\documentclass[12pt, a4paper]{report}
\usepackage[utf8]{inputenc}
\usepackage{amsmath}
\usepackage{amsfonts}
\usepackage{amssymb}
\usepackage{graphicx}
\usepackage{natbib}
\setcitestyle{authoryear}
\usepackage[colorlinks=true,linkcolor=blue,citecolor=blue, urlcolor=blue]{hyperref}
\usepackage{import} 
\usepackage{enumerate}

\newcommand{\R}{\mathbb R}
\usepackage{enumitem}
\usepackage{subcaption}
\usepackage[toc,page]{appendix}
\usepackage{rotating}
\usepackage{bigints}
\usepackage{sectsty} 
\usepackage[lithuanian,greek,english]{babel}
\usepackage{pdfpages}
\allsectionsfont{\raggedright}

\usepackage{caption}	

\begin{document}
\pagenumbering{alph}	
\thispagestyle{empty}	
\vspace*{2mm}	
\begin{center}
\huge\textbf{\textsf{Testing and Emulating Modified Gravity on Cosmological Scales}}\\  
\vspace{7mm}
\includegraphics[width=0.65\columnwidth]{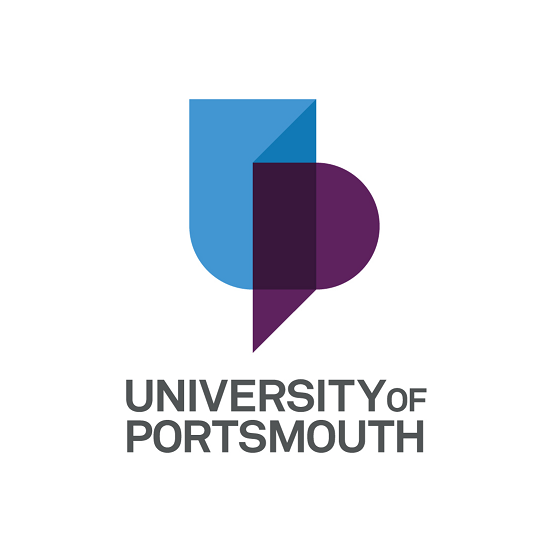}

\vspace{6mm}
\normalsize by\\
\vspace{1mm}
\large \textbf{Andrius Tamošiūnas}\\
\vspace{2mm}

\vspace{2mm}
\normalsize This thesis is submitted in partial fulfilment of\\
the requirements for the award of the degree of\\
Doctor of Philosophy of the University of Portsmouth.\\
\vspace{5mm}
\textbf{Supervisors}:\\
Prof. Bob Nichol, Prof. David Bacon, Prof. Kazuya Koyama\\
\vspace{3mm}
\today	
\end{center}
\newpage
\pagenumbering{roman}	
\phantomsection	
\addcontentsline{toc}{chapter}{Abstract}	
\chapter*{Abstract}	

This thesis explores methods and techniques for testing and emulating models of modified gravity. In particular, the thesis can be split into two parts. The first part corresponding to chapters \ref{ch:galaxy_clusters}, \ref{ch:modified_gravity} and \ref{ch:testing_EG} introduces a method of testing modified gravity models on galaxy cluster scales. In more detail, chapter \ref{ch:galaxy_clusters} introduces the main concepts from galaxy cluster physics, which are important in the context of testing modified gravity. I present a summary of the different probes of mass in galaxy clusters. More specifically, the properties of the X-ray-emitting intracluster medium are discussed in the context of measuring hydrostatic masses. In addition, the key equations for weak lensing of background galaxies by galaxy clusters are summarized in terms of their importance for measuring cluster lensing masses. 

Chapter \ref{ch:modified_gravity} introduces a technique for detecting the modifications of gravity using a combination of X-ray and weak lensing data obtained by stacking multiple galaxy clusters. This technique first discussed in \citet{Terukina2013} and \citet{Wilcox2015} allows us to put some of the most competitive constraints on scalar-tensor theories with chameleon screening as well as the closely related $f(R)$
gravity models. Chapter \ref{ch:modified_gravity} also contains a discussion of the key theoretical concepts of scalar-tensor models and their relationship to $f(R)$ gravity. Finally, the chapter is concluded by introducing novel results which update the tests done in \citet{Wilcox2015} with an improved dataset consisting of 77 galaxy clusters from the XCS and CFHTLenS surveys. The updated dataset containing less noisy tangential shear data allows us to put tight constraints on the chameleon field background value and the related $f_{R0}$ parameter:  $\phi_{\infty}<8 \times 10^{-5} M_{\mathrm{pl}}$ and $\left|f_{R 0}\right|<6.5 \times 10^{-5}$.

Chapter \ref{ch:testing_EG} expands the mentioned techniques for testing a different type of a model. In particular, the model of emergent gravity (introduced in \citet{Verlinde2017}) is tested by using a variation of the techniques introduced in chapter \ref{ch:modified_gravity}. The key prediction of Verlinde's emergent gravity is a scaling relation similar to the baryonic Tully-Fisher relation, which allows us to determine the dark matter distribution in a cluster directly from the baryonic mass distribution. The mentioned scaling relation was tested by determining the baryonic mass from the X-ray surface brightness data and calculating the predicted weak lensing tangential shear profile, which was then compared against the $\Lambda$CDM prediction based on the Navarro-Frenk-White profile. The test was performed for the Coma Cluster using data from \citet{Terukina2013} and for the 58 galaxy cluster stack from \citet{Wilcox2015}. The obtained results indicate that according to the Coma Cluster data, the emergent gravity predictions agree with the $\Lambda$CDM predictions only in the range of $r \approx$ $250$-$700$ kpc. Outside the mentioned radial range the standard model results are preferred according to the Bayesian information criterion analysis (despite needing two extra free parameters). The same general conclusion can be drawn from the 58 cluster stack data, which indicates a good agreement between the models only for $r \approx 1$-$2$ Mpc. Outside of that radial range the standard model is strongly preferred.

The second part of the thesis, referring specifically to chapters \ref{ch:machine_learning} and \ref{ch:GAN_emulators}, contains a study of machine learning techniques for emulating cosmological simulations. More specifically, generative adversarial networks are studied as an effective tool for emulating $N$-body simulation data quickly and efficiently. Chapter \ref{ch:machine_learning} contains a brief discussion of the different machine learning algorithms in the context of emulators. Specifically, artificial neural networks are introduced along with gradient boosting algorithms. Chapter \ref{ch:GAN_emulators} introduces a generative adversarial network algorithm for emulating cosmic web data along with weak lensing convergence map data coming from $N$-body simulations. The presented approach is based on the \textit{cosmoGAN} algorithm first described in \citet{Mustafa2019}, which allows us to generate thousands of realistic weak lensing convergence maps in a matter of seconds. The mentioned approach is then modified to allow emulating cosmic web and weak lensing data for $\Lambda$CDM and $f(R)$ gravity with different cosmological parameters and redshifts. In addition, a similar approach was used to simultaneously emulate dark matter and baryonic simulation data coming from the Illustris simulation. The obtained results indicate a 1-20\% difference between the power spectra of the emulated and original ($N$-body simulation) datasets depending on the training data used. Finally, the chapter contains an in-depth study of the technique of latent space interpolation and how it can be applied to control the cosmological/modified gravity parameters during the emulation procedure. The obtained results illustrate that such machine learning algorithms will play an important role in producing accurate mock data in the era of future large scale observational surveys.

\newpage
\renewcommand{\contentsname}{Table of Contents}	
\pdfbookmark{Table of Contents}{contents}	
\tableofcontents

\newpage
\pdfbookmark{List of Tables}{tables}	
\listoftables

\newpage
\pdfbookmark{List of Figures}{figures}	
\listoffigures


\newpage
\phantomsection
\addcontentsline{toc}{chapter}{Declaration}
\chapter*{Declaration}
Whilst registered as a candidate for the above degree, I have not been registered for any other research award. The results and conclusions embodied in this thesis are the work of the named
candidate and have not been submitted for any other academic award.\\
\vspace{50pt}
\begin{center}
Word count: 41,724 words.
\end{center}

\newpage
\phantomsection
\addcontentsline{toc}{chapter}{Acknowledgements}
\chapter*{Acknowledgements}

I would like to use this opportunity to express my sincere gratitude to the people that made this thesis possible. Firstly, I would like to thank my lovely family for their never-ending love and support. I am grateful to my Dad, who has sparked in me the love for science and education. I am grateful to my Mom for always being there for me. And I am grateful to my brother for being someone I can look up to.

I would also like to thank my dear supervisors Prof. Bob Nichol, David Bacon and Kazuya Koyama for their infinite patience, kindness and keeping my passion for learning new things aflame. Thanks for teaching me so much about myself and the Universe. In turn, I would like to extend my sincere gratitude to all my teachers and mentors (from primary school to university), who have taught me so much. Likewise I am grateful to the people who have influenced me in many different ways throughout the years including: Prof. Andrew Liddle, Bruce Bassett and Kathy Romer. 

Finally, my gratitude goes to my friends and colleagues in the ICG. Thanks to my academic brothers and sisters -- Michael, Mike, Maria, Sam Y., Sam L. and Natalie, for sharing the academic path and for the fun times and the endless nights in the pubs of Portsmouth. I am also sincerely grateful to everyone in the ICG for making the institute a warm and welcoming place to work and study in.

\newpage
\phantomsection
\addcontentsline{toc}{chapter}{Dissemination}
\chapter*{Dissemination}

Chapter \ref{ch:modified_gravity} contains work in preparation for publication. Chapters \ref{ch:testing_EG} and \ref{ch:machine_learning} contain original work described in the following publications:
\\
\\
A. Tamosiunas, H. A. Winther, K. Koyama, D. J. Bacon, R. C. Nichol, B. Mawdsley. \textit{Towards Universal Cosmological Emulators with Generative Adversarial Networks}. April 2020. Submitted for publication in the Journal of Computational Astrophysics and Cosmology. arXiv: \href{https://arxiv.org/abs/2004.10223}{arXiv:2004.10223 [astro-ph.CO]}.
\\
\\
A. Tamosiunas. D. J. Bacon, K. Koyama, R. C. Nichol. \textit{Testing Emergent Gravity on Galaxy Cluster Scales}. January 2019. Published in JCAP. DOI:  \href{https://iopscience.iop.org/article/10.1088/1475-7516/2019/05/053}{10.1088/1475-7516/2019/05/053}. arXiv: \href{https://arxiv.org/abs/1901.05505}{arXiv:1901.05505 [astro-ph.CO]}

\newpage
\phantomsection
\addcontentsline{toc}{chapter}{Notation}
\chapter*{Notation and Abbreviations}

\begin{table}[h!]
\begin{flushright}
\def\arraystretch{1.0}%
\begin{tabular}{l|l}
\textbf{Description:} & \textbf{Abbreviation:} \\ \hline
Active galactic nuclei & \textbf{AGN} \\
Bayesian information criterion & \textbf{BIC} \\
Canada-France-Hawaii Telescope Lensing Survey & \textbf{CFHTLenS} \\
The cosmic microwave background & \textbf{CMB} \\
Convolutional neural networks & \textbf{CNN} \\
Deep convolutional generative adversarial networks & \textbf{DCGAN} \\
The theory of  emergent gravity & \textbf{EG} \\
The Friedman-Lema\^itre-Robertson-Walker metric & \textbf{FLRW} \\
Generative adversarial network & \textbf{GAN} \\
Graphical processing unit & \textbf{GPU} \\
The theory of general relativity & \textbf{GR} \\
Intracluster medium in galaxy clusters & \textbf{ICM} \\
The Markov chain Monte Carlo algorithm & \textbf{MCMC} \\
Modified Newtonian dynamics & \textbf{MOND} \\
The Navarro-Frenk-White density profile & \textbf{NFW} \\ 
The Sunyaev-Zeldovich effect  & \textbf{SZ} \\
The XMM Cluster Survey & \textbf{XCS} \\
The standard model of cosmology & \textbf{$\Lambda$CDM} 
\end{tabular}%
\end{flushright}
\end{table}

\newpage
\pagenumbering{arabic}	



\chapter{Introduction}
\label{ch:introduction}

The science of cosmology dates back to ancient times. When defined in a broad sense cosmology is inseparable from the earliest inquiries into how nature works by ancient cultures as described in the historical records \citep{Kragh2017}. Initially such inquiries were tightly intertwined with religion and superstition. The principles of what is now known as modern cosmology were arguably first combined into a consistent framework in ancient Greece. As an example, the great philosopher Plato in his works \textit{Timaeus} and \textit{Republic} introduced the two-sphere model, which placed Earth at the centre of the Universe, surrounded by a celestial sphere holding the stars and other heavenly bodies \citep{Evans1998}. Another school of philosophy, the Pythagoreans, sought to build models of the celestial motion based on known mathematical principles. A key stepping stone to mention here is that the Pythagoreans treated astronomy as one of the key mathematical arts (along with arithmetic, geometry and music). Such a high regard for natural philosophy eventually led to the formulation of the first known heliocentric model of the solar system by the mathematician and astronomer Aristarchus of Samos \citep{Heath1991}. The model of Aristarchus placed the Sun at the centre of the known Universe with the Earth and the other known planets orbiting around it. Another visionary insight by Aristarchus was that the stars were in fact objects analogous to the Sun at such great distances from Earth that no parallax was observed \citep{Wright1995}. The original texts describing Aristarchus' models were later lost for over a millennium with only references in contemporary texts surviving.        

Another visionary work that came from this historical period is \textit{The Sand Reckoner} (Gr: \textgreek{Ψαμμίτης}) by Archimedes \citep{Hirshfeld2009}. In this work Archimedes sets out to calculate the number of grains of sand that fit into the Universe. In order to do this, Archimedes had to estimate the size of he Universe based on Aristarchus' model. In addition, Archimedes had to invent new mathematical notation for dealing with large numbers. The obtained results estimated the diameter of the Universe to be no more than $10^{14}$ stadia or roughly 2 light-years in contemporary units. Equivalently, a Universe of such size could fit $10^{63}$ grains of sand. The importance of this work lies in the fact that it is likely the first known systematic estimate of the size of the Universe based on mathematics and the known principles of astronomy.

Major leaps in understanding of cosmology and astronomy were made during the Renaissance. The works of Nicolaus Copernicus reintroduced the heliocentric model from relative obscurity due to the original works of Aristarchus being mostly unknown. The works of Copernicus, when combined with astronomical observations, allowed predictions of planetary motions and orbital periods as well as experimental comparison of the two competing theories of geocentrism and heliocentrism. The observational tradition of Copernicus was carried on by later astronomers including Tycho Brahe and Johannes Kepler, the work of whom led to the three laws of Kepler. Another key discovery from this historical period came from Galileo, who is traditionally credited as one of the discoverers of the telescope. The mentioned theoretical and observational efforts culminated in the work of Newton and in particular Newton's law of universal gravitation, which formed the core of our understanding of gravity for over two centuries after its discovery \citep{Taton2003,Curley2012}.

Arguably the most important theoretical development in the modern era of cosmology came with Einstein's theory of general relativity (GR) \citep{Einstein1916}. GR revolutionized our understanding of gravity by promoting space and time from a mere stage in which events take place to a 4-dimensional dynamical canvas that interacts with matter and energy in intricate ways (\textit{spacetime tells matter how to move; matter tells spacetime how to curve} according to John A. Wheeler). 

After more than a century GR has been extensively confirmed observationally and now forms the basis of our understanding of how gravity behaves in a wide range of systems starting with the solar system and galaxies and ending with the Universe as a whole. For this reason GR is also the theoretical basis behind the currently most complete model of standard cosmology -- the $\Lambda$CDM model. The $\Lambda$CDM model has been extremely successful in explaining structure formation in the Universe along with the anisotropies of the cosmic microwave background (CMB). However, as the name implies, in order to make accurate predictions, the theory requires two extra components -- the cosmological constant ($\Lambda$) or some other form of dark energy along with non-luminous non-baryonic matter (CDM). Dark matter, in particular, was first inferred to exist by Fritz Zwicky by studying the mass distribution in the Coma Cluster in 1933 \citep{Zwicky1933}. Now we know that some form of dark matter is crucial for explaining galaxy rotation curves and large scale structure formation in general. Another key discovery came in 1998, when the Supernova Cosmology Project and the High-Z Supernova Search Team found evidence for the accelerating expansion of the Universe using data from type Ia supernovae \citep{Riess1998}. To explain the accelerating expansion, some form of dark energy is required.

Today we know a lot about dark energy and dark matter, but the physical origin of them still eludes astronomers and particle physicists. This is one of the key motivations for developing models that modify GR. Since the publication of the original theory in 1915 a plethora of modified gravity models have been proposed. These models can be generally classified based on the type of modification they introduce to the original GR framework. Namely, modified models can introduce extra scalar, vector and tensor fields, extra spatial dimensions and higher order derivatives. In addition, certain assumptions that exist in the original model can be relaxed (e.g. non-local theories). These approaches form a complex family of modified gravity models, each of which comes with unique observational signatures. A need to discriminate between the different families of modifications of gravity has led to a variety of observational tests on scales ranging from the laboratory, to the solar system and all the way to cosmological scales \citep{Koyama2016}. In this thesis special emphasis will be put on galaxy cluster-related methods for testing for such modification of gravity. In addition, various machine learning techniques will be explored as tools for emulating modified gravity simulations. The rest of chapter \ref{ch:introduction} will introduce the relevant basic concepts in GR and cosmology. In addition, a brief overview of the current theoretical and observational developments in the field of modified gravity will be given. 

\section{General Relativity}
\label{introduction:GR}

Einstein's theory of general relativity published in 1915 forms the basis of the modern understanding of gravity. One of the key postulates of the theory is the equivalence principle, which equates the gravitational and inertial masses of a given body. This at first glance inconsequential idea, with its roots dating back to the observations of Galileo, has led Einstein on a path towards finding deep connections between gravity and the geometry of spacetime. Namely, by demonstrating the equivalence of the forces felt by a body in an accelerated frame and those felt in a gravitational field, Einstein was able to generalize the tools and techniques first developed for his theory of special relativity \citep{Wald2010}. 

GR relates the energy-momentum contents of a given gravitational system to the geometric effects on spacetime. Hence, if the mass/energy distribution in a given system is known, accurate predictions can be made about the resulting dynamics of the system. A key equation in this regard is the Einstein-Hilbert action. Describing GR in terms of an action has a number of advantages. In particular, it allows to describe the theory following a similar formalism as in the other classical field theories (e.g. Maxwell theory). Varying the action allows a straightforward way for deriving the field equations. In addition, the effects of other fields (e.g. matter fields) can be easily added to the total action. The Einstein-Hilbert action is given by: 

\begin{equation}
S_{\mathrm{EH}}=\int d^{\mathrm{4}} x \frac{\sqrt{-g}}{2 \kappa}(R-2 \Lambda)+S_{m}\left[\psi_{M},g_{\mu \nu}\right],
\label{eq:EH_action}
\end{equation}

\noindent where $g_{\mu \nu}$ is the spacetime metric, $g$ is the determinant of the metric, $\kappa = 8\pi G$, $G$ is the gravitational constant, $R$ is the Ricci scalar, $\Lambda$ is the cosmological constant and $S_{m}$ is the matter action governed by the matter field $\psi_{M}$. The Ricci scalar can be obtained by contracting the indices of the Ricci tensor $R \equiv g^{\mu \nu} R_{\mu \nu}$, which, in turn, can be derived from the Riemann curvature tensor: 

\begin{equation}
R_{\sigma \mu \nu}^{\rho}=\partial_{\mu} \Gamma_{\sigma \nu}^{\rho}-\partial_{\nu} \Gamma_{\sigma \mu}^{\rho}+\Gamma_{\lambda \mu}^{\rho} \Gamma_{\sigma \nu}^{\lambda}-\Gamma_{\lambda \nu}^{\rho} \Gamma_{\sigma \mu}^{\lambda}.
\end{equation}

\noindent The Riemann curvature tensor quantifies the amount of curvature in the 4-D spacetime manifold. Here $\Gamma$ refers to the Christoffel symbols, given by:

\begin{equation}
\Gamma_{\sigma \nu}^{\rho}=\frac{1}{2} g^{\rho \gamma}\left(\frac{\partial g_{\gamma \nu}}{\partial x^{\sigma}}+\frac{\partial g_{\gamma \sigma}}{\partial x^{\nu}}-\frac{\partial g_{\sigma \nu}}{\partial x^{\gamma}}\right).
\end{equation}

By varying the Einstein-Hilbert action, w.r.t. the spacetime metric the Einstein field equations are obtained: 

\begin{equation}
G_{\mu \nu}=\kappa T_{\mu \nu}-\Lambda g_{\mu \nu},
\end{equation}

\noindent where $G_{\mu \nu} = R_{\mu \nu} - Rg_{\mu \nu}/2$ is the Einstein tensor. $T_{\mu \nu}$ refers to the energy-momentum tensor, defined by: 

\begin{equation}
T^{\mu \nu} = - \frac{2}{\sqrt{-g}}\frac{\delta S_{m}}{\delta g_{\mu \nu}}.
\end{equation}

In the case of a perfect fluid with density $\rho$, pressure $p$ and the four-velocity $U^{\mu}$:

\begin{equation}
T^{\mu \nu}=\left(\rho c^{2}+p\right) U^{\mu} U^{\nu}+p g^{\mu \nu}.
\end{equation}

In this framework, the dynamics of bodies can be deduced from the geodesic equation, which generalizes the notion of a straight line to curved spaces: 

\begin{equation}
\frac{d^{2} x^{\mu}}{d s^{2}}+\Gamma^{\mu}_{\alpha \beta} \frac{d x^{\alpha}}{d s} \frac{d x^{\beta}}{d s}=0.
\label{eq:geodesic}
\end{equation}

\noindent Hence, if the metric $g_{\mu \nu}$ describing a given gravitational system is known, one can solve eq. \ref{eq:geodesic} to obtain the trajectory of a body in terms of the four spacetime coordinates and some affine parameter $x^{\mu}(\lambda)$.

The key significance of GR in the context of cosmology comes from its ability to relate mass/energy distributions to the corresponding effects on spacetime and ultimately the resulting motion of bodies. This makes GR one of the key foundations of the standard model of cosmology. 

\section{The Standard Model of Cosmology}
\label{introduction:lambda-cdm}

The $\Lambda$CDM model is currently the most well-tested framework capable of describing a wide range of phenomena, such as the anisotropies of the CMB and the underlying large-scale structure formation. The standard model is based on three key assumptions \citep{Peebles1993, Li2019}: 

\begin{enumerate}[label=\roman*),align=left] 
  \item The cosmological principle. This principle refers to the matter distribution, on large scales, being homogeneous and isotropic. 
  \item The known laws of gravity are universal. Or, more specifically, in the context of cosmology, gravity is described by GR everywhere in the Universe. 
  \item The matter-energy budget of the Universe contains a significant contribution from some form of non-luminous, non-baryonic matter (dark matter) along with the usual baryonic matter and radiation.
\end{enumerate}

The first assumption can be expressed mathematically by choosing the most general metric fulfilling the needed conditions of isotropy and homogeneity -- the Friedman-Lema\^itre-Robertson-Walker (FLRW) metric \citep{Friedmann1922, Lemaitre1931}. The FLRW metric is obtained by starting with the most general metric in 4-D and constraining the form of the metric to account for isotropy, homogeneity and the different types of the spatial curvature of the Universe. This leads to the following form:

\begin{equation}
\mathrm{d} s^{2}=\mathrm{d} t^{2}-a(t)^{2}\left(\frac{\mathrm{d} r^{2}}{1-k r^{2}}+r^{2} \mathrm{d} \theta^{2}+r^{2} \sin ^{2} \mathrm{d} \phi^{2}\right),
\label{eq:FLWR}
\end{equation}

\noindent where $t$ is proper time, $r,\theta,\phi$ are the usual spherical coordinates, $a(t)$ is the scale factor and $k$ is a constant related to spatial curvature. The numerical values of $k = \{-1, 0, 1 \}$ (in the units of $\text{length}^{-2}$) refer to an open, flat and closed spatial curvature of the Universe correspondingly. 

Applying the FLRW metric to the Einstein field equations results in the two equations that govern the evolution of the scale factor $a(t)$ known as the Friedmann equations: 

\begin{equation}
H^{2}=\left(\frac{\dot{a}}{a}\right)^{2}=\frac{8 \pi G}{3} \rho-\frac{k}{a^{2}}+\frac{\Lambda}{3},
\end{equation}

\begin{equation}
\frac{\ddot{a}}{a}=-\frac{4 \pi G}{3}(\rho+3 p)+\frac{\Lambda}{3},
\end{equation}

\noindent where $\ddot{a}$ and $\dot{a}$ correspond to the time derivatives of the scale factor, $\rho$ is the density, $p$ is the pressure and $\Lambda$ is the cosmological constant. The Friedmann equations are profound as they describe the expansion of space and relate it to the matter content. Hence, assuming that the underlying density distribution can be determined, one can deduce the future evolution of the Universe. 

The $\Lambda$CDM model is based on 6 main parameters that are needed to fit the key observational datasets such as the CMB anisotropies, large scale galaxy clustering and the redshift/brightness relation for supernovae. These parameters are the baryon density parameter $\Omega_{B}h^{2}$ (with $h=\mathrm{H}_{0} /\left(100 \mathrm{km} \mathrm{s}^{-1} \mathrm{Mpc}^{-1}\right)$ as the dimensionless Hubble parameter), the dark matter density parameter $\Omega_{c}h^{2}$, the angular scale of the sound horizon at the last scattering $\theta_{*}$, the scalar spectral index $n_{s}$, the initial super-horizon curvature fluctuation amplitude (at $k_{0} = 0.05$ $\text{Mpc}^{-1}$) $A_{s}$ and the reionization optical depth $\tau$. 

Physically, the $\Omega_{B}$ and the $\Omega_{c}$ parameters quantify the amount of baryonic and dark matter relative to the critical density. The $\theta_{*}$ parameter quantifies the ratio between the sound horizon (i.e. the distance sound waves could have traveled in the time before recombination) and the distance to the surface of last scattering. The spectral index $n_{s}$ quantifies the scale dependence of the primordial fluctuations (with $n_{s} = 1$ referring to scale invariant case). Finally, in the context of the CMB observations, the optical depth to reionization, $\tau$, is a unitless quantity which provides a measure of the line-of-sight free-electron opacity to CMB radiation. This is the case as Thomson scattering of the CMB photons by the free electrons produced by reionization serves as an opacity source that suppresses the amplitude of the observed primordial anisotropies.

Table \ref{table:plank_results} lists the values of the 6 key parameters according to the recent Planck results. Knowing these values with sufficient accuracy allows us to determine other parameters of interest, such as the Hubble parameter and the dark energy density. More generally, being able to measure these parameters with accuracy leads to the most detailed picture of the Universe we have as of yet: a spatially flat Universe expanding at an accelerated rate.

\begin{table}[ht!]
\centering
\begin{tabular}{ |c|c| } 
\hline
 \textbf{Parameter:} & \textbf{Constraint:} \\ 
 \hline \hline
 $\Omega_{B}h^{2}$ & $0.02233$ $\pm$ $0.00015$  \\ 
 $\Omega_{c}h^{2}$ & $0.1198$ $\pm$ $0.0012$ \\
 $100\theta_{*}$ & $1.04089$ $\pm$ $0.00031$ \\
 $ln(10^{10}A_{s})$ & $3.043$ $\pm$ $0.014$ \\
 $n_{s}$  &  $0.9652$ $\pm$ $0.0042$ \\
 $\tau$  & $0.0540$ $\pm$ $0.0074$ \\
 \hline
\end{tabular}
\caption[Base-$\Lambda$CDM cosmological parameters from Planck 2018]{Base-$\Lambda$CDM cosmological parameters from Planck 2018 TT,TE,EE + LowE + lensing results \citep{Aghanim2018}. }
\label{table:plank_results}
\end{table}

The standard model is successful not only in being able to fit the observational data, but also in terms of making testable predictions. Namely, the polarization of the CMB, predicted by the model has been discovered in 2002 \citep{Kovacs2002}. Similarly, the prediction and detection of the baryon acoustic oscillations is another recent success of the model \citep{Cole2005}. 

Despite the great successes of the $\Lambda$CDM model, a number of challenges remain. Starting with the validity of the outlined key assumptions and ending with the reliance on the existence of dark energy and dark matter, the issues facing the standard model must be discussed in greater detail.

\section{The Standard Model: Problems and Challenges}
\subsection{The Validity of the Cosmological Principle}

The key assumptions of the $\Lambda$CDM framework have been criticized thoroughly ever since the inception of the standard model. Namely, it is clear that the cosmological principle, i.e. the homogeneity and isotropy of the structure in the Universe, does not hold on some scales (e.g. the Local Group with its complex structure is far from being homogeneous and isotropic). Multiple observational tests have been performed to test the cosmological principle, generally confirming it on large scales \citep{Lahav2001, Bengaly2019}. However, on smaller scales multiple questions remain, such as what effects do local deviations from isotropy and homogeneity have on our measurements of the accelerating expansion of the Universe. More specifically, different models of inhomogeneous cosmology argue that inhomogeneities on different scales affect the local gravitational forces leading to skewed measurements of the expansion of the Universe. However, these models also suffer from various issues (see \citet{Bolejko2017} for an overview).       

\subsection{The Validity of GR on Different Scales}

The second key assumption of the standard model, i.e. GR being valid on all scales, can be challenged as well. Firstly, it is known that the theory is incomplete in terms of not being able to describe systems where quantum effects have to be fully taken into account. This implies that the very early Universe along with some astrophysical systems, such as black holes, cannot be fully described by the theory. This touches a more fundamental problem in theoretical physics of not being able to reconcile GR with quantum field theory. GR is thought to be an effective theory only valid up to around the Planck scale. This has led to a search for a complete quantum gravity theory resulting in multiple prominent approaches, such as string theory, the theory of loop quantum gravity and a plethora of modified gravity models \citep{Mukhi2011, Agullo2016}. 

\begin{figure}
\centering
\includegraphics[width=0.85\columnwidth]{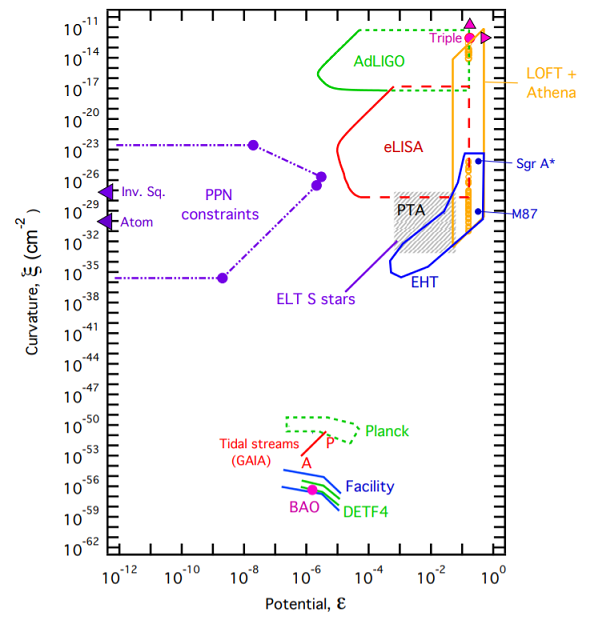}
\caption[Observational tests of modified gravity]{The observational parameter space of gravity tests from \citet{Baker2015}. The meaning of the key abbreviations is as follows: PPN = Parameterized Post-Newtonian region, Inv. Sq. = laboratory tests of the inverse square law of gravity, Atom = atom interferometry experiments, EHT = the Event
Horizon Telescope, Facility = a futuristic large radio telescope such as the Square Kilometre Array, DETF4 = a hypothetical "stage 4" experiment according to the classification scheme of the Dark Energy Task Force. The other abbreviations correspond to observational mission names or specific objects.    
\label{figure:modified_gravity_tests}}
\end{figure}

In a more observational context, the assumption of the validity of GR has been tested exquisitely, but only on certain scales. Figure \ref{figure:modified_gravity_tests} summarizes the current state of tests of gravity on various scales, with curvature and potential referring to $\xi = GM/c^{2}r^{3}$ and $\varepsilon = GM/c^{2}r$ (for a spherical object of mass $M$ and radius $r$) correspondingly.

A key takeaway from figure \ref{figure:modified_gravity_tests} is that even though modern observational missions have explored a wide variety of scales, there is still a large section of the parameter space that remains unexplored. Specifically, gravity is well-tested in the solar system and binary pulsars, however the low curvature regime remains to be explored and is of special interest for understanding various relevant phenomena such as that associated to dark matter. Similarly, tests of gravity in the strong curvature regime could improve our understanding of systems where both gravitational and quantum effects are important (e.g. black holes). Overall, testing gravity in low and high curvature regimes will likely provide a fuller understanding of how gravity works, which in turn will improve our understanding of cosmology and astrophysics on all scales. 
 
\subsection{Issues Related to Dark Matter}
 
\begin{figure}[ht!]
\centering
\makebox[\textwidth][c]{\includegraphics[width=1.2\textwidth]{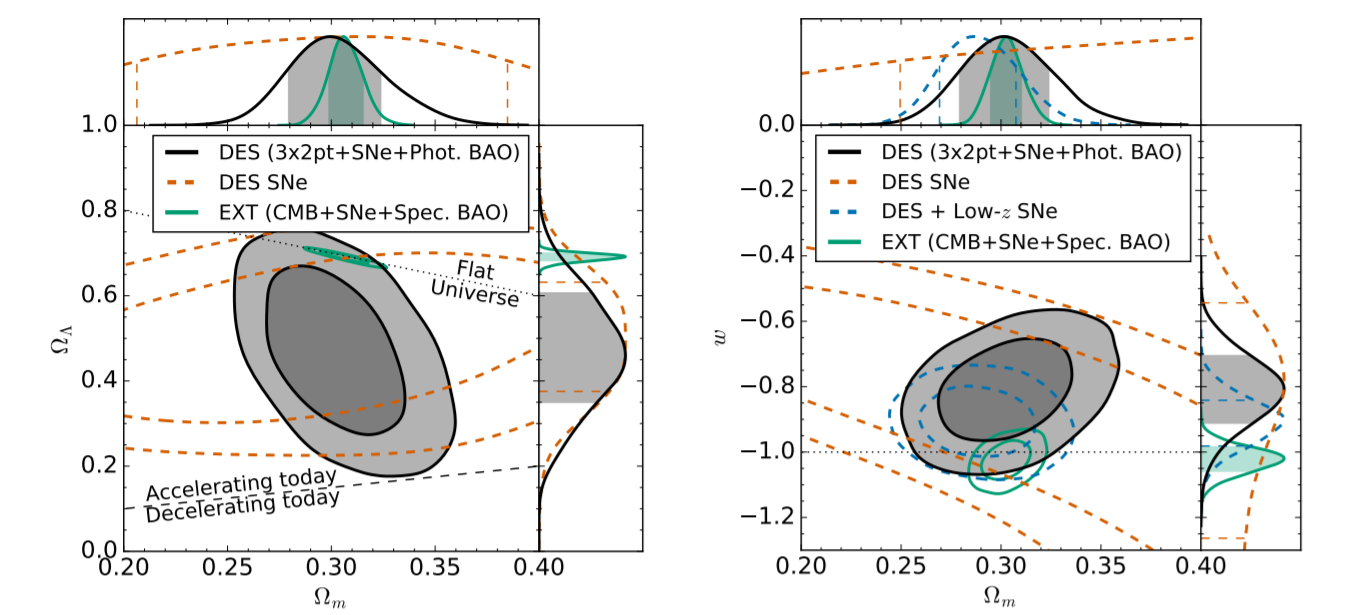}}%
\caption[Recent cosmological constraints from multiple probes in the Dark Energy Survey]{Recent cosmological constraints from the DES survey. \textbf{Left:} constraints on the present-day dark energy density $\Omega_{\Lambda}$ and matter density $\Omega_{m}$. Black contours correspond to DES data alone (including information from weak lensing, large scale structure, type Ia supernovae and BAO data); green contours correspond to best available constraints from external data; orange contours correspond to DES supernovae constraints alone. \textbf{Right:} equivalent constraints on the dark energy equation of state $w$ and matter density $\Omega_{m}$. The dashed blue contours show the low redshift supernovae constraints. The contours in both plots correspond to 68\% and 95\% confidence limits. External data specifically refers to Planck, Pantheon and BOSS DR12 datasets \citep{DES2018}. Note that the significant tension between the DES and the external data is related to the known tension in the measurements of the $S_{8}$ parameter in DES and the Planck tomographic weak lensing data as discussed in \citet{Joudaki2020}.   
\label{figure:omega_m_vs_omega_lambda}}
\end{figure}

The third base assumption that the $\Lambda$CDM model is based on is related to the existence of dark matter. Historically some form of dark matter was hypothesized to exist in order to explain the rotation curves of galaxies. Most recent observational evidence indicates that dark matter is crucial for explaining the formation and evolution of galaxy clusters and large scale structure as well \citep{Freese2017}. Other key evidence comes from weak lensing surveys, CMB anisotropies and baryon acoustic oscillations \citep{Roos2010}. Figure \ref{figure:omega_m_vs_omega_lambda} shows the combined constraints on $\Omega_{\Lambda}$ and $\Omega_{m}$ coming from the weak lensing, large scale structure, supernovae and baryon acoustic oscillation data. These results clearly illustrate a need for some form of dark energy and non-baryonic matter to explain the currently available observational data.

Despite the great success of the cold dark matter paradigm, certain questions remain unanswered. This is especially clear in the context of galaxy formation where a number of challenges to the $\Lambda$CDM model have emerged in recent years. These include the missing satellites problem, which indicates a mismatch between the observed dwarf galaxy numbers and the corresponding prediction from numerical simulations. Similarly, the cusp/core problem indicates a mismatch between the predicted and observed cuspiness and density of the dark matter dominated galaxies. Another inconsistency comes in the form of the \textit{too big to fail} problem, which states that the observed satellites in the Milky Way are not massive enough to be consistent with the $\Lambda$CDM predictions \citep{Bullock2017}. These issues can be viewed in a wider context of reconciling theoretical predictions with the cosmological simulations and observational data. Inconsistencies could originate due to the lack of understanding of the galaxy formation processes, difficulty of building realistic simulations of such processes or a lack of understanding of the fundamental nature of dark matter.

\subsection{Issues Related to Dark Energy}

Another key challenge that the standard model of cosmology is facing at the moment is explaining the nature of dark energy. As illustrated by figure \ref{figure:omega_m_vs_omega_lambda}, $\Omega_{\Lambda}$ dominates the total energy budget of the Universe. Some form of dark energy is required to account for the accelerated expansion of the Universe. The energy scale for the cosmological constant deduced from the available observational data is of the order of: $\rho_{\Lambda} \equiv \Lambda/8 \pi G \approx ( 10^{-3}$  $\mathrm{eV})^{4}$ \citep{Koyama2016}. However, arguments in quantum field theory and semi-classical gravity suggest existence of vacuum energy $T^{\rm vac}_{\mu\nu} \equiv -\rho_{\rm vac} g_{\mu\nu}$, which should contribute to the total energy budget of the Universe. Calculations in quantum field theory suggest $\left|\rho_{\mathrm{vac}}\right| \approx 2 \times 10^{8}$      $\mathrm{GeV}^{4}$, which is a huge value comparable to $2 \times 10^{11} \rho_{\text {nucl }}$, where $\rho_{\text {nucl }}$ is the density of atomic nuclei \citep{Weinberg1989}. Such a major contribution to the total energy budget is clearly not observed in the available data, which leads to the \textit{old} cosmological constant problem (\textit{why doesn't the vacuum energy gravitate as expected?}). In addition, a related problem arises when trying to explain the observed accelerating expansion of the Universe. Namely, extreme fine tuning is required between the value of the cosmological constant and the predicted vacuum energy in order to explain the observed cosmological expansion. This is referred to as the \textit{new} cosmological constant problem. 

Other conundrums include the \textit{why now?} problem, as in why is the current vacuum energy density of similar magnitude to the matter energy density at this particular cosmic epoch \citep{Lombriser2019}? These issues have been studied extensively and various possible solutions have been proposed in the context of different models of dark energy and modified gravity (e.g. see \citet{Li2011}).

\subsection{Tensions in the Cosmological Parameters}

\begin{figure}[t!]
\centering
\includegraphics[width=0.90\columnwidth]{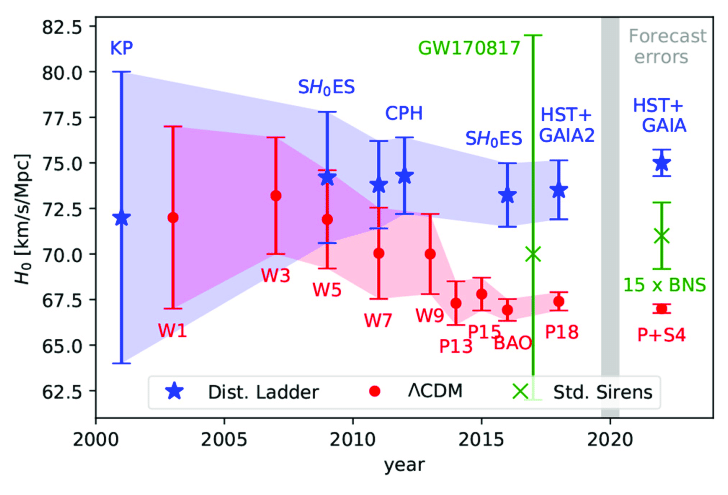}
\caption[The $H_{0}$ tension]{Measured $H_{0}$ value as a function of time from \citet{Ezquiaga2017}. The blue contour indicates the local measurements with calibration based on Cepheids. The red contour shows the CMB measurements done under the assumption of $\Lambda$CDM. The green results correspond to the most recent direct measurements of $H_{0}$ with standard sirens. The forecast errors refer to CMB stage IV experiments \citep{Abazajian2016}, standard sirens \citep{Nissanke2013} and the distance ladder with full GAIA and HST \citep{Casertano2016, Riess2016}. The error bars show the 1-$\sigma$ error. 
\label{figure:H_0_tension}}
\end{figure}

Another key contemporary challenge to the standard model is the existence of the various tensions between the different observables. A prime example of this is the tension between the early and late Universe measurements for the expansion rate parameter $H_{0}$. In more detail, the local measurements of $H_{0}$ using the distance ladder indicate a significantly higher value when compared to the Planck CMB measurements (at around 3.5-$\sigma$ level) \citep{Riess2018}. Such a tension could indicate various systematic problems both with the early and the late Universe measurements or, alternatively, it could indicate new physics. Figure \ref{figure:H_0_tension} summarizes some of the recent measurements of $H_{0}$. Various ways of relieving the tension have been proposed, such as independent ways of measuring $H_{0}$ through gravitational wave measurements or through the calibration of the tip of the red giant branch \citep{Abbott2017, Freedman2019}. These measurements relieve the tension to some extent, however more accurate observational data might be needed to fully account for the discrepancy.

Another example of a tension between the different types of cosmological measurements is the $S_{8}$ tension, where $S_{8}=\sigma_{8} \sqrt{\Omega_{\mathrm{m}} / 0.3}$. The $\sigma_{8}$ parameter here refers to the amplitude of the linear power spectrum on the scale of 8 $h^{-1}$ Mpc. It is one of the key cosmological parameters due to being related to the growth of the fluctuations in the early Universe. As described in \citet{Joudaki2020}, there is a $2.5$-$\sigma$ tension between the combined Kilo Degree Survey (KV450), DES-Y1 and the Planck weak lensing measurements of the $S_{8}$ parameter. This tension is also likely one of the key reasons behind the significant difference in the cosmological parameter constraints observed in figure \ref{figure:omega_m_vs_omega_lambda}. As is the case with the $H_{0}$ tension, it is not exactly clear what is the root cause for such a divergence of measurements. As illustrated by the results in \cite{Joudaki2020}, the DES measurements reduce but do not solve the tension observed between the KV450 and the Planck datasets. Data from the surveys in the upcoming decade will likely give additional clues about the nature of the $S_{8}$ and other related tensions.  

The outlined problems indicate that despite the great success of the $\Lambda$CDM model many issues remain. It is possible that these issues could be resolved rather naturally with high quality observational data from the upcoming surveys along with more realistic simulations and a better understanding of the properties of dark matter and dark energy. However, it could also indicate a need for new physics. In either case, all the discussed phenomena are intimately related to our understanding of how gravity works on different scales. Starting with the intricacies of galaxy formation and ending with the issues related to the accelerated expansion, a better understanding of gravity could help resolve some of the key issues outlined above. Because of this, modifying GR has been proposed as a possible solution to the many conundrums facing the standard model of cosmology. 

\section{Modified Gravity: Tests and Current Developments}
\raggedbottom
\label{introduction:modified_gravity}

The motivations for modifying GR are generally trifold: accounting for the accelerated expansion of the Universe, explaining the nature of the missing mass on cosmological scales and giving a deeper understanding of how gravity relates to quantum field theory. These are all goals of key importance and making significant progress in any of these directions could account for the various shortcomings of the $\Lambda$CDM model. For these reasons, a vast family of modified gravity models has been developed. 

One rather natural way of classifying modifications to GR can be defined in the context of Lovelock's theorem. Lovelock's theorem states the following: \textit{in 4-D the only divergence-free symmetric rank-2 tensor constructed from only the metric $g_{\mu \nu}$ and its derivatives up to second order, and preserving diffeomorphism invariance, is the Einstein tensor with a cosmological constant term}. In slightly simpler words, Einstein field equations are unique equations of motion for a single metric derivable from a covariant action in 4-D \citep{Berti2015, Li2019}. This theorem is profound as it shows that GR in this context is the simplest theory of gravity with the outlined properties. Hence, if one was to modify GR, some of the outlined conditions would necessarily have to be broken. In fact, Lovelock's theorem gives a recipe on how to generate modified gravity theories: a modification of gravity will have one (or multiple) of the following features:

\begin{enumerate}[label=\roman*), align=left]
  \item Extra degrees of freedom. This refers to extra scalar, vector and tensor fields introduced to the action. This class of models includes the Horndeski theory, which is the most general scalar-tensor theory in 4 dimensions leading to second order equations of motion \citep{Horndeski1974}. Horndeski theory includes many familiar theories such as Brans-Dicke gravity, chameleon gravity and quintessence. This class also contains models such as massive gravity and bi-gravity \citep{Kenna-Allison2019-1, Kenna-Allison2019-2}. 
  \item Lorentz Violations. These models break the Lorentz invariance. Examples models include Ho\v{r}ava gravity, Einstein-Aether theory and n-DBI gravity \citep{Blas2014}.  
  \item Higher spacetime dimensionality. Early models including extra spacetime dimensions, such as the Kaluza-Klein theory, have inspired a number of contemporary models such as string theory. Other prominent models in this class include braneworld models \citep{Maartens2010}. 
  \item Non-locality. Non-local models contain terms of the form of $Rf(\Box^{-1}R)$ or $m^{2} R \Box^{-2} R$ in the Einstein-Hilbert action. More generally, various string-inspired non-local models have gained popularity in recent years. Such models have been used in the context of dark energy, inflation and bouncing cosmology scenarios \citep{Koshelev2011}.  
  \item Higher derivatives. These models introduce higher degree derivatives to the action. Such theories are difficult to construct as higher derivatives can lead to Ostrogradsky instability. However, there are ways to avoid such instabilities, as shown in beyond Horndeski models \citep{Langlois2016}.  
\end{enumerate}

\begin{figure}[t!]
\centering
\includegraphics[width=0.92\columnwidth]{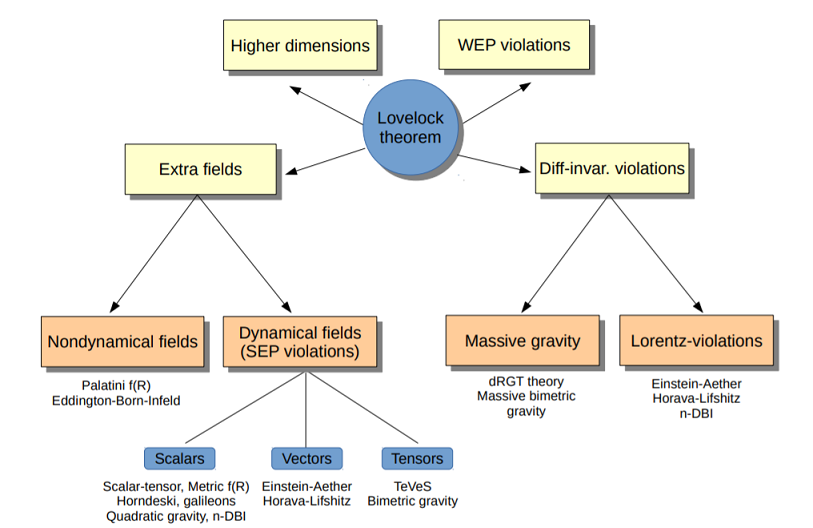}
\caption[Modified gravity models]{A classification of modified gravity models based on Lovelock's theorem \citep{Berti2015}. The abbreviations refer to the weak and the strong equivalence principles (WEP and SEP) and to diffeomorphism invariance.  
\label{figure:modified_gravity_models}}
\end{figure}

Figure \ref{figure:modified_gravity_models} shows some of the more popular models classified according to Lovelock's theorem. It is important to note that there are many models that do not easily fit into such classification. A prime example of this in the context of this thesis refers to various emergent/entropic gravity models. Emergent gravity can refer to a wide class of not necessarily related theories that describe gravity as an emergent phenomenon. Such theories combine ideas from black hole thermodynamics and condensed matter physics in order to explore the possible emergence of gravity with prime examples being approaches described in \citet{Padmanabhan2015} and \citet{Verlinde2017}.

The mentioned models can give insight into the various conundrums of the standard model. In particular, the mentioned classes of models can explain the accelerating expansion with various degrees of success. Or, additionally, some of the models can give insights into the problem of dark matter and shine light on the various incompatibilities between GR and quantum physics. However, as of yet, there is no single framework that fully accounts for the effects associated with dark energy and dark matter while also fitting all the key observational datasets. Observational constraints, in particular, play a crucial role in exploring the space of the allowed theories. There is a plethora of astrophysical and cosmological tests on scales ranging from laboratory and interferometry tests all the way to large scale structure tests of modified gravity. Here we will review the main types of observational and experimental tests in a rough order of scale. A deeper discussion of the cluster scale tests will be given in chapters \ref{ch:galaxy_clusters} and \ref{ch:modified_gravity}.

\subsection{Laboratory Tests}

Laboratory tests aim to detect fifth force effects on the smallest scales accessible by the currently available instruments ($\mu$m and larger). A key challenge for these types of experiments is reducing the Newtonian force effects from the environment. This can be done by using vacuum chambers and optimizing the geometry of the experiment (i.e. the geometry of the mass/density distribution). 

At sub-mm scales one could in principle detect the Casimir force effects, which are predicted by quantum electrodynamics, manifesting as an interaction between two parallel uncharged plates. At these scales one can also detect chameleon forces, which would dominate over the Casimir force. Hence, the deviation from the predicted Casimir force can be used as a probe for chameleon force effects. Chameleon models refer to a class of scalar-tensor theories that avoid the solar system constraints by employing a special form of a non-linear potential. A common general choice for chameleon models is of the following inverse power law form \citep{Burrage2018}:

\begin{equation}
V(\phi)=\tilde{\Lambda}_{0}^{4}+\frac{\Lambda_{0}^{4+n}}{\phi^{n}},
\label{eq:chameleon_potential}
\end{equation}

\noindent where $\phi$ is the scalar field, and the different choices of the $\{ \tilde{\Lambda}_{0}, \Lambda_{0}, n \}$ parameters corresponds to different models. $\tilde{\Lambda}_{0}$ can be set to $\approx 10^{-3}$ eV to account for the accelerating expansion (discussed further in chapter \ref{ch:modified_gravity}). 
 
When it comes to Casimir force experiments, the most precise measurements are achieved by measuring the force between a plate and a sphere rather than two plates, which leads to the chameleon force scaling with the distance between the sphere and the plate, $d$ as follows: 

\begin{equation}
F_{\phi} \sim d^{\frac{2-n}{n+2}},
\label{eq:chameleon_force_scaling}
\end{equation}

\noindent with $F_{\phi}$ as the chameleon force and $n$ as a constant that dictates the scaling. Stringent constraints can be put for $n=-4$ and $n=-6$ models \citep{Burrage2016}.  

Other experiments that probe the Casimir force effects include optically levitated dielectric spheres with radii ranging around $r \sim \mathcal{O}(\mu \mathrm{m})$. In these types of experiments laser beams are used to counter the Earth's Newtonian gravity effects. Such an approach can put constraints on the $n = 1$ models \citep{Burrage2018}.

Atom interferometry is another powerful technique that can be used for constraining chameleon models. These experiments employ interferometers, which allow probing the acceleration experienced by atoms due to chameleon forces. In particular, atoms are put into a superposition of states related to the two different paths that can be taken (the two arms of the interferometer). The two paths are later recombined and a measurement is made that allows to put constraints on the acceleration of the atoms with precisions of around $10^{-6}g$, with $g \approx 9.8$ $\mathrm{m/s^{2}}$ \citep{Elder2016}.

\begin{figure}[ht!]
\centering
\includegraphics[width=0.97\columnwidth]{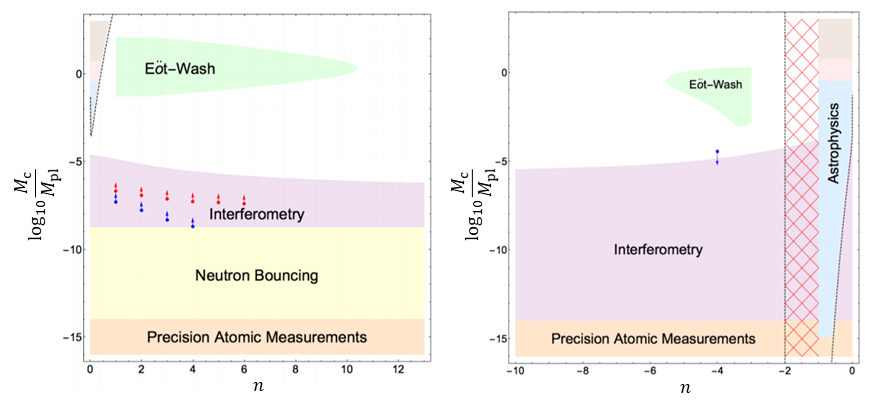}
\caption[Summary of laboratory tests of gravity]{A summary of the observational and laboratory tests constraining chameleon models. The different regions mark the excluded subsets of the parameter space. The black, blue and red dots show the lower bounds (indicated by the arrow) on the coupling strength $M_{c}$ at the dark energy scale coming from neutron bouncing and interferometry experiments respectively. The dark energy scale refers to $\Lambda_{0} = 2.4 \times 10^{-3}$ $\mathrm{eV}$ and is marked by the dotted lines. The two plots refer to the constraints with $\Lambda_{0}$ fixed to the dark energy scale and positive values of $n$ (left figure) and negative values (right figure). The red hashed area refers to regions where the model does not possess chameleon screening. Finally, the brown subsets correspond to parameter space regions accessible by cosmological observations \citep{Hamilton2015, Lemmel2015, Li2016, Burrage2016}.  
\label{figure:modified_gravity_laboratory_tests}}
\end{figure}

Another class of laboratory tests that is worth mentioning is precision neutron tests. Neutrons, being electrically neutral particles, are perfect for isolating the fifth force effects from the gravitational and electromagnetic forces due to the environment. Different experiments using neutrons place constraints on the chameleon coupling strength $M_{c}$. For instance, using ultra cold neutrons interacting with a mirror one can put a constraint in the range of $M_{c} > 1.7 \times 10^{6}$ $\mathrm{TeV}$ \citep{Jenke2014}. Figure \ref{figure:modified_gravity_laboratory_tests} summarizes some the currently available laboratory constraints on chameleon models.

\subsection{Solar System Tests}

Solar system tests of GR date back to the very beginnings of Einstein's revolutionary theory. In fact, long before the development of GR, deviations of the perihelion precession of Mercury from the Newtonian gravity prediction were known. This observation later led to one of the key tests confirming the validity of GR.

Another early test confirming the validity of GR was performed by measuring the deflection of light by the Sun. The observations of Arthur Eddington and collaborators during the solar eclipse of 1919 measured the displacement of the position of stars behind the sun proving one of the key tenets of the theory.

Modern tests put some of the tightest constraints on the deviations from GR. Experiments, such as the Shapiro time delay measurements, which give the relativistic time day experienced by radar signals in a round trip to Mercury and Venus, agree with the theoretical GR prediction at 5\% level \citep{Shapiro1971}. More recently measurements based on the same basic principle were performed using the data from the Cassini spacecraft, which measured the frequency shift of radio photons to and from the spacecraft. This experiment constrains the parametrized post-Newtonian formalism Eddington parameter $\gamma$ (which quantifies the deflection of light by a gravitational source) with high precision: $\gamma = 1+(2.1 \pm 2.3) \times 10^{-5}$ \citep{Bertotti2003}.    

Tests of the strong equivalence principle (laws of gravity are independent of velocity and location) are of special importance in the context of modified gravity models. A wide class of theories predict violations to the strong equivalence principle on some level. In general, tests of the strong equivalence principle test the universality of free fall, which is measured by comparing accelerations $a_{1}$ and $a_{2}$ of two different bodies:

\begin{equation}
\frac{\Delta a}{a}=\frac{a_{1}-a_{2}}{\frac{1}{2}\left(a_{1}+a_{2}\right)}=\left(\frac{M_{G}}{M_{I}}\right)_{1}-\left(\frac{M_{G}}{M_{I}}\right)_{2},
\label{eq:test_of_sep}
\end{equation}

\noindent with $M_{G}$ and $M_{I}$ as gravitational and inertial masses correspondingly. In the case of the solar system tests of the equivalence princple, the two bodies are the Earth and the Moon as measured in the lunar laser ranging experiments. These experiments put strong constraints on the anomalous perihelion angular advance of the Moon: $|\delta \theta|<2.4 \times 10^{-11}$ \citep{Williams2004, Li2019B}. Such experiments also constrain the time variation of Newton's constant: $\dot{G}/G=(2 \pm 7) \times 10^{-13}$ per year \citep{Williams2009}. Finally, the constraints from the lunar laser ranging experiments can be combined with the E\"{o}t-Wash torsion balance measurements to provide a confirmation for the strong equivalence principle at 0.04\% \citep{Merkowitz2010}. 

The solar system constraints have had a profound influence on the theoretical development of modified gravity models. The outlined constraints clearly indicate that GR is valid in the solar system, leaving nearly no space for even miniscule modifications of the model. This has led to the development of various screening mechanisms, which suppress the fifth force effects in the solar system, while still allowing interesting effects on cosmological scales. 
  
\subsection{Gravitational Wave Tests} In terms of observational constraints, one of the key developments at the time of writing this thesis has been the detection of the gravitational wave and gamma ray burst signals from a neutron star merger event GW170817/GRB 170817A. The event resulted in a 100 second gravitational wave signal and a corresponding 2 second duration gamma-ray burst caused by the merger \citep{Ligo2017}. The optical counterpart of the event has subsequently been observed by over 70 observatories marking the beginning of this type of multi-messenger astronomy \citep{Nichol2017}.

The key significance of the mentioned gravitational wave observation in the context of this thesis comes in terms of the constraints on modified gravity models. In general, introducing new fields coupled to gravity in modified models of gravity affects the propagation speed of gravitational waves. Hence, the speed of the propagation of gravitational waves can be used as reliable probe of modified gravity. Probing modified gravity models with gravitational waves has a number of advantages, such as the fact that gravitational waves can be used to test theories with screening mechanisms (given that the signals come from extragalactic sources). In addition, even small deviations from the speed of light in gravitational wave propagation can accumulate over large distances, making such a probe extremely sensitive. In particular, the observed event GW170817 allowed putting extremely tight constraints on the speed of the gravitational waves: $\left| c_{\mathrm{GW}}/c -1  \right| \leq 5 \times 10^{-16}$ \citep{Abbott2017_2}. This result has single-handedly ruled out a wide subset of modifications of gravity. More specifically, such a strong constraint practically rules out any model that predicts variation of the gravitational wave propagation speed with respect to the speed of light. It is useful at this point to discuss some of the effects of the gravitational wave results on the various classes of models discussed previously without going into great detail. In this regard, it is useful to introduce Horndeski theory, which contains several models important to this thesis as subsets of the theory. Models of special importance to this thesis will be discussed further in chapters \ref{ch:modified_gravity} and \ref{ch:testing_EG}.  

As previously mentioned, Horndeski theory refers to the most general scalar-tensor theory with 2nd degree equations of motion. The theory can be described by the following Lagrangian: 

\centerline{\noindent\begin{minipage}{1.1\textwidth}
\begin{equation}
\begin{split}
\mathcal{L_{\rm H}}= & G_{2}(\phi, X)-G_{3}(\phi, X) \square \phi +G_{4}(\phi, X) R+G_{4,X}(\phi, X)\left[(\square \phi)^{2}-\left(\nabla_{\mu} \nabla_{\nu} \phi\right)^{2}\right] + \\
&G_{5}(\phi, X) G^{\mu \nu} \nabla_{\mu} \nabla_{\nu} \phi-\frac{1}{6} G_{5, X}(\phi, X)\left[(\square \phi)^{3}-3 \square \phi\left(\nabla_{\mu} \nabla_{\nu} \phi\right)^{2}+2\left(\nabla_{\mu} \nabla_{\nu} \phi\right)^{3}\right],
\end{split}
\label{eq:horndeski_lagrangian}
\end{equation}
\end{minipage}}

\noindent where $G_{2}$, $G_{3}$, $G_{4}$ and $G_{5}$ are free functions of the scalar field $\phi$ and $X \equiv -1/2 g^{\mu \nu} \partial_{\mu} \phi \partial_{\nu} \phi$, $G^{\mu \nu}$ is the Einstein tensor, $R$ is the Ricci scalar, $\square \phi=\nabla_{\mu} \nabla^{\mu} \phi$ and the subscript commas denote derivatives \citep{Horndeski1974}. The different choices for the set of functions $\{ G_{2}, G_{3}, G_{4}, G_{5} \}$ represent different scalar-tensor models. 

The gravitational wave speed in Horndeski theory can be deduced via the tensor sound speed $\alpha_{\rm T}$ by noting that $c_{\mathrm{\rm  GW}}^{2}=1+\alpha_{\rm T}$. The tensor sound speed has been shown to have the following form \citep{Kobayashi2011, Li2019}:

\begin{equation}
\alpha_{\rm T}=\frac{2 X\left[2 G_{4,X}+G_{5,\phi}-(\ddot{\phi}-H \dot{\phi}) G_{5,X}\right]}{2\left[G_{4}-2 X G_{4,X}-\frac{1}{2} X G_{5, \phi}-\dot{\phi} H X G_{5,X}\right]}.
\label{eq:horndeski_c_gw}
\end{equation}

\noindent The observational requirement of $c_{\mathrm{GW}} \approx c$ (or equivalently $\alpha_{\rm T} \approx 0$) can be satisfied by setting $G_{5} =0$ and $G_{4} = G_{4}(\phi)$. This ultimately results in only the following class of Lagrangians surviving: 

\begin{equation}
\mathcal{L_{\rm H}}=G_{4}(\phi) R+G_{2}(\phi, X)-G_{3}(\phi, X) \square \phi.
\label{eq:surviving_L_horndeski}
\end{equation}

The effect of this is that a wide class of Horndeski and beyond Horndeski models are ruled out (see \citet{Sakstein2017} and \citet{Baker2017} for a wider discussion). This includes some important models in the context of the accelerating expansion of the Universe. In particular, a subclass of Galileon models, which can account for the accelerated expansion without a need for a cosmological constant are ruled out. Similarly, a wide subset of degenerate higher order scalar-tensor theories (DHOST) has been ruled out. The same can be said about the Fab Four models that contain interesting cosmological solutions. 

\begin{figure}[t!]
\centering
\includegraphics[width=0.99\columnwidth]{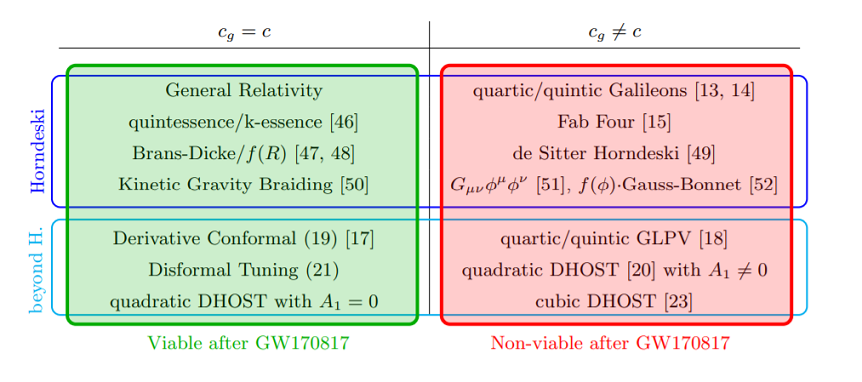}
\caption[Summary of modified gravity models after GW170817]{Summary of the state of modified gravity models after the GW170817 gravitational wave results \citep{Ezquiaga2017}. The models are classified according to the predicted value of the gravitational wave speed $c_{g}=c_{\mathrm{GW}}$. 
\label{figure:gravitational_wave_constraints}}
\end{figure}

The surviving models include a class of theories where gravity is minimally coupled like the kinetic gravity braiding models and quintessence. The k-essence models are also still valid. So are the models relevant to this thesis, such as the $f(R)$ and Brans-Dicke theories. Figure \ref{figure:gravitational_wave_constraints} summarizes the current state of the various modified gravity models after the release of the gravitational wave results.    

\subsection{Galaxy Scale Tests}

Observations of galaxies have historically played an important role in the theoretical development of dark matter and modified gravity models. Namely, galaxy rotation curve measurements acted as one of the initial pieces of evidence for the existence of dark matter. More generally, the complex morphology of galaxies allows testing modified gravity models with different screening mechanisms along with alternative models of dark matter. 

Theories with screening mechanisms predict different effects on the gas and the stars that make up galaxies. This is the case, as stars are generally screened, while the diffuse gas is not. Hence, comparing the rotation curves of stars and gas allows putting constraints on theories with screening. As an example, this method has been used to constrain the $f_{R0}$ parameter to values of $f_{R0} < 10^{-6}$ in $f(R)$ models (see chapter \ref{ch:modified_gravity} for a wider discussion of these models) \citep{Vikram2018}. More generally, for theories with screening, the self-screening parameter has been constrained to values of: $\chi_{c} < 10^{-6}$.

In addition, screening can lead to morphological and kinematical distortions of galaxies. In this case the stellar
component of a dwarf galaxy is self-screened while the surrounding dark matter halo and gaseous component are unscreened. Different fifth force effects experienced by the different parts of galaxies lead to an offset of stellar disks from the HI (neutral atomic hydrogen) gaseous components. In addition, galactic disks are warped in a way whereby the screened stars are displaced from the principal axis. A recent example of such measurements includes \citet{Desmond2018}, where offsets between the optical and HI centroids were constrained. The mentioned measurements also put a constraint on the $f(R)$ theories: $3\left|f_{R 0}\right| / 2< 1.5 \times 10^{-6}$.

Galaxies also offer ways of testing gravity via gravitational lensing. A recent example of such a measurement comes from the ESO 325-G004 elliptical galaxy. Comparing the mass estimates from the stellar motion and weak lensing data coming from the Hubble Space Telescope and the Very Large Telescope indicated no significant deviation from GR with $\gamma=0.97 \pm 0.09$ with 1-$\sigma$ confidence \citep{Collett2018}. 

The mentioned techniques are only a small subset of the tests performed on galaxy scales in recent years. For a more systematic review see \citet{Jain2011,Vikram2013,Koyama2016}.

\subsection{Galaxy Cluster Tests}

Galaxy clusters and superclusters, being the largest gravitationally bound structures, offer a multitude of ways of testing the effects of gravity on large scales. Galaxy clusters contain anywhere from hundreds to thousands of galaxies, with total masses in the range of $10^{14}-10^{15}$ M\textsubscript{\(\odot\)}. The mass distribution of galaxy clusters is dominated by galaxies and the lower density intracluster medium (ICM) with temperatures ranging between 2-15 keV \citep{Kravtsov2012}. This combination of high density regions (where the fifth force would be screened) and lower density intracluster gas, especially in the outskirts of clusters (where there would be no screening), makes clusters great for testing modified gravity theories.

Various modified gravity models with screening mechanisms can leave imprints in the observational properties of galaxy clusters. More specifically, modifications of GR can affect cluster density profiles and correspondingly X-ray surface brightness and weak lensing profiles. As an example, recent work in \citet{Schmidt2009} and \citet{Cataneo2016} investigated the abundance of massive halos as a tool for detecting $f(R)$ gravity effects. Both studies found similar constraints for $f(R)$ models: $\left|f_{R_{0}}\right|\lesssim 10^{-4}$. 

As discussed, the effects of modified gravity with chameleon screening would not be detectable in the high density galaxy cluster cores, however, the fifth force would have an effect in the outskirts of clusters. This introduces a deviation between the hydrostatic and lensing masses, which, in principle, can be observed by combining X-ray and weak lensing measurements. Using this technique, the constraints of $\left|f_{R 0}\right| \lesssim 6 \times 10^{-5}$ at 95\% confidence were obtained in \cite{Terukina2013} and \cite{Wilcox2015}. 

These and other cluster scale constraints are discussed in greater detail in chapter \ref{ch:modified_gravity}.

\subsection{Large Scale Structure Tests}

Large scale structure formation is sensitive to the underlying model of gravity. Most types of deviations from GR should in principle be detectable in the CMB anisotropy data. Furthermore, measurements of the CMB power spectrum and the secondary bispectrum have some sensitivity to modified gravity growth of structure effects through the large-scale integrated Sachs-Wolfe
effect and weak lensing. Another method of constraining gravity is via redshift space distortions. This refers to the spatial distribution of galaxies appearing distorted when their positions are plotted as a function of their redshift rather than as a function of their distance. Comparing these effects against the theoretical GR predictions places stringent constraints on modifications of the standard laws of gravity.

As a concrete example, competitive constraints on $f(R)$ models were obtained in \citet{Lombriser2012}, where a subclass of $f(R)$ models designed to reproduce the $\Lambda$CDM expansion history was tested. In the context of the expansion history, such models can be parametrized by  $B_{0}$, which corresponds to the Compton wavelength parameter. By combining the data from supernovae distances, baryon acoustic oscillations and the CMB, constrains of $B_{0} < 1.1 \times 10^{-3}$ at 95\% confidence were determined.

\begin{figure}[ht!]
\centering
\includegraphics[width=0.80\columnwidth]{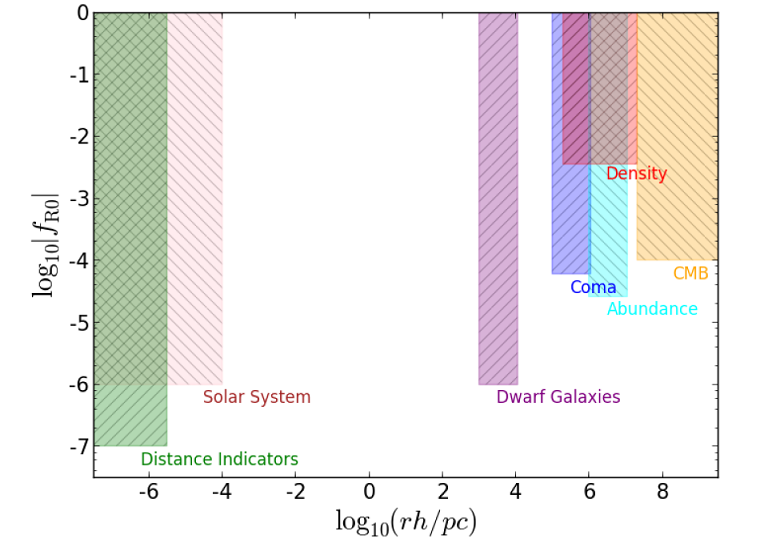}
\caption[Summary of the observational constraints on $f(R)$ gravity]{Summary of different observational constraints on the $f_{R0}$ parameter as a function of scale. Figure from \citet{Wilcox2016}, originally adapted from \citet{Terukina2013}. 
\label{figure:f_R_constraints}}
\end{figure}

Needless to say, the outlined list of the observational probes is far from complete. A number of techniques will be left undiscussed due to being out of scope of this thesis. In addition, a much deeper discussion of the $f(R)$ model and the corresponding constraints is given in chapter \ref{ch:modified_gravity}. Finally, figure \ref{figure:f_R_constraints} shows a summary of the relevant constraints on $f(R)$ models.

\chapter{Galaxy Clusters}
\label{ch:galaxy_clusters}

Chapter \ref{ch:galaxy_clusters} introduces the key ideas from galaxy cluster physics. In particular, the basic structure of galaxy clusters is introduced and discussed in the context of the underlying astrophysics. In addition, the properties of the intracluster medium are discussed in the context of measuring X-ray surface brightness as one of the key probes in our tests of modified gravity. Similarly, the key features of the SZ effect are introduced. Galaxy kinematics is discussed as an important technique for measuring cluster masses. Finally, weak lensing by galaxy clusters is summarized as a key tool for testing modified gravity as discussed in chapters \ref{ch:modified_gravity} and \ref{ch:testing_EG}.  

\section{The Structure and Basic Properties of Galaxy Clusters }

Historically the observational studies of galaxy clusters date back to the work of Herschel and Messier, who were the first to notice the tendency of galaxies (then only known as \textit{galactic nebulae}) to cluster \citep{Kravtsov2012}. Later work by Hubble in 1926 showed that galactic nebulae are in fact galaxies, which in turn resulted in a better understanding of the nature of galaxy clusters \citep{Heilbron2005}. In 1933, under the assumption of virial equilibrium, Zwicky made a crucial discovery that the visible mass in the Coma Cluster is not enough to account for the motion of galaxies in the cluster \citep{Zwicky1933}. In particular, Zwicky calculated the dispersion of radial velocities of 8 galaxies in the Coma Cluster and found the value of $\sigma=1019 \pm 360$ km/s \citep{Figueras2007}. Comparing this result against the prediction derived using hydrostatic equilibrium equations Zwicky found that the Coma Cluster had to be over 400 times more massive than the mass contained in the visible parts of galaxies in the cluster. This marks the beginning of the observational studies of dark matter. 

Modern multi-wavelength studies of galaxy clusters allow us to draw a detailed picture of the physical properties of these objects. Galaxy clusters, being among the largest gravitationally bound structures, contain from hundreds to thousands of galaxies. Typical masses of galaxy clusters fall in the range of $10^{14}-10^{15}$ M\textsubscript{\(\odot\)} \citep{Sarazin1988}. A key feature of galaxy clusters is the high energy intracluster medium (ICM), consisting of heated, X-ray emitting gas with temperatures of around $2-15$ $\mathrm{keV}$ \citep{Fabian1992}. Measuring the composition of galaxy clusters is difficult, as it varies significantly among individual clusters. However, as a guideline, dark matter makes up $\sim 84 \%$ of the mass budget in clusters, with the leftover $\sim 16 \%$ corresponding to the high-energy ICM and stars \citep{Rosati2002}. 

\begin{figure}[ht!]
\centering
\includegraphics[width=0.99\columnwidth]{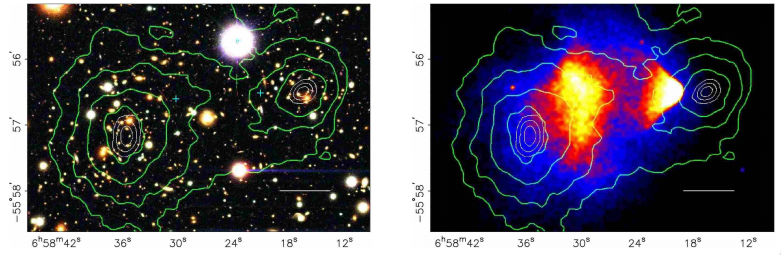}
\caption[The mass distribution in the Bullet Cluster]{\textbf{Left}: an optical image of the Bullet Cluster taken by the Magellan telescope with overplotted contours representing the total mass distribution inferred by gravitational lensing. \textbf{Right}: the hot plasma distribution (in red) in the Bullet Cluster inferred by the Chandra X-ray data with the same mass contours plotted over. \citep{Clowe2006}.  
\label{figure:bullet_cluster}}
\end{figure}

Modern studies of galaxy clusters have played a crucial role in understanding the properties of dark matter. A prime example of this is the case of merging galaxy clusters. During a merger the different components that make up a cluster interact differently. In particular, visible matter located in stars and galaxies is mostly not affected by the collision. High energy ICM as detected by X-ray observations, however, is slowed down significantly due to the electromagnetic interactions. Finally, the major mass component in the form of dark matter passes through the baryonic matter with no interaction. This results in a mass distribution where the bulk of the mass resides in regions different to those dominated by the X-ray emitting ICM. The most well-known system of merging clusters is the Bullet Cluster, which provides some of the best existing evidence for the existence of dark matter on galaxy cluster scales \citep{Clowe2006}. Figure \ref{figure:bullet_cluster} illustrates the total and the ICM mass distributions in the Bullet Cluster clearly showing that the two mass components appear in different locations in the merging cluster system. In the case of modified gravity models, such as modified Newtonian dynamics (MOND), which assume no existence of dark matter, most of the mass in merging clusters should coincide with the visible baryonic mass distribution. Hence merging clusters, such as the Bullet Cluster, give important evidence against modified gravity models of such kind. 

Other properties that are worthy of discussion are related to the formation and the spatial distribution of galaxy clusters. It is widely accepted that the likely cause of the observed large scale structure in the Universe is the seed density fluctuations that were formed during the period of cosmic inflation. Due to gravitational instability these initial perturbations have been amplified and eventually collapsed to form an intricate web of filaments and voids. Observing galaxy clusters allows us to deduce the various properties of the structure of the Universe on the largest accessible scales. In addition, cluster observations give information about the underlying cosmological model. The simplest models of inflation assume the primordial density fluctuations to be Gaussian. However, more complicated models predict varying amounts of non-Gaussianity. Detecting such non-Gaussianity via the CMB or other means would be of great interest in the early Universe studies. A number of recent studies have been dedicated to investigating the viability of using galaxy clusters as an alternative probe for non-Gausianities \citep{Mana2013, Trindade2017}. 

Clusters are known to be much more strongly clustered than galaxies \citep{Bahcall1988}. This can be expressed in terms of the two-point correlation function that follows a power law of the form $\xi_{c c}(r)=\left(r / R_{0}\right)^{-1.8}$. More specifically, $\xi_{c c}(r)$ refers to the two-point spatial correlation function which is related to the joint probability of finding two objects at separation $r$. $R_{0}$ has been shown to obey the following scaling relation: $R_{0} \approx 0.4 D_{c}$ for $20 h^{-1} \mathrm{Mpc}<D_{c}<100 h^{-1} \mathrm{Mpc}$ and $D_{c} \equiv n_{c}^{-1 / 3}$, with $n_{c}$ as the mean space density of clusters \citep{Bahcall1992}. This agrees well with the more recent observational results described in \citet{Basilakos2004, Balaguera2014}. Recent studies also reinforce the conclusion that the spatial distribution of the tracers of the large scale structure, such as galaxies and clusters of galaxies, are a powerful probe for the early Universe physics. In particular, combining cluster and galaxy power spectra along with the cross power spectrum offers a way of accessing a plethora of information about the underlying cosmology. Hence future surveys such as Euclid will place unprecedented constrains on primordial non-Gaussianity \citep{Euclid2018}. 

Another key quantity to discuss is the cluster mass function, which is of special importance when studying large scale structure formation. In particular, the cluster mass function quantifies the number of clusters of a given mass at a given redshift: $n(M,z)$. The mass function can be estimated using the Press-Schechter formalism, which assumes that the fraction of matter that ends up in objects with mass $M$ can be deduced from the portion of the initial density field (smoothed on the mass scale $M$) lying at an overdensity exceeding a given critical threshold value $\delta_{c}$. The gradient of the mass function can be shown to take the following form \citep{Press1974, Borgani2006}:

\begin{equation}
\frac{d n(M, z)}{d M} =\sqrt{\frac{2}{\pi}} \frac{\bar{\rho}_{c}}{M^{2}} \frac{\delta_{c}}{\sigma_{M}(z)}\left|\frac{d \log \sigma_{M}(z)}{d \log M}\right| \exp \left(-\frac{\delta_{c}^{2}}{2 \sigma_{M}(z)^{2}}\right),
\label{eq:press_schechter}
\end{equation}

\noindent with $\bar{\rho}_{c}$ as the mean cluster density and $\sigma_{M}(z)$ is the variance at mass scale $M$ linearly extrapolated to redshift $z$. The key takeaway from equation \ref{eq:press_schechter} is that there is an intimate link between the early Universe primordial perturbations and the late Universe structure. Through the linear perturbation growth factor the value of $\sigma_{M}$ can be directly related to the power spectrum and the cosmological density parameters. This further illustrates the value of galaxy clusters as probes for the formation and evolution of large scale structure and, in turn, for the underlying model of cosmology. 

There are other important galaxy cluster probes of the underlying cosmology. These include the mass-to-light ratio and the baryon fraction. The mass-to-light ratio quantifies the ratio between the total mass in a given volume versus the corresponding luminosity. This ratio can be used to deduce the matter density $\Omega_{m}$. The baryon fraction also provides a constraint on the matter density parameter (assuming that the cosmic baryon density parameter is known). With an additional assumption that the baryon fraction does not evolve in galaxy clusters, one can constrain the dark energy equation of state parameters \citep{Borgani2006}.

\section{Intracluster Medium and X-ray Observations}
\label{section:X-ray_masurements}

The ICM is composed of high-energy superheated X-ray emitting plasma. The ICM mainly consists of ionized helium and hydrogen, which dominates the total baryonic content of galaxy clusters. Heavier elements, such as iron, can also be found as quantified by the ratio to hydrogen known as metallicity. Average values of metallicity range from one third to a half of the value observed in the Sun \citep{Mantz2017}. Studying the chemical structure of the ICM and its evolution with redshift offers a record of the overall element production and evolution throughout the history of the Universe. 

The high temperature of the ICM leads to X-ray emission via the process known as bremsstrahlung radiation. Bremsstrahlung radiation refers to the braking radiation produced by deceleration of charged particles when deflected by other charges. A typical example of such a process is the deflection of electrons by atomic nuclei leading to X-ray emission with a frequency proportional to the energy change. The emissivity at frequency $\nu$ for an ion of charge $Z$ in a plasma with an electron temperature $T_{e}$ is \citep{Sarazin1988}:

\begin{equation}
\epsilon_{v}^{\rm ff}=\frac{32 \pi e^{6} Z^{2} n_{e} n_{i}}{3 m_{e} c^{3}} \sqrt{\frac{2 \pi}{3 k_{B} T_{e} m_{e}}} g_{\rm ff}(T_{e}, v) e^{-h v / k_{B} T_{e}} ,
\label{eq:bremsstrahlung}
\end{equation}

\noindent with $e$ as the elementary charge, $n_{i}$ and $n_{e}$ as the number densities of ions and electrons, $m_{e}$ as the electron mass, $g_{\rm ff}$ as the Gaunt factor which corrects for quantum effects, $k_{B}$ as the Boltzmann constant and $h$ as the Planck constant. The Gaunt factor in this case is given by \citep{Nozawa1998}: 

\begin{equation}
g_{\rm ff} \approx \frac{3}{\sqrt{\pi}} \ln \left(\frac{9 k_{B} T_{e}}{4 h v}\right), 
\label{eq:gaunt_factor}
\end{equation}

\noindent The total emission including all the other components (such as line emission) is then given by:

\begin{equation}
\epsilon_{\nu}=\int_{0}^{\infty} e_{v}^{\rm ff} d v \approx 3.0 \times 10^{-27} \sqrt{\frac{T_{e}}{1 \text{ K}}}\left(\frac{n_{e}}{1 \text{ cm}^{-3}}\right)^{2} \mathrm{erg} \cdot \mathrm{cm}^{-3} \mathrm{s}^{-1}.
\label{eq:total_emissivity}
\end{equation}

\noindent The different emission processes described in equation \ref{eq:total_emissivity} can be written as follows: 

\begin{equation}
\epsilon_{v}=\sum_{i} n_{i} n_{e} \lambda_{c}\left(n_{i}, T_{e}\right),
\label{eq:emissivity_cooling}
\end{equation}

\noindent where $\lambda_{c}\left(n_{i}, T_{e}\right)$ is the temperature and ion dependent cooling function that is related to the emission mechanism. Observationally a more natural quantity to work with is the surface brightness, which is equal to the integral of $\epsilon_{\nu}$ \citep{Terukina2013, Wilcox2015}:

\begin{equation}
S_{\mathrm{B}}\left(r_{\perp} \right)=\frac{1}{4 \pi(1+z_{cl})^{4}} \int n_{\mathrm{e}}^{2}\Big(\sqrt{r_{\perp}^{2}+z^{2}}\Big) \lambda_{\mathrm{c}}\left(n_{e},T_{e}\right) d z.
\label{eq:surface_brightness2}
\end{equation}

\noindent  Here we assumed spherical symmetry and switched to projected coordinates, such that a point at some radius $r$ from the centre of the cluster is given by $r = \sqrt{r_{\perp}^{2} + z^{2}}$, with $r_{\perp}$ as the perpendicular radial distance and $z$ as the distance from the centre in the direction parallel to the line of sight. In addition, we assumed that the gas within a given cluster at redshift $z_{cl}$ is dominated by hydrogen, i.e. $n_{i} = n_{e}$. The $4\pi$ factor comes from the assumption that the emissivity is isotropic, while the $(1 + z_{cl})^{4}$ term accounts for the cosmological transformations of spectral surface brightness and energy. The electron number density $n_{e}$ is clearly related to the gas distribution in a cluster which allows us to use surface brightness as a probe for the underlying mass distribution.

Another cluster scale observable worthy of discussion is the Sunyaev-Zeldovich (SZ) effect. The SZ effect refers to the distortion of the CMB through inverse Compton scattering by high-energy electrons from the ICM. The SZ effect, more specifically, is a combination of multiple primary and secondary effects. These include thermal interactions between the CMB photons and the high-energy electrons as well as secondary kinematic and polarization effects. This method does not depend on redshift and provides a way of measuring cluster masses as well as detecting clusters at great distances. In addition, it is possible to use a combination of the SZ effect and X-ray measurements to accurately deduce distances to clusters. 

\begin{figure}[ht!]
\centering
\includegraphics[width=0.50\columnwidth]{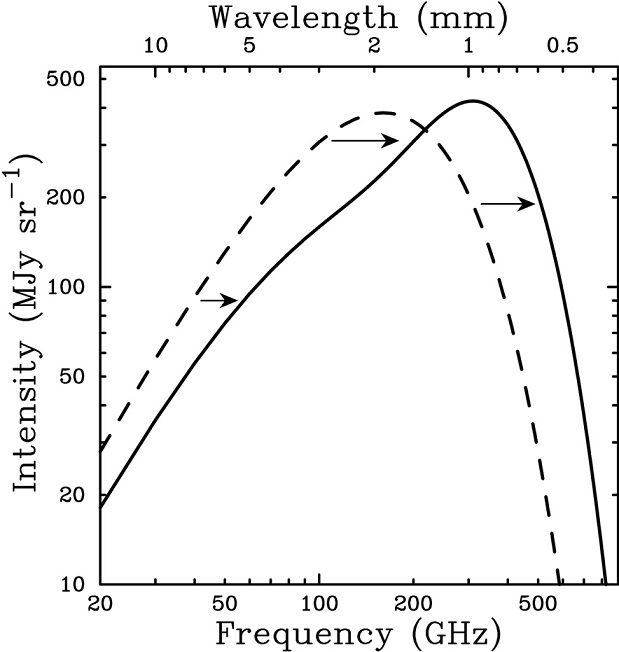}
\caption[Sunyaev-Zel'dovich effect]{The CMB spectrum, undistorted (dashed line) and distorted by the SZ effect. The SZ distortion is shown for a fictional cluster 1000 times more massive than an average cluster in order to illustrate the effect \citep{Sunyaev1980, Zeldovich_review2002}.  
\label{figure:SZ_effect}}
\end{figure}

As CMB photons pass through massive clusters, there is around 1\% probability of interacting with a high-energy ICM electrons \citep{Birkinshaw1999}. This results in a boost of energy of the photons by $k_{\mathrm{B}} T_{\mathrm{e}} / m_{\mathrm{e}} c^{2}$, where $T_{e}$ and $m_{e}$, as before, are the temperature and the mass of the electrons correspondingly. This results in distortion of $\lesssim  1$ mK in the CMB spectrum (see figure \ref{figure:SZ_effect}). More accurately, the SZ effect spectral distortions can be expressed as follows \citep{Rephaeli1995}: 

\begin{equation}
\frac{\Delta T_{\rm SZ}}{T_{\rm CMB}}=f(x) \int n_{e} \frac{k_{B} T_{e}}{m_{e} c^{2}} \sigma_{T} d \ell,
\label{eq:SZ_effect}
\end{equation}

\noindent where $f(x)$ is a function of a dimensionless frequency $x = h\nu/k_{B}T_{\rm CMB}$, $n_{e}$ is the electron number density, $m_{e}$ is the associated mass and $\sigma_{T}$ is the Thomson cross-section. 

\section{Methods To Estimate Cluster Masses}

\subsection{Hydrostatic Equilibrium}

A key notion when it comes to measuring galaxy cluster masses is that of hydrostatic equilibrium. The hydrostatic equilibrium equation relates the pressure gradient with the gravitational force in a galaxy cluster. Under spherical symmetry, the hydrostatic equilibrium equation is given by \citep{Terukina2013, Wilcox2015}: 

\begin{equation}
\frac{d P_{\mathrm{\rm total}}}{d r}=-\rho_{\rm gas} \frac{G M(<r)}{r^{2}},
\label{eq:hydrostatic_equilibrium}
\end{equation}

\noindent where $P_{\rm total}$ and $\rho_{\rm gas}$ are the pressure and the density of the gas and $r$ is the radial coordinate. $P_{\rm total}$ refers to the total gas pressure, including the thermal and non-thermal pressure contributions: $P_{\rm total} = P_{\rm thermal} + P_{\rm non-thermal}$ (discussed in more detail in chapter \ref{ch:modified_gravity}). Similarly, $M(<r)$ is the mass enclosed in radius $r$ and can also be split into the two contributions \citep{Lagan2010}: 

\begin{equation}
M(<r) = -\frac{r^{2}}{G \rho_{\rm gas}(r)} \Big( \frac{d P_{\text {thermal }}(r)}{d r} + \frac{d P_{\mathrm{non}-\mathrm{thermal}}(r)}{d r} \Big).
\label{eq:cluster_mass}
\end{equation}

Intuitively, equation \ref{eq:hydrostatic_equilibrium} represents a balance between the pressure and the gravitational force in a galaxy cluster. This indicates that the galaxy cluster is not undergoing formation processes or is not taking part in a merger. 

Using the equation of state for gas with a number density $n_{\rm gas}$ and temperature $T_{gas}$: $P_{\text {thermal }}=k_{B} n_{\mathrm{gas}} T_{\mathrm{gas}}$, one can rewrite the thermal mass component in a more useful form:

\begin{equation}
M_{\text {thermal}}(r)=-\frac{k_{B} T_{\text {gas}}(r) r}{\mu m_{\mathrm{p}} G}\left(\frac{d \ln \rho_{\rm gas}(r)}{d \ln r}+\frac{d \ln T_{\text {gas}}(r)}{d \ln r}\right).
\label{eq:thermal_mass}
\end{equation}

\noindent Here the identity $\rho_{gas} = \mu m_{p} n_{gas}$ was used, with $\mu$ as the mean molecular weight and $m_{p}$ as the proton mass. The mean molecular weight for the fully ionised gas is given by $\mu(n_{\rm e} + n_{\rm H} + n_{\rm He})m_{\rm p} = m_{\mathrm{p}} n_{\mathrm{H}}+4 m_{\mathrm{p}} n_{\mathrm{He}}$, where $n_{\mathrm{H}}$ and $n_{\mathrm{He}}$ are the number densities of hydrogen and helium correspondingly \citep{Ettori2013}.  

The fraction of the non-thermal contribution to the total pressure is given by \citep{Shaw2012}: 

\begin{equation}
P_{\text {non-thermal }}(r) = 
\alpha_{\mathrm{nt}}(1+z)^{\beta_{\mathrm{nt}}}\left(\frac{r}{r_{500}}\right)^{n_{\mathrm{nt}}}\left(\frac{M_{200}}{3 \times 10^{14} M_{\odot}}\right)^{n_{\mathrm{M}}} P_{\text {total}}(r),
\label{eq:non_thermal_pressure}
\end{equation}

\noindent where $\alpha_{\mathrm{nt}}, \beta_{\mathrm{nt}}, n_{\mathrm{nt}}$ and $n_{\mathrm{M}}$ are parameters determined by hydrodynamical simulations. The $r_{500}$ parameter corresponds to the radius at which the dark matter halo average density is equal to five hundred times the critical density. Analogously, $M_{200}$ corresponds to the mass at $r_{200}$.

Equation \ref{eq:hydrostatic_equilibrium} is of key significance as it allows us to relate the mass distribution in galaxy clusters to the observed pressure/temperature distribution, which can be inferred via X-ray surveys. Such a way of measuring masses is based on two key assumptions: that the majority of observed clusters are in fact in hydrostatic equilibrium and that most clusters are on average spherical. In general, effects of non-spherical geometries of clusters can be averaged out by stacking a large numbers of clusters. The hydrostatic equilibrium equation can also be tested by comparing independent measurements of cluster masses. More concretely, one can compare the X-ray determined masses to those deduced by weak lensing. As an example, recent measurements by \citet{Smith2016} indicate that the mean ratio of X-ray to lensing masses for 50 LoCuSS clusters at $0.15<z<0.3$ is $\beta_{X} = 0.95 \pm 0.05$, hence showing no significant deviation from the hydrostatic equilibrium assumption. Note, however, that the results of such measurements strongly depend on the method and the dataset used. For instance, the results in \cite{Biffi2016} derived using simulated galaxy clusters show variations up to 10-20\% up to the virial radius.

At this stage it is important to discuss the various astrophysical effects that can lead to biases when estimating cluster masses. In particular, an important concept in the context of galaxy cluster formation is that of virialization. Theoretically, the cluster merging and formation processes cease once virialization is reached (i.e. when the forces acting on the cluster are in balance and the potential energy is twice the negative kinetic energy as described by the virial theorem). Many real-world clusters, however, are not fully virialized, with the inner regions being more relaxed than the outer parts of the cluster. In addition, there is an ongoing accretion of gas and dark matter in the outer parts of the cluster. These effects complicate the mass estimates using the SZ effect and the galaxy kinematics. In addition, these effects could introduce extra bias when constraining models of modified gravity using the methods described in chapters \ref{ch:modified_gravity} and \ref{ch:testing_EG}. However, it is important to note that the mentioned effects have been investigated in the previous studies and found to be subdominant when compared to the non-thermal pressure effects and the predicted deviation between the different mass estimates due to modified gravity (for a wider discussion see \cite{Wilcox2016, Rumbaugh2018,Walker2019}). Nonetheless, effects such as these are important to understand and to quantify using both observational data and simulations. Further analysis of some of these systematics is given in chapters \ref{ch:modified_gravity} and \ref{ch:testing_EG}.

\subsection{Galaxy kinematics}
Another important method of estimating cluster masses uses the kinematics of the member galaxies. Under the assumption of the virial theorem, one can express the mass as follows \citep{Borgani2006}:

\begin{equation}
M_{\rm kin}=\frac{\pi}{2} \frac{3 \sigma_{v}^{2} R_{V}}{G},
\label{eq:kinematic_mass}
\end{equation}

\noindent where $\sigma_{v}$ is the line-of-sight velocity dispersion and $R_{V}$ is the viral radius. The virial radius can be estimated if a sufficient sample of member galaxies is available: 

\begin{equation}
R_{V}=N^{2}\left(\sum_{i>j} r_{i j}^{-1}\right)^{-1},
\label{eq:virial_radius}
\end{equation}

\noindent with $N$ as the number of galaxies and $r_{ij}$ as the separation between the $i$-th and the $j$-th galaxies.

Such a method of determining cluster mass comes with a set of challenges. Namely, as previously discussed, the assumption of the virial theorem can be valid to varying degrees in different populations of galaxies (for instance late and early types of galaxies). Another challenge in the context of observational data, comes in terms of non-member (foreground or background) galaxies that can bias the mass measurements. Algorithms for filtering out such interloper galaxies are of special importance for accurate mass estimates \citep{Girardi1993, Van_haarlem1997}. 

A key issue when it comes to such virial theorem-based approaches is not knowing the full underlying dark matter distribution. Such approaches are based on the assumption that, in general, dark matter follows the visible mass distribution. However, if one were to relax this assumption, the under/over-estimation of the cluster mass is given by \citep{Sadat1997}: 

\begin{equation}
\mu_{cl}=\frac{\left[1+2 C_{c} R_{\rm true}+C_{\lambda} R_{\rm true}^{2}\right]}{\left[1+C_{v} R_{\rm true}\right]},
\label{eq:cluster_mass_bias}
\end{equation}

\noindent where $R_{\rm true}$ is the true ratio of the dark matter and galaxy masses, while the $C_{c}, C_{\lambda}, C_{v}$ are the relative concentration parameters. 

A related method of determining the underlying mass distribution in clusters is via the orbits of member galaxies. In particular, assuming hydrostatic equilibrium, one can show that the mass distribution is given by: 

\begin{equation}
M(<r)=\frac{-G n_{\mathrm{gal}}(r)}{r^{2}}\left[\frac{d n_{\mathrm{gal}}(r)}{d r} \sigma_{r}(r)^{2}+\frac{2 n_{\mathrm{gal}}(r)}{r}\left[\sigma_{r}(r)^{2}-\sigma_{t}(r)^{2}\right]\right],
\label{eq:cluster_mass_orbits}
\end{equation}

\noindent where $n_{gal}$ is the number density of galaxies, and the dispersion parameters $\sigma_{t}$ and $\sigma_{r}$ control the shape of the orbit ($\sigma_{t} = \sigma_{r}$ for isotropic orbits). The drawback of this particular method is that the velocity dispersion profiles are not well known, leading to bias in the mass estimates \citep{Wojtak2010}.

\section{Weak Lensing in Galaxy Clusters}
\label{section:cluster_weak_lensing}
\subsection{The Basics of Gravitational Lensing}

One of the most important astronomical probes used to track the underlying cluster mass distribution is gravitational lensing. Gravitational lensing effect refers to the bending of light by large distributions of matter. The existence of such an effect was known long before the development of GR. In fact, if one allows the possibility of light having even a minuscule mass, Newtonian physics predicts bending of light rays by massive bodies \citep{Yajnik2019}. As later shown and calculated by the German astronomer Johann Georg von Soldner, the deflection angle of light due to a massive body is proportional to the gradient of the gravitational potential \citep{Gine2008}. This turned out to be a surprisingly accurate prediction that agreed with the initial calculations of Einstein as of 1911. Only in the final version of GR in 1915 Einstein managed to obtain the correct result, which was equal to twice the predicted Newtonian value. More specifically, Einstein predicted a deflection of 1.7 arc seconds for light passing the Sun \citep{Einstein1916}.  

Here we will lay out some of the equations for a single lens system as well as extended mass distributions e.g. galaxy clusters. The derivations are based primarily on \citet{Wright1999} and \citet{Bartelmann2017}. 

To estimate the deflection angle our starting point is assuming that the Newtonian potential in cosmological systems is small, i.e. $|\Psi| / c^{2} \ll 1$. In addition, the peculiar velocities that mass distributions have on cosmological scales are relatively small. These assumptions allow us to describe gravitational lensing by using the perturbed Minkowski metric: 

\begin{equation}
\mathrm{d} s^{2}=-c^{2}\left(1+\frac{2 \Psi}{c^{2}}\right) \mathrm{d} t^{2}+\left(1-\frac{2 \Phi}{c^{2}}\right) \mathrm{d} \vec{x}^{2}.
\label{eq:perturbed_minkowski}
\end{equation}

\noindent For propagating light $ds = 0$, which gives: 

\begin{equation}
c^{\prime}=\left|\frac{\mathrm{d} \vec{x}}{\mathrm{d} t}\right|=c\left(1+\frac{2 \Phi_{L}}{c^{2}}\right),
\label{eq:effective_light_speed}
\end{equation}

\noindent where $c^{\prime}$ is the effective light speed (note that the gravitational potential is negative) and $\Phi_{L} = (\Psi + \Phi)/2$. Note that in GR (in the absence of anisotropic stress) $\Phi = \Psi$, however, this is not generally the case in modified gravity models. To be consistent with the analysis in the later parts of this chapter, we are using a more general notation here. Also note here that the form of equation \ref{eq:effective_light_speed} allows us to define the refraction index in the usual manner: $c^{\prime} = c/n$, giving $n = 1 - 2\Phi_{L}/c^{2}$. Here, by analogy, we are treating a spacetime region with a gravitational potential $\Phi_{L}$ present as a material of refractive index $n$. One can then apply Fermat's principle, which states that the path taken by light rays minimizes the time of travel, i.e.:

\begin{equation}
\delta \tau=\delta \int_{A}^{B} \frac{c}{n} \mathrm{d} t=0,
\label{eq:fermats_principle}
\end{equation}

\noindent where $\tau$ is the path of the photon, $A$ and $B$ are the initial and final points and $\delta$ stands for a variation. Varying equation \ref{eq:fermats_principle} w.r.t. the light path leads to the identity for the deflection angle: 

\begin{equation}
\hat{\vec{\alpha}}=-\frac{2}{c^{2}} \int \vec{\nabla}_{\perp} \Phi_{L} \mathrm{d} l,
\label{eq:deflection_angle}
\end{equation}

\noindent where the gradient is taken perpendicular to the line of sight. Note that this calculation gives the correct result as predicted by GR and is twice larger than the corresponding Newtonian result. 

\begin{figure}[ht!]
\centering
\includegraphics[width=0.70\columnwidth]{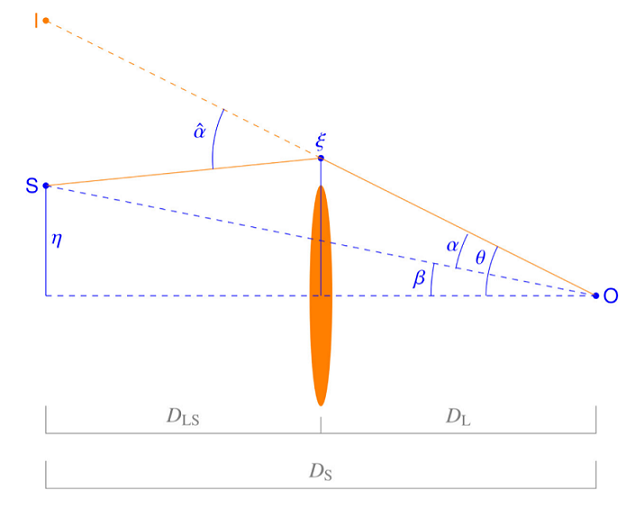}
\caption[The geometry of gravitational lensing]{The geometry of gravitational lensing. The image of a source $S$ as observed by an observer $O$ is displaced by the deflection angle $\alpha$ due to the gravitational potential of the lensing mass in the centre \citep{Bartelmann2017}.   
\label{figure:weak_lensing_geometry}}
\end{figure}

The integral in equation \ref{eq:deflection_angle} is not easy to evaluate, however a simplified result can be obtained by using the Born approximation. In particular, for small deflection angles (of the order arc seconds or smaller), the integration path can be approximated by straight lines: 

\begin{equation}
\hat{\alpha}=-\frac{2}{c^{2}} \frac{\partial}{\partial b} \int_{-\infty}^{\infty} \mathrm{d} z \frac{G M}{\sqrt{b^{2}+z^{2}}}=\frac{4 G M}{b c^{2}}=\frac{2 R_{\mathrm{S}}}{b},
\label{eq:born_approximation}
\end{equation}

\noindent where the deflection angle was calculated for a point mass $M$ at the origin with the light ray propagating parallel to the $z$ axis with an impact parameter $b$. $R_{s}$ here refers to the Schwarzschild radius. For the Sun, $M_{\odot} \approx 2 \cdot 10^{33}$ g, resulting in a deflection angle of $\hat{\alpha}_{\odot} \approx 8.6 \cdot 10^{-6} \approx 1.7^{\prime \prime}$ at the solar radius $R_{\odot}=7 \cdot 10^{5}$ km. This is the famous result confirmed by the observational data collected by Dyson, Eddington and Davidson in 1919 \citep{Dyson1920}. 

More complicated mass distributions require a more complex treatment. However, if the lensing mass distribution is thin compared to the distances in the lens system, the light ray paths between the source, the lens and the observer can be approximated as straight lines as shown in figure \ref{figure:weak_lensing_geometry}. This is known as the \textit{thin-lens approximation} and is sufficient to describe the basic lensing properties of isolated masses such as galaxy clusters. 

Following the geometry in figure \ref{figure:weak_lensing_geometry} one can define the reduced deflection angle $\vec{\alpha}$:

\begin{equation}
\vec{\alpha}\equiv\frac{D_{\mathrm{LS}}}{D_{\mathrm{S}}} \hat{\vec{\alpha}},
\label{eq:reduced_angle}
\end{equation}

\noindent with $D_{LS}$ and $D_{S}$ as the (angular diameter) distance between the lens and the source and the distance to the source correspondingly. This allows us to relate the angles shown in figure \ref{figure:weak_lensing_geometry} in the following way: $\vec{\beta}=\vec{\theta}-\vec{\alpha}$. Equation \ref{eq:reduced_angle} can also be expressed as:

\begin{equation}
\vec{\alpha}=\vec{\nabla}_{\perp}\left[\frac{2}{c^{2}} \frac{D_{\mathrm{LS}}}{D_{\mathrm{S}}} \int \Phi_{L} \mathrm{d} z\right],
\label{eq:reduced_angle2}
\end{equation}

\noindent where equation \ref{eq:deflection_angle} was used and the  $\vec{\nabla}_{\perp}$ refers to the perpendicular gradient. The perpendicular gradient can be replaced with the angular gradient w.r.t. angle $\theta$: $\vec{\nabla}_{\perp}=D_{\mathrm{L}}^{-1} \vec{\nabla}_{\theta}$. This finally allows writing $\vec{\alpha}$ in terms of the quantity $\psi$, which refers to the lensing potential: $\vec{\alpha}=\vec{\nabla}_{\theta} \psi$, with:

\begin{equation}
\psi \equiv \frac{2}{c^{2}} \frac{D_{\mathrm{LS}}}{D_{\mathrm{L}} D_{\mathrm{S}}} \int \Phi_{L} \mathrm{d} z.
\label{eq:lensing_potential}
\end{equation}

\noindent The lensing potential captures the key imaging properties of a gravitational lens.

Given equation \ref{eq:lensing_potential}, one can define two quantities: $\kappa$ (convergence) and $\gamma$ (shear), such that: 

\begin{equation}
\kappa(\vec{\theta})=\frac{1}{2}\left(\frac{\partial^{2} \psi}{\partial \theta_{1}^{2}}+\frac{\partial^{2} \psi}{\partial \theta_{2}^{2}}\right),
\label{eq:convergence}
\end{equation}

\begin{equation}
\gamma_{1}(\vec{\theta})=\frac{1}{2}\left(\frac{\partial^{2} \psi}{\partial \theta_{1}^{2}}-\frac{\partial^{2} \psi}{\partial \theta_{2}^{2}}\right),
\label{eq:shear1}
\end{equation}

\begin{equation}
\gamma_{2}(\vec{\theta})=\frac{\partial^{2} \psi}{\partial \theta_{1} \partial \theta_{2}}=\frac{\partial^{2} \psi}{\partial \theta_{2} \partial \theta_{1}}.
\label{eq:shear2}
\end{equation}

\begin{figure}[ht!]
\centering
\includegraphics[width=0.75\columnwidth]{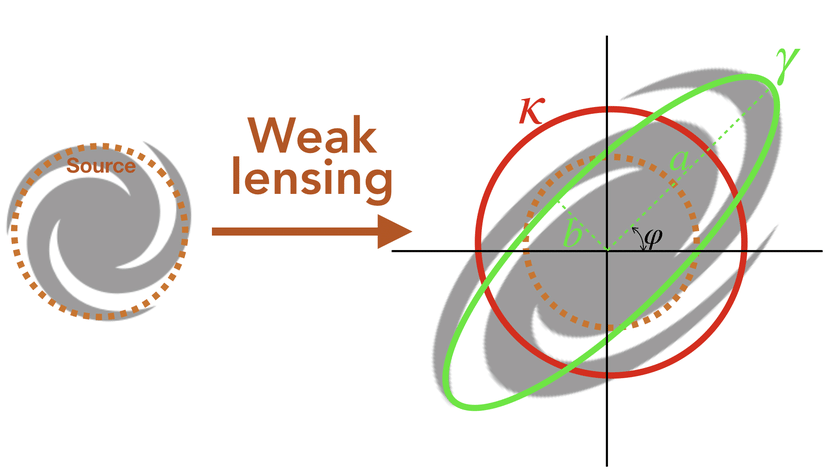}
\caption[The weak lensing effects on an image of a background galaxy]{The weak lensing effects on an image of a background galaxy as quantified by convergence $\kappa$ and shear $\gamma$. Image from \citet{Shuntov2019}.  
\label{figure:weak_lensing_effects}}
\end{figure}

\noindent The magnitude of shear is then simply given by: $\gamma=|\gamma|=\left(\gamma_{1}^{2}+\gamma_{2}^{2}\right)^{1 / 2}$. Figure \ref{figure:weak_lensing_effects} illustrates how a background source is deformed due to weak lensing and how these effects are related to quantities $\kappa$ and $\gamma$. More specifically, ellipticity of a deformed source galaxy can be defined as:

\begin{equation}
\varepsilon \equiv \frac{a-b}{a+b}=\frac{\gamma}{1-\kappa},
\label{eq:ellipticity}
\end{equation}

\noindent where $a$ and $b$ are the semi-major and semi-minor axes as illustrated in figure \ref{figure:weak_lensing_effects}. For most weak lensing systems $\kappa \ll 1$, resulting in $\varepsilon \approx \gamma$. 

An archetypal example of a weak lensing system is that of galaxy clusters distorting the shapes of the background galaxies. More specifically, galaxy clusters imprint a coherent distortion pattern onto the distant background galaxies as measured by their ellipticities. Hence statistically studying the distortions of the ellipticity of the background galaxies allows us to ultimately deduce the mass distribution of the lens cluster. Such measurements, however, are highly complicated by the fact that the background galaxies have an intrinsic ellipticity $\varepsilon_{\mathrm{S}}$, which is generally not known. A crucial assumption taken in weak lensing studies is that the average intrinsic ellipticity, for a sufficiently large sample of galaxies, is expected to be: $\left\langle\varepsilon_{\mathrm{S}}\right\rangle \approx 0$ \citep{Hirata2007}. As discussed in \citet{Bartelmann2017}, the standard deviation of the intrinsic ellipticity is measured to be $\sigma_{\varepsilon} \approx 0.2$ and averaging over $N$ faint galaxy images reduces the scatter of the intrinsic ellipticity to:

\begin{equation}
\Delta\left\langle\varepsilon_{\mathrm{S}}\right\rangle \approx \frac{\sigma_{\varepsilon}}{\sqrt{N}}.
\label{eq:ellipticity_scatter}
\end{equation}

\noindent As illustrated by the description above, the distortion of the ellipticities of the background sources by a foreground lens mass ultimately depends on the lensing mass distribution. Hence, studying the weak lensing effects we can infer the underlying distribution of the lensing system. In fact, weak lensing offers one of the most powerful probes for studying the properties of galaxy clusters. 

\subsection{Weak Lensing by NFW Halos}

Recent observational evidence along with evidence from numerical simulations strongly supports the idea that there is a universal density profile for dark matter haloes \citep{Navaro1997, Bartlemann1998,Young2016}. In fact, the mentioned evidence shows that systems ranging from globular clusters to large galaxy clusters can be described by the same universal density profile. That, of course, refers to the well-studied Navarro, Frenk, White (NFW) profile given by \citep{Navarro1996}:

\begin{equation}
\rho(r)=\frac{\delta_{c} \rho_{c}}{\left(r / r_{s}\right)\left(1+r / r_{s}\right)^{2}},
\label{eq:NFW_profile}
\end{equation}

\noindent where $\rho_{c}=(3 H^{2}(z)) /(8 \pi G)$ is the critical density, and $r_{s} = r_{200}/c_{v}$ is the scale radius. The virial radius term $r_{200}$ refers to the radius inside which the mass density of the halo is equal to $200\rho_{c}$. Here $\delta_{c}$ refers to the characteristic overdensity of the halo, given by:

\begin{equation}
\delta_{c}=\frac{200}{3} \frac{c_{v}^{3}}{\ln (1+c_{v})-c_{v} /(1+c_{v})}.
\label{eq:characteristic_overdensity}
\end{equation}

\noindent Equation \ref{eq:NFW_profile} describes the underlying dark matter distribution in galaxy clusters. Hence, assuming the NFW profile, one can derive an analytic expression for the radial dependence of the convergence and the shear due to the dark matter halos in galaxy clusters \citep{Wright1999}. 

The local value of convergence can be described by:

\begin{equation}
\kappa(\vec{\theta})=\frac{\Sigma(\vec{\theta})}{\Sigma_{c}},
\label{eq:convergence2}
\end{equation}

\noindent where $\Sigma(\vec{\theta})$ refers to the surface mass density and $\Sigma_{c}$ is the critical surface mass density given by:

\begin{equation}
\Sigma_{c} \equiv \frac{c^{2}}{4 \pi G} \frac{D_{S}}{D_{L} D_{LS}}.
\label{eq:critical_surface_density}
\end{equation}

\noindent Assuming spherical symmetry, the surface mass density is simply given by: 

\begin{equation}
\Sigma(R)=2 \int_{0}^{\infty} \rho(R, z) d z,
\label{eq:surface_mass_density_rho}
\end{equation}

\noindent where the integral is evaluated over the coordinate $z$ along the line of sight and $R=D_{L}\left(\theta_{1}^{2}+\theta_{2}^{2}\right)^{1 / 2}$ is the projected radius relative to the center of the lens. Equation \ref{eq:NFW_profile} can then be integrated along the line of sight to give: 

\begin{equation}
    \Sigma_{\rm NFW}(x) = \begin{cases} 
                                    \frac{2 r_{s} \delta_{c} \rho_{c}}{\left(x^{2}-1\right)}\left[1-\frac{2}{\sqrt{1-x^{2}}} \operatorname{arctanh} \sqrt{\frac{1-x}{1+x}}\right] & (x < 1) \\ 
                                    
                                    \frac{2 r_{s} \delta_{c} \rho_{c}}{3} & (x = 1)\\ 
                                    
                                    \frac{2 r_{s} \delta_{c} \rho_{c}}{\left(x^{2}-1\right)}\left[1-\frac{2}{\sqrt{x^{2}-1}} \arctan \sqrt{\frac{x-1}{1+x}}\right]  & (x > 1),
                      \end{cases}
                      \label{eq:nfw_simga}
\end{equation}

\noindent where a dimensionless radial distance was defined as $x = R/r_{s}$. The tangential shear for an NFW density distribution is then given by:

\begin{equation}
\gamma_{\mathrm{NFW}}(x)=\frac{\bar{\Sigma}_{\mathrm{NFW}}(x)-\Sigma_{\mathrm{NFW}}(x)}{\Sigma_{c}},
\label{eq:nfw_shear}
\end{equation}

\noindent where $\bar{\Sigma}_{\mathrm{NFW}}(x)$ refers to the mean surface density inside radius $x$. More specifically, the mean surface density is given by the following integral: 

\begin{equation}
\bar{\Sigma}_{\mathrm{NFW}}(x)=\frac{2}{x^{2}} \int_{0}^{x} x^{\prime} \Sigma_{\mathrm{NFW}}\left(x^{\prime}\right) d x^{\prime}.
\label{eq:sigma_nfw_mean}
\end{equation}

\noindent Putting everything together and evaluting the integrals gives the following result:

\begin{equation}
    \gamma_{\rm NFW}(x) = \begin{cases} 
                                    \frac{r_{s}\delta_{c}\rho_{c}}{\Sigma_{c}} g_{<}(x) & (x < 1) \\ 
                                    \frac{r_{s}\delta_{c}\rho_{c}}{\Sigma_{c}} \Big( \frac{10}{3} + 4\ln{(\frac{1}{2})} \Big) & (x = 1)\\ 
                                    \frac{r_{s}\delta_{c}\rho_{c}}{\Sigma_{c}} g_{>}(x)  & (x > 1),
                      \end{cases}
                      \label{eq:nfw_shear_final}
\end{equation}

\noindent where the two functions $g_{<}(x)$ and $g_{>}(x)$ were defined for convenience. The two functions are given explicitly by:

\begin{equation}
\centering
\begin{split}
    g_{<}(x) ={} & \frac{8 \arctan(\sqrt{(1-x)/(1+x)})}{x^{2}\sqrt{1-x^2}} + \frac{4}{x^2} \ln{(x/2)} \\
    & -\frac{2}{(x^{2}-1)} + \frac{4 \arctan(\sqrt{(1-x)/(1+x)})}{(x^{2} -1)(1-x^{2})^{1/2}     }, 
\label{eq:g_<}
\end{split}
\end{equation}

\begin{equation}
\centering
\begin{split}
    g_{>}(x) ={} & \frac{8 \arctan(\sqrt{(1-x)/(1+x)})}{x^{2}\sqrt{x^2-1}} + \frac{4}{x^2} \ln{(x/2)} \\
    & - \frac{2}{(x^{2}-1)} + \frac{4 \arctan(\sqrt{(1-x)/(1+x)})}{(x^{2}-1)^{3/2}     }.
\label{eq:g_>}
\end{split}
\end{equation}

In summary, the concepts introduced in this chapter clearly illustrate how various observational probes can be used to measure the underlying mass distribution of galaxy clusters. In addition, astrophysics in galaxy clusters is shown to be tightly related to the properties of dark matter and the underlying model of gravity. For this reason galaxy clusters have been extremely important in testing models of modified gravity on cosmological scales. Chapter \ref{ch:modified_gravity} delves deeper into the effects of modified gravity on galaxy clusters. In addition, a technique for testing chameleon gravity using combined X-ray and weak lensing data from stacked galaxy clusters is introduced.

\chapter{Modified Gravity on Galaxy Cluster Scales}
\label{ch:modified_gravity}

This chapter introduces chameleon and $f(R)$ gravity models along with an effective technique for testing modified gravity on galaxy cluster scales. More specifically, the chapter starts by introducing the relationship between the scalar-tensor models and $f(R)$ gravity. In addition, a technique of testing models of modified gravity with chameleon gravity using cluster X-ray and weak lensing data based on the previous work in \citet{Terukina2013} and \citet{Wilcox2015} is introduced. Original results reproducing the tests described in \citet{Wilcox2015} with an updated dataset are presented. Finally, the implications of the results for model-independent tests of gravity are discussed in the last section of the chapter. The original results presented in this chapter were produced in collaboration with Carlos Vergara and Kathy Romer, as described in \citet{Vergara2019}. 

\section{Scalar-Tensor Gravity with Chameleon Screening}
\subsection{The Action}

As discussed in chapter \ref{ch:introduction}, modified gravity models offer a novel approach in tackling some of the key issues in modern cosmology. However, a major shortcoming of such modified gravity approaches comes in the context of the stringent observational constraints in the solar system. In this respect, models with different types of screening mechanisms are of special importance as they can avoid the rigid solar system constraints while still possessing a cosmologically interesting phenomenology. A natural question to ask, however, is how natural and fine-tuned such models are? Undeniably, most models that allow screening behaviour are fine-tuned to turn off the fifth force on the scales of the solar system to avoid the strict constraints. However, it should be noted that similar screening behaviour can be observed in various scenarios in electromagnetism and hence it is not entirely unnatural to expect a scalar field to posses screening. Here the key features of the scalar-tensor models with chameleon screening are summarized based primarily on \citet{Khoury2004, Waterhouse2006, Burrage2018}.

The chameleon model can be described by introducing a scalar field $\phi$ with a potential $V(\phi)$. The dynamics of the theory can then be captured by the action, which, as usual, refers to a functional that, when varied w.r.t. the metric and the scalar field, gives the set of equations of motion. In this case the action for a scalar field $\phi$ is given by:

\begin{equation}
S_{\phi}=-\int d^{4} x \sqrt{-g}\left\{\frac{1}{2}(\partial \phi)^{2}+V(\phi)\right\}.
\label{eq:chameleon_action}
\end{equation}

\noindent This can be combined with the standard Einstein-Hilbert action with a term describing the matter fields $\psi_{m}$ (equation \ref{eq:EH_action}) giving the following combined action: 

\begin{equation}
\begin{gathered}
    S = S_{EH} + S_{\phi} + S_{m} = \\
     =  \int d^{4} x \sqrt{-g}\left\{\frac{M_{\mathrm{pl}}^{2}}{2} R-\frac{1}{2} \nabla_{\mu} \phi \nabla^{\mu} \phi-V(\phi)-\frac{1}{\sqrt{-g}} \mathcal{L}_{\mathrm{m}}\left(\psi_{\mathrm{m}}^{(i)}, g_{\mu \nu}^{(i)}\right)\right\}.
\end{gathered}
\label{eq:chameleon_action_full}
\end{equation}

\noindent Note that the last term is generalized to allow multiple matter species, while $g_{\mu \nu}^{(i)}$ refers to the Jordan frame metric, that is conformally related to the Einstein frame metric $g_{\mu \nu}$ by:

\begin{equation}
g_{\mu \nu}^{(i)} \equiv e^{2 \beta_{i} \phi / M_{\rm pl}} g_{\mu \nu}. 
\label{eq:conformal_metrics}
\end{equation}

\noindent Note that here we allow for different coupling constants $\beta_{i}$ for different matter species. Jordan and Einstein frames refer to the two different ways of expressing the scalar-tensor action. In particular, in the Jordan frame the scalar field (or some function of it) is multiplied by the Ricci scalar, while in the Einstein frame it is not. More formally, the Jordan frame refers to the frame in which the matter is minimally coupled to the metric. The equations appearing in this section can be translated between the different frames by using the conformal transformation defined in equation \ref{eq:conformal_metrics}. Also, it is important to note that the Jordan frame metric $g_{\mu \nu}^{(i)}$ is the metric that the matter experiences. 

As usual, varying the action w.r.t. $\phi$ allows us to obtain the equations of motion for the scalar field:

\begin{equation}
    \delta S =\int d^{4} x \sqrt{-g}\left\{\nabla^{2} \phi-V_{, \phi}(\phi)-\sum_{i} \frac{1}{\sqrt{-g}} \frac{\partial \mathcal{L}_{m}}{\partial g_{\mu \nu}^{(i)}} \frac{2 \beta_{i}}{M_{\mathrm{pl}}} g_{\mu \nu}^{(i)}\right\} \delta \phi = 0. 
\label{eq:action_variation}
\end{equation}

\noindent The terms in the brackets give the equation of motion for the field $\phi$. The last term can be expressed more explicitly by noting that energy density for matter species $i$ in the Einstein frame is given by:

\begin{equation}
\rho_{i} = e^{-\left(1-3 w_{i}\right) \beta_{i} \phi / M_{\mathrm{pl}}} \frac{1}{1-3 w_{i}} \frac{2}{\sqrt{-g}} \frac{\partial \mathcal{L}_{\mathrm{m}}}{\partial g_{\mu \nu}^{(i)}} g_{\mu \nu}^{(i)},
\label{eq:Einstein_frame_density}
\end{equation}

\noindent where $w_{i}$ relates the pressure and the density in the equation of state and the index $i$ refers to the $i$-th species of matter as before. Using this expression the equation of motion can then be written as:

\begin{equation}
\nabla^{2} \phi=V_{, \phi}(\phi)+\sum_{i}\left(1-3 w_{i}\right) \frac{\beta_{i}}{M_{\mathrm{pl}}} \rho_{i} e^{\left(1-3 w_{i}\right) \beta_{i} \phi / M_{\mathrm{pl} }}.
\label{eq:eom_chameleon}
\end{equation}

\noindent The shape of equation \ref{eq:eom_chameleon} allows us to conveniently define an effective potential $V_{\rm eff}(\phi)$, such that:

\begin{equation}
V_{\mathrm{eff}}(\phi) \equiv V(\phi)+\sum_{i} \rho_{i} e^{\left(1-3 w_{i}\right) \beta_{i} \phi / M_{\rm pl}}.
\end{equation}

\noindent The equation of motion can then be written succinctly:

\begin{equation}
\nabla^{2} \phi=V_{\mathrm{eff}, \phi}(\phi).
\label{eq:eom_effective}
\end{equation}

\subsection{Properties of the Effective Potential}

The behaviour of the chameleon field can be controlled by choosing a particular form of the bare potential $V(\phi)$. As discussed in \citet{Waterhouse2006}, as a starting point, one might choose a potential such that it can give rise to cosmic acceleration via slow roll. In addition, we also want it to have the screened behaviour, such that the fifth force effects are suppressed in high density regions. It is important to note, however, that there are certain no-go theorems that prohibit scalar-tensor models, which possess both a screening mechanism and self-acceleration \citep{Wang2012B}. In other words, if our theory possesses a screening mechanism it will still require some form of dark energy to account for the accelerating expansion.  

In order to possess screening, the potential $V(\phi)$ has to be continuous and bounded from below, while also strictly decreasing. In addition, its first derivative $V_{,\phi}$ should be negative and increasing. The second derivative $V_{,\phi \phi}$ should be positive and decreasing. Finally, the potential should have the following behaviour for vanishing $\phi$ values: $\lim _{\phi \rightarrow 0} V(\phi)=\infty$.

The two often-used potentials possessing the outlined properties are an exponential potential of the form:

\begin{equation}
V(\phi)=M^{4} \exp \left(\frac{M^{n}}{\phi^{n}}\right),
\label{eq:exponential_potential}
\end{equation}

\noindent and the inverse power-law potential of the form:

\begin{equation}
V(\phi)=\frac{M^{4+n}}{\phi^{n}}.
\label{eq:ratra_peebles_potential}
\end{equation}

\noindent $M$ here refers to a constant with a dimension of mass while $n$ is a positive constant. 

The effective potential has an important feature such that if the coupling $\beta_{i}$ is positive, there exists a minimum at $\phi = \phi_{min}$:

\begin{equation}
V_{, \phi}\left(\phi_{\min }\right)+\sum_{i}\left(1-3 w_{i}\right) \frac{\beta_{i}}{M_{\mathrm{pl}}} \rho_{i} e^{\left(1-3 w_{i}\right) \beta_{i} \phi_{\min } / M_{\mathrm{pl}}}=0.
\label{eq:eff_pot_minimum}
\end{equation}

\noindent In addition, one can define a mass $m$ associated with the field $\phi$:

\begin{equation}
m^{2} \equiv V_{\mathrm{eff}, \phi \phi}(\phi)=V_{, \phi \phi}(\phi)+\sum_{i}\left(1-3 w_{i}\right)^{2} \frac{\beta_{i}^{2}}{M_{\mathrm{pl}}^{2}} \rho_{i} e^{\left(1-3 w_{i}\right) \beta_{i} \phi / M_{\mathrm{pl} }}.
\label{eq:chameleon_mass}
\end{equation}

\begin{figure}[!ht]
\centering
\captionsetup[subfigure]{justification=centering}
  \begin{subfigure}[b]{0.475\textwidth}
    \includegraphics[width=\textwidth]{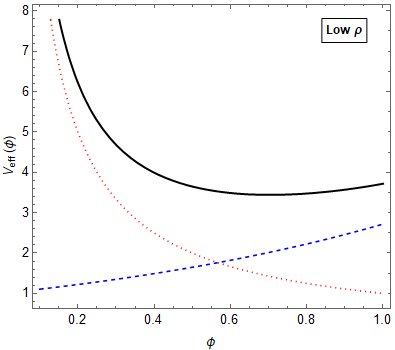}
    \label{hameleon_potential1}
  \end{subfigure}
  \begin{subfigure}[b]{0.49\textwidth}
    \includegraphics[width=\textwidth]{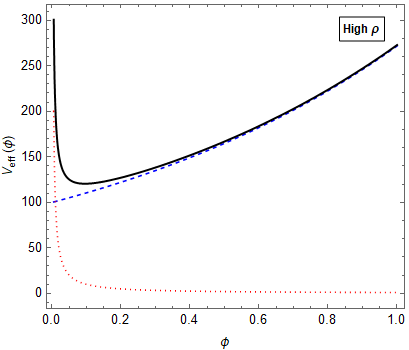}
    \label{chameleon_potential2}
  \end{subfigure}
  \caption[The effects of the local density on the shape of the chameleon potential]{The effects of a local mass density $\rho$ on the shape of the effective potential $V_{\rm eff}(\phi)$ with all the constants suppressed. The red dotted line corresponds to the $\phi^{-1}$ term. The dashed blue line corresponds to the $\rho e^{\phi}$ term. The black line corresponds to the sum of the two terms. The density is set to a numerical value of $\rho = 1$ in the figure on the left and $\rho = 100$ in the figure on the right. The figure is author's own. }
  \label{figure:chameleon_potential}
\end{figure}

\noindent Setting $\phi = \phi_{min}$ in equation \ref{eq:chameleon_mass}
gives $m^{2} = m^{2}_{min}$. The minimum mass $m_{min}$ is of special importance as it is equal to the inverse of the characteristic range of the chameleon force. Figure \ref{figure:chameleon_potential} illustrates the behaviour of the effective potential and the minimum mass for different values of the local density. More specifically, as illustrated by equation \ref{eq:chameleon_mass}, when the density $\rho_{i}$ increases, the minimum value $\phi_{min}$ decreases while the $m_{min}$ value increases. In other words, for larger density regions, such as the solar system, the characteristic range of the chameleon force becomes very short and hence the modified gravity effects are suppressed. This is true as $V_{,\phi}$ and $e^{\left(1-3 w_{i}\right) \beta_{i} \phi / M_{\mathrm{pl}}}$  are increasing functions of $\phi$, while $V_{,\phi \phi}$ is a decreasing function of $\phi$.

The interaction between the chameleon field and matter can be determined by the geodesic equation: 

\begin{equation}
\ddot{x}^{\rho}+\tilde{\Gamma}_{\mu \nu}^{\rho} \dot{x}^{\mu} \dot{x}^{\nu}=0,
\label{eq:jordan_frame_geodesic}
\end{equation}

\noindent where $\tilde{\Gamma}_{\mu \nu}^{\rho}$ is the Christoffel symbol corresponding to the Jordan frame metric ${g}^{(i)}_{\mu \nu}$, while the dot is the derivative w.r.t. the proper time $\tilde{\tau}$. Remembering that the Jordan frame metric is related to the Einstein frame metric via a conformal transformation (equation \ref{eq:conformal_metrics}), one can evaluate the needed metric derivatives: 

\begin{equation}
g^{(i)}_{\mu \nu, \sigma}=\left(\frac{2 \beta_{i}}{M_{\mathrm{pl}}} \phi_{, \sigma} g_{\mu \nu}+g_{\mu \nu, \sigma}\right) e^{2 \beta_{i} \phi / M_{\mathrm{pl}}}.
\label{eq:metric_derivatives}
\end{equation}

\noindent This expression can be used to evaluate the Jordan frame Christoffel symbols in terms of the Einstein frame metric. The geodesic equation can then be expressed as: 

\begin{equation}
\ddot{x}^{\rho}+\tilde{\Gamma}_{\mu \nu}^{\rho} \dot{x}^{\mu} \dot{x}^{\nu}=\ddot{x}^{\rho}+\Gamma_{\mu \nu}^{\rho} \dot{x}^{\mu} \dot{x}^{\nu}+\frac{\beta_{i}}{M_{\mathrm{pl}}}\left(2 \phi_{, \mu} \dot{x}^{\mu} \dot{x}^{\rho}+g^{\sigma \rho} \phi_{, \sigma}\right) = 0,
\end{equation}

\noindent where $\Gamma_{\mu \nu}^{\rho}$ corresponds to the Einstein frame Christoffel symbols. The second term in the equation above is the familiar gravitational term, while the last term corresponds to the force due to the chameleon field. More specifically, in the non-relativistic limit, a test mass of matter species $i$ experiences a force $F_{\phi}$, which can simply be expressed as:

\begin{equation}
\frac{\vec{F}_{\phi}}{m}=-\frac{\beta_{i}}{M_{\mathrm{pl}}} \vec{\nabla} \phi.
\label{eq:chameleon_force}
\end{equation}

\section{$f(R)$ Gravity}

Another type of modified gravity theory important in the context of this chapter is $f(R)$ gravity. This refers to a family of theories in which the Ricci scalar is replaced by a general function $f(R)$. Depending on the function used, such theories can modify GR in a way that accounts for the accelerated expansion and offers possible solutions to some of the other contemporary issues in cosmology. However, many functional forms are ruled out by theoretical arguments and observational constraints (e.g. see \citet{Felice2010,Jain2013,Dombriz2016}). Such theories can also contain a time and scale dependent gravitational constant and also exhibit massive gravitational waves. Finally, an important feature of $f(R)$ theories is that performing a certain conformal transformation allows us to write them in a form equivalent to scalar-tensor theories. This means that the observational constraints on scalar-tensor theories with chameleon screening can be converted to the equivalent constraints on a subset of $f(R)$ theories.

The action for $f(R)$ gravity takes the following form:

\begin{equation}
S_{f(R)}=\int \mathrm{d}^{4} x \sqrt{-\tilde{g}} \frac{M_{\mathrm{pl}}^{2}}{2} f(\tilde{R})+S_{\mathrm{matter}}\left[\psi,\tilde{g}_{\mu \nu}\right],
\label{eq:fR_action}
\end{equation}

\noindent where we simplified the notation for clarity by setting $g_{\mu \nu}^{(i)} = \tilde{g}_{\mu \nu}$. Here the tilde denotes quantities in the Jordan frame. Also note that now we assume a single matter species. The equation of motion in $f(R)$ theories takes the following form:

\begin{equation}\begin{aligned}
\tilde{R}_{\mu \nu} f^{\prime}(\tilde{R}) &-\frac{1}{2} f(\tilde{R}) \tilde{g}_{\mu \nu}=\frac{ T_{\mu \nu}^{\text {matter }}}{M_{\rm pl}^{2}} +\nabla_{\mu} \nabla_{\nu} f^{\prime}(\tilde{R})-\tilde{g}_{\mu \nu} \square f^{\prime}(\tilde{R}),
\end{aligned}\end{equation}

\noindent where $T_{\mu \nu}^{\text {matter }}$ is the energy-momentum tensor and $\square=\tilde{g}^{\mu \nu} \nabla_{\mu} \nabla_{\nu}$, while the prime symbol denotes a derivative w.r.t. $\tilde{R}$.

The action described above can be recast as a scalar-tensor theory by defining a scalar field $\phi$ as follows:

\begin{equation}
\exp \left(-\frac{2 \beta \phi}{M_{\mathrm{pl} }}\right)=f^{\prime}(R),
\label{eq:fR_transformation}
\end{equation}

\noindent where $\beta = \sqrt{1/6}$. Using equation \ref{eq:fR_transformation} and switching to Einstein frame (equation \ref{eq:conformal_metrics}) allows to rewrite the action \ref{eq:fR_action} in the exact same form as the scalar-tensor action in equation \ref{eq:chameleon_action_full}, if we express the potential as:

\begin{equation}V(\phi)=\frac{M_{\mathrm{pl}}^{2}\left(\tilde{R} f^{\prime}(\tilde{R})-f(\tilde{R})\right)}{2 f^{\prime}(\tilde{R})^{2}}.
\label{eq:fR_potential}
\end{equation}

\noindent In other words, a subset of $f(R)$ theories defined via equation \ref{eq:fR_transformation} are equivalent to a scalar-tensor theory with the potential defined above. The possibility of recasting the $f(R)$ model as a scalar-tensor model where a new scalar field sources the accelerated expansion (e.g. quintessence) also leads to a nuanced question of what is the difference between models of modified gravity and dark energy. As illustrated above, in the case of the $f(R)$ model, it can be interpreted as both a modified gravity and a dark energy model depending on the chosen frame. Similarly many other models, such as Brans-Dicke theory can be transformed to the familiar scalar-tensor form (eq. \ref{eq:chameleon_action_full}) by choosing a suitable conformal rescaling. It should be noted, however, that not every model can be recast in such a way. Also, there has been a long-running debate on whether the Einstein and the Jordan frames are equivalent or alternatively, which of the frames is the physical one (see \cite{Faraoni1999} for a more in-depth discussion).         
 
A concrete example of an $f(R)$ model with an interesting phenomenology is the Hu-Sawicki model, with:

\begin{equation}
f(\tilde{R})=-m_{1}^{2} \frac{c_{1}\left(\tilde{R}/ m_{1}^{2}\right)^{n}}{c_{2}\left(\tilde{R} / m_{1}^{2}\right)^{n}+1},
\label{eq:hu_sawicki}
\end{equation}

\noindent with $m_{1},n,c_{1}$ and $c_{2}$ as constants. The constants can be chosen such that the accelerated expansion can be accounted for while also mimicking the expansion history of the standard concordance model. 

More specifically, in the high curvature regime when $\tilde{R} \gg m_{1}^{2}$, equation \ref{eq:hu_sawicki} can be expanded \citep{Hu2007}:

\begin{equation}
\lim _{m_{1}^{2} / \tilde{R} \rightarrow 0} f(\tilde{R}) \approx-\frac{c_{1}}{c_{2}} m_{1}^{2}+\frac{c_{1}}{c_{2}^{2}} m_{1}^{2}\left(\frac{m_{1}^{2}}{\tilde{R}}\right)^{n}.
\label{eq:hu_sawicki_expansion}
\end{equation}

\noindent In the limiting case of $c_{1} / c_{2}^{2} \rightarrow 0$, the $c_{1}/c_{2}$ term acts as the cosmological constant. Furthermore, at finite $c_{1}/c_{2}^{2}$, the curvature freezes to a fixed value and stops declining with the matter density resulting in a class of models which accelerate in a manner similar to $\Lambda$CDM. Finally, the constants can be chosen such that the potential has a form that exhibits the chameleon mechanism as shown in figure \ref{figure:chameleon_potential}. Note, however, that the previous comments regarding the no-go theorems that prohibit models with both screening and self-acceleration apply here as well.   

\section{Testing Modified Gravity on Galaxy Cluster Scales}
\subsection{Non-Thermal Pressure and the Modified Hydrostatic Equilibrium Equation}

\label{section:hydrostatic_equilibrium}

Galaxy clusters, as discussed in the previous chapter, being among the largest gravitationally bound structures in the Universe with regions of high and low densities, offer a plethora of ways to test modifications of gravity. In this section, a specific approach first introduced in \citet{Terukina2013} and later extended in \citet{Wilcox2015} is discussed and summarized. In particular, it is an approach based on combining multiple galaxy cluster probes, such as X-ray surface brightness and weak lensing data, in order to constrain modifications of gravity predicted by chameleon scalar-tensor and the related $f(R)$ models. As discussed, a key feature of such models is the suppression of the fifth force effects in the high density regions. In the context of galaxy clusters, such suppression would manifest in the fifth force being screened in the dense cluster cores, but not in the outskirts of clusters. In the outskirts of clusters the intracluster gas would be affected by the usual force of gravity plus an additional fifth force, which would then result in the gas being slightly more compact than predicted by GR. This, in turn, would lead to a slightly higher temperature and the corresponding X-ray surface brightness. Analogously, the hydrostatic mass inferred using X-ray measurements would be affected as well. In contrast, the weak gravitational lensing profile is not affected in chameleon gravity models (discussed later in the section). Hence by comparing the X-ray and the weak lensing measurements, the fifth force effects can be constrained observationally. 

A key assumption when comparing the hydrostatic and the weak lensing masses is that of the hydrostatic equilibrium (equation \ref{eq:hydrostatic_equilibrium}). In particular, it describes the balance between the gas pressure gradient and the gravitational force in the cluster. The total pressure described in equation \ref{eq:hydrostatic_equilibrium} can be split into thermal and non-thermal contributions: $P_{\text {total }}=P_{\text {thermal }}+P_{\text {non-thermal}}$. The non-thermal pressure component here is related to a variety of effects such as the bulk motion and turbulence of the ICM gas along with the effects of the cosmic rays and magnetic fields. Such effects are important to account for when estimating the hydrostatic mass \citep{Lagan2010}. More specifically, observational evidence and hydrostatic simulations indicate a common trend of the non-thermal fraction increasing towards large radii and becoming comparable to the thermal pressure at around the virial radius \citep{Shi2014}. The hydrostatic equilibrium assumption then allows us to define the mass components corresponding to thermal and non-thermal pressure: $M(<r)=M_{\text {thermal }}(r)+M_{\text {non-thermal }}(r)$, where:

\begin{equation}
M_{\text {thermal }}(r) \equiv-\frac{r^{2}}{G \rho_{\text {gas }}(r)} \frac{d P_{\text {thermal }}(r)}{d r},
\label{eq:thermal_mass}
\end{equation}

\begin{equation}
M_{\text {non-thermal }}(r) \equiv-\frac{r^{2}}{G \rho_{\text {gas }}(r)} \frac{d P_{\text {non-thermal }}(r)}{d r}.
\label{eq:non_thermal_mass}
\end{equation}

\noindent The thermal mass component can be re-expressed in terms of the density and temperature distributions by using the equation of state: $P_{\text {thermal }}=k_{B} n_{\text {gas}} T_{\text {gas}}$ and $\rho_{\mathrm{gas}}=\mu m_{\mathrm{p}} n_{\mathrm{gas}}$:

\begin{equation}
M_{\text {thermal }}(r)=-\frac{k_{B} T_{\mathrm{gas}}(r) r}{\mu m_{\mathrm{p}} G}\left(\frac{d \ln \rho_{\mathrm{gas}}(r)}{d \ln r}+\frac{d \ln T_{\mathrm{gas}}(r)}{d \ln r}\right).
\label{eq:hydrostatic_mass}
\end{equation}

\noindent Here $\mu$ and $m_{p}$ refer to the mean molecular weight and the proton mass correspondingly. The mean molecular weight  for a fully ionised cluster gas can be defined as: $\mu(n_{e} + n_{H} + n_{He})m_{p} = m_{\mathrm{p}} n_{\mathrm{H}}+4 m_{\mathrm{p}} n_{\mathrm{He}}$ with $n_{e} = n_{H} + 2n_{He}$, where $n_{e}$, $n_{H}$, $n_{He}$ refer to the number density of the electrons, hydrogen and helium respectively \citep{Terukina2013}. Adopting the mass fraction of hydrogen of $n_{\mathrm{H}} /\left(n_{\mathrm{H}}+4 n_{\mathrm{He}}\right)=0.75$ leads to $\mu = 0.59$.

The non-thermal pressure effects can be redefined as a fraction $g(r)$ of the total pressure: 

\begin{equation}
    P_{\text {non-thermal }}(r) \equiv g(r) P_{\text {total }}(r).
    \label{eq:non_thermal_pressure2}
\end{equation}

\noindent And hence $P_{\text {total }}=g^{-1} P_{\text {non-thermal }}=(1-g)^{-1} P_{\text {thermal }}$, allowing us to write the non-thermal component as: 

\begin{equation}
    P_{\text {non-thermal }}(r)=\frac{g(r)}{1-g(r)} n_{\text {gas }}(r) k_{B} T_{\text {gas }}(r).
    \label{eq:non_thermal_pressure3}
\end{equation}

\noindent The functional shape of $g(r)$ has been studied using hydrodynamical simulations \citep{Shaw2010, Battaglia2012}. In particular, the cited works show that the non-thermal pressure fraction can be represented by:

\begin{equation}
g(r)=\alpha_{\mathrm{nt}}(1+z)^{\beta_{\mathrm{nt}}}\left(\frac{r}{r_{500}}\right)^{n_{\mathrm{nt}}}\left(\frac{M_{200}}{3 \times 10^{14} M_{\odot}}\right)^{n_{\mathrm{M}}}.
\label{eq:non_thermal_fraction}
\end{equation}

\noindent with $\alpha_{nt}$, $\beta_{nt}$, $n_{nt}$ and $n_{M}$ as constants. The set of values of $\left(\alpha_{\mathrm{nt}}, \beta_{\mathrm{nt}}, n_{\mathrm{nt}}, n_{\mathrm{M}}\right)=(0.3,0.5,0.8,0.2)$ was determined in \citet{Shaw2012}. These are also the values used in the related works in the literature \citep{Terukina2013, Wilcox2015, Vergara2019}.

Having discussed the thermal and the non-thermal pressure terms, we can now write down the modified hydrostatic equilibrium equation that takes into account the fifth force effects:

\begin{equation}
\frac{1}{\rho_{\rm gas}(r)} \frac{P_{\rm total}(r)}{d r}=-\frac{G M(r)}{r^{2}}-\frac{\beta}{M_{\rm pl}} \frac{\mathrm{d} \phi(r)}{\mathrm{d} r},
\label{eq:modified_hydrostatic_equilibrium}
\end{equation}

\noindent where the last term is due to the chameleon force. The last term can also be used to define a mass corresponding to the chameleon gravity effects: 

\begin{equation}M_{\phi}(r) \equiv-\frac{r^{2}}{G} \frac{\beta}{M_{\mathrm{pl}}} \frac{d \phi(r)}{d r}.
\label{eq:chameleon_mass2}
\end{equation}

\noindent The $M_{\phi}$ then modifies the mass inferred by using the hydrostatic equilibrium equation, such that the total mass is given by:

\begin{equation}
    M(<r)=M_{\text {thermal }}(r)+M_{\text {non-thermal }}(r)+M_{\phi}(r).
\label{eq:total_hydro_mass}
\end{equation}

At this point it's worthwhile to summarize the underlying assumptions that allow us to calculate the cluster mass using the equations above. In particular, the key assumptions that lead the equations to have the form laid out above are those of hydrostatic equilibrium and spherical symmetry. In addition, to get the correct mean molecular weight, a good knowledge of the intracluster gas composition is assumed. Finally, the non-thermal pressure effects are based on studies that come from hydrodynamical simulations, which are assumed to be sufficiently realistic to approximate the real cluster astrophysics. As previously discussed, all these assumptions can be challenged to some degree and, as always, further work is need both in the context of simulations and observational data. These assumptions will be further examined in the rest of this chapter and chapter \ref{ch:testing_EG}.

\subsection{Weak Lensing in Chameleon Gravity}

A key aspect of the tests of the chameleon gravity described throughout this chapter is that the weak lensing effects are not affected by the fifth force in such models of modified gravity. This is the case as the chameleon field is coupled to the trace of the energy-momentum tensor. More specifically, as shown in \citet{Arnold2014}, if one adopts the Newtonian gauge in a spatially flat background:

\begin{equation}
\mathrm{d} s^{2}=a(\eta)^{2}\left[-(1+2 \Psi) \mathrm{d} \eta^{2}+(1-2 \Phi) \mathrm{d} \mathbf{x}^{2}\right],
\label{eq:newtonian_gauge}
\end{equation}

\noindent then the gravitational lensing potential is given by: $\Phi_{L}=(\Phi+\Psi) / 2$. Note that $\eta$ here denotes conformal time, which is related to cosmic time $t$ via the scale factor $a(\eta)$: $a(\eta) \mathrm{d} \eta \equiv \mathrm{d} t$. The equations for the two potentials $\Phi$ and $\Psi$ can then be derived in $f(R)$ gravity under the assumptions of weak field limit ($|\Phi| \ll 1$ and $|\Psi| \ll 1$), quasi-static approximation and the energy-momentum being described by the pressureless perfect fluid, such that  $T_{00}=\rho a^{2}$. Then it can be shown that the modified Poisson equation is given by: 

\begin{equation}
\frac{1}{a^{2}} \nabla^{2} \Psi=\nabla_{\mathrm{phys}}^{2} \Psi=\frac{16 \pi G}{3} \delta \rho-\frac{1}{6} \delta R,
\label{eq:modified_poisson}
\end{equation}

\noindent where $\nabla_{\mathrm{phys}}$ denotes the Laplace operator with respect to the physical coordinates (rather than the comoving coordinates). Similarly, for the $\Phi$ potential:

\begin{equation}
\frac{1}{a^{2}} \nabla^{2} \Phi=\nabla_{\mathrm{phys}}^{2} \Phi=\frac{8 \pi G}{3} \delta \rho+\frac{1}{6} \delta R.
\label{eq:modified_psi}
\end{equation}

\noindent Hence, for the lensing potential $\Phi_{\rm L}$ we have:

\begin{equation}
\nabla_{\mathrm{phys}}^{2} \Phi_{\rm L}=\frac{\nabla_{\mathrm{phys}}^{2} \Phi+\nabla_{\mathrm{phys}}^{2} \Psi}{2}=4 \pi G \delta \rho=\nabla_{\mathrm{phys}}^{2} \phi_{\mathrm{N}},
\label{eq:lensing_potential_poisson}
\end{equation}

\noindent which has the usual form for the Newtonian gravitational potential $\phi_{N}$. This means that we can use the familiar equations described in section \ref{section:cluster_weak_lensing} to described lensing in case of chameleon gravity as well. Note that in the context of $f(R)$ models this argument is true only for $|f_{R0}| \ll 1$.

Assuming the NFW profile (equation \ref{eq:NFW_profile}), the mass inferred from weak lensing can then be expressed as follows:

\begin{equation}
M_{\rm WL}(<r)=4 \pi \int_{0}^{r} d r r^{2} \rho(r)=4 \pi \rho_{\mathrm{s}} r_{\mathrm{s}}^{3}\left[\ln \left(1+r / r_{\mathrm{s}}\right)-\frac{r / r_{\mathrm{s}}}{1+r / r_{\mathrm{s}}}\right],
\label{eq:weak_lensing_mass}
\end{equation}

\noindent where $\rho_{s}$ and $r_{s}$ as before are the characteristic density and the characteristic scale -- the two parameters used in the NFW profile. Note that the $r_{s}$ term here can be expressed as follows: 
\begin{equation}
r_{\mathrm{s}}=\frac{1}{c}\left(\frac{3 M_{200}}{4 \pi \rho_{\mathrm{c}} \delta_{\mathrm{c}}}\right)^{1 / 3},
\label{eq:r_s}
\end{equation}

\noindent where $\delta_{c}$ is given in equation \ref{eq:characteristic_overdensity} and $M_{200}$ refers to the mass enclosed by $r_{200}$, i.e. the radius at which the average density of the halo is equal to $200 \rho_{c}$, where $\rho_{c} = 3 H^{2}(z) / 8 \pi G$. In summary, this means that the NFW profile can be characterised by two free parameters, the concentration parameter $c_{v}$ and the mass parameter $M_{200}$.

Given the assumption of the hydrostatic equilibrium along with the fact that the weak lensing mass is not affected by the fifth force effects we can then relate all the defined masses as follows: 

\begin{equation}
M_{\text {thermal }}+M_{\text {non-thermal }}+M_{\phi} = M_{\mathrm{WL}}.
\label{eq:wl_hydro_masses}
\end{equation}

\noindent Hence, assuming that we can measure the $M_{\text{thermal}}$ and the $M_{\text{non-thermal}}$ terms using X-ray data along with  $M_{\mathrm{WL}}$ using the corresponding shear data, the chameleon mass term can be constrained. 

\subsection{X-ray Surface Brightness}

The hydrostatic equilibrium equation for the thermal pressure component can be integrated to obtain an expression for gas pressure:

\begin{equation}
P_{\mathrm{thermal}}(r)=P_{\mathrm{thermal}, 0}+\mu m_{\mathrm{p}} \int_{0}^{r} n_{\rm gas}(r) \left(-\frac{G M(<r)}{r^{2}}-\frac{\beta}{M_{\mathrm{pl} }} \frac{d \phi(r)}{d r}\right) d r,
\label{eq:pressure}
\end{equation}

\noindent where $P_{\mathrm{thermal}, 0}$ is the central pressure and we used: $\rho_{\rm gas} = \mu m_{p} n_{\rm gas}$. Note that this expression can also be written in terms of the electron pressure $P_{e}$ and the electron number density $n_{e}$ by noting that:

\begin{equation}
n_{\mathrm{e}}=\frac{2+\mu}{5} n_{\mathrm{gas}},
\label{eq:electron_gas_number_density}
\end{equation}

\begin{equation}
P_{\mathrm{e}}=n_{\mathrm{e}} k_{B} T_{\mathrm{gas}}=\frac{2+\mu}{5} P_{\text {thermal }},
\label{eq:electron_thermal_pressure}
\end{equation}

\noindent where $\mu$ is the mean molecular weight as before. This then gives an expression for the electron pressure:

\begin{equation}
P_{\mathrm{e}}(r)=P_{\mathrm{e}, 0}+\mu m_{\mathrm{p}} \int_{0}^{r} n_{e}(r) \left(-\frac{G M(<r)}{r^{2}}-\frac{\beta}{M_{\mathrm{pl}}} \frac{d \phi(r)}{d r}\right) d r.
\label{eq:electron_pressure}
\end{equation}

\noindent The electron distribution in a cluster dictates the form of the $n_{e}(r)$ function. A standard choice to parametrize it adapted in \citet{Terukina2013} and \citet{Wilcox2015} is the isothermal beta model: 
\begin{equation}
n_{\mathrm{e}}(r)=n_{0}\left[1+\left(\frac{r}{r_{1}}\right)^{2}\right]^{b_{1}},
\label{eq:isothermal_beta_profile}
\end{equation}

\noindent with $n_{0}$, $r_{1}$ and $b_{1}$ as the free parameters.

By using the equation of state of gas in equation \ref{eq:electron_pressure}, the temperature of gas in the cluster can be directly related to the X-ray surface brightness:

\begin{equation}
S_{\mathrm{B}}\left(r_{\perp} \right)=\frac{1}{4 \pi(1+z_{\rm cl})^{4}} \int n_{\mathrm{e}}^{2}\Big(\sqrt{r_{\perp}^{2}+z^{2}}\Big) \lambda_{\mathrm{c}}\left(T_{\mathrm{gas}}\right) d z,
\label{eq:surface_brightness3}
\end{equation}

\noindent where $\lambda_{\rm c}$ is the cooling function and $z_{\rm cl}$ is the cluster redshift. The form of the cooling function was obtained in \citet{Wilcox2015} by using the XSPEC software and the APEC model over the range of $0.5$ - $2.0$ keV \citep{Arnaud1996, Smith2001}. The mentioned model takes gas temperature, the cluster redshift and the cluster metallicity and outputs X-ray cluster flux for a range of temperatures. The metallicity value of $Z=0.3 Z_{\odot}$ was adopted following \citet{Sato2011}. Fitting equation \ref{eq:surface_brightness3} to X-ray data, allows us to determine the free parameters in the isothermal beta profile in equation \ref{eq:isothermal_beta_profile}, which can then be used to calculate the thermal and the non-thermal masses. 

\subsection{X-ray and Weak Lensing Datasets}

Here the key datasets used to compare our results against the previous results in the literature are discussed. The general technique of combining multiple observational probes on galaxy cluster scales in order to constrain chameleon gravity was first described in \citet{Terukina2013}. More specifically, in the mentioned work the modified gravity constraints were obtained by performing a multi-dataset MCMC analysis. In particular, this was done by using a combined dataset consisting of the X-ray temperature data from \citet{Snowden2008} and \citet{Wik2009}, X-ray surface brightness profile data from \citet{Churazov2012}, SZ effect data  from \citet{Ade2016} and the tangential shear data from \citet{Okabe2010B}. The mentioned datasets are described in more detail in the later parts of this chapter and chapter \ref{ch:testing_EG}. 

The techniques described in \citet{Terukina2013} were later expanded \citet{Wilcox2015} where the surface brightness and tangential shear profiles were produced by stacking data from 58 galaxy clusters. In particular, the mentioned dataset contains data from 58 galaxy clusters, at redshifts $0.1<z<1.2$ from the XMM Cluster Survey (XCS) and the Canada France Hawaii Telescope Lensing Survey (CFHTLenS). The clusters were stacked in order to improve the signal to noise ratio and to remove various irregularities that individual clusters posses. Combining observations of multiple clusters is a complicated procedure, as the cluster images have to be rescaled in a consistent matter while also taking into account the fact that each observation comes with different background properties and flare corrected exposure times. In order to produce a single stack, the 58 individual cluster images were rescaled to a common projected size by estimating the $M_{500}$ and $M_{200}$ masses using the approach described in \citet{Hu2003,Sahlen2009}. Subsequently, the $r_{200}$ radius was calculated for each cluster, which in turn allowed rescaling each image to a $500 \times 500$ pixel format, such that each cluster had an $r_{200}$ equivalent to 125 pixels. Each of the images was then centered on the source centroid given in the XCS data. The final stacked surface brightness map was produced by taking the mean value for each pixel across all the images. 

The tangential shear profiles were calculated using the ellipticity components and the photometric redshifts for each source galaxy, as given in the CFHTLenS catalogue. In particular, for each galaxy the tangential and the cross-shear components $\gamma_{\mathrm{t}}, \gamma_{\mathrm{x}}$  were calculated as a function of their position relative to the cluster position (as measured by an angle $\phi_{g}$ relative to the baseline of zero declination). The tangential shear around each XCS cluster centroid was then binned into 24 equally spaced logarithmic annuli reaching $10 \times r_{200}$. The shear values were then stacked by summing the profiles of each cluster and calculating an average shear value in each radial bin. 

The dataset was also split into two bins based on the X-ray temperature. The temperature for each cluster was determined by using a mass-temperature relation following the procedure laid out in \citet{Stott2010}. In particular, the cluster stack was cut into a low temperature bin ($T < 2.5$ keV) with a median redshift $z=0.32$ and a high temperature bin ($T > 2.5$ keV) with $z = 0.34$. This roughly corresponds to splitting the dataset into galaxy groups and galaxy clusters. The logic for such a split was based on tests, where different splits were considered with a goal of producing the tightest possible constraints of the modified gravity parameters. More specifically, having multiple temperature bins were shown not to have a significant effect on the modified gravity constraints, hence two bins were used. 

\begin{figure}[ht!]
\centering
\includegraphics[width=0.70\columnwidth]{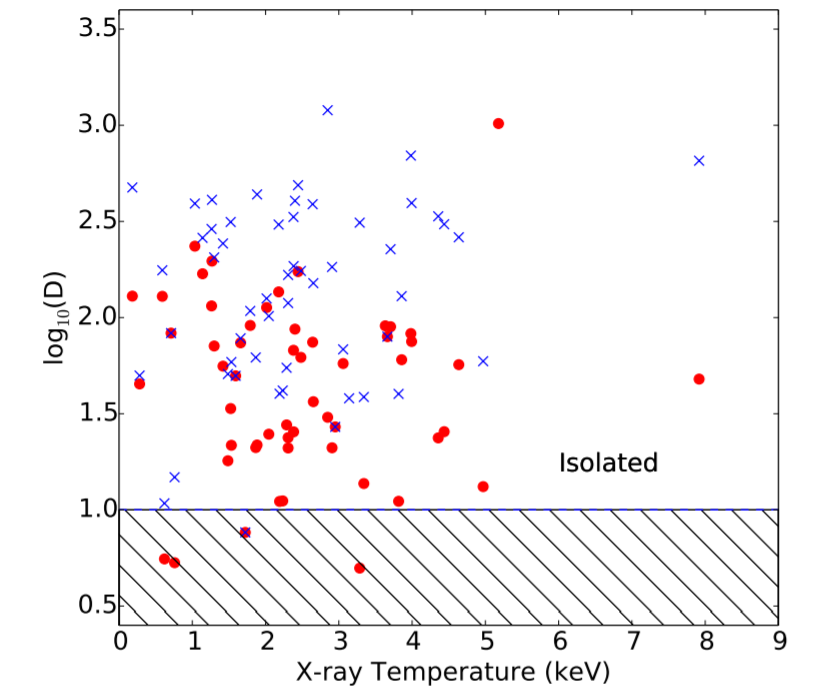}
\caption[Analysis of the cluster separation]{Analysis of the 58 cluster stack dataset in terms of the separation parameter $D$, which is a measure of the distance between a given cluster and the nearest overdensity in the top 30\% and 10\% overdensity values in the given dataset correspondingly shown as red dots and blue crosses \citep{Wilcox2015}. 
\label{figure:cluster_separation}}
\end{figure}

An important aspect of the dataset described in \citet{Wilcox2015} is that the majority of the mentioned 58 clusters are sufficiently isolated from the neighbouring clusters. This is of key importance, as if clusters are not sufficiently isolated, they might be screened by the neighbouring clusters essentially suppressing any fifth force effects. In order to measure the separation of individual clusters in the dataset, the separation parameter $D$ which quantifies the separation between a given cluster and the nearest cluster scaled by the $r_{200}$ for each given cluster was calculated \citep{Zhao2011}. In such a parametrization, $D>1$ corresponds to a well isolated cluster. Figure \ref{figure:cluster_separation} shows the $D$ values for each cluster in the dataset. Only 5 clusters are found to be not sufficiently isolated from the local environment.

\begin{figure}[ht!]
\centering
\includegraphics[width=0.99\columnwidth]{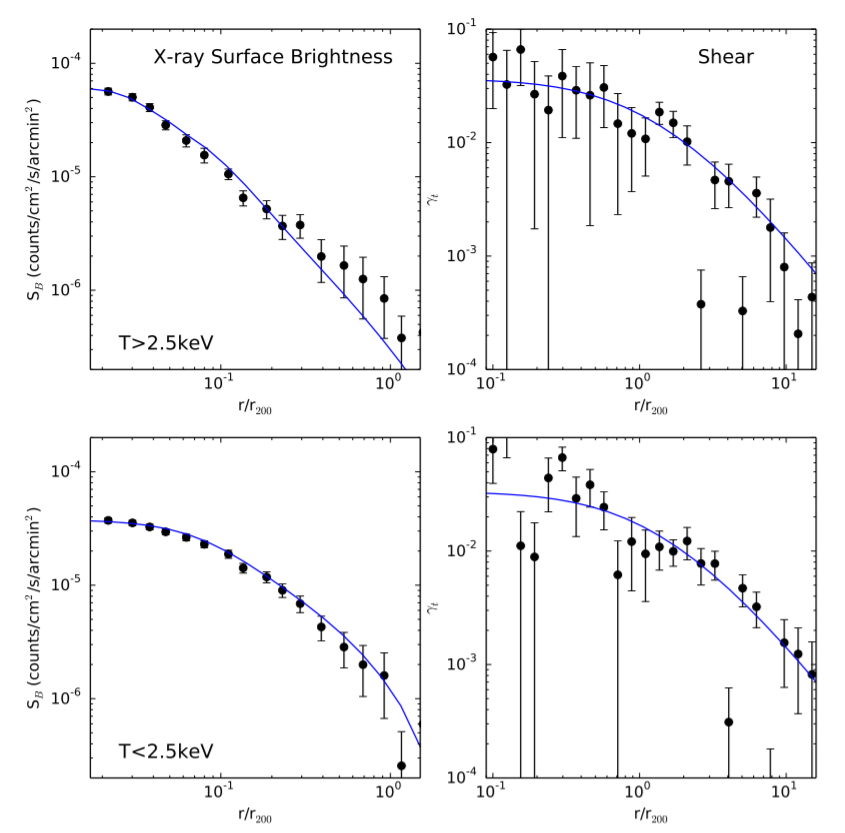}
\caption[X-ray and tangential shear profile dataset from \citet{Wilcox2015}]{The X-ray surface brightness and the tangential shear profiles for the 58 galaxy cluster dataset from \citet{Wilcox2015}. The blue lines correspond to the best-fit results from the MCMC analysis. The best-fit values are given in table \ref{table:old_vs_new_results}. 
\label{figure:HW_dataset}}
\end{figure}

The stacked X-ray surface brightness and the tangential shear profiles for both temperature bins are shown in figure \ref{figure:HW_dataset}. The best-fit results and the corresponding modified gravity constraints were obtained using an MCMC analysis (described at the end of this section).

\begin{figure}[ht!]
\centering
\includegraphics[width=0.85\columnwidth]{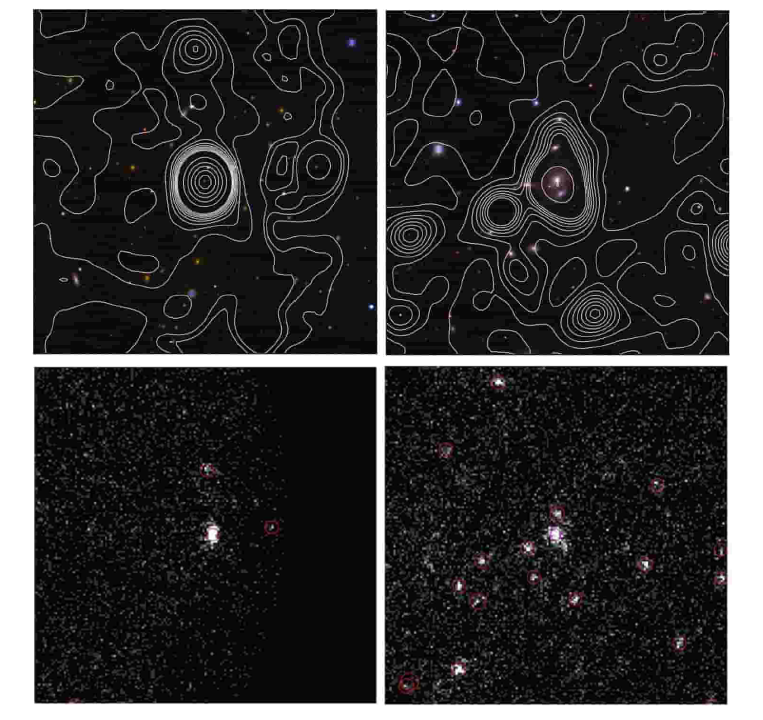}
\caption[A sample of sources misclassified as galaxy clusters in the original XCS-CFHTLenS dataset in \citet{Wilcox2015}]{A sample of sources misclassified as clusters in the original XCS-CFHTLenS dataset from \citet{Wilcox2015}. The top row corresponds to the optical images with the X-ray contours while the bottom row shows the corresponding X-ray images. Image adapted from figure 4.7 in \citet{Vergara2019}. 
\label{figure:misclassified_sources}}
\end{figure}

A recent analysis of the original 58 cluster dataset, as described in great detail in section 3.2 in \citet{Vergara2019}, indicated that a number of sources were possibly misclasified as galaxy clusters. In particular, the newest XCS master source list indicated that the XCS automated pipeline algorithm (XAPA), that is used to  detect X-ray sources, flagged 8 of the objects as point sources and 11 as extended sources with a point spread function warning flag (indicating a need for further investigation). After further investigation, a total of 27 sources were removed as a precaution due to being possibly misclassified. More specifically, the mentioned 27 sources were found to resemble AGN sources rather than multiple galaxies with an extended X-ray emission (see a sample of the misclassified sources in figure \ref{figure:misclassified_sources}). The removed sources can be split into bins in terms of the photon counts of over and less than 200. For the case of $\leq 200$ the removed sources had 
$\langle z\rangle = 0.459$ and $\left\langle T\right\rangle = 2.0854$ keV. While, for the case of $\geq 200$ photons, the mean redshift and X-ray temperature were correspondingly $\langle z\rangle = 0.4842$ and $\left\langle T\right\rangle = 2.1458$ keV.

\begin{figure}[ht!]
\centering
\includegraphics[width=0.90\columnwidth]{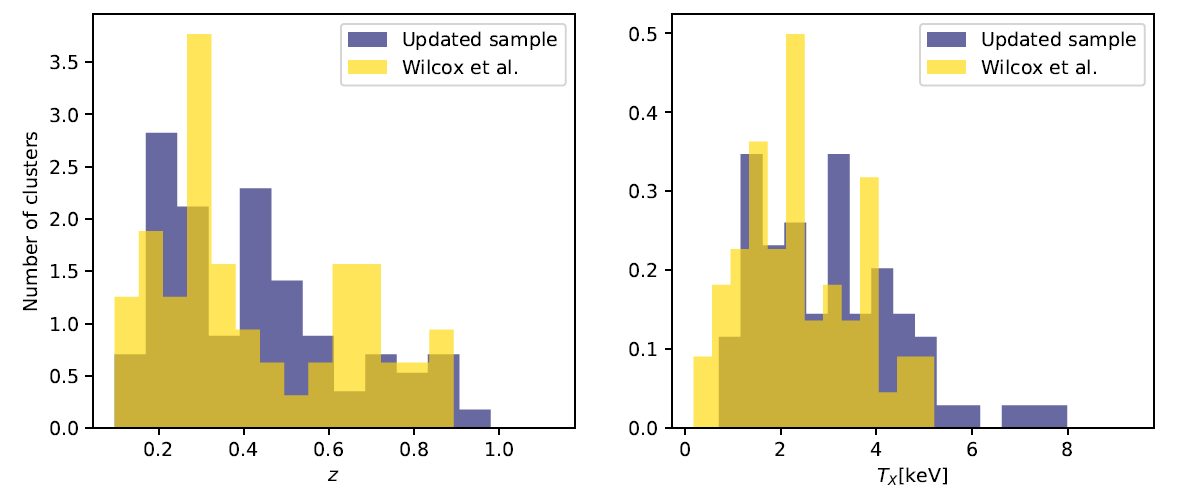}
\caption[A comparison of the original and the updated XCS-CFHTLenS datasets]{A comparison of the original and the updated XCS-CFHTLenS datasets \citep{Vergara2019}. 
\label{figure:original_new_dataset}}
\end{figure}

In addition to a significant part of the original dataset being removed, new clusters were added to the original dataset as described in detail in \citet{Vergara2019}. More specifically, new XCS cluster candidates in the CFHTLenS footprint were analyzed. A cross-match between the latest XCS master source catalogue and the CFHTLenS 3-D matched-filter catalogue was performed in order to find clusters with both the X-ray and the corresponding weak lensing data. Extended sources with a photon count of  $\geq 200$ were chosen for the updated dataset. Combining the new cluster samples with the existing correctly classified samples from the original dataset resulted in a dataset of 77 X-ray selected, optically confirmed clusters in the CFHTLenS footprint. Note that the new clusters were also chosen to satisfy the $\log_{10}(D) > 1$ condition in the same fashion as the majority of the clusters in the original dataset (see figure \ref{figure:cluster_separation}). Figure \ref{figure:original_new_dataset} shows a comparison between the original and the updated datasets in terms of the redshift and the X-ray temperature distribution.

The updated dataset of the 77 clusters was then used to produce stacked X-ray surface brightness and weak lensing profiles using an identical procedure to the one used to produce the original 58 cluster stack. 

\begin{figure}[ht!]
\centering
\includegraphics[width=0.95\columnwidth]{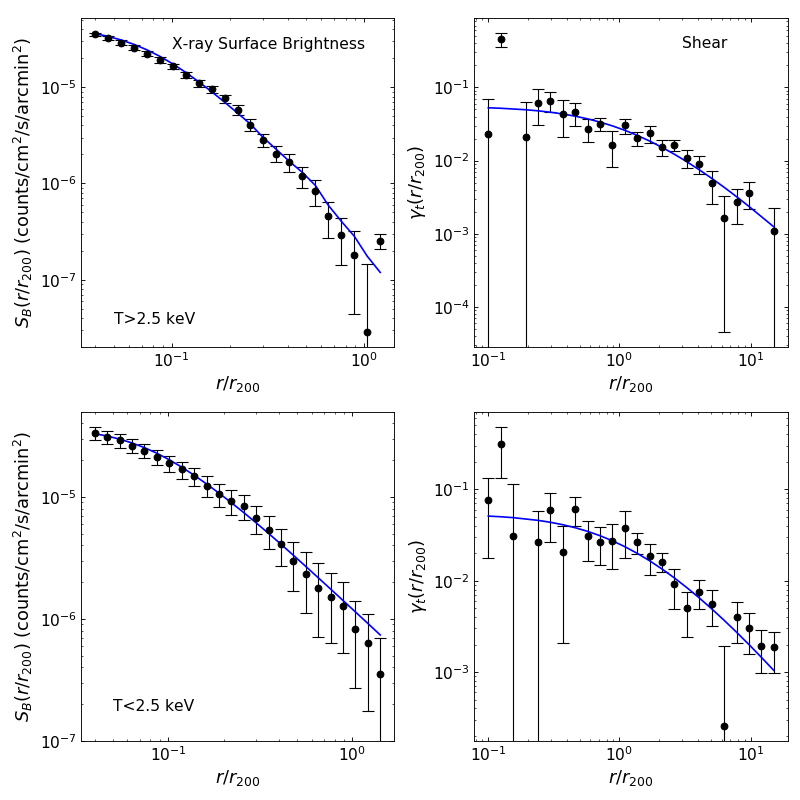}
\caption[The stacked X-ray surface brightness and tangential shear profiles for the updated 77 cluster dataset]{The stacked tangential shear and X-ray surface brightness profiles produced using the updated XCS-CFHTLenS 77 cluster dataset as described in \citep{Vergara2019}. The blue lines correspond to the best-fit results from the MCMC analysis. The best-fit values are given in table \ref{table:old_vs_new_results}. 
\label{figure:updated_dataset_profiles}}
\end{figure}

Combining the non-cluster sources along with the genuine galaxy clusters has multiple effects on the X-ray surface brightness and tangential shear profiles. Specifically, a lot of the misclassified sources were AGNs, which have different surface brightness profiles when compared to clusters. In addition, such sources would not produce a shear signal comparable to that of galaxy clusters. These two factors affect the shape and the errors of the resulting stacked profiles, however it is important to emphasize that the magnitude of such effect is limited due to the averaging procedure. In other words, irregularities of the individual sources are mostly averaged out during the stacking procedure as long as the number of the misclassified sources is not dominant in the dataset. In our case the resulting updated dataset profiles are generally similar to the original dataset profiles. However, as expected, removing the misclassified sources resulted in lower tangential shear errors. Nonetheless, the updated dataset had less clusters being stacked for the lower X-ray temperature bin $T < 2.5$ keV, which results in slightly higher error bars for the corresponding surface brightness profile. Figures \ref{figure:HW_dataset} and \ref{figure:updated_dataset_profiles} show the X-ray surface brightness and the tangential shear profiles for the original and the updated datasets. 

\begin{figure}[ht!]
\centering
\includegraphics[width=0.98\columnwidth]{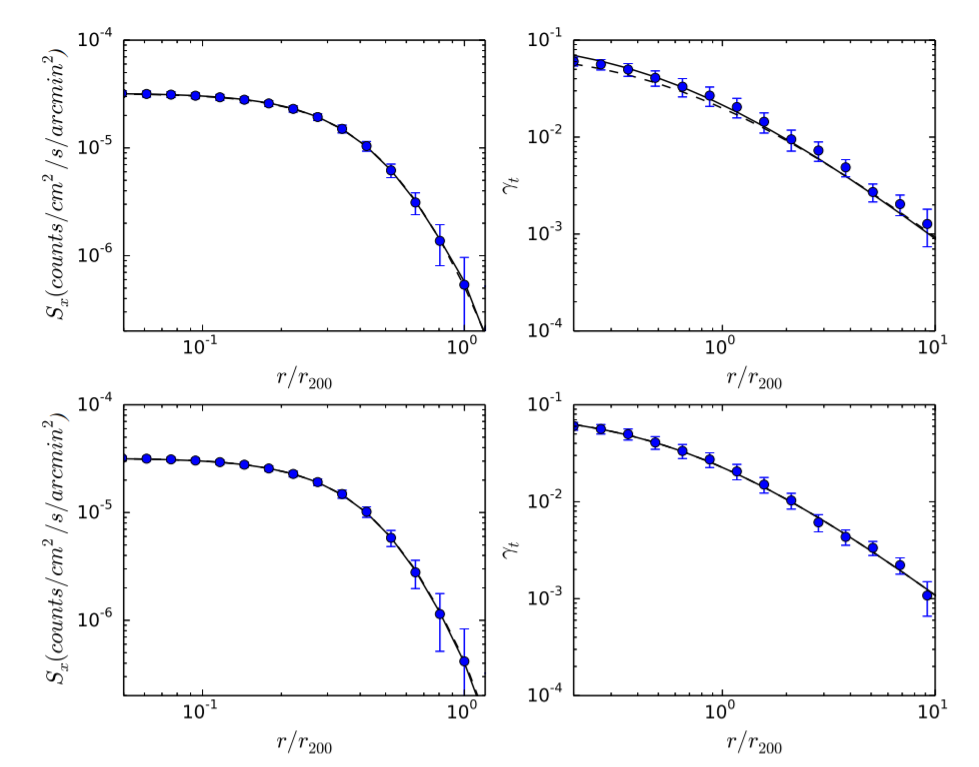}
\caption[X-ray surface brightness from two types of simulations from \citet{Wilcox2016}]{The X-ray surface brightness and weak lensing profiles from clusters produced using two types of MGENZO simulations: $\Lambda$CDM (top) and $f(R)$ with $\left|f_{\mathrm{R} 0}\right|=10^{-5}$ (bottom) \citep{Wilcox2016}. In the case of the $\Lambda$CDM simulation, 103 clusters were stacked; in the case of the $f(R)$ simulation, 99 clusters were stacked. The best-fit lines correspond to the best-fitting analytical model with (dashed line) and without (solid line) non-thermal pressure component added. The best-fit values are given in table \ref{table:terukina_wilcox2016}. 
\label{figure:simulation_dataset}}
\end{figure}

Another dataset that is important to discuss for comparison purposes is the dataset produced in \citet{Wilcox2016}. In particular, this dataset includes simulated galaxy clusters produced using the MGENZO simulation, which is an extension of the ENZO code allowing hydrodynamical simulations with $f(R)$ and scalar-tensor gravity \citep{Bryan2014}. Two types of simulations were produced, one with the standard $\Lambda$CDM parameters (103 clusters) and the second one with an $f(R)$ gravity (99 clusters) with $\left|f_{\mathrm{R} 0}\right|=10^{-5}$. Both simulations were run with $2 \times 128^{3}$ particles with $4 \times 10^{11} \mathrm{M}_{\odot}$ mass and $128$ $\mathrm{Mpc} / \mathrm{h}$ box size. The \textit{Rockstar} Friends-of-Friends (FOF) algorithm  was then used to locate the main dark matter haloes \citep{Behroozi2013}. The X-ray images were created using the \textit{PHOX} software, which is designed to obtain synthetic observations from hydro-numerical simulations \citep{Biffi2011}. Since the used simulation does not simulate the effects of lensing, the expected convergence $\kappa$ was estimated using the following equation: 

\begin{equation}
\kappa=\frac{3 H_{0}^{2} \Omega_{m}}{2 c^{2}} \sum_{i} \Delta_{\chi i} \chi_{i} \frac{\left(\chi_{c l u s t}-\chi_{i}\right)}{\chi_{c l u s t}} \frac{\delta_{i}}{a_{i}},
\label{eq:convergence_estimate}
\end{equation}

\noindent where the Born approximation was used and the sumation is over the co-moving distance $\chi_{i}$, using bins of width $\Delta_{\chi i}$ and $\delta_{i}$ is the overdensity and $a_{i}$ as the scale factor. Once the X-ray and the weak lensing images were produced, they were stacked using an analogous procedure to the one in \citet{Wilcox2015} to allow a detailed comparison of the results. Figure \ref{figure:simulation_dataset} shows the described datasets.

\subsection{MCMC Fitting}

The same general procedure was used to constrain the modified gravity parameters in \citet{Terukina2013}, \citet{Wilcox2015} and our approach described in this thesis. More specifically, the only free parameters appearing in the surface brightness and the weak lensing equations are $n_{0}, b_{1}, r_{1}$ (from the electron number density profile), $T_{0}$ (the central temperature value) and $M_{200}$ and $c$ characterizing the NFW density profile. Given that the dataset is split into two bins in terms of X-ray temperature, the total set of free parameters includes: $\{ T_{0}^{\mathrm{I}}, n_{0}^{\rm I}, b_{1}^{\rm I}, r_{1}^{\rm I}, M_{200}^{\rm I}, c^{\rm I}_{v}, T_{0}^{\mathrm{II}}, n_{0}^{\rm II}, b_{1}^{\mathrm{II}}, r_{1}^{\mathrm{II}}, M_{200}^{\mathrm{II}},c^{\mathrm{II}}_{v}, \beta_{2}, \phi_{\infty,2}  \}$. The superscript notation here refers to the two temperature bins of $T < 2.5$ and $T > 2.5$ keV respectively. The last two parameters are used to parametrize the modifications of gravity and refer to the rescaled coupling constant and the field value at $r \rightarrow \infty$, such that: $\beta_{2}=\beta /(1+\beta)$ and   
$\phi_{\infty, 2}=1-\exp \left(-\phi_{\infty} / 10^{-4} M_{\mathrm{pl} }\right)$. The following priors were used when exploring the parameter space: $T_{0}^{\rm I} = [0.1,50]$ keV, $n_{0}^{\rm I} = [1 \times 10^{-5},10]$ $\rm cm^{-2}$, $b_{1}^{\rm I} = [-10, -1 \times 10^{-7}]$, $r_{1}^{\rm I} = [1 \times 10^{-3}, 1.0]$ Mpc, $M_{\rm 200}^{\rm I} = [1 \times 10^{-13}, 4 \times 10^{15}]$ $\rm M_{\odot}$, $c_{v}^{\rm I} = [0.1, 40]$, $T_{0}^{\rm II} = [0.1, 50]$ keV, $n_{0}^{\rm II} = [1 \times 10^{-5}, 10]$ $\rm cm^{-2}$, $b_{1}^{\rm II} = [-10, -1 \times 10^{-7}]$, $r_{1}^{\rm II} = [1 \times 10^{-4}, 1.0]$ Mpc, $M_{200}^{\rm II} = [1 \times 10^{13}, 4 \times 10^{15}]$ $\rm M_{\rm \odot}$, $c_{v}^{\rm II} = [0.1, 40]$, $\beta_{2} = [0.0,1.0]$, $\phi_{\infty,2} = [0.0,1.0]$.  

The goodness of fit can then be quantified as follows: 

\begin{equation}
\begin{gathered}
    \chi^{2}(T_{0}^{\mathrm{I}}, n_{0}^{\rm I}, b_{1}^{\rm I}, r_{1}^{\rm I}, M_{200}^{\rm I}, c^{\rm I}_{v}, T_{0}^{\mathrm{II}}, n_{0}^{\rm II}, b_{1}^{\mathrm{II}}, r_{1}^{\mathrm{II}}, M_{200}^{\mathrm{II}}, c^{\mathrm{II}}_{v}, \beta_{2}, \phi_{\infty,2} )= \\
    =\chi^{\rm I \; 2}_{\rm WL} + \chi^{\rm II \; 2}_{\rm WL} + \chi^{\rm I \; 2}_{\rm SB} + \chi^{\rm II \; 2}_{\rm SB},
    \end{gathered}
    \label{eq:chi_squared_total}
\end{equation}

\noindent where the total $\chi^{2}$ is split into contributions due to the two temperature bins for the weak lensing and surface brightness datasets. The goodness of fit components can be quantified by evaluating the squared residuals between the predicted and the observed values divided by the corresponding error values:

\begin{equation}
\chi_{\mathrm{WL}}^{\mathrm{I} \; 2}=\sum_{i} \frac{\left(\gamma_{\rm t}\left(r_{\perp, \mathrm{i}}^{\mathrm{I}}\right)-\gamma_{\mathrm{t,i}}^{\mathrm{obs}, \mathrm{I}}\right)^{2}}{\left(\Delta \gamma_{\mathrm{t,i}}^{\mathrm{obs}, \mathrm{I}}\right)^{2}},
\label{eq:chi_wl_1}
\end{equation}

\begin{equation}
\chi_{\mathrm{WL}}^{\mathrm{II} \; 2} =\sum_{i} \frac{\left(\gamma_{\rm t}\left(r_{\perp, \mathrm{i}}^{\mathrm{II}}\right)-\gamma_{\mathrm{t,i}}^{\mathrm{obs}, \mathrm{II}}\right)^{2}}{\left(\Delta \gamma_{\mathrm{t,i}}^{\mathrm{obs}, \mathrm{II}}\right)^{2}},
\label{eq:chi_wl_2}
\end{equation}

\begin{equation}
\chi_{\mathrm{SB}}^{\mathrm{I} \; 2}=\sum_{i, j}\left(S_{\mathrm{B}}\left(r_{\perp, \mathrm{i}}^{\mathrm{I}}\right)-S_{\mathrm{B}, \mathrm{i}}^{\mathrm{obs}, \mathrm{I}}\right) C_{i, j}^{-1}\left(S_{\mathrm{B}}\left(r_{\perp, \mathrm{j}}^{\mathrm{I}}\right)-S_{\mathrm{B}, \mathrm{j}}^{\mathrm{obs}, \mathrm{I}}\right),
\label{eq:chi_SB_1}
\end{equation}

\begin{equation}
\chi_{\mathrm{SB}}^{\mathrm{II} \; 2} =\sum_{i, j}\left(S_{\mathrm{B}}\left(r_{\perp, \mathrm{i}}^{\mathrm{II}}\right)-S_{\mathrm{B}, \mathrm{i}}^{\mathrm{obs}, \mathrm{II}}\right) C_{i, j}^{-1}\left(S_{\mathrm{B}}\left(r_{\perp, \mathrm{j}}^{\mathrm{II}}\right)-S_{\mathrm{B}, \mathrm{j}}^{\mathrm{obs}, \mathrm{II}}\right).
\label{eq:chi_SB_2}
\end{equation}

\noindent Here $S_{B}(r_{\perp})$ is the X-ray surface brightness at a perpendicular radial distance from the cluster centre, $\gamma_{\rm t}(r_{\perp})$ refers to tangential shear, $\Delta \gamma$ is the corresponding error and $C_{i,j}$ refers to the components of the covariance matrix. The covariance, in particular, is a measure of how changes in one surface brightness bin affect the values in the other bins. Following the assumption in \citet{Terukina2013} and \citet{Wilcox2015}, the covariance matrix for the weak lensing dataset was approximated as diagonal. More specifically, this choice was based on the correlation matrices having dominant diagonal terms in all the described weak lensing datasets.  

The $\chi^{2}$ in equation \ref{eq:chi_squared_total} was then optimized using an MCMC sampler. In particular, the \textit{Zeus} sampler was used to find the optimal values of the outlined free parameters \citep{karamanis2020}. The sampler was run using 42 walkers for 10000 steps with 4000 steps removed as burn-in.

\subsection{Results}

Table \ref{table:old_vs_new_results} summarizes the best-fit parameter values corresponding to the fits in figures \ref{figure:HW_dataset} and \ref{figure:updated_dataset_profiles}. Similarly, the best-fit results from \citet{Terukina2013} and \citet{Wilcox2016} are given for comparison in table \ref{table:terukina_wilcox2016}. The best-fit parameter values were generally found to be degenerate in the sense that multiple combinations of the parameters can lead to equivalent best-fit results.

\begin{table}[!ht]
\centering
\begin{tabular}{lcc}
\textbf{Parameters:} & Terukina et al. (2014)  & Wilcox et al. (2016) \\ \hline
$T_{0}$ (keV) & 11.3  & 26.5      \\
$n_{0}$ ($\rm cm^{-2}$) & $2.34 \times 10^{-3}$  & $1.10 \times 10^{-3}$  \\
$r_{1}$ (Mpc) & $0.30$  & 0.63  \\
$M_{200}$ ($\mathrm{M}_{\odot}$) & $24.6 \times 10^{14}$  &  $10.0 \times 10^{14}$  \\
$b_{1}$ & -0.915 & -2.0    \\
$c_{v}$ & 2.64 & 9.0    \\
$\beta_{2}$ &  0.94  & 0.75     \\
$\phi_{\infty,2}$ & 0.98   & 0.50     \\ \hline
\end{tabular}%
\caption[Comparison of the best-fit parameters in \citet{Terukina2013} and \citet{Wilcox2016}.]{Comparison of the best-fit parameters in \citet{Terukina2013} and \citet{Wilcox2016}. In the case of the \citet{Wilcox2016} results the $\Lambda$CDM best-fit values are shown. For the information about the likelihood and the errors of the best-fit parameters see figure \ref{MCMC_contours_terukina}. }
\label{table:terukina_wilcox2016}
\end{table}

\begin{figure}[ht!]
\centering
\makebox[\textwidth][c]{\includegraphics[width=1.2\textwidth]{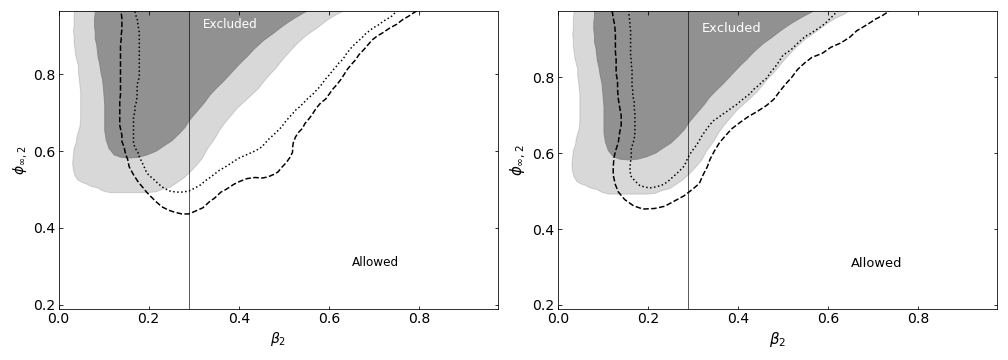}}%
\caption[A comparison of the modified gravity constraints from the Coma Cluster \citep{Terukina2013}, the 58 galaxy cluster stack from \citep{Wilcox2015} and the 77 cluster stack described in this work]{A comparison of the results presented in this work against the previous work in the literature. \textbf{Left:} modified gravity constraints from the 77 cluster stack with the 95\% confidence limits in light grey and the 99\% confidence limits in dark grey. The dash and the dotted lines are the equivalent constraints from the 58 galaxy cluster stack as described in \citet{Wilcox2015}. \textbf{Right:} the 77 cluster stack constraints (in light and dark grey as before) compared against the constraints calculated using the data from the Coma Cluster (shown as contours in dashed and dotted lines) as described in \citet{Terukina2013}. The vertical lines correspond to $\beta = \sqrt{1/6}$ and allows us to put the following constraints on the modifications of gravity: $\phi_{\infty} < 8 \times 10^{-5}$ $M_{\rm pl}$ or equivalently $|f_{R0}| < 6.5 \times 10^{-5}$ at 95\% confidence.
\label{figure:new_vs_old_vs_coma}}
\end{figure}

\begin{figure}[ht!]
\centering
\makebox[\textwidth][c]{\includegraphics[width=1.2\textwidth]{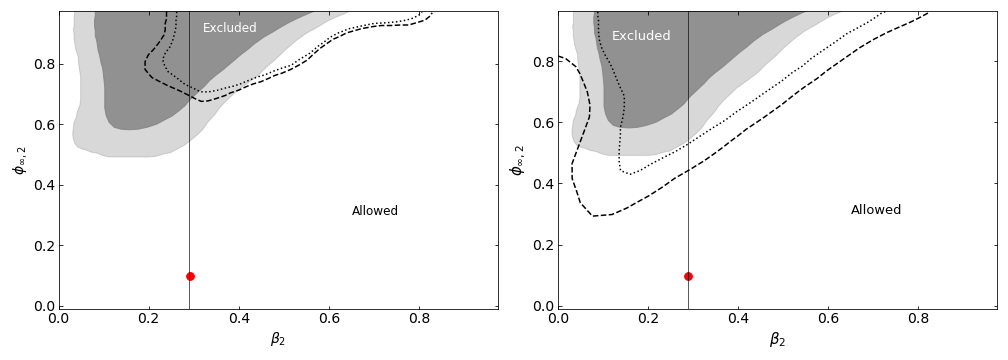}}%
\caption[A comparison of the modified gravity constraints from the $\Lambda$CDM and $f(R)$ simulations with 103 and 99 clusters stacked correspondingly and the 77 cluster stack described in this work]{A comparison of the results presented in this work against the previous work in the literature. \textbf{Left:} modified gravity constraints from the 77 cluster stack with the 95\% confidence limits in light grey and the 99\% confidence limits in dark grey. The dashed and dotted lines are the corresponding results derived from the 103 cluster stack produced by the $\Lambda$CDM simulation (with the non-thermal pressure effects included) as described in \citet{Wilcox2016}. \textbf{Right:} same as the left, but compared against the $f(R)$ simulation 99 cluster stack results (with the non-thermal pressure effects included) in \citet{Wilcox2016}. The vertical lines correspond to $\beta = \sqrt{1/6}$ and give the corresponding $f_{R0}$ parameter constraints. The red points corresponds to fiducial model value of $|f_{R0}| = 10^{-5}$  that was used in the $f(R)$ simulations. 
\label{figure:new_vs_hw2016_results}}
\end{figure}

The results for the constraints on the modified gravity parameters are shown in figures \ref{figure:new_vs_old_vs_coma} and \ref{figure:new_vs_hw2016_results}. These figures compare the modified gravity constraints derived in this work against the previous works in the literature. More specifically, figure \ref{figure:new_vs_old_vs_coma} shows the comparison against the results described in \citet{Terukina2013} and \citet{Wilcox2015}. The contours in light and dark grey correspond to the parameter space regions that are ruled out at 95\% and 99\% confidence correspondingly, while the dashed and dotted contours are the corresponding results from the previous work in the literature. The vertical lines in all the plots correspond to the value of $\beta = \sqrt{1/6}$ and hence allow converting the constraints on the $\phi_{\infty,2}$ parameter to the constraints on the $f_{R0}$ parameter. More specifically, the portion of the line that is in the allowed region of the parameter space gives the allowed values of the $\phi_{\infty,2}$ parameter, which can be converted back to $\phi_{\infty}$ and then to the $f_{R0}$ by noting that: $f_{R 0}=-\sqrt{2 / 3}\left(\phi_{\infty} / M_{\mathrm{pl} }\right)$. Figure \ref{figure:new_vs_hw2016_results} shows the corresponding comparison against the $\Lambda$CDM and $f(R)$ simulation results from \citet{Wilcox2016}. The contours correspond to the ruled-out parameter space regions coming from the 103 and 99 cluster stacks produced in $\Lambda$CDM and $f(R)$ simulations correspondingly. The red points in both plots refer to the fiducial $f(R)$ simulation value of $|f_{R0}| = 10^{-5}$.

In general, the constraints derived in this work using the updated 77 cluster stack are similar to the previous results in the literature. More specifically, comparing the new results against the constraints derived using the original dataset of 58 clusters in \citet{Wilcox2015} shows that both results are capable of ruling out a region of the parameter space of nearly identical size. This is somewhat expected as, even though the tangential shear errors are smaller in the updated dataset, the surface brightness errors for the $T < 2.5$ keV bin are significantly larger due to a different number of clusters being stacked in that bin. The main difference between our results and the original 58 cluster results is the fact that the triangular area of the ruled out parameter values is shifted towards lower values of $\beta_{2}$. The triangular shape of the contours originates from the relationship between the critical radius $r_{\rm crit}$ and the values of the coupling constant $\beta$ and the $\phi_{\infty}$, which is connected to the effectiveness of the screening mechanism. More specifically, the critical radius is given by \citep{Wilcox2015}:

\begin{equation}
r_{\text {crit }}=\frac{\beta \rho_{\mathrm{s}} r_{\mathrm{s}}^{3}}{M_{\mathrm{\rm pl}} \phi_{\infty}}-r_{\mathrm{s}},
\label{eq:critical_radius}
\end{equation}

\noindent where $\rho_{s}$ is the density at this particular radius. Hence for very small values of $\beta$, the deviations from GR are too insignificant to be observed given the observational data errors. Similarly, as $\beta$ increases, a lower value of $\phi_{\infty}$ is required to obtain $r_{\rm crit}$ that is inside the cluster. This sets an upper limit on $\beta/\phi_{\infty}$ and results in the triangular shape seen in all the result plots.

\begin{table}[!ht]
\centering
\begin{tabular}{lcc}
\textbf{Parameters:} & Wilcox et al. (2015)  & This work \\ \hline
$T_{0}^{\rm I}$ (keV) & 12.6  & 20.8     \\
$T_{0}^{\rm II}$ (keV) & 7.8   & 9.1    \\
$n_{0}^{\rm I}$ ($\rm cm^{-2}$) & $2 \times 10^{-2}$  & $1.1 \times 10^{-2}$    \\
$n_{0}^{\rm II}$ ($\rm cm^{-2}$) & $4.90 \times 10^{-2}$  & $1.38 \times 10^{-2}$    \\
$r_{1}^{\rm I}$ (Mpc) & $6.0 \times 10^{-2}$  & $8.9 \times 10^{-2}$    \\
$r_{1}^{\rm II}$ (Mpc) & $5.0 \times 10^{-2}$  & $9.0 \times 10^{-2}$    \\
$M_{200}^{\rm I}$ ($\mathrm{M}_{\odot}$) & $12.2 \times 10^{14}$  & $21.2 \times 10^{14}$    \\
$M_{200}^{\rm II}$ ($\mathrm{M}_{\odot}$) & $13.7 \times 10^{14}$  & $25.8 \times 10^{14}$    \\
$b_{1}^{\rm I}$ & -0.42 &  -0.6   \\
$b_{1}^{\rm II}$ & -0.89 & -0.74    \\
$c^{\rm I}_{v}$ & 3.5 &  4.7   \\
$c^{\rm II}_{v}$ & 3.8  &  4.5   \\
$\beta_{2}$ &  0.67  &   0.57 \\
$\phi_{\infty,2}$ & 0.88  &  0.39  \\ \hline
\end{tabular}%
\caption[Comparison of the best-fit parameters in \citet{Wilcox2015} and this work]{Comparison of the best-fit parameters in \citet{Wilcox2015} and this work. The bin notation of I and II here refers to the $T < 2.5$ keV and $T > 2.5$ keV bins correspondingly. For the information about the likelihood and the best-fit parameter errors, see figures \ref{MCMC_contours_wilcox} and \ref{MCMC_contours_77_clusters}.}
\label{table:old_vs_new_results}
\end{table}

Our results derived from the updated cluster stack profiles offer some of the most competitive constraints on cosmological scales. Due to the contours being shifted towards the lower $\beta_{2}$ values, when compared to the results for the Coma Cluster and the 58 cluster stack, the constraints on the $f_{R0}$ parameter are slightly weaker. Table \ref{table:fR0_constraints} summarizes the constraints on the $f_{R0}$ parameter from all the previously mentioned works. In summary, our results agree well with the previous work with the $|f_{R0}|$ constraints being slightly weaker that those in \citet{Terukina2013} and \citet{Wilcox2015}, but stronger than those in \citet{Wilcox2016} in the case of $\Lambda$CDM. 

The outlined results show that combining cluster X-ray and weak lensing data and, in particular, stacking cluster profiles offers a reliable technique of putting some of the strongest constraints on galaxy cluster scales. It is, however, important to discuss the validity of the key assumptions taken in this work. One of such assumptions was that galaxy clusters are spherically symmetric. This, of course, is not valid for real clusters, however, in our work we stack multiple clusters, which averages out the deviations from spherical symmetry. A natural question to ask then is how our results would be affected by introducing small deviations from spherical symmetry. This was investigated in \citet{Terukina2013}, where a small perturbation $\delta_{n_{e}}$ in the electron number density profile is introduced:

\begin{equation}
n_{\mathrm{e}}(r, \theta, \varphi)=\bar{n}_{\mathrm{e}}(r)\left[1+\delta_{n_{e}}(r, \theta, \varphi)\right],
\label{eq:spherical_perturbation}
\end{equation}

\noindent where $\bar{n}_{e}$ is the mean electron number value. Assuming that $\left\langle\delta_{n_{e}}\right\rangle=0$ and $\left\langle\delta_{n_{\mathrm{e}}}^{2}\right\rangle \neq 0$, allows us to express the effect on the X-ray surface brightness:

\begin{equation}
S_{\mathrm{B}} \propto \int n_{\mathrm{e}}^{2} d z=\left(1+\left\langle\delta_{n_{\mathrm{e}}}^{2}\right\rangle\right) \int \bar{n}_{\mathrm{e}}^{2} d z.
\label{eq:non_spherical_SB}
\end{equation}

\begin{table}[]
\centering
\begin{tabular}{lcc}
 & $\phi_{\infty}$ (95\% CL): & $|f_{R0}|$ (95\% CL): \\ \hline
Terukina et al. (2014): & $\lesssim 7 \times 10^{-5}$ $M_{\rm pl}$  &  $\lesssim 6 \times 10^{-5}$ \\
Wilcox et al. (2015): & $< 5.8 \times 10^{-5}$ $M_{\rm pl}$  & $<6 \times 10^{-5}$  \\
Wilcox et al. (2016) $\Lambda$CDM: & $<1.1 \times 10^{-4}$ $M_{\rm pl}$  & $< 1.1 \times 10^{-4}$  \\
Wilcox et al. (2016) $f(R)$: & $ <5.7 \times 10^{-5}$ $M_{\rm pl}$ & $< 5.5 \times 10^{-5}$ \\
This work: & $< 8 \times 10^{-5}$ $M_{\rm pl}$  & $<6.5 \times 10^{-5}$    \\ \hline
\end{tabular}%
\caption[Modified gravity constraints from previous works in the literature]{Modified gravity constraints from previous works in the literature compared with the constraints derived in this work. Note that constraints in \citet{Terukina2013} are rounded to a different significant figure, which gives the same general result to the one in \citet{Wilcox2015} despite the slightly different $\phi_{\infty} $values.}
\label{table:fR0_constraints}
\end{table}

\noindent Here $1+\left\langle\delta_{n_{\mathrm{e}}}^{2}\right\rangle$ is usually referred to as the clumping factor and it can be estimated observationally. As an example for the cluster Abell 1835, the clumping factor is $\sim 1.5$ \citep{Morandi2013}. This can then be used to calculate the effect on the estimates of the hydrostatic mass resulting in a factor of $\sim 1.2$. In summary, the systematics from the clumpiness of typical galaxy clusters can then be estimated to be of order of a few $\times 10 \%$. For a more accurate estimate a detailed study is required using observational and simulation data, which is out of the scope of this work. 

Another key point to discuss is how the quality of the data affects our constraints. More specifically, it is clear that in all of the discussed datasets the weak lensing data is the dominant source of uncertainty. This is the case, as measuring weak lensing is complicated and, even after stacking a significant number of clusters, the errors are relatively large when compared to the corresponding surface brightness errors. In addition, even after stacking, multiple outlier points remain. In order to investigate how these points affect our results we tested removing the first few lensing data points that are closest to the cluster center. We found that the outlier points did not have a significant effect on the best-fit parameters and the related constraints. To understand why, it is important to emphasize that we are fitting all the four datasets simultaneously rather than each dataset individually. Hence a small change in the best-fit profile of the shear data does not have a significant effect on the corresponding constraints. 

\section{Implications for the Gravitational Slip Parameter}

\subsection{Gravitational Slip in Galaxy Clusters}

The modified gravity tests described in the previous sections are model dependent -- i.e. the results depend on the model-specific assumptions and generally cannot be easily converted to the corresponding constraints on other models (except the $f(R)$ model, which is directly related to the chameleon scalar-tensor model via a conformal transformation). Recently there has been a lot of interest in exploring model-independent tests of modified gravity. These are tests that do not depend on a specific model and can be used to constrain deviations from GR in a way that allows to apply the constraints to a wide class of models. One such way of testing modifying gravity in a model-independent way is by measuring the gravitational slip parameter. This section discusses how our techniques can be adapted to calculated the gravitational slip parameter. In addition, the estimation of the constraints on the gravitational slip is calculated using different DES datasets.  

As mentioned, one way to parametrize deviations from GR is via the so-called gravitational slip parameter. The gravitational slip parameter is defined as the ratio of the two gravitational potentials appearing in equation \ref{eq:newtonian_gauge}, $\eta_{s} \equiv \Phi/\Psi$. More specifically, the gravitational slip parameter can be interpreted as the ratio between the effective gravitational coupling of light to the coupling of matter. In GR, $\eta_{s} = 1$ (in the absence of anisotropic stress), however, in a large class of modified gravity models the gravitational slip parameter deviates from unity.

A detailed study of constraining the gravitational slip parameter using simulated galaxy cluster data was done in \citet{Pizzuti2019}. This work, in particular, studied the viability of constraining $\eta_{s}$ using a combination of simulated strong and weak gravitational lensing data along with the data from galaxy dynamics. The key point presented in \citet{Pizzuti2019} is that deviation from $\eta_{s} = 1$ in the context of galaxy clusters is equivalent to the deviation between the dynamical and the lensing cluster masses. The key concepts discussed in \citet{Pizzuti2019} are summarized here in the context of our results presented in the previous sections.

The two gravitational potentials used in the definition of the gravitational slip parameter can be related to the properties of galaxy clusters. For instance, the cluster galaxy dynamics can be described by the Jeans equation:

\begin{equation}
\frac{\partial\left(\nu \sigma_{r}^{2}\right)}{\partial t}+2 \beta(r) \frac{\nu \sigma_{r}^{2}}{r}=-\nu(r) \frac{\partial \Psi}{\partial r},
\label{eq:jeans_equation}
\end{equation}

\noindent where $\nu(r)$ is the number density of tracers, $\sigma_{r}^{2}$ is the velocity dispersion along the radial direction and $\beta \equiv 1-\left(\sigma_{\theta}^{2}+\sigma_{\phi}^{2}\right) / 2 \sigma_{r}^{2}$, with $\sigma_{\theta}$ and $\sigma_{\phi}$ as the velocity dispersion along the angular directions. The potential $\Psi$ is given by the Poisson equation:

\begin{equation}
\nabla^{2} \Psi=4 \pi G \rho_{m},
\label{eq:poisson_equation}
\end{equation}

\noindent with $\rho_{m}$ as the total mass density in a cluster (dominated by the dark matter, gas and galaxy mass components). The total mass enclosed in some radius $R$ is simply:

\begin{equation}M_{\mathrm{tot}}(R)=4 \pi \int^{R}_{0} r^{2} \rho_{m}(r) \mathrm{d} r.
\label{eq:total_mass}
\end{equation}

\noindent The gravitational potential $\Psi$ can then be expressed as: 

\begin{equation}
\Psi(R)=G \int_{R_{0}}^{R} \frac{d s}{s^{2}} M_{\mathrm{dyn}}(s),
\label{eq:psi_potential}
\end{equation}

\noindent where we identified the total mass as the mass measured by galaxy kinematics. 

A similar expression can be found for the lensing mass. In this case, the geodesics of light respond to the sum of the two mentioned potentials: $\Psi + \Phi$, such that:

\begin{equation}
\nabla^{2}(\Phi+\Psi)=8 \pi G \rho_{\text {lens }}.
\label{eq:poisson_WL}
\end{equation}

\noindent The $\rho_{\rm lens}$ term here is the density corresponding to the lensing mass, which is given by:

\begin{equation}
M_{\mathrm{lens}}=\frac{r^{2}}{2 G} \frac{d}{d r}(\Phi+\Psi).
\label{eq:lensing_mass}
\end{equation}

\noindent Equations \ref{eq:poisson_WL} and \ref{eq:psi_potential} allow us to express the $\Phi$ potential as:

\begin{equation}
\Phi(r)=G \int^{r}_{0} \frac{d s}{s^{2}}\left[2 M_{\mathrm{lens}}(s)-M_{\mathrm{dyn}}(s)\right].
\label{eq:phi_potential2}
\end{equation}

\noindent Finally, this allows expressing the gravitational slip parameter in terms of the lensing and dynamical mass of a galaxy cluster:

\begin{equation}
\eta_{s}(r)=\frac{\bigintsss_{\, 0}^{r} \frac{d s}{s^{2}}\left[2 M_{\mathrm{lens}}(s)-M_{\mathrm{dyn}}(s)\right]}{\bigintsss_{\, 0}^{r} \frac{d s}{s^{2}} M_{\mathrm{dyn}}(s)}.
\label{eq:gravitational_slip}
\end{equation}

\noindent This expression shows that the gravitational slip essentially quantifies the deviation between the lensing and the dynamical masses at different radii. Note that the gravitational slip parameter can be easily related to the hydrostatic mass inferred from the intra-cluster gas studies, by noticing that the hydrostatic equilibrium equation (eq. \ref{eq:hydrostatic_equilibrium}) is related to the gravitational potential $\Psi$. In other words, the same potential that dictates the cluster galaxy kinematics also affects the intra-cluster gas, allowing us to treat the dynamical mass in equation \ref{eq:gravitational_slip} as equivalent to the hydrostatic mass, which is equal to the sum of $M_{\rm thermal} + M_{\rm non-thermal}$. Replacing $M_{\rm dyn}$ with $M_{\rm thermal} + M_{\rm thermal}$ in equation \ref{eq:gravitational_slip} allows us to calculate the gravitational slip parameter following an approach similar to the one described in the previous sections in this chapter. More specifically, the gravitational slip parameter is directly related to the validity of the hydrostatic equilibrium, i.e. the equivalence of the lensing and the hydrostatic masses in a galaxy cluster.

\subsection{Constraining the Deviations from the Hydrostatic Equilibrium and the Gravitational Slip Parameter}

\begin{figure}[ht!]
\centering
\makebox[\textwidth][c]{\includegraphics[width=1.1\textwidth]{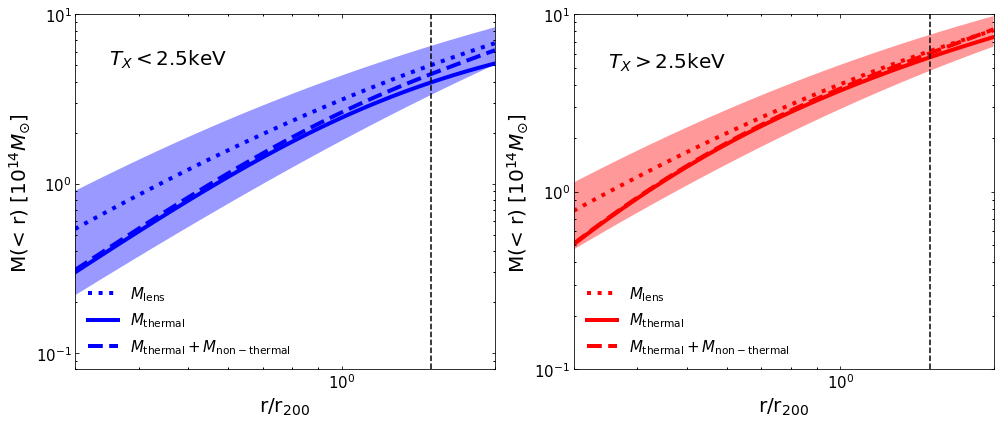}}%
\caption[Checking the validity of the assumption of hydrostatic equilibrium]{ A comparison of the hydrostatic and weak lensing masses from \citet{Wilcox2015}. The coloured bands in both plots correspond to the lensing mass uncertainty ($1$-$\sigma$) region. The mass profiles were extrapolated above the radii marked by the vertical dashed line to show that the masses also agree well in the outskirts of clusters if the non-thermal term is added.  
\label{figure:thermal_vs_lensing_mass}}
\end{figure}

The question of the validity of the hydrostatic equilibrium assumption has been investigated in detail in \citet{Terukina2013} and \citet{Wilcox2015}. In particular, cluster masses inferred by the weak lensing data were compared against the masses inferred from the X-ray data. The results are shown in figure \ref{figure:thermal_vs_lensing_mass}. The results show a general agreement between the gas and the weak lensing masses for the full radial range covered by the dataset. An important conclusion that can be drawn from figure \ref{figure:thermal_vs_lensing_mass} is that the non-thermal pressure effects are increasingly more important at large radii. However, if the non-thermal term is added, given the weak lensing errors, there is a good agreement between the lensing and the gas masses, justifying the hydrostatic equilibrium assumption.  

\begin{figure}[ht!]
\centering
\includegraphics[width=0.70\columnwidth]{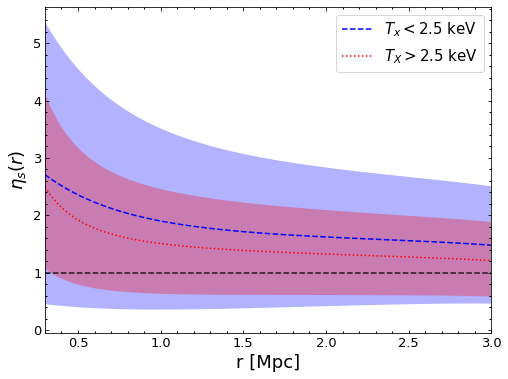}
\caption[The gravitational slip parameter calculated using the X-ray and the weak lensing data from \citet{Wilcox2015}.]{The gravitational slip parameter calculated using the mass profiles described in figure \ref{figure:thermal_vs_lensing_mass}. The dotted lines correspond to the mean values for each temperature bin. The x-axis variable was converted from $r/r_{200}$ to $r$ (in order to allow an easier comparison with the other results in the literature) by noting that $r_{200} \approx 2.2$ Mpc for this particular dataset. 
\label{figure:gravitational_slip}}
\end{figure}

The mass profiles shown in figure \ref{figure:thermal_vs_lensing_mass} can be used to calculate the gravitational slip parameter. In particular, $\eta_{s}$ is proportional to the integrated difference between the two mass profiles, leading to the results shown in figure \ref{figure:gravitational_slip}. The results indicate that the mean value of $\eta_{s}$ approaches 1 only at high $r$ values. However, given the dominant weak lensing errors, the value of $\eta_{s} = 1$ is well within the allowed region. This illustrates the two key issues when measuring $\eta_{s}$ using galaxy cluster data. First, the available weak lensing data has very high error bars, which result in poor constraints for the gravitational slip. In addition, the complex astrophysics happening in the core regions of clusters complicate the different mass estimates in those regions, hence the validity of hydrostatic equilibrium in the inner region of clusters has been debated extensively (eg. see \citet{Fabian1992, Peterson2006, Fujita2011}).

An interesting question to ask is whether stacking more clusters for the X-ray and the weak lensing profiles would bring the gravitational slip constraint errors close to those predicted by simulated data. This, however, is complicated by the lack of access to cluster data that contains both high-quality X-ray and weak lensing information. Hence, in order to estimate the errors on the gravitational slip resulting from a larger weak lensing dataset (which is dominates the errors in figure \ref{figure:gravitational_slip}), a significantly higher number of clusters were stacked when calculating the tangential shear profile. For this estimation, instead of the CFHTLenS data, the DES year 1 cluster catalogue was used hoping to get smaller weak lensing errors. In particular, DES Y1 Gold catalogue was used, which includes measurements of 137 million objects in 5 filters over 1800 square degrees of the sky as described in full detail in \citet{DES2018_Y1_GOLD}. The clusters were chosen such that the resulting dataset would be as similar as possible to the original 58 XCS-CFHTLenS clusters. In particular, as shown in figure \ref{figure:original_new_dataset}, the original 58 cluster dataset span redshifts between $0.1 \lesssim z \lesssim 0.9$ and temperatures between $0.1 \lesssim T \lesssim 5.0$, hence the new clusters were chosen from the DES data to have a similar redshift and temperature distribution. In addition, cluster richness was taken into account, by removing clusters that had richness significantly higher/lower than the mean richness of the original 58 cluster stack. More specifically, values larger or smaller than the mean value by 50\% were removed from the list. Other values, such as 25\% and 75\% were tried, but the effect on the resulting weak lensing profiles was not significant. A list of $\sim 100$ and $ \sim 1000$ clusters was then chosen in a way that the mean redshift and temperature values were as close as possible to $\langle z \rangle = 0.33$  and $\langle T \rangle = 2.3$ keV (i.e. the median values of the original dataset). Finally the results were split into two temperature bins as before. 

Once a list of clusters was obtained, the tangential shear profiles were obtained by using \textit{xpipe}, which is a software package that automates the pipeline for producing weak lensing profiles using DES data \citep{xpipe_github2020}. More specifically, \textit{xpipe} automates the stacking procedure by calculating the $r_{200}$ and $M_{200}$, rescaling each cluster to the same size, calculating the tangential shear and averaging the results in each radial bin. This resulted in two tangential shear profiles for $\sim$$100$ and $\sim$$1000$ clusters with errors significantly lower than those shown in figure \ref{figure:HW_dataset}. 

\begin{figure}[ht!]
\centering
\makebox[\textwidth][c]{\includegraphics[width=1.1\textwidth]{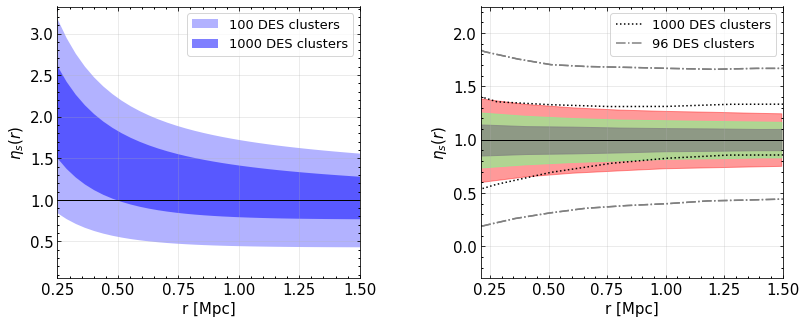}}%
\caption[Estimate of the gravitational slip constraints using $\sim$$100$ and $\sim$$1000$ stacked galaxy clusters.]{Analysis of the estimated gravitational slip constraints using stacked galaxy clusters. \textbf{Left:} the estimated galaxy cluster constraints calculated using the original 58 cluster dataset (figure \ref{figure:HW_dataset}, $T>2.5$ keV bin) with the weak lensing error bars rescaled according to the errors calculated by stacking 100 and 1000 DES year 1 clusters. \textbf{Right:} the estimated galaxy cluster constraints calculated using the original 58 cluster dataset X-ray data along with the 96 and 1000 DES year 1 cluster weak lensing data. The results are compared against the simulation results for 15 (red), 30 (green) and 75 (grey) clusters from \citet{Pizzuti2019} (adapted from figure 5). 
\label{figure:grav_slip_comparison}}
\end{figure}

Finally, the gravitational slip parameter was estimated in two different ways. In the first instance, we used the X-ray data from the 58 cluster stack ($T>2.5$ keV bin) and the corresponding 58 cluster weak lensing data, however, the lensing error bars were rescaled by a factor determined from the $100$ and $1000$ cluster stacks from the DES data. In more detail, this was done by splitting the 100 and the 1000 cluster stacks into radial bins and calculating the mean error bar size in each bin, which was then compared against the original 58 cluster stack error bars. The original dataset error bars were then rescaled by that factor to estimate how the errors would be reduced by stacking a significantly higher number of clusters. The second calculation was done by using the original 58 cluster X-ray data, but the weak lensing data was replaced entirely by the new weak lensing shear profiles. The obtained results were compared against the previously mentioned simulation results from \citet{Pizzuti2019}, where simulated data was used to derive the gravitational slip estimate from the $M_{\rm dyn}$  and $M_{\rm lens}$ masses. Figure \ref{figure:grav_slip_comparison} summarizes the results. In summary, the results indicate that stacking 1000 galaxy clusters significantly improves the gravitational slip constraints. Here it is important to emphasize that these constraints should not be taken as a rigid estimation of $\eta_{s}$, but rather as a \textit{back of an envelope} estimation of the associated errors, as we are using different clusters for the X-ray and the shear data. The results however are a good estimate of how competitive the constraints can become once high quality X-ray and weak lensing data becomes available from DES and future surveys. In comparison, the simulation results from \citet{Pizzuti2019} are much stronger; however, it is important to note that these results were produced using different methodology and using simulated data that is significantly less noisy than observational data. In order to improve our estimates in figure \ref{figure:grav_slip_comparison}, more high-quality X-ray data with the weak lensing counterpart is required from the newest DES data and future X-ray surveys.

\chapter{Testing Emergent Gravity on Galaxy Cluster Scales}
\label{ch:testing_EG}

Chapter \ref{ch:testing_EG} introduces a novel test of the theory of emergent gravity using the data described in chapter \ref{ch:modified_gravity}. The methods and techniques described in chapter \ref{ch:modified_gravity} are adapted to test a different type of a model hence illustrating that our tests can be generalized quite easily. In addition, a more detailed discussion of the various assumptions and systematics is given. 

The material in this chapter is primarily based on the results obtained in \citet{Tamosiunas2019}. Note that at the time of writing the paper, the newest 77 cluster stack data was not yet available. Hence, the results presented in this chapter were produced using the Coma Cluster data as described in \citet{Terukina2013}. The cluster stack results were produced using the original 58 cluster stack produced using CFHTLenS and XCS data as described in \citet{Wilcox2015}. Reproducing the tests with the newest 77 cluster stack data is left for future work.

\section{Motivations for Emergent Gravity}

Emergent gravity (EG) refers to a family of theories united by a common principle that gravity can be described as an emergent phenomenon. Ideas of such kind date back to the early work of Jacob Bekenstein, who, in a great stroke of ingenuity in 1973, demonstrated that black holes are thermodynamical objects with an entropy proportional to the area of the event horizon: 

\begin{equation}
S_{\mathrm{BH}}=k_{\mathrm{B}} \eta_{c} \frac{A}{\ell_{\mathrm{pl}}^{d-2}},
\label{eq:black_hole_entroypy}
\end{equation}

\noindent with $k_{B}$ as the Boltzmann constant, $\eta_{c}$ as a numerical constant, $A$ as the area of the event horizon, and $\ell_{\mathrm{pl}}^{d-2} \equiv \left(\hbar G / c^{3}\right)^{\frac{1}{d-2}}$ as the $d$-dimensional Planck length \citep{Bekenstein1973}. Soon after the results by Bekenstein were published the numerical constant was determined by Hawking to have the value of $\eta_{c} = 1/4$ \citep{Hawking1974, Hawking1974B}. This insight was shown to be far from just a mere analogy, as illustrated by the later work of Bekenstein, Hawking and others, which established a clear link between black hole physics and classical thermodynamics. In fact, one can define consistent laws of black hole mechanics which are analogous to the four laws of thermodynamics. The four laws can be summarized as follows \citep{Wald2010, Carlip2014}:

\begin{enumerate}[align=left] 
\addtocounter{enumi}{-1}
  \item The event horizon has a constant surface gravity $\kappa_{g}$ (for stationary black holes). In this respect a stationary black hole is comparable to a body in thermal equilibrium and $\kappa_{g}$ is comparable to temperature $T$ in classical thermodynamics. 
  \item The change in energy of a black hole is proportional to the changes to the area of the event horizon, the angular momentum and the electric charge, i.e.:  
  \begin{equation}
  d E=\frac{\kappa_{g}}{8 \pi} d A+\Omega d J+\Phi_{e} d Q,
  \label{eq:black_hole_thermodynamics_laws}
  \end{equation}
  
   \noindent where $\kappa_{g}$ is the surface gravity, $A$ is the area of the event horizon, $\Omega$ is the angular velocity, $J$ is the angular momentum, $\Phi_{e}$ is the electrostatic potential and $Q$ is the charge. As before, this law can be related to the first law of classical thermodynamics, which stems from the conservation of energy. 
  
  \item The area of the event horizon does not decrease with time: 
  \begin{equation}
  \frac{d A}{d t} \geq 0.
  \label{eq:second_law_bh_thermo}
  \end{equation}
  
  \noindent Hence the event horizon area is analogous to entropy in the second law of thermodynamics. It is important to note that later work by Hawking showed that black holes radiate, which over long enough periods of time can eventually lead to the decrease of the event horizon area.
  \item Black holes cannot have $\kappa_{g} = 0$. This indicates that one cannot produce black holes with naked singularities. The analogy with the classical thermodynamics here is a bit more subtle, but it turns out that this is indeed equivalent to the third law in classical thermodynamics, which states that as $T \rightarrow 0$ entropy is a well defined constant. 
\end{enumerate}

\noindent The exact nature of this correspondence between black hole and classical thermodynamics has been a matter of an ongoing debate since the formulation of the 4 laws. One fascinating possibility is that the analogies laid out in the laws above hint towards certain fundamental properties of gravity and spacetime. For instance, in classical thermodynamics it has been long known that macroscopic quantities ultimately have a microscopic nature. As an example, the temperature of a body can be directly related to the energy stored in the discrete microscopic degrees of freedom. Realizations of such kind historically led to the discovery of the discrete atomic nature of matter. Given the deep connections between gravitational systems and thermodynamics, as outlined above, a natural question to ask is whether spacetime itself has an underlying microscopic structure (sometimes referred to as \textit{atoms of spacetime}). Suspicions of such kind have only been strengthened by the discovery of the Fulling-Davies-Unruh effect, which allows accelerating observers to observe the vacuum as having a well-defined temperature \citep{Fulling1973, Davies1975,Unruh1976}. These observations have inspired a family of different approaches, which treat gravity as a phenomenon that emerges from the underlying microscopic dynamics that obeys the laws of thermodynamics. 

The research program of EG has resulted in a number of important breakthroughs in our understanding of the fundamental nature of gravity. A prime example of this is the result by Jacobson, which demonstrates that the Einstein field equations can be derived starting from general considerations related to the entropy-area relation (equation \ref{eq:black_hole_entroypy}) \citep{Jacobson1995}. More generally, later results by Padmanabhan showed that it is possible to derive the field equations for a large class of gravitational theories from the thermodynamic extremum principle \citep{Padmanabhan2007, Padmanabhan2008}. These and other recent successes of the emergent paradigm are discussed in full detail in \citet{Padmanabhan2015}.   

\section{Verlinde's Emergent Gravity}

\subsection{The Predictions of the Model}

One of the most recent additions to the family of emergent theories has been proposed by Eric Verlinde. Here we will lay out some of the key results of this approach primarily based on \citet{Verlinde2017, Brouwer2017, Tamosiunas2019}. From hereon, EG will refer specifically to the approach introduced in \citet{Verlinde2011} and \citet{Verlinde2017}. The figures and the results described in this chapter are the author's own (as described in \citet{Tamosiunas2019}) unless specified otherwise.  

In \citet{Verlinde2011} Newton's laws are derived starting from general considerations in statistical mechanics and the holographic principle. In particular, the mentioned work shows that gravitation can be described as an entropic force arising from the changes in the information associated with the positions of material bodies in a gravitational system. In addition, Verlinde specifies a method for deriving Einstein's field equations from general considerations along the same lines.  

A key notion in such an emergent description of gravity is the holographic principle, which states that the information in a volume of space can be thought of as encoded on a lower-dimensional boundary to the region. This principle has been originally inspired by the insight in black hole thermodynamics, that the information about the objects that have fallen into the hole might be stored entirely in the surface fluctuations of the event horizon \citep{Susskind1995}. A prime example of an application of the holographic principle is the AdS/CFT correspondence, which refers to a certain duality between string theory models described in anti-de Sitter space and conformal field theories \citep{Maldacena1999}. In his work Verlinde uses the holographic principle as a tool for relating changes in the configuration of masses to the corresponding change in entropy. In particular, the holographic principle allows us to generalize the ideas used in black hole thermodynamics to other gravitational systems. As a concrete example, one of the main results in \citet{Verlinde2011} shows how changes in the entropy of a gravitational system can be related to the changes in the gravitational potential acting on a test mass near a spherical mass distribution enclosed by a holographic screen:

\begin{equation}
    \frac{\Delta S}{n} = -k_{B}\frac{\Delta \Phi_{N}}{2c^{2}},
    \label{eq:EG_entropy_change}
\end{equation}

\noindent where $\Delta S$ is the change in entropy, $n$ is the number of bits of information stored on the holographic screen bounding the system, $\Delta \Phi_{N}$ is the change in the gravitational potential and $k_{B}$ and $c$ are the Boltzmann constant and the speed of light.

The more recent proposal by Verlinde extends these ideas in an attempt to describe gravity as an emergent force in cosmological scenarios \citep{Verlinde2017}. As previously discussed, the entropy-area relationship is of monumental importance in the EG model and can be used to derive the familiar laws of gravity (e.g. the Einstein field equations). However, in \citet{Verlinde2017} the author argues that due to the presence of positive dark energy in our Universe an extra contribution to the total entropy\footnote{Technically, this is the entanglement entropy of the underlying microscopic degrees of freedom (see \citet{Verlinde2017} for a more detailed explanation). From hereon the terms entropy and entanglement entropy are used interchangeably.} in the form of a volume law must exist. In particular, Verlinde argues that modifying the entropy-area relationship leads to extra gravitational effects that become important on scales set by the Hubble acceleration scale: $a_{0} = cH_{0}$. Another key achievement in \citet{Verlinde2017} is extending a number of important ideas in the EG paradigm, which are best described in anti-de Sitter space, to a more cosmologically realistic de Sitter space.

\begin{figure}[ht!]
  \centering
    \includegraphics[width=0.55\columnwidth]{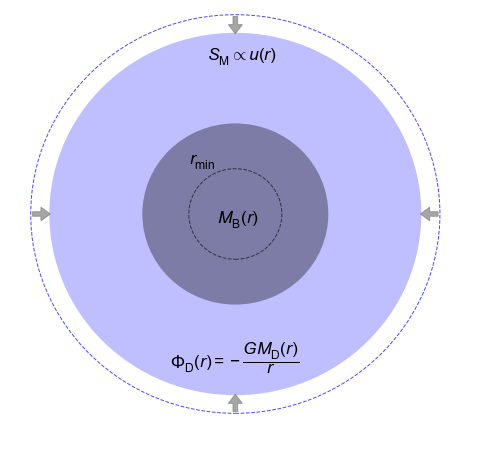}
    \caption[Emergent gravity effects]{Diagram illustrating the physical effects predicted by EG. Introducing a central baryonic distribution $M_{B}(r)$ (for instance a galaxy cluster here shown in dark purple) causes a reduction of the entanglement entropy $S_{M}$, which is quantified by the displacement field $u(r)$. This results in a central force and a potential $\Phi_{D}(r)$, which can be calculated using the scaling relation (equation \ref{eq:EG_main_prediction}). These effects are expected to become significant at radii larger than $r_{\rm min}$ (determined from equation \ref{eq:r_min}).  }
\label{figure:emergent_gravity_effects}
\end{figure}

In terms of extra gravitational effects, Verlinde shows that introducing a central baryonic mass distribution on galaxy and galaxy cluster scales results in the reduction of the total entanglement entropy of the system, which is equivalent to extra gravitational effects (i.e. a force pointing towards the matter distribution (see figure \ref{figure:emergent_gravity_effects})). These extra gravity effects are comparable in size to the effects usually associated with those of cold dark matter. 

\noindent The entropy change in a spherical system caused by introducing a spherical central distribution of baryonic matter $M_{B}$ can be expressed through the displacement field $u(r)$, such that:

\begin{equation}
    S_{M}(r) = \frac{u(r)A(r)}{V_{0}^{*}}
    \quad\mathrm{with}\quad
    V_{0}^{*}= \frac{2G \hbar}{cH_{0}},
    \label{eq:entropy displacement}
\end{equation}

\noindent where $S_{M}$ is the amount of displaced entropy, $A(r) = 4 \pi r^{2}$ is the surface area of the system, $G$ is the gravitational constant, $c$ is the speed of light, $\hbar$ is the reduced Planck constant and $H_{0}$ is the current value of the Hubble parameter.

To fully describe the entropy displacement effect by baryonic matter Verlinde draws a useful analogy with the effects of inclusions in elastic materials as described by the linear theory of elasticity. This seemingly random connection turns out surprisingly useful in calculating the changes in entropy caused by \textit{inclusions} of baryonic matter in gravitational systems. Namely, introducing inclusions into elastic materials causes strain $\epsilon$, which can be related to the change in entropy of the system. In \cite{Verlinde2017} the author notices that the effects of inclusions in elastic materials share certain similarities with the effects of baryonic matter distributions on the entanglement entropy in de Sitter space. An elasticity/gravity correspondence\footnote{For a better understanding of this correspondence see table 1 in \cite{Verlinde2017}} is then established to derive the exact result for the extra gravitational effects due to entropy displacement. Note that similarities between the theory of elasticity and gravity has been studied previously in some detail in the literature. As an example, in 1967 Sakharov introduced the idea of \textit{induced gravity}, which argues that gravitation emerges from quantum field theory in roughly the same sense that hydrodynamics or continuum elasticity theory emerges from molecular physics \citep{Visser2002}. More recently, Padmanabhan discusses a similar approach treating gravity as elasticity of spacetime \citep{Padmanabhan2004}. 

If spacetime in our system is mathematically treated as an incompressible elastic medium, the strain caused by the baryonic matter $\epsilon_{D}(r) = u'(r)$ is then given by: 

\begin{equation}
\int^{r}_{0}\epsilon_{D}^{2}(r')A(r')dr' = V_{M_{B}},
\label{eq:EG_strain}
\end{equation}

\noindent where $A$ is the area of a sphere we are integrating over and $V_{M_{B}}$ is a quantity related to the amount of entropy displaced by the baryonic matter distribution $M_{B}$\footnote{Note that $V_{M_{B}}$ is equal to the volume that would contain the amount of entropy that is removed by a mass $M_{B}$ inside a sphere of radius $r$, if that volume was filled with the average entropy density of the universe (see \citet{Brouwer2017} for a wider discussion).} and is given below in equation \ref{eq:EG_volume}. In \cite{Verlinde2017} it is shown that in de Sitter space $S_{M}(r) = (-2 \pi Mr)/\hbar$, which leads to $\epsilon_{D}(r)$ being given by: 

\begin{equation}
\epsilon_{D}(r) = \frac{8 \pi G}{cH_{0}}\frac{M_{D}(r)}{A(r)},
\label{eq:EG_strain2}
\end{equation}

\noindent where $M_{D}(r)$ refers to the \textit{apparent dark matter} distribution\footnote{Here we want to emphasize that in EG, there is only baryonic matter. However, gravity acts differently on large scales, which can be modeled as a consequence of an effective extra mass distribution, here called $M_{D}$. The effects of $M_{D}$ can then be compared against those of dark matter in standard cosmology.}. The $V_{M_{B}}$ term is given by:

\begin{equation}
V_{M_{B}} = \frac{8 \pi G}{3cH_{0}}M_{B}(r)r.
\label{eq:EG_volume}
\end{equation}

Substituting equations (\ref{eq:EG_strain2}) and (\ref{eq:EG_volume}) into (\ref{eq:EG_strain}) and integrating leads to the main result which is tested in this chapter:

\begin{equation}
M_{D}^{2}(r) = \frac{cH_{0}r^{2}}{6G} \frac{d(M_{B}(r)r)}{dr},
\label{eq:EG_main_prediction}
\end{equation}

\noindent where $M_{D}(r)$ is the apparent dark matter mass enclosed in $r$ and $M_{B}(r)$ is the baryonic mass. This can be interpreted as an effective dark matter distribution caused by gravity acting differently on large scales, rather than a new form of matter as in the $\Lambda$CDM framework. Hence Verlinde's EG offers an alternative solution to the problem of dark matter.

This result has a number of interesting consequences. For instance, computing the total acceleration due to $M_{B}$ and $M_{D}$, assuming that the baryonic mass is concentrated in the centre, leads to the result below that agrees well with the baryonic Tully-Fisher relation as seen in MOND-like theories:

\begin{equation}
\frac{GM_{D}(r)}{r^{2}} = \sqrt{\frac{a_{0}GM_{B}(r)}{6r^{2}}}, 
\label{eq:mond_limit}
\end{equation}

\noindent with $a_{0} = cH_{0}$ as the scale familiar from modified Newtonian dynamics \citep{Milgrom2016}. Similarly, applying equation \ref{eq:EG_main_prediction}, for extended mass distributions in galaxy clusters, highly reduces the missing mass problem, hence possibly offering an alternative to dark matter on galaxy and cluster scales.

The points outlined above illustrate that the EG model is of special interest in the context of dark matter on galaxy and galaxy cluster scales. In the point mass approximation limit, EG reproduces the original MOND predictions, while still leading to unique results for more general mass distributions. This is an attractive feature of the model, as it could potentially resolve some of the issues of the non-relativistic MOND framework, such as generally poor fit to data on galaxy cluster scales.   

The result in equation \ref{eq:EG_main_prediction} offers a testable prediction for a ratio between the dark matter and baryonic matter mass distributions with no free parameters. The rest of this chapter is dedicated to introducing a novel test of this relation on galaxy cluster scales. In addition, we will review the current theoretical criticisms and observational constraints of Verlinde's theory. However, before introducing the methods for testing this relation, it is worthwhile to lay out all the key assumptions under which equation \ref{eq:EG_main_prediction} is valid.  

\subsection{The Main Assumptions}

The current predictions of EG are valid only under a certain set of assumptions. Here I will list those key assumptions in the context of the observational data that is used to test the model: 

\begin{itemize}
    \item The EG predictions are only applicable for approximately spherically symmetric, sufficiently isolated and non-dynamic mass distributions. This means that, for instance, the Bullet Cluster would not be a valid test case for EG. This is rather unfortunate, as the Bullet Cluster offers a perfect test case for theories that predict a scaling relationship between the dark matter and baryonic matter distributions. Figure \ref{figure:bullet_cluster} clearly shows the bulk of the baryonic matter being centered in different parts of the merging cluster system when compared against the total dark matter distribution. This is clear evidence against scaling relations of the form of equation \ref{eq:EG_main_prediction}. However, given that the EG prediction was derived assuming spherical symmetry and the mass distributions being approximately static, merging cluster systems, such as the Bullet Cluster, cannot be used to test this particular model.  
    
    \item Since there is no rigid description of cosmology in EG yet, all the equations are only valid for the current value of the Hubble parameter, $H_{0}$ (i.e. $H(z)$ will be approximated as $H_{0}$ and only small redshift clusters will be considered). This also implies that Verlinde's theory is not capable of addressing such phenomena as the CMB and structure formation. However, note that more recent EG approaches, such as Hossenfelder's covariant approach (see \citet{Hossenfelder2017} and section \ref{section:covariant_EG}), could in principle address the CMB.  
    \item There is also no geodesic equation in EG as of yet, so a crucial assumption will be made that weak lensing works in EG the same way as in GR. In particular, following the work in \citet{Brouwer2017}, it will be assumed that the extra gravity effects predicted by the model affect the paths of photons in the same way as dark matter does in GR. In turn, this implies that dark matter is distributed according to equation \ref{eq:EG_main_prediction}, which allows us to derive the weak lensing predictions. Future theoretical and observational work will be required to test the validity of this assumption. 
    \item As discussed in \citep{Brouwer2017}, the effects of EG are only expected to become important in the regime where the volume law contribution to the total entropy ($S \propto V$) is significantly larger than the entropy displaced by baryonic matter $M_{B}$. This, following equation 18 in \citet{Brouwer2017}, is expressed by introducing a minimal radius, $r_{\rm min}$, above which we expect the EG effects to become noticeable, as described by the following inequality:

\begin{equation}
r > \sqrt{\frac{2M_{B}(r)G}{cH_{0}}}.
\label{eq:r_min}
\end{equation}

    Solving equation \ref{eq:r_min} gives the value for $r_{\rm min}$. Table \ref{table:r_min_table} lists the typical values for $r_{\rm min}$ for various systems of different sizes and masses. 
    
    \begin{table}[h!]
        \centering
        \begin{tabular}{lllll}
         \textbf{Scale}& \textbf{Typical mass ($\boldsymbol{\mathrm{M\textsubscript{\(\odot\)}}}$)} & \textbf{Typical size (Mpc)}  & \textbf{$\boldsymbol{r_{\rm min}}$ (Mpc)} &  \\ \hline \hline
         Solar system& 1.0014   & $5.8 \times 10^{-10}$ & $2 \times 10^{-8}$  &  \\
         Galaxy& $10^{10} - 10^{11}$ & $3-6 \times 10^{-2}$  & $2-6 \times 10^{-3}$ &  \\
         Galaxy cluster&$10^{14} - 10^{15}$& $2-10$  & 0.2 - 0.64&  \\ 
         Coma Cluster & $2.2 \times 10^{13}$ & $6$ & 0.094 &  \\ 
         Cluster stack& $1.3 \times 10^{13}$  &  4  &  0.073  &  \\ \hline
        \end{tabular}
        \caption[The typical values of the $r_{\rm min}$ parameter]{ The typical sizes and average masses of different objects along with the values of $r_{\rm min}$ assuming the point mass approximation. For the Coma Cluster and the cluster stack (see section \ref{section:EG_stacked_clusters}), we have chosen the mean mass in the region of interest covered by our data rather than the full mass (see sections \ref{section:EG_coma_test} and \ref{section:EG_stacked_clusters} for more details). The typical sizes were chosen in the same manner.}
        \label{table:r_min_table}
    \end{table}

\end{itemize}

\section{Testing Emergent Gravity}

\subsection{Testing Emergent Gravity with the Coma Cluster}
\label{section:EG_coma_test}

Galaxy clusters, being the largest gravitationally bound systems, offer a natural setting for testing models of gravity. Having regions of high and low density as well as a mass distribution dominated by dark matter, clusters have been used extensively for testing models with screening mechanisms and comparing the predictions with general relativity. In this section an approach similar to the one developed in \citet{Terukina2013} and \citet{Wilcox2015} is used, where chameleon and $f(R)$ gravity models were tested in the Coma Cluster as well as a 58 cluster stack coming for CFHTLenS and XCS surveys. More specifically, in these works multiple probes are used to constrain the modified gravity effects in the outskirts of galaxy clusters under the assumption of hydrostatic equilibrium. 

Here we use the intracluster gas temperature profile to determine the baryonic mass distribution in the Coma Cluster and to calculate the predicted weak lensing signal, which is then compared with the actual weak lensing data. The same procedure is done for the standard model (GR + cold dark matter described by a Navarro-Frenk-White profile) and the EG model. The results are then compared in terms of the $\chi^{2}$ and the Bayesian information criterion (BIC) values. 

\begin{figure}[ht!]
  \centering
    \includegraphics[width=0.80\columnwidth]{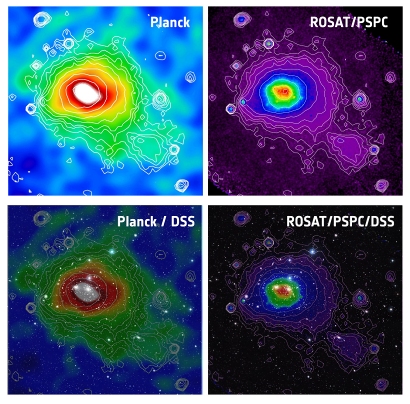}
    \caption[The Coma Cluster]{The Coma Cluster (Abell 1656). \textbf{Top left:} the cluster as seen by using the Planck data for the SZ effect. \textbf{Top right:} the cluster as seen by using the X-ray data from the ROSAT PSPC survey. \textbf{Bottom left:} the contours from the figure on the top left superimposed on a wide-field optical image of the Coma Cluster from the Digital Sky Survey. \textbf{Bottom right:} the contours from the figure on top right superimposed on a wide-field optical image from the Digital Sky Survey. The colours in all images correspond to the intensity of the signal, while the contours are the X-ray signal overlaid. Image courtesy of the ESA/LFI \& HFI Consortia, the Max Planck Institute and the Digital Sky Survey \citep{Coma2020}.    }
\label{figure:coma_cluster}
\end{figure}

The Coma Cluster (Abel 1656) is a large well-studied nearby ($z =0.0231$) galaxy cluster with over 1,000 identified galaxies \citep{Gavazzi2009}. The cluster has an extensively-studied mass distribution and has been the subject of numerous weak lensing and X-ray studies. Figure \ref{figure:coma_cluster} shows the underlying distribution of the Coma Cluster as seen in SZ and X-ray data. It is important to note that the cluster is not entirely spherical and evidence of substructures of various sizes can be seen distributed around the main mass distribution of the cluster. 

The equations below illustrate how the temperature profile of the Coma Cluster can be used to determine the total mass distribution and, in turn, to calculate the predicted weak lensing signal. Assuming hydrostatic equilibrium we can relate pressure to the mass using the same equation as described in chapter \ref{ch:modified_gravity}:

\begin{equation}
\frac{1}{\rho_{\rm gas}(R)} \frac{dP_{\rm total}}{dr}  = -\frac{GM(<r)}{r^{2}},
\label{eq:hydrostatic_equilibrium2}
\end{equation}

\noindent where $\rho_{\rm gas}$ is the gas density, $P_{\rm total}$ is the total pressure and $M(<r)$ is the mass enclosed in radius $r$. This allows us to calculate the gas temperature corresponding to the thermal pressure term by using the ideal gas law: $P_{\rm thermal} = n_{\rm gas}kT_{\rm gas}$. Integrating equation \ref{eq:hydrostatic_equilibrium2} gives: 

\begin{equation}
T_{\rm gas}(r) =    -\frac{m_{p}\mu}{n_{e}(r)k}\Big( \int^{r}_{0} n_{e}(r') \frac{GM(<r')}{r'^{2}}  + P_{\rm thermal,0} \Big) dr',
\label{eq:gas_temperature}
\end{equation}

\noindent where we switched to electron number density and $\mu$ is the mean molecular weight, $m_{p}$ is the proton mass, $n_{e}(r)$ is the electron number density and the last term, the central pressure, is an integration constant. As before, for a fully ionised gas, the mean molecular weight is given by $\mu = 0.59$. The non-thermal pressure terms can be derived in the same manner (as already discussed in section \ref{section:hydrostatic_equilibrium}). Equation (\ref{eq:gas_temperature}) then allows us to determine the underlying mass distribution in the Coma Cluster, given that we have a way to measure the temperature accurately. 

In this work we adopted the standard beta-model electron density profile \citep{Cavaliere1976}. The baryonic mass distribution is then given by:

\begin{equation}
M_{B}(<r) = M_{\rm gal}(<r) + 4 \pi m_{a} \int^{r}_{0} n_{e}(r') r'^{2}dr',
\label{eq:baryonic_mass}
\end{equation}

\noindent where we summed the total stellar galaxy mass with the intracluster gas mass and $m_{a}$ is the average mass of an atom in the cluster gas, given by $2m_{H}/(1 + X)$ where $m_{H}$ is the Hydrogen mass and $X$ is the mass fraction of the Hydrogen atoms. 

\begin{figure}[ht!]
  \centering
    \includegraphics[width=0.75\columnwidth]{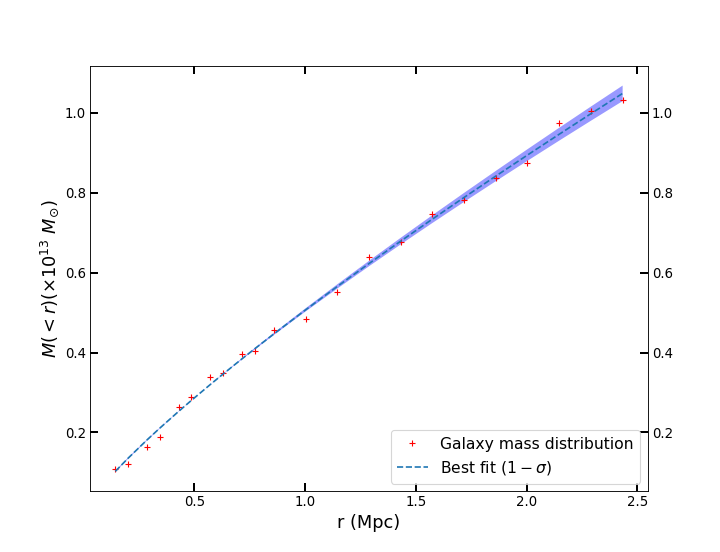}
    \caption[The Coma Cluster galaxy mass distribution]{Galaxy mass distribution of the Coma Cluster with the 1-$\sigma$ errors shown as the blue band. }
\label{figure:coma_cluster_galaxy_masses}
\end{figure}

In order to estimate the galaxy mass distribution in the Coma Cluster, we queried the SDSS data catalogue (Data Release 14) for the median estimate of the total stellar masses of galaxies located within the 180 arcminute diameter around the central point of the cluster for $0.01<z<0.05$ \citep{SDSS2018}. This region was then split into radial bins of 5 arcminutes, and for each cylindrical shell we summed the stellar masses for all the detected galaxies. This results in a galaxy mass distribution in a spherical region of $r \simeq 2.5$ Mpc around the centre of the cluster. Figure \ref{figure:coma_cluster_galaxy_masses} shows the results for the galaxy mass distribution. Summing the stellar galaxy and the X-ray emitting gas mass distributions gives a good measure of the total baryonic mass distribution, which can then be used to calculate the total mass distribution using eq. (\ref{eq:EG_main_prediction}). Finally, having obtained the total mass distribution for the cluster, we have all that is needed to compute the weak lensing predictions.

In order to compare the predictions from EG with those from standard cosmology (GR + dark matter), we chose to describe the dark matter distribution in the cluster by the NFW profile:

\begin{equation}
M_{\rm NFW}(<r) = 4 \pi \rho_{s} r_{s}^{3} \Big( \ln(1 + r/r_{s}) - \frac{r/r_{s}}{1+r/r_{s}} \Big),
\label{nfw1}
\end{equation}

\noindent where $\rho_{s}$ is the characteristic density and $r_{s}$ is the characteristic scale \citep{Wright1999}. This was then used to calculate the total mass in the cluster and, in turn, to predict the weak lensing profile. Note that in the case of EG, the model assumes only the existence of baryonic matter and the \textit{apparent} dark matter effects are fully described by equation \ref{eq:EG_main_prediction}. 

Following the approach taken by \citet{Brouwer2017} and using the equations described in \citet{Wright1999}, we calculated the weak lensing profiles as follows:

\begin{equation}
\gamma_{t}(r) = \frac{\bar{\Sigma}(r) - \Sigma(r)}{\Sigma_{c}},
\label{eq:tangential_shear_coma}
\end{equation}

\noindent where $\gamma_{t} (r)$ is the tangential shear, while $\Sigma (r)$ and $\Sigma_{c}(r)$ are correspondingly the surface density and critical surface density (see section \ref{section:cluster_weak_lensing} for more information). The surface density of a given radial density distribution is given by:

\begin{equation}
\Sigma(r) = \int^{\infty}_{-\infty} \rho(r) dr = \int^{\infty}_{-\infty} \rho(R,z) dz,
\label{eq:surface_density_coma}
\end{equation}

\noindent where we switched to cylindrical coordinates ($R$, $\phi$, $z$) centered on the central point of our cluster. $\Delta \Sigma(R)$ for both baryonic and apparent dark matter can be calculated using the general expression:

\begin{equation}
\Delta \Sigma(R) = \bar{\Sigma}(<R) - \Sigma(R) = \frac{2\pi \bigintsss_{\;0}^{R} R' \Sigma(R') dR'}{\pi R^{2}} - \Sigma(R).
\label{eq:delta_Sigma_coma}
\end{equation}

\noindent In the case of EG, the shear equations are then given by:

\begin{equation}
\gamma_{t}(r) = \frac{\Delta \Sigma_{EG}(R)}{\Sigma_{c}} = \frac{\Delta \Sigma_{B}(R) + \Delta \Sigma_{D}(R)}{\Sigma_{c}},
\label{eq:tantential_shear_EG_coma}
\end{equation}

\noindent where we have split $\Delta \Sigma$ into contributions from baryonic and apparent dark matter for the surface density.

Having laid out the main equations at this point it is worth noticing that in order to derive the total mass distribution of the cluster we need to choose a way of parametrizing the electron number density $n_{e}(r)$. As previously, this is done by using the simple isothermal beta profile of the following form: $n_{e} = n_{0}(1 + (r/r_{1})^{2})^{b_{1}}$. The only free parameters for the EG model appearing in the equations above are then $n_{0}$, $r_{1}$, $b_{1}$ and $T_{0}$ (central temperature). On the other hand (given our assumption that dark matter is distributed according to the NFW profile), for the GR model we have the following free parameters: $n_{0}$, $r_{1}$, $b_{1}$ and $T_{0}$, $c_{v}$ and $M_{200}$ (where the last two parameters refer to concentration and the mass enclosed by $r_{200}$). The values for the free parameters were then obtained by looking for solutions that fit the temperature profile data and, at the same time, produce weak lensing predictions which agree well with the observational data (in other words, both datasets were fit simultaneously by minimizing the combined value of $\chi^{2}_{T_{\rm gas}} + \chi^{2}_{\gamma_{t}}$). The data used included the X-ray temperature profile (combined from \citet{Snowden2008} and \citet{Wik2009}) and the weak-lensing profile (\citet{Gavazzi2009}, \citet{Okabe2010}) of the Coma Cluster.

\begin{figure}
\begin{subfigure}{.5\linewidth}
\centering
\includegraphics[width=.99\textwidth]{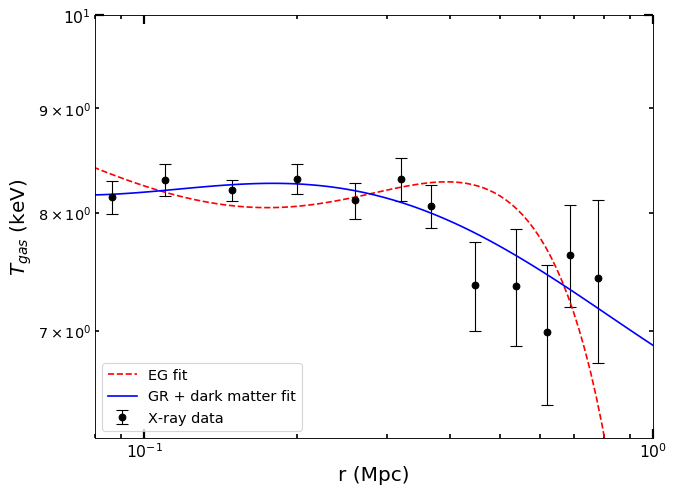}
\caption{\small Gas temperature fit in EG and GR}
\label{fig:mean and std of net14}
\end{subfigure}%
\begin{subfigure}{.5\linewidth}
\centering
\includegraphics[width=.95\textwidth]{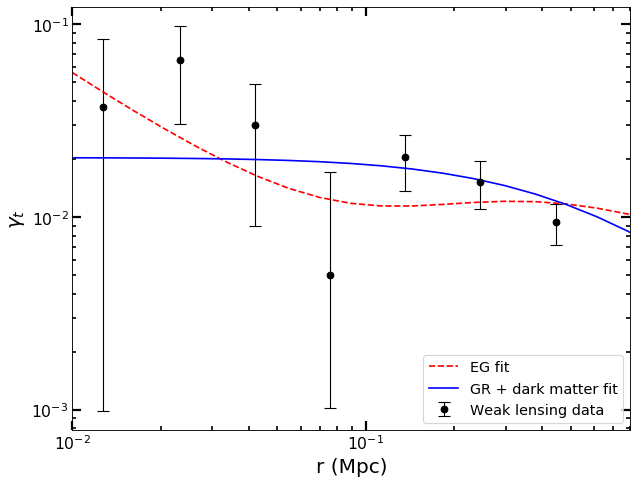}
\caption{\small Weak-lensing fit in EG and GR}
\label{fig:mean and std of net24}
\end{subfigure}\\[1ex]
\begin{subfigure}{0.5\linewidth}
\centering
\includegraphics[width=.97\textwidth]{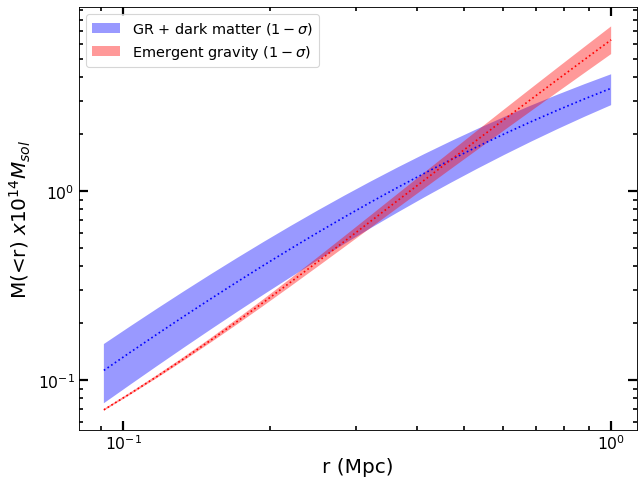}
\caption{\small Total mass profiles in the two models}
\label{fig:fig:mean and std of net44}
\end{subfigure}
\begin{subfigure}{.5\linewidth}
\centering
\includegraphics[width=.99\textwidth]{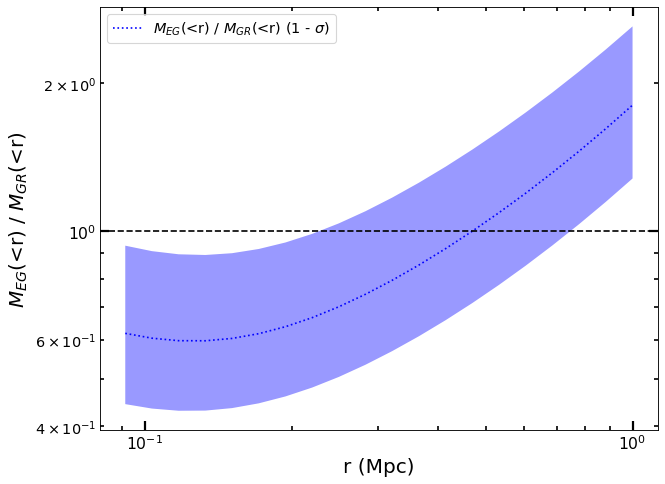}
\caption{\small Ratio of the mass profiles}
\label{fig:mean and std of nets}
\end{subfigure}
\caption[The results for the test of emergent gravity using the Coma Cluster]{A comparison of the EG and the standard model (GR + dark matter described by the NFW profile) results. Figure $a$ shows the gas temperature fit for both models. Figure $b$ shows the weak lensing fits. Figure $c$ shows the mass distributions calculated using the best-fit parameters for both models (with the contours corresponding to the 1-$\sigma$ confidence intervals). Figure $d$ shows the ratio of the two mass distributions. The goodness of fit statistics are summarized in table \ref{table:coma_goodness_of_fit}. }
\label{fig:coma_main_results}
\end{figure}

The data fitting was performed by minimizing the combined residuals using the limited memory Broyden–Fletcher–Goldfarb–Shanno (L-BFGS) algorithm available from the \textit{SciPy} python library \citep{Scipy2020}. The 1-$\sigma$ confidence intervals were determined using the in-built features of the \textit{SciPy.optimize} library, which use the estimated inverse Hessian matrix to calculate the standard deviation of each best-fit parameter. The $\chi^{2}$ values were calculated using the standard formula: $\chi^{2} = \sum_{i} (C_{i} - O_{i})^{2}/\sigma_{i}^{2}$, where $C_{i}$ refers to the calculated values, $O_{i}$ to the observed values, $\sigma_{i}$ to the variance at a given data point. The covariance matrix here was assumed to be diagonal, however, in the case of the cluster stack data, we used the full covariance matrix. The best-fit results for the standard model (GR + dark matter) and EG results are summarized in table \ref{table:coma_results} and figure \ref{fig:coma_main_results}. The goodness of fit statistics are given in table \ref{table:coma_goodness_of_fit}.

\newcommand\T{\rule{0pt}{3.0ex}}       
\newcommand\B{\rule[-2.0ex]{0pt}{0pt}} 

\begin{table}[ht!]
\noindent\makebox[\textwidth]{%
\begin{tabular}{ccccccc}
\centering
& $\boldsymbol{n_{0}}$ \textbf{($\mathrm{\mathbf{cm}^{-3}}$)} & $\boldsymbol{r_{1}}$ \textbf{(Mpc)} & $\boldsymbol{b_{1}}$  & $\boldsymbol{T_{0}}$ \textbf{(keV)} & $\boldsymbol{M_{200}}$ ($\boldsymbol{\mathrm{M\textsubscript{\(\odot\)}}}$)  & $\boldsymbol{c_{v}}$  \\ \hline \hline
\textbf{GR:}                   &  $4.2^{+0.21}_{-0.17} \times 10^{-3}$ \T \B  &$0.07^{+0.04}_{-0.04}$    & $-0.201^{+0.512}_{-0.512}$    &  $8.77^{+0.60}_{-0.61}$   & $2.39^{+1.18}_{-1.16} \times 10^{14}$                        &     $ 4.68^{+1.37}_{-1.36}$                   \\
\textbf{EG:}        &  $3.2^{+0.21}_{-0.20} \times 10^{-3}$ \T \B   &  $0.26^{+0.026}_{-0.025}$   &  $-0.615^{+0.056}_{-0.062}$   &  $9.18^{+0.13}_{-0.14}$  &            n/a            &         n/a               \\ \hline
\end{tabular}}
\caption[The best-fit parameters for the Coma Cluster test of emergent gravity]{The best-fit parameters for the standard model (GR + dark matter) and the model with gravity behaving according to EG fitted to the Coma Cluster data (see figure \ref{fig:coma_main_results}). }
\label{table:coma_results}
\end{table}

\begin{table}[ht!]
\centering
\begin{tabular}{lcccccc}
                    & \textbf{$\chi_{T_{gas}}^{2}$} & \textbf{$\chi_{\gamma_{t}}^{2}$}  & $N_{\rm d.o.f.}^{T_{gas}}$ & $N_{\rm d.o.f.}^{\gamma_{t}}$ & \textbf{$\mathrm{BIC}_{ T_{gas}}$} & \textbf{$\mathrm{BIC}_{\gamma_{t}}$}  \\ \hline \hline
\textbf{GR:\,\,\,\,\,\,\,\,\,\,\,\,} &   0.9   &  1.3                                             &    8.0                    &  5.0               &   0.8 & 16.7            \\
\textbf{EG:\,\,\,\,\,\,\,\,\,\,\,\,} &      3.0  &                                 1.7              &   9.0                     &    4.0            & 18.1  &    6.2         \\ \hline
\end{tabular}
\caption[Goodness of fit statistics for the Coma Cluster test of emergent gravity]{Goodness of fit statistics. BIC is the Bayesian information criterion statistic. $N_{\rm d.o.f.}$ corresponds to the number of degrees of freedom, i.e. the difference between the number of data points and the number of free parameters being fit (i.e. parameters seen in table \ref{table:coma_results}). The $\chi^{2}$ values here correspond to the reduced chi-squared statistic (i.e. chi-squared per d.o.f.). Note that the rather high $\chi_{\gamma_{t}}^{2}$ for the GR fit originates due to the outlier point at around $8 \times 10^{-2}$ Mpc. Removing the outlier point reduces the chi-squared statistic to $\chi^{2}_{\gamma_{t}} = 0.89$. Also note that the combined BIC values are $\mathrm{BIC}_{EG} = 21.8$ and $\mathrm{BIC}_{GR} = 20.3$ with no outlier points removed. With the outlier point removed, $\mathrm{BIC}_{EG} = 20.1$ and $\mathrm{BIC}_{GR} = 20.8$. Note, however, that the BIC difference in both cases is too small to be conclusive in determining the preferred model for this dataset. } 
\label{table:coma_goodness_of_fit}
\end{table}

The Coma Cluster results above indicate that EG is capable of producing fits that are generally comparable to the standard model fits and are in agreement with the observational data (within the shown uncertainties). The best-fit parameters from the gas temperature and the weak lensing data then result in mass distributions for the two models that are in agreement for $250$ kpc $< r <$ 700 kpc. The calculated mass distributions can be compared with the other results in the literature, such as \citet{Brownstein2006}, where the total mass profile was determined using X-ray data or \citet{Lokas2003}, where elliptical galaxy velocity moments were used instead. In general, our GR + dark matter profile, within the given uncertainties, is in good agreement with the mentioned results from the literature, with the exception of around $r\sim 1$ Mpc, where the profiles in the mentioned papers fall between our EG and GR results. Overall this indicates that the EG result underestimates the total mass distribution for $r \lesssim 250$ kpc and overestimates it for $r \gtrsim 800$ kpc given 1-$\sigma$ confidence. The values in table \ref{table:coma_goodness_of_fit} indicate that, despite requiring more free parameters, GR is still the preferred model according to the  $\chi^{2}$ analysis. The BIC analysis is not conclusive, with both models having very similar BIC values. This is the case, as, even though EG has a poorer fit to the temperature data, it has significantly better fit to the weak lensing data and less free parameters. 

There are, however, a number of important issues that need to be discussed in terms of using the Coma Cluster data for testing EG. In particular, as is shown in figure \ref{figure:coma_cluster}, the cluster is not exactly spherical and is not completely isolated for external mass distributions. Recent investigations in the structure of the Coma Cluster show that Coma is a typical example of a $z = 0$ cluster in terms of its internal kinematics. However, the X-ray temperature in the cluster has been shown to be significantly higher than in a sample of clusters of similar masses \citep{Pimbblet2014}. In addition, as shown in figure \ref{fig:coma_main_results}, the data for the Coma Cluster is limited, especially for the weak lensing profile. This is generally true when using single galaxy cluster data, as acquiring high accuracy weak lensing data is difficult. Hence, in order to avoid various biases and problems due to non-spherical symmetry, a test of the EG model with the data coming from 58 stacked galaxy clusters is introduced in the next section.

\subsection{Testing Emergent Gravity with Stacked Galaxy Clusters}
\label{section:EG_stacked_clusters}

\begin{figure}[ht!]
\centering
  \includegraphics[width=0.95\linewidth]{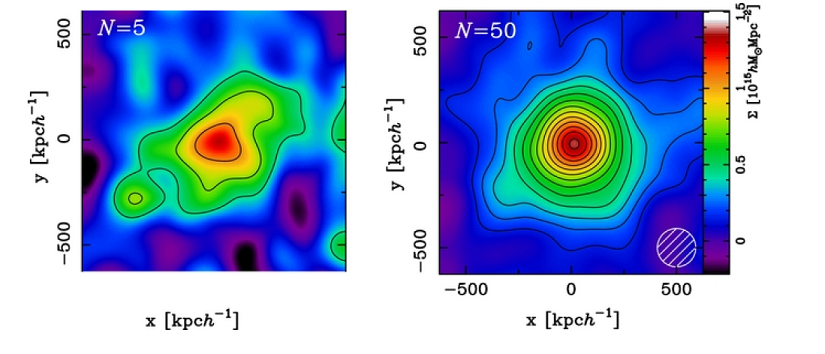}  		  
  \caption[An illustration of the effects of stacking galaxy clusters]{An illustration of the effects of stacking galaxy clusters (starting with a sample of 5 and ending with 50) on the density distribution and shape. Adapted from \citet{Okabe2013}.}
  \label{fig:stacking_clusters}
\end{figure}

Stacking multiple galaxy clusters allows us to form a dataset that is representative of typical galaxy cluster properties at a given redshift. Figure \ref{fig:stacking_clusters} illustrates the effects on the projected mass distribution due to stacking 50 galaxy clusters.   

To mitigate some of the mentioned issues with using the data from the Coma Cluster and to test the effects of EG with a larger sample of galaxy clusters we followed an approach similar to that taken in chapter \ref{ch:modified_gravity}, where 58 clusters with redshifts ranging between $0.1 < z < 1.2$ were stacked using X-ray (from the XMM Cluster Survey) and weak lensing data (from the Canada France Hawaii Telescope Lensing Survey) \citet{Mehrtens2012, Erben2013}. Stacking clusters in such a way averages away most irregularities in shape and density and provides an approximation to an average galaxy cluster. In addition, the signal to noise ratio is improved. More importantly, stacking multiple well-isolated low redshift clusters produces a perfect dataset to test the predictions of EG.  

The dataset used to test EG consisted of the original 58 cluster stack, as described in detail in chapter \ref{ch:modified_gravity} and in \citet{Wilcox2015}. The cluster stack has a number of important properties in the context of the assumptions under which EG predictions are significant. In particular, most galaxy clusters in the dataset are isolated from the other nearby mass distributions (see figure 4 in \citet{Wilcox2015}). In addition, our dataset consists of clusters with a mean redshift of $z \approx 0.33$, justifying the assumption that we can neglect the effects of varying the Hubble parameter $H(z)$ in our test. Finally, the cluster stack has been binned in terms of temperature, to approximately separate it into galaxy groups and galaxy clusters. This allows us to investigate how well the theory in question works for objects of significantly different masses.  

\begin{figure}[!ht]
    \centering
    \subfloat{{\includegraphics[width=6.2cm]{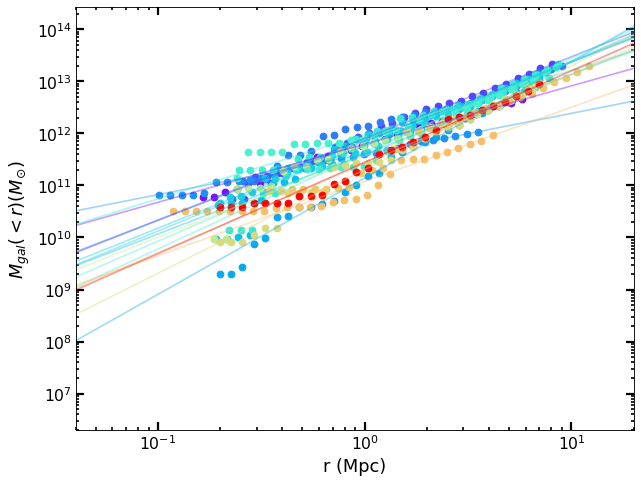}} }
    \qquad
    \subfloat{{\includegraphics[width=6.3cm]{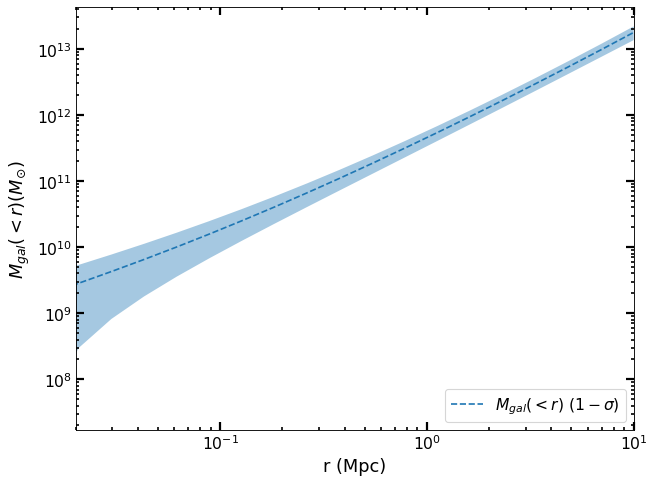}} }
    \caption[The galaxy mass distribution for the cluster stack]{Results for the galaxy mass distribution of the 58 cluster stack. The result on the left contains galaxy masses $M_{gal}(<r)$ at different $r$ for each cluster in the stack along with the extrapolated best-fit lines to bring the measurements to the same scale. The figure on the right shows the mean and the standard deviation of the cluster stack galaxy masses.}
    \label{fig:cluster_galaxy_masses}
\end{figure}

In order to determine the galaxy mass distribution for our cluster stack, we queried the CFHTLenS survey catalog (as described in \citet{Erben2013}) for each individual cluster following a similar procedure as before for the Coma Cluster (the main difference being that we adjusted the angular region that we queried based on the distance to each cluster). In particular, for each cluster the galaxy stellar masses were summed in concentric cylindrical shells (allowing the redshift to vary in each direction by $\sim z/4$). The results for each cluster were then linearly fitted and extrapolated to cover the same range. Finally, we then averaged over the masses for each cluster for each value of radii to determine the mean galaxy mass and the corresponding uncertainty. Figure \ref{fig:cluster_galaxy_masses} illustrates the procedure and the obtained results.

As previously, we used equation \ref{eq:tangential_shear_coma} to calculate the tangential shear profiles. In order to determine the underlying baryonic mass distribution the projected surface brightness data was fit by using equation \ref{eq:surface_brightness3}. This then results in the same free parameters of $n_{0}$, $r_{1}$, $b_{1}$ and $T_{0}$ for EG plus and extra two free parameters $c_{v}$ and $M_{200}$ for the standard model (GR + dark matter) due to the assumption of the NFW profile. The free parameters were determined by simultaneously fitting the surface brightness and the weak lensing datasets. The total mass profiles were then calculated using the obtained best-fit parameters. The results were compared with the analogous results calculated in GR. The best-fit was performed using the non-linear least-squares minimization using the python LmFit library \citep{Newville2014}. In particular, the Levenberg–Marquardt algorithm was used to determine the best-fit parameter values along with the corresponding confidence limits. 

The goodness of fit was evaluated by following the approach taken in chapter \ref{ch:modified_gravity} and the appendix A in \cite{Wilcox2015}. In particular, for the weak lensing data we approximated the covariance matrix as diagonal. For the surface brightness data the covariance matrix was included in the $\chi^{2}$ calculations to account for the correlations between the surface brightness radial bins. The results (split into two temperature bins) are summarized in table \ref{table:cluster_stack_params}. The goodness of fit statistics are summarized in table \ref{table:cluster_stack_goodness_of_fit}.

\begin{table}[ht!]
\noindent\makebox[\textwidth]{%
\begin{tabular}{ccccccc}
\centering
& $\boldsymbol{n_{0}}$ \textbf{($\mathrm{\mathbf{cm}^{-3}}$)} & $\boldsymbol{r_{1}}$ \textbf{(Mpc)} & $\boldsymbol{b_{1}}$  & $\boldsymbol{T_{0}}$ \textbf{(keV)} & $\boldsymbol{M_{200}}$ ($\boldsymbol{\mathrm{M\textsubscript{\(\odot\)}}}$)  & $\boldsymbol{c_{v}}$  \\ \hline \hline
\textbf{GR} (Bin 1):  &$5.5^{+1.8}_{-1.8} \times 10^{-3}$ \T \B  &$0.023^{+0.009}_{-0.009}$    & $-0.59^{+0.04}_{-0.04}$    &  $7.7^{+5.5}_{-5.3}$   &$4.0^{+2.2}_{-2.0} \times 10^{14}$                        &     $7.00^{+1.37}_{-1.49}$                  \\
\textbf{GR} (Bin 2):        &  $8.9^{+1.8}_{-1.8} \times 10^{-3}$ \T \B   &  $0.021^{+0.008}_{-0.009}$   &  $-0.57^{+0.04}_{-0.04}$   &  $6.6^{+5.5}_{-5.3}$  &           $9.61^{+2.52}_{-2.47} \times 10^{14}$            &         $ 4.95^{+1.42}_{-1.51}$             \\
\textbf{EG} (Bin 1):        &  $5.5^{+0.4}_{-0.3} \times 10^{-3}$ \T \B   &  $0.096^{+0.03}_{-0.02}$    &  $-1.00^{+0.04}_{-0.03}$   &  $8.3^{+0.95}_{-0.92}$  &            n/a            &         n/a               \\ 
\textbf{EG} (Bin 2):        &  $3.2^{+0.4}_{-0.4} \times 10^{-3}$ \T \B   &  $0.062^{+0.03}_{-0.02}$   &  $-0.70^{+0.03}_{-0.03}$   &  $7.0^{+0.93}_{-1.01}$  &            n/a            &         n/a               \\ \hline

\end{tabular}}
\caption[Best-fit parameters for the EG and the standard model fits using the cluster stack data]{The best-fit parameters for the standard model (GR + dark matter) and the EG model for the 58 cluster stack data. Bins 1 and 2 refer to the $T > 2.5$ keV and $T < 2.5$ keV temperature bins correspondingly. }
\label{table:cluster_stack_params}
\end{table}


\begin{figure}[ht!]
\begin{subfigure}{.5\linewidth}
\centering
\includegraphics[width=.95\textwidth]{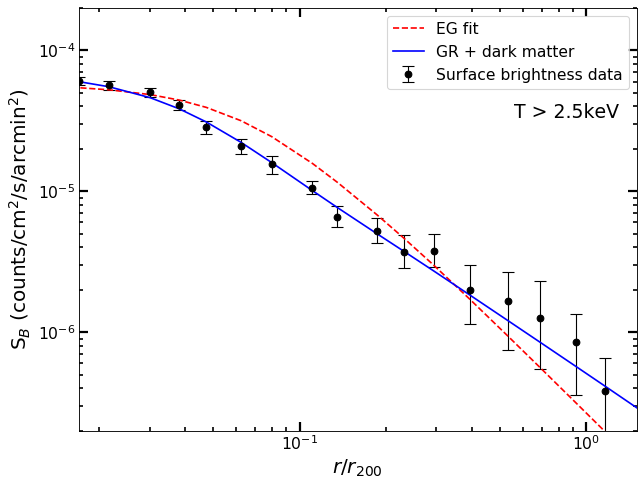}
\caption{X-ray surface brightness fits}
\label{fig:mainA}
\end{subfigure}%
\begin{subfigure}{.5\linewidth}
\centering
\includegraphics[width=.95\textwidth]{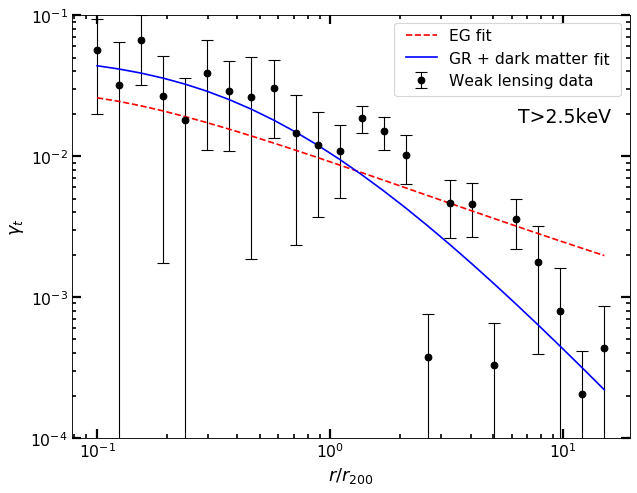}
\caption{Weak lensing fits}
\label{fig:mainB}
\end{subfigure}\\[1ex]
\begin{subfigure}{0.5\linewidth}
\centering
\includegraphics[width=.95\textwidth]{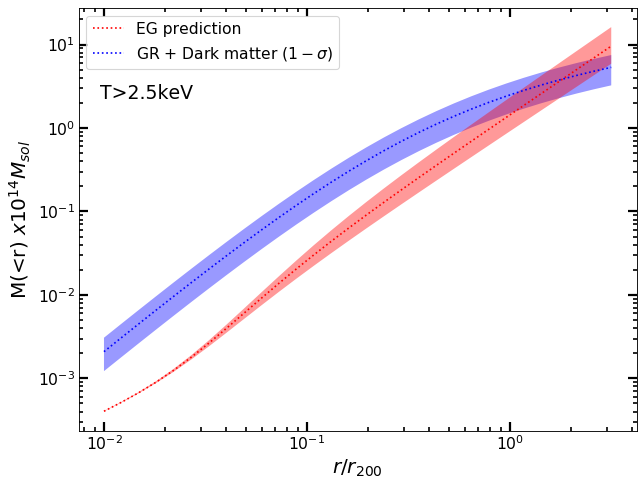}
\caption{Mass distributions in the two models}
\label{fig:mainC}
\end{subfigure}
\begin{subfigure}{.5\linewidth}
\centering
\includegraphics[width=.95\textwidth]{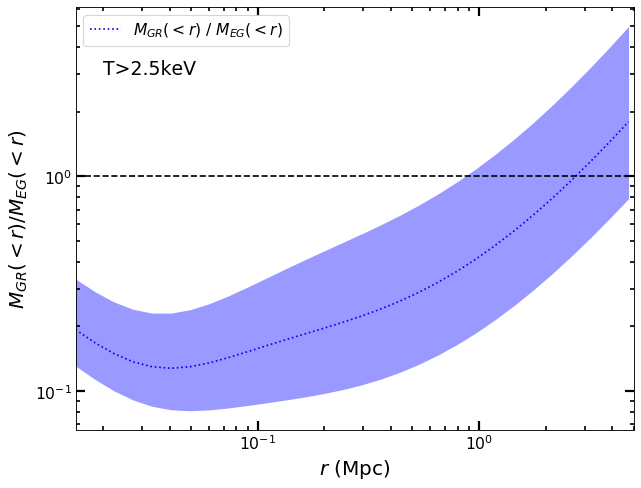}
\caption{Ratio of the mass distributions}
\label{fig:mainD}
\end{subfigure}
\caption[The results of testing emergent gravity with stacked galaxy clusters ($T > 2.5$ keV bin)]{A comparison of the EG and the standard model (GR + dark matter described by the NFW profile) results (for the $T > 2.5$ keV bin roughly corresponding to more massive galaxy clusters). Figure $a$ shows the surface brightness fit for both models. Figure $b$ shows the weak lensing (tangential shear) fit for both models. Figures $c$ and $d$ show the corresponding masses calculated using the best-fit parameters from figures $a$ and $b$. The blue and red bands correspond to the 1-$\sigma$ confidence intervals. The goodness of fit statistics are summarized in table \ref{table:cluster_stack_goodness_of_fit}. Note that the rather poor EG fit to the lensing data indicates that the model is not capable of simultaneously fitting X-ray and shear data due to the form of equation \ref{eq:EG_main_prediction}.}
\label{fig:main_results_stack}
\end{figure}


\begin{figure}[ht!]
\begin{subfigure}{.5\linewidth}
\centering
\includegraphics[width=.95\textwidth]{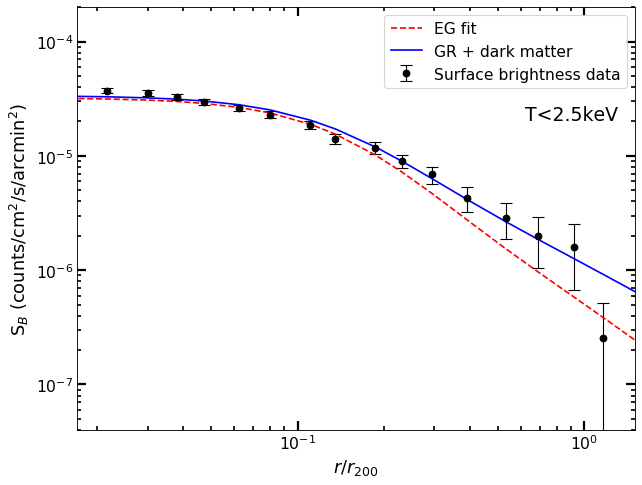}
\caption{X-ray surface brightness fits}
\label{fig:mainA2}
\end{subfigure}%
\begin{subfigure}{.5\linewidth}
\centering
\includegraphics[width=.95\textwidth]{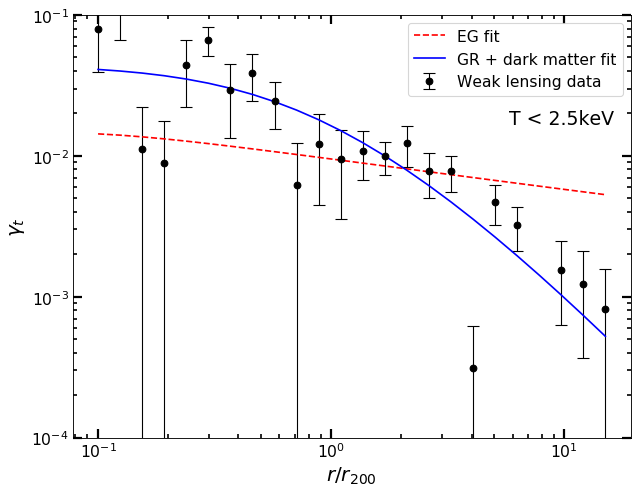}
\caption{Weak lensing fits}
\label{fig:mainB2}
\end{subfigure}\\[1ex]
\begin{subfigure}{0.5\linewidth}
\centering
\includegraphics[width=.95\textwidth]{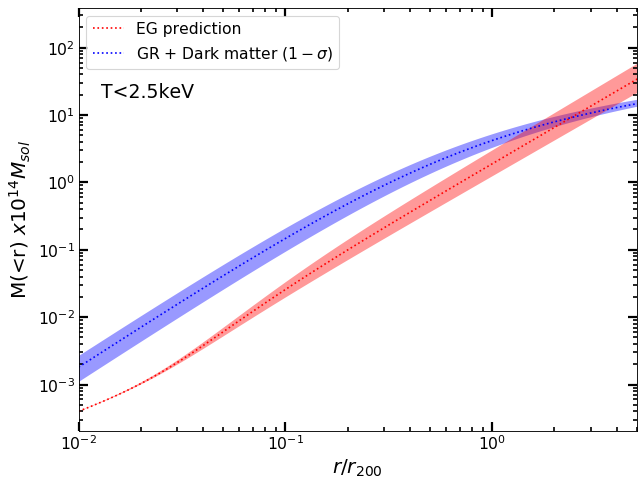}
\caption{Mass distributions in the two models}
\label{fig:mainC2}
\end{subfigure}
\begin{subfigure}{.5\linewidth}
\centering
\includegraphics[width=.95\textwidth]{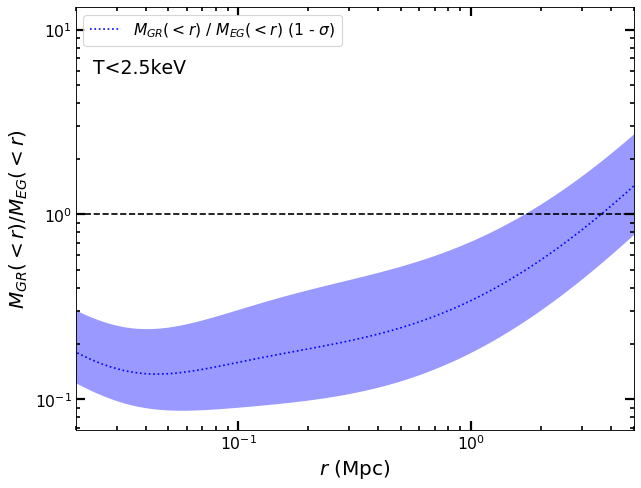}
\caption{Ratio of the mass distributions}
\label{fig:mainD2}
\end{subfigure}
\caption[The results of testing emergent gravity with stacked galaxy clusters ($T < 2.5$ keV bin)]{A comparison of the EG and the standard model (GR + dark matter described by the NFW profile) results (for the $T < 2.5$ keV bin roughly corresponding to galaxy groups). Figure $a$ shows the surface brightness fit for both models. Figure $b$ shows the weak lensing (tangential shear) fit for both models. The mass profiles in figures $c$ and $d$ were calculated using the best-fit parameters from figures $a$ and $b$. The blue and red bands correspond to the 1-$\sigma$ confidence intervals. The goodness of fit statistics are summarized in table \ref{table:cluster_stack_goodness_of_fit}. Note that the rather poor EG fit to the lensing data indicates that the model is not capable of simultaneously fitting X-ray and shear data due to the form of equation \ref{eq:EG_main_prediction}. }
\label{fig:main_results2_stack}
\end{figure}


Figure \ref{fig:main_results_stack} shows the results for the clusters with temperatures higher than 2.5 keV, which roughly corresponds to galaxy clusters (rather than galaxy groups). In this case the surface brightness fits are comparable for both models. However, the tangential shear profile fit in EG is significantly worse than the corresponding GR result. In general we found that EG could not simultaneously fit both datasets with accuracy. In other words, if we want to fit the surface brightness profiles accurately for $r/r_{200} \lesssim 2 \times 10^{-1}$, the resulting tangential shear profile will have a gradient that is too large to agree with the observational data for large values of $r$. This results in the total mass distributions in EG and GR that agree only at around $r \gtrsim  800$ kpc. 

In figure \ref{fig:main_results2_stack}, for the $T < 2.5$ keV bin (roughly corresponding to galaxy groups) a similar trend emerges. In this case, the EG tangential shear fits are even poorer resulting in total mass distributions that agree well only for $r \gtrsim 1.5$ Mpc. Otherwise, for $r \gtrsim 10$ Mpc, the EG fits are in strong tension with the data. The values in table \ref{table:cluster_stack_goodness_of_fit} also indicate that GR is strongly preferred. Also, it is important to note that the rather high values of the $\chi^{2}$ are dominated by the contribution from the outlier points at $r/r_{200} \approx 2.7$, $r/r_{200} \approx 5.0$ for bin 1 and $r/r_{200} \approx 2 \times 10^{-1}$ as well as $r/r_{200} \approx 4.0$ for bin 2. However, removing the outlier points does not lead to a different conclusion regarding the preferred model. 

\begin{table}[ht!]
\centering
\begin{tabular}{lcccccc}
                    & \textbf{$\chi^{2}_{ S_{B}}$} & $\chi^{2}_{ \gamma_{t}}$ & $N_{\rm d.o.f.}^{S_{B}}$ & $N_{\rm d.o.f.}^{ \gamma_{t}}$ & $\mathrm{BIC}_{S_{B}}$ & $\mathrm{BIC}_{\gamma_{t}}$\\ \hline \hline
\textbf{GR (bin 1):\,\,\,\,\,\,\,\,\,\,\,\,} &   0.7  & 4.1                                                 &    13.0                    &  22.0               &   4.1  &  3.1          \\
\textbf{GR (bin 2):\,\,\,\,\,\,\,\,\,\,\,\,} &      1.2  & 2.0                                              &   12.0                     &    21.0            & 13.2  &   19.2        \\  
\textbf{EG (bin 1):\,\,\,\,\,\,\,\,\,\,\,\,} &      6.0  &    16.4                                           &   14.0                     &    21.0            & 37.2 & 48.5              \\ 
\textbf{EG (bin 2):\,\,\,\,\,\,\,\,\,\,\,\,} &      1.84     &    6.3
&   13.0                     &    20.0            &  21.8     &  46.6        \\ \hline
\end{tabular}
\caption[Goodness of fit statistics for the cluster stack results]{Goodness of fit statistics. BIC is the Bayesian information criterion statistic. $N_{\rm d.o.f.}$ corresponds to the number of degrees of freedom, i.e. the difference between the number of data points and the number of free parameters being fit. The $\chi^{2}$ values here correspond to the reduced chi-squared statistic (i.e. chi-squared per d.o.f.). Note that the rather high $\chi_{\gamma_{t}}^{2}$ for the GR fit originates due to the outlier points. For bin 1, removing the two outlier points at $r/r_{200} = 2.7$ and $r/r_{200} = 5$ reduces the chi-squared statistic to $\chi^{2}_{\gamma_{t}} = 1.1$. For bin 2, removing points at around $r/r_{200} = 2 \times 10^{-1}$ and $r/r_{200} = 4$ reduces the chi-squared statistic to $\chi^{2}_{\gamma_{t}} = 1.3$. Likewise, removing the outliers also reduces the EG fit $\chi^{2}_{\gamma_{t}}$, however it does not change the preferred model, hence the outliers were not removed in the final analysis.  }
\label{table:cluster_stack_goodness_of_fit}
\end{table}

\begin{figure}[!ht]
\centering
\captionsetup[subfigure]{justification=centering}
  \begin{subfigure}[b]{0.49\textwidth}
    \includegraphics[width=\textwidth]{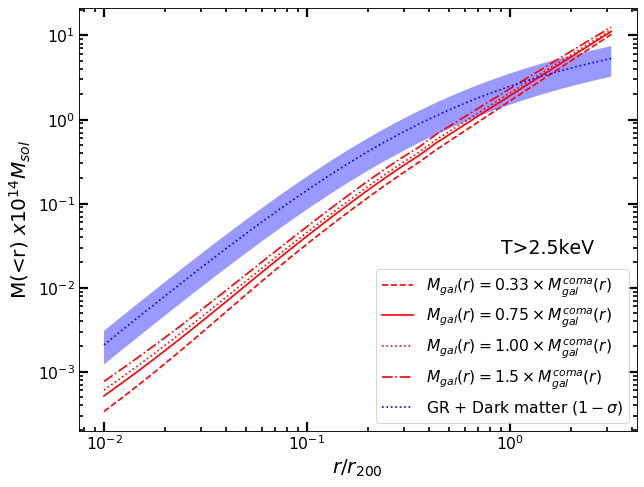}
    \caption{\small Results with different $M_{\rm gal}(r)$}
    \label{fig:a1}
    
  \end{subfigure}
  \begin{subfigure}[b]{0.49\textwidth}
    \includegraphics[width=\textwidth]{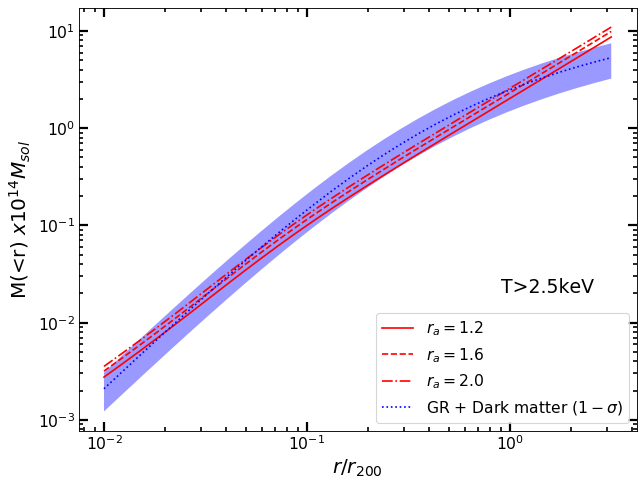}
    \caption{Results with modified $M_{D}(r)$}
    \label{fig:a2}
  \end{subfigure}
  \caption[Analysis of the various systematics]{Analysis of how the main results are affected by varying the galaxy mass function (compared to the Coma Cluster) and modifying the scaling relation for $M_{D}(r)$.}
  \label{fig:a}
\end{figure}

Comparing the results for the galaxy clusters, groups and the Coma Cluster indicates that, in general, EG seems to work better for massive clusters. This in turn means that accurate measurements of the total galaxy mass distribution (which dominates over the intracluster gas mass at low radii and hence could push the predicted mass profiles closer to those predicted in GR), are of special importance. In order to test the importance of the stellar galaxy mass measurements on our final results, we repeated the analysis outlined above, for the 58 cluster stack with various $M_{\rm gal}(r)$ distributions (which were compared against the galaxy mass distribution of the Coma Cluster). In particular the total mass distributions were deduced in the same way as in figure \ref{fig:main_results_stack}, but now with a galaxy mass distribution closer to that of the Coma Cluster (i.e. being equal to $0.33\mathrm{-}1.5 \times M_{\rm gal}^{\rm Coma}(r)$, where $M_{\rm gal}^{\rm Coma}(r)$ is the Coma galaxy mass distribution determined from the SDSS data). As figure \ref{fig:a1} illustrates, having significantly larger galaxy masses (while keeping the intracluster gas component unchanged) results in a better agreement between the standard model (GR + cold dark matter), EG and the observational data. 

Finally, we investigated how the results were affected by relaxing some of the assumptions in the derivation of the scaling relation in equation \ref{eq:EG_main_prediction}. In particular, as the authors point out in \citet{Halenka2018}, the mentioned scaling relation originally comes from the inequality, which is ultimately set to be equal (this results in equation \ref{eq:EG_strain}; for more information see section 7.1 in \citep{Verlinde2017}). Hence, they propose a phenomenological model, in which the $r^{2}$ term in the numerator of equation \ref{eq:EG_main_prediction} is replaced by $r_{a}r$, where $r_{a}$ is a constant, leading to:

\begin{equation}
    M_{D}^{2}(r) = \frac{cH_{0}r_{a}r}{6G} \frac{d(M_{B}(r)r)}{dr}.
\label{eq:mod_EG}
\end{equation}

\noindent The $r_{a}$ parameter, more specifically, describes how the elastic medium in Verlinde's theory is affected by the baryonic matter. Analysis in \citet{Halenka2018} shows that $r_{a} = 1.2$ Mpc leads to a good agreement between the EG prediction and the data. 

If we use the modified scaling relation and carry out our analysis again, the results in figure \ref{fig:a2} are obtained. In agreement with the results in \citet{Halenka2018} and \citet{Tortora2018}, for values of $r_{a} \approx 1.2$ we find a good agreement between GR and EG.

The results in figures \ref{fig:main_results_stack} and \ref{fig:main_results2_stack} are in agreement with most of the other results in the literature.  Specifically, in \citet{Ettori2017} X-ray and SZ effect data is used to deduce the baryonic and, in turn, the total mass distributions for EG and the standard model, resulting in distributions very similar to ours (see figure 3 in \citet{Ettori2017} in particular). More recently, in \citet{Ettori2019} the same approach was extended for a larger sample of clusters, once again resulting in mass distributions that agree only at around 1 Mpc radial scales. In \citet{Hodson2017} the EG scaling relation is used to calculate the acceleration radial distributions again resulting in profiles for GR and EG, that only become comparable for $r \gtrsim 2$ Mpc. Finally, the results reported in \citet{Halenka2018} (without modifying the original EG scaling relation) are closer to our results for the Coma Cluster. Note, however, that such comparisons with other results in the literature should be treated with caution, as the methods and the datasets used to derive them are in general distinct and are affected by different systematics. 

To conclude, it should be noted that it is impressive that a prediction derived from considerations in black hole thermodynamics and information theory leads to a result (with no free parameters) for the dark matter distribution that is of the right magnitude. However, as indicated by our results and the results from the literature, it is now clear that the scaling relation in equation \ref{eq:EG_main_prediction} cannot fully account for the effects associated with non-baryonic matter on galaxy cluster scales. To some degree this was expected, as equation \ref{eq:EG_main_prediction} was derived based on a number of simplifying assumptions, such as the mass distributions being spherical, non-dynamic and isolated. All of these assumptions are broken in real clusters to varying degrees, hence a natural question to ask is what effect does breaking these assumptions have on the main predictions of the model. In this regard, the covariant EG model introduced in \citet{Hossenfelder2017} offers a way forward by introducing more general covariant equations that could in principle be solved for non-spherical mass distributions (see section \ref{section:covariant_EG} for a brief review of the key features on the covariant EG model). In addition, the covariant approach also allows for calculating the weak lensing predictions, which do not rely on simplifying assumptions, such as that the lensing works in the same way as it does in GR (assumption made in this work). Interestingly, as pointed out in the literature, the covariant EG approach shares many mathematical similarities with the model of superfluid dark matter introduced in \citet{Berezhiani2015}. More specifically, both models predict MOND-like effects in the non-relativistic limit similar to those predicted in Verlinde's EG. These similarities could ultimately hint at some deeper connection between the phenomena described in Verlinde's work and the different phases of dark matter. Such considerations are outside the scope of this thesis, however, they point out some interesting directions for future work. 

The next two sections contain a brief review of some of the theoretical criticisms and other observational tests of the EG model along with a summary of the key features of the covariant EG formulation. 

\section{The Current State of the Model}

As previously mentioned, the theory has already been tested using several methods and on a range of scales. Here for completeness the key observational tests and their results are summarized with an emphasis on their significance to the validity of Verlinde's model on different scales. 

In \citet{Brouwer2017} the average surface mass density profiles of isolated central galaxies were used to perform the first known test the model. In particular, the average surface mass density profiles of 33,613 isolated central galaxies were compared against the theoretical predictions. The study found that the predictions of the model are in good agreement with the measured galaxy-galaxy lensing profiles in four different stellar mass bins. 

More recently the predictions of EG were compared against the predictions from a list of selected modified gravity theories for a single galaxy cluster, showing that the model only approaches the measured acceleration profile data at the outskirts of the cluster \citet{Hodson2017}. Similarly, the X-ray and weak lensing data from the A1689 cluster 
was used in \citet{Nieuwenhuizen2017} to compare the predictions of EG with some selected modified gravity models, also finding that model fails to account for the missing mass in the mentioned cluster.

In \citet{Halenka2018} the authors used mass densities from a sample of 23 galaxy clusters to test the predictions of EG on galaxy cluster scales. They found that EG could only correctly predict the baryon and dark matter mass profiles at around the virial radius, while being ruled out at a 5-$\sigma$ level in the other parts of the clusters. However, as the authors pointed out, fully accounting for the systematic uncertainties and modifying certain assumptions in the model leads to a much better agreement between GR and EG. 

A similar study in \citet{ZuHone2019} used matter densities of relaxed, massive clusters of galaxies using a combination of X-ray and weak lensing data. A key improvement in this work was to include the baryon mass contribution of the brightest cluster galaxy in each system along with the total mass profiles from gravitational lensing. The results indicate that the EG predictions for the mass profiles and baryon mass fractions disagree with the observational data by a factor of up to $\sim 2$-$6$ for radii in the range of $100$-$200$ kpc.

In a more theoretical context, Verlinde's theory has received numerous criticisms. In \citet{Dai2017} the authors argue against the idea of treating gravity as a force of entropic origin. Namely, Newtonian gravitational force is conservative, which implies that the dynamics of bodies in gravitational systems as well as the action should be reversible. This implies that gravitational dynamics cannot be caused purely by the increase of entropy in the gravitational system. More specifically the authors point out that the equation $F \Delta x=T \Delta S$ in \citet{Verlinde2011}, where $F$ is the force, $\Delta x$ is the change in position, $T$ is the temperature and $\Delta S$ is the change in entropy, is missing a term corresponding to the change of kinetic energy of the system $\Delta E_{k}$. Including the missing term leads to a conclusion that gravity cannot be an entropic force arising solely from the change in entropy in the system. 

A further criticism in \citet{Dai2017} points out a flaw with the averaging procedure used in the derivation of the MOND scaling relations (equations at the end of the section 4.4 in \citet{Verlinde2017}). This point, however, was later criticised in \citet{Yoon2020}, where it was shown that the criticism in \citet{Dai2017} stemmed from a misunderstanding of the details of the gravity-elasticity correspondence in the Verlinde's original argument. 

More generally, it has been shown that fully accounting for energy-momentum conservation and cosmological homogeneity and isotropy conditions severely restricts a wide class of possible entropic gravity models \citep{Wang2012}. 

\section{Covariant Emergent Gravity}
\label{section:covariant_EG}

Another recent development that is important to discuss in more detail is the covariant EG model introduced by Hossenfelder \citep{Hossenfelder2017}. In particular, Hossenfelder introduces a model in which a vector field fills the de Sitter space and via interactions with baryonic matter results in effects similar to those of to dark matter. In the non-relativistic limit this model reproduces the main predictions of Verlinde's EG along with correction terms. In addition, the covariant formulation also demonstrates that the introduced vector field can mimic the effects of dark energy. 

In Hossenfelder's covariant formulation, the displacement field $u(r)$ (introduced in equation \ref{eq:entropy displacement}) is treated as an extra vector field that originates from the volume term in the total entropy equations in Verlinde's theory. The vector field couples to baryonic matter and drags on it to create an effect similar to dark matter. The proposed Langragian is given by: 

\begin{equation}
    \mathcal{L} = M_{\rm pl}^{2}R + \mathcal{L}_{M} - \frac{u^{\mu}u^{\nu}}{Lu}T_{\mu \nu} + \frac{M_{\rm pl}^{2}}{L^{2}} \chi^{3/2} - \frac{\lambda^{2}M_{\rm pl}^{2}}{L^{4}} (u_{\kappa}u^{\kappa})^{2},
    \label{eq:covariant_EG_lagrangian}
\end{equation}

\noindent with $R$ as the Ricci scalar, $\mathcal{L}_{M}$ as the matter Lagrangian, $L$ as the Hubble radius, $T_{\mu \nu}$ as the stress-energy tensor, $\chi$ as the kinetic term of the vector field and $\lambda$ as the mass term.

Noting that the gravitational potential is given by $\phi_{u} = \sqrt{-u^{\alpha}u_{\alpha}}/L$, equation (\ref{eq:covariant_EG_lagrangian}) can used to derive an equation of motion for the field and solve it for $\phi_{u}(r)$. For a covariant, spherically symmetric case the solution is given by: 

\begin{equation}
    \phi_{u}(r) = \sqrt{\frac{M}{LM_{\rm pl}^{4}}} \Biggl[   \gamma\Bigg(-C + \ln{\Bigg(\frac{rM_{\rm pl}^{2}}{M}\Bigg)} \Bigg) + \frac{3M}{M_{\rm pl}^{2}r} + \mathcal{O} \Bigg( \Bigg(\frac{M}{M_{\rm pl}^{2}r} \Bigg)^{2} \Bigg)        \Biggr] ,
    \label{eq:covariant_EG_potential}
\end{equation}

\noindent where $\gamma(r) = 1 - 2M/M_{\rm pl}^{2}r$ and $C$ is an integration constant that needs to be set using boundary or initial conditions. As discussed in \cite{Hossenfelder2017}, $C$ should be set by taking the limit $r \xrightarrow{} \infty$, however, given the assumptions used to derive \ref{eq:covariant_EG_potential} this is not possible; the other option is to treat it as a free parameter and to deduce it numerically given some baryonic mass distribution. In general the result given in equation \ref{eq:covariant_EG_potential} is expected to be different from the potential due to apparent dark matter in Verlinde's original formulation (i.e. it will contain correction terms). Hence, solving equation \ref{eq:covariant_EG_potential} is the natural next step in both the theoretical development and the observational tests of the model. 

More generally, the form of the Lagrangian in equation \ref{eq:covariant_EG_lagrangian} indicates some clear differences between the Verlinde's and Hossenfelder's formulations. In particular, the extra terms in the Lagrangian indicate that even when no baryonic mass is present in the system, the field $u_{\alpha}$ does not vanish. Or, in other words, stress-energy conservation would require the field $u_{\alpha}$ to be a source of gravity as well. This means that the solutions for the total potential for general gravitational systems will not be identical to those derived by Verlinde and will contain correction terms. Another interesting feature of the Lagrangian is the 2/3 power of the kinetic term. There have been multiple modified gravity approaches that have a similar kinetic term, most notably \citet{Berezhiani2015}, where a theory of dark matter superfluidity is proposed. 

As discussed in \citet{Hossenfelder2017}, the Langrangian above can be solved for $\phi_{u}$, however the solution contains an integration constant that cannot be determined analytically and would require numerical solutions. Finding these solutions is out of the scope of this thesis. However, further exploration of the covariant formulation of EG for spherical and non spherical mass distributions, and comparison of the results with the predictions from Verlinde's original formulation will be an interesting direction for future work. 

It is also important to discuss some general criticisms to the covariant formalism. Namely, as pointed out in \citet{Dai2017B}, small perturbations around the de Sitter space in \citet{Hossenfelder2017} grow rapidly indicating unstable cosmology. However the authors point out that adding matter and radiation to the model could in principle provide stability. 

More generally, models with fractional powers of the kinetic term (such as described in equation \ref{eq:covariant_EG_lagrangian} and by the model of superfluid dark matter in \citet{Berezhiani2015}) have been criticised in \citet{Zatrimaylov2020}. In this work the author investigates the effects of enforcing certain theoretical and observational constraints on the family of models described above. More specifically, \citet{Zatrimaylov2020} imposes the constraints of the energy density being bounded from below, superluminal propagation being absent in relativistic settings and the models being able to account for gravitational lensing effects. The conclusions indicate that scalar, vector and tensor theories with fractional kinetic terms in generally struggle to satisfy the mentioned energy density conditions while also abiding by the observational constraints (for instance, the LIGO results for the speed of gravitational waves). This also applies to $f(R)$ models with MOND-like potentials, which reproduce MOND effects on galaxy and cluster scales, similar to those predicted by the superfluid dark matter and covariant EG models.

\chapter{A Brief Introduction to Machine Learning}
\label{ch:machine_learning}

Chapter \ref{ch:machine_learning} marks the beginning of the second part of the thesis. The main focus of the second part is on machine learning techniques in the context of $N$-body simulation emulators. Chapter \ref{ch:machine_learning} contains an overview of basic machine learning techniques and their relevance to natural sciences. This includes a more in-depth look at decision tree algorithms, artificial neural networks, generative adversarial networks (GANs) and gradient boosting. Chapter \ref{ch:GAN_emulators} consists of a novel technique for emulating $N$-body simulation data using a GAN algorithm. In particular, an algorithm capable of efficiently emulating cosmic web and weak lensing convergence maps is introduced. 

The key goal of the algorithms introduced in the upcoming chapters is to produce realistic mock data quickly and efficiently. In particular, emulating $N$-body simulation data from $\Lambda$CDM and modified gravity simulations is of great importance for survey mock data generation as well as modified gravity tests. More concretely, such emulators could be used to generate mock weak lensing and galaxy cluster data without resorting to computationally expensive hydrodynamic simulations. In this respect, the topics discussed in chapters \ref{ch:machine_learning} and \ref{ch:GAN_emulators} are nicely linked with the topics discussed in the preceding chapters.

\section{Machine Learning and Artificial Intelligence}

The field of machine learning dates back to the beginning of the 20th century and is intimately linked with the studies of the human brain and the field of neuroscience. In fact, the theoretical basis for the studies of the human brain in this context dates back even earlier to the work by Alexander Bain and William James, who independently proposed a model of the brain as a network of neurons \citep{Bain1873, James2012}. Later, in 1943 Warren McCulloch and Walter Pitts created a computational model for neural networks \citep{McCulloh1943}. In 1958 this culminated in the invention of the perceptron algorithm by Frank Rosenblatt, which is a precursor to modern artificial neural networks \citep{Rosenblatt1958}.

The term \textit{machine learning} itself dates back to 1959 and refers to the study of techniques and algorithms that make decisions and predictions without having been programmed to do so explicitly \citep{Samuel1959}. In this regard, machine learning is closely related to the fields of computational statistics, automation, data science, mathematical optimization and robotics.  The studied algorithms can be broadly classified into supervised learning, unsupervised learning and reinforcement learning. Supervised learning algorithms are programmed to deduce a rule that maps a certain set of inputs to a set of outputs based on a \textit{training} dataset, which contains data split into categories. An archetypal example of such a machine learning task is image classification, often done using artificial neural networks, decision forests and other commonly used algorithms. Unsupervised learning algorithms, on the other hand, deduce patterns and correlations in a given dataset without it being explicitly classified into categories before the training procedure. Algorithms of such type generally work based on principle component and cluster analysis. Reinforcement learning algorithms, on the other hand, are trained based on their interactions with a dynamic environment with the aim of performing a specified goal. Reinforcement learning algorithms are often applied to solve problems in robotics and gaming (e.g. the AlphaZero algorithm) \citep{Silver2017}. 

Another class of models that does not easily fit into the classification outlined above (and often contains a combination of supervised and unsupervised techniques) contains generative models. Generative models refer to a class of algorithms that aim to generate statistically realistic mock data based on a training dataset. More specifically, given a set of data instances $X$ and the corresponding set of labels $Y$, a generative model is trained to capture the joint probability $P(X,Y)$. The two prime examples of generative algorithms are variational autoencoders (VAEs) and generative adversarial networks (GANs) \citep{Zamorski2019}. The latter will be discussed in greater detail at the end of this chapter.

\section{Machine Learning in Cosmology and Astrophysics}

The key goal of machine learning (and data science more generally) is to extract useful information from data. This makes machine learning techniques an important tool in the natural sciences where the key goal is to build physical models based on observational and experimental data. Naturally, throughout the last few decades, machine learning techniques have become an important tool in the toolset of astrophysicists and cosmologists. Here we overview some of the key machine learning techniques used in cosmology. 

Recently a combination of techniques (naive Bayes, k-nearest neighbours, support vector machines and neural networks) have been studied as a tool for photometric supernova classification \citep{Lochner2016}. Machine learning techniques are also key for the photometric LSST astronomical time-series classification challenge (PLAsTiCC) \citep{Plasticc2018}.

In the field of CMB studies, extracting constraints on cosmological parameters is of key importance. In this regard, machine learning has been shown to provide competitive techniques for calculating these constraints quickly and efficiently. These techniques are of special importance when studying the non-Gaussian foreground contributions in particular. An example of machine learning used for such data is the DeepCMB algorithm, which uses deep convolutional neural networks (CNN) for cosmological parameter estimation and lensing reconstruction \citep{Caldeira2019}. Similarly, a 3-D CNN algorithm has been used to extract cosmological paramters from large scale structure data in \citet{Ravanbakhsh2017}.

Galaxy cluster mass estimation is another important task that has greatly benefited from using different machine learning approaches. Recently, it has been shown that machine learning techniques (support distribution machines, support vector regression, decision trees, CNNs and others) allow significant reduction in the scatter in cluster mass estimates when compared to the more traditional statistical methods \citep{Ntampaka2015, Armitage2019, Ho2019}.  

In weak lensing CNNs have been used as quick and efficient tools for discriminating between different models of modified gravity especially in the context of non-Gaussian information encoded in the weak lensing maps \citep{Gupta2018, Ribli2019}. Machine learning has also been used with strong lensing data, where it was found to be significantly faster and more efficient in identifying strong lensing arcs \citep{Lanusse2017}.

It is also important to mention generative models, which have found great use in emulating cosmological simulation data. GANs \citep{Goodfellow2014}, in particular, have been employed to produce statistically realistic mock data for both weak lensing convergence maps and cosmic web slices \citep{Rodriguez2018, Mustafa2019}. 

The methods and techniques mentioned in this section clearly illustrate the effectiveness of machine learning when applied to a variety of problems in astrophysics and cosmology. However, it is also important to discuss some of the common drawbacks that a lot of the mentioned models have. In particular, many machine learning models suffer from being difficult to interpret (i.e. the "black box" problem). This is is especially true in the case of neural networks. Similarly, when it comes to most algorithms, it is generally difficult to introduce prior physics knowledge that would be used when making predictions. With some models one could introduce priors in a Bayesian fashion, however, with many models that might not be possible. In addition, another important issue with many models is that it is not easy to implement physical constraints (i.e. conservation of energy etc.). These and similar issues might not be problematic depending on the application at hand, but nonetheless should be taken into consideration when choosing an algorithm to tackle a specific problem.  

The rest of this chapter is dedicated to introducing some of the key machine learning algorithms in terms of relevance for this thesis. In particular, decision trees and gradient boosting is introduced as a tool for accurate classification. Similarly, different types of neural networks are reviewed. Finally the GAN algorithm is discussed as a tool for emulating cosmological data.

\section{Decision Trees and Gradient Boosting}

An algorithm that has recently gained significant popularity in the literature is the \textit{XGBoost} algorithm \citep{Chen2016}. \textit{XGBoost} refers to extreme gradient boosting, which is a technique that produces a prediction based on an ensemble of weak prediction models that are optimized during the training procedure. The employed prediction models are usually modeled using decision trees. Here we overview some of the main features of the gradient boosting procedure more generally and in the context of decision trees. 

Generally speaking gradient boosting algorithms work by iteratively improving the prediction of a model by fitting the residual points and adding extra terms to the model to account for those residuals. This leads to such models being very successive in approximating complex functions, as during the training procedure gradient boosting allows the algorithm to focus on the data points that the initial model struggled to fit and to incrementally improve this. Here this procedure will be described mathematically based primarily on \citet{Friedman2001}.

The goal of most supervised machine learning models is to produce accurate predictions based on a training dataset: $\left\{\left(x_{1}, y_{1}\right), \ldots,\left(x_{n}, y_{n}\right)\right\}$. Here $x_{i}$ is a vector corresponding to the training data, while $y_{i}$ is either a class in a classification task or a value that the model tries to predict in a regression task. More specifically, a machine learning algorithm aims to find an accurate approximation $\hat{F}(x)$ of the function $F(x)$ that minimizes some cost function $E(y,F(x))$. More formally, the function $\hat{F}(x)$ is determined by evaluating the following: 

\begin{equation}
\hat{F}= \underset{F}{\arg \min} \mathbb{E}_{x, y}[E(y, F(x))],
\label{eq:gradient_optimization}
\end{equation}

\noindent where $\mathbb{E}$ refers to the expectation function. Algorithms, such as \textit{XGBoost}, determine the function $\hat{F}(x)$ by expressing it as a sum of weighted functions $h_{i}(x)$:

\begin{equation}
\hat{F}(x)=\sum_{i=1}^{M} \gamma_{i} h_{i}(x)+\text {const},
\label{eq:gradient_boosting_function}
\end{equation}

\noindent where $\gamma_{i}$ are the weight parameters and $M$ is the number of iterations. In case of the \textit{XGBoost} algorithm, the functions $h_{i}$ represent decision trees that fit the residuals between the prediction of the model and the training data at each iteration of training. 

Here a brief overview of the gradient boosting training algorithm is given in pseudo-code without going into full detail (for more information see \citet{Friedman2001,Hastie2013}):

\begin{enumerate}
    \item The model is initialized with a constant value. This can be an average value based on the data or a simple fit based on the cost function (note that $\gamma$ here should not be confused with the previously mentioned weight parameters, as it simply denotes the initial fit that minimizes the cost function): 
    
    \begin{equation}
    F_{0}(x)=\underset{\gamma}{\arg \min } \sum_{i=1}^{n} E\left(y_{i}, \gamma\right).
    \label{eq:initial_cost}
    \end{equation}
    
    \item For iterations $m = 1$ to $M$ \textbf{do:}
        \begin{enumerate}[label=\roman*)]
            \item evaluate the pseudo-residuals (for $i = 1,...,n$): 
            
            \begin{equation}
            r_{i m}=-\left[\frac{\partial E\left(y_{i}, F\left(x_{i}\right)\right)}{\partial F\left(x_{i}\right)}\right]_{F(x)=F_{m-1}(x)}.
            \label{eq:pseudo_residuals}
            \end{equation}
            
            \item Fit a weak learner (e.g. a decision tree algorithm) $h_{m}(x)$ to the residuals. This is done by training $h_{m}(x)$ on the residual dataset: $\left\{\left(x_{i}, r_{i m}\right)\right\}_{i=1}^{n}$.
            
            \item Evaluate the weight parameter $\gamma_{m}$ via optimization:
            
            \begin{equation}
            \gamma_{m}=\underset{\gamma}{\arg \min } \sum_{i=1}^{n} E\left(y_{i}, F_{m-1}\left(x_{i}\right)+\gamma h_{m}\left(x_{i}\right)\right).
            \label{eq:weight_optimization}
            \end{equation}
            
            \item Update the model: 
            
            \begin{equation}
            F_{m}(x)=F_{m-1}(x)+\gamma_{m} h_{m}(x).
            \label{eq:updating_the_model}
            \end{equation}
            
        \end{enumerate}

    \textbf{end for}

    \item Output the final result $ \hat{F}(x) = F_{M}(x)$.
    
\end{enumerate}

\begin{figure}[ht!]
  \centering
    \includegraphics[width=0.95\columnwidth]{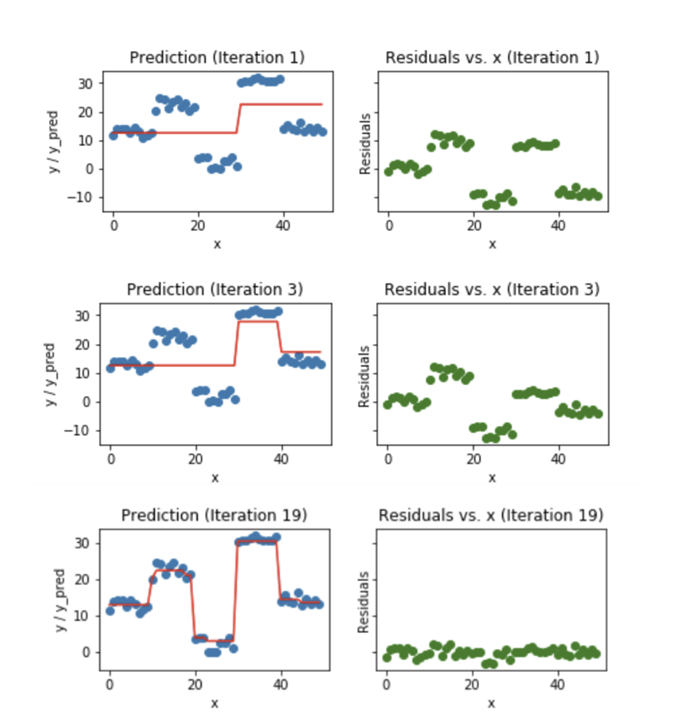}
    \caption[Gradient boosting training procedure]{Illustration of the gradient boosting procedure. A sample dataset $x$ (in blue) is fit by a model $F_{m}(x)$ (red line). After each iteration, the residual points (in green) are calculated and fit using a decision tree algorithm $h_{m}(x)$, which is then added to the original model. After several iterations, the total model fits the data nearly perfectly. Image credit: \citet{Grover2017}. }
\label{figure:gradient boosting}
\end{figure}

\noindent Figure \ref{figure:gradient boosting} illustrates the gradient boosting procedure on a sample dataset.

\begin{figure}[ht!]
  \centering
    \includegraphics[width=0.65\columnwidth]{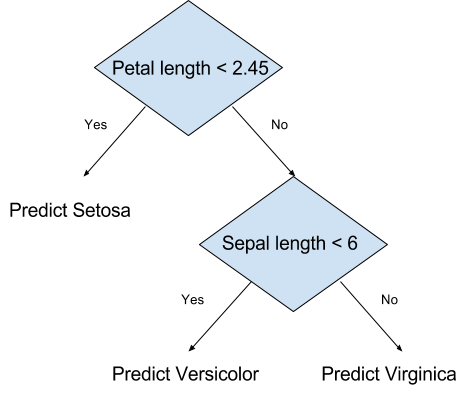}
    \caption[An example of a decision tree]{An example of a decision tree based on the Iris flower dataset. Image from \citet{Anam2018}. }
\label{figure:decision_tree_example}
\end{figure}

The final point to discuss is how the $h_{m}(x)$ functions are actually determined. As mentioned, this is usually done by using decision trees, however it could be any algorithm capable of fitting the residual data. Decision trees refer to a technique of splitting a dataset in a way that allows making accurate predictions (in a classification or a regression task). The technique is easiest to understand by referring to a simple example. A classic dataset used to illustrate machine learning classification problems is the Fisher-Anderson Iris flower dataset \citet{Anderson1936,Dua2019}. This dataset contains 50 data samples of the length and the width of the sepals and the petals for 3 different species of Iris flowers (\textit{Iris Setosa}, \textit{Iris Virginica} and \textit{Iris Versicolor}). The dataset is often used as a pedagogical example on building machine learning classification algorithms with the goal of using the petal and sepal features in order to predict the flower type. With decision trees this can be done by finding an optimal way of splitting the dataset into categories based on the values of the mentioned features. In particular, decision tree algorithms are optimized to find the optimal way of splitting the training data into categories in a way that allows predicting the flower type as accurately as possible. Figure \ref{figure:decision_tree_example} illustrates such a decision tree.

A natural question to ask is how does one quantify which way of splitting the data leads to the most accurate predictions. This is done by measuring the \textit{pureness} of a subset of the decision tree. A commonly used measure for this is entropy $I_{H}$. For a particular node of the decision tree the entropy can be calculated as follows:

\begin{equation}
I_{H}(t)=-\sum_{i=1}^{C} p(i \mid t) \log _{2} p(i \mid t),
\label{eq:entropy_decision_trees}
\end{equation}

\noindent where $p(i \mid t)$ is the proportion of the samples that belong to class $i$ for a particular node, $t$ is the training subset of the parent node and $C$ refers to the number of unique class labels. The entropy can then be used to estimate the information gain due to a given split in the data, as quantified by the information gain parameter $G_{s}$:

\begin{equation}
G_{s}\left(D_{p}, x_{i}\right)=I_{H}\left(D_{p}\right)-\frac{N_{\text {left}}}{N_{p}} I_{H}\left(D_{\text {left}}\right)-\frac{N_{\text {right}}}{N_{p}} I_{H}\left(D_{\text {right}}\right),
\label{eq:entropy_gain}
\end{equation}

\noindent where $D_{p}$ is the training subset of the node under consideration, $x_{i}$ is the feature that the split is being performed for, $N_{p}$ is the number of samples in the parent node, $N_{\rm left}$ is the number of samples in the left child node, $N_{\rm right}$ is the number of samples in the right child node, $D_{\rm left}$ is the training subset of the left child node and $D_{\rm right}$ is the training subset of the right child node. Note that this equation is only correct for decision trees where only two splits are available at each node, however, it is easy to generalize the equation above for further possible splits by adding analogous further terms. 

To put it simply, equation \ref{eq:entropy_gain} is the measurement of the difference in entropy before and after the split. In particular, equation \ref{eq:entropy_gain} can be better understood by looking at a concrete example, i.e. the decision tree shown in figure \ref{figure:decision_tree_example}. For instance, the information gain parameter corresponding to the right-hand side node (sepal length of $<6$) can be calculated by considering the number of the data samples that correspond to Versicolor and Virginica Iris flowers. In this case $N_{\rm left}$ corresponds to the size of the subset of data in the left child note, $N_{\rm right}$ is analogous and $D_{p}$ corresponds to the subset of data that has sepal length of $< 6$ and is used to calculate the entropy of the parent node. Analogously, the entropy of child nodes can be calculated by using the corresponding data subsets in those nodes.  

Given these ways of quantifying the change in entropy due to any split in the dataset, decision trees can then be optimized to maximise the purity (information gain) of those splits. Decision trees are \textit{greedy} algorithms in the sense that they optimize the information gain for each split sequentially. 

In summary, decision tree algorithms can be used for both regression and classification and are relatively easy to interpret (the decision splits can be plotted graphically). However, the models tend to be prone to overfitting, i.e. performing poorly when making predictions on unseen data. There are a number of remedies for overfitting, for instance carefully choosing the hyperparameters corresponding to the maximum allowed depth and number of leaves of the decision tree during the optimization procedure.  

The described gradient boosting techniques and decision forests are combined in the \textit{XGBoost} algorithm that offers cutting edge accuracy when it comes to classification and regression tasks. In the upcoming chapter \textit{XGBoost} will be used for classifying the produced weak lensing and overdensity field data based on cosmological parameters. This will be crucial for proper analysis of the emulated data sets.

\section{Artificial Neural Networks}

\begin{figure}[ht!]
  \centering
    \includegraphics[width=0.9\columnwidth]{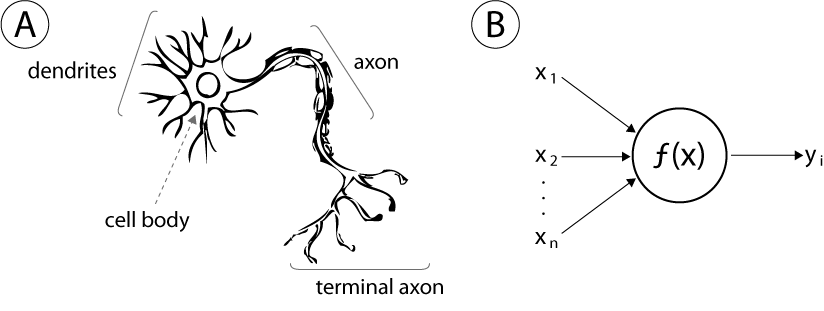}
    \caption[Biological and artificial neurons]{\textbf{A:} a biological neuron. \textbf{B:} an artificial neuron (perceptron). Diagram adapted from figure 1 in \citet{Maltarollo2013}. }
\label{figure:biological_vs_artificial_neurons}
\end{figure}

The main building block of artificial neural networks is the previously mentioned Rosenblat's perceptron which mimics the key features of biological neurons. Figure \ref{figure:biological_vs_artificial_neurons} illustrates the key difference and similarities of the biological and artificial neurons. In addition, the figure also summarizes the key components of the biological neurons. The signals received from the neighbouring neurons are delivered via protoplasmic nerve extensions called dendrites and sent to the cell body, where the signals are processed. If sufficient input signal is received, the neuron generates an action potential. The action potential is then transmitted via longer cytoplasmic protrusions (known as axons) to the other neighbouring neurons. If sufficiently strong input is not received, the signal quickly decays and no action potential is generated.  

Artificial neurons are built to mimic the key elements of their biological counterparts. As illustrated by figure \ref{figure:biological_vs_artificial_neurons}, they contain a node that is equivalent to the cell body, which receives multiple inputs, processes them and produces an output. The received inputs are usually weighted and processed by a non-linear activation function. This process can be expressed as follows:

\begin{equation}
y_{i}=f\left(\sum_{j=0}^{n} w_{i j} x_{j}\right),
\label{eq:artificial_neuron}
\end{equation}

\noindent where $w_{ij}$ refers to the set of the weight parameters, $f$ is the activation function, $x_{j}$ is the set of inputs and $y_{i}$ refers to the output(s). The activation function improves the training procedure for multi-layered neural networks and allows the network to approximate non-linear functions easier. A commonly used function is the sigmoid logistic function: 

\begin{equation}
f(x)=\frac{1}{1+e^{-x}}=\frac{e^{x}}{e^{x}+1}.
\label{eq:sigmoid_function}
\end{equation}

\noindent The sigmoid function has a characteristic \textit{S} shape and the following asymptotic behaviour: $f(x) \rightarrow 1$ for $x \rightarrow \infty$ and $f(x) \rightarrow 0$ for $x \rightarrow -\infty$. Many other functions can be used, which all share the common \textit{S}-like shape: $tanh(x)$, $arctan(x)$, $erf(x)$ (the error function), $f(x) = x/\sqrt{1+x^{2}}$ etc. The output values (and the asymptotic behaviour) can be controlled by normalizing the mentioned functions to the needed range of $[y_{min},y_{max}]$ (which is usually $[0,1]$ or $[-1,1]$ depending on the value range of the training data).

\begin{figure}[ht!]
  \centering
    \includegraphics[width=0.7\columnwidth]{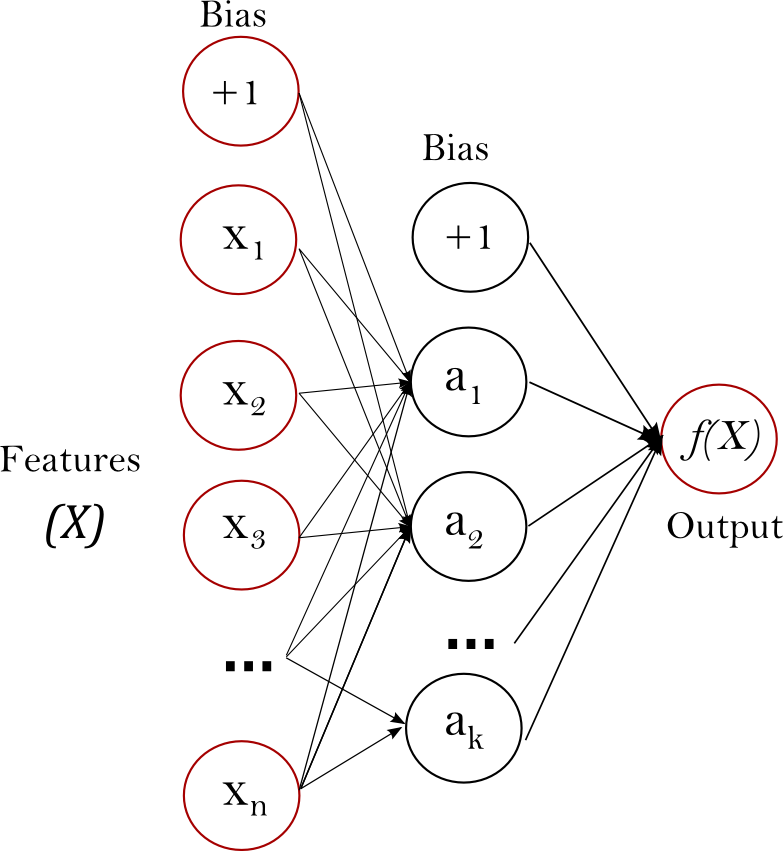}
    \caption[A multilayer perceptron]{The multilayer perceptron architecture. Image from \citet{scikit-learn2020}.}
\label{figure:multilayer_perceptron}
\end{figure}

Joining multiple artificial perceptrons into a layered structure results in the familiar multilayer perceptron architecture shown in figure \ref{figure:multilayer_perceptron}. In a multilayer perceptron each node is a single artificial neuron, with multiple inputs coming in, being processed by an activation function and sent out as output signals to be received by other neurons.

\noindent One minor difference in this architecture, is the addition of bias nodes, which add a certain value to the output of each layer in the network. The bias term introduces a slight shift to the activation function, which has been shown to significantly improve the performance of artificial neural networks \citep{Hellstrom2020}. A succinct way of representing the multilayered structure of a neural network is by using a function composition notation, where a neural network $N$ maps an input vector $x_{i}$ to an output in each layer as follows:

\begin{equation}
    N = l^{1} \circ l^{2} \circ ... \circ l^{n}
    \quad\text{with}\quad
    l_{k}^{i}(x^{i}) = f(W_{k}^{i}x^{i} + b^{i}).
\end{equation}

\noindent Here each layer $l^{i}$ maps from an input $x^{i}$ to an output as shown above with $l_{k}^{i}$ as the $k$-th element of the $i$-th layer, $f$ as an activation function, $b_{i}$ as the bias term, $W^{i}_{k}$ as the weight matrix. The function composition is an operation for two functions $g(x)$ and $h(x)$ that can be defined as $(g \circ h)(x) = g(h(x))$.

Hence an artificial neural network $N$ can be treated as a complicated non-linear function that maps an input $x_{i}$ to an output and the goal of the training procedure is to find the optimal set of the weight parameters in the weight matrix $W^{i}_{k}$. The training procedure for multilayered neural networks as described above is usually done using the backpropagation algorithm with gradient descent. 

A key quantity when evaluating the performance of an artificial neural network during the training procedure is the cost (error) function $E$. For an input-output data pair $(x,y)$ the cost function can be something as simple as the square difference between the output of the network and the true value corresponding to a given input (e.g. the correct class of an image in an image classification task or the correct value in a regression task): $E(t,y) = (t - y)^{2}$. More sophisticated cost functions are usually used in modern neural networks, such as the cross-entropy function: 

\begin{equation}
E_{\rm CE}(t,y) = -(t \log (y)+(1-t) \log (1-y)).
\label{eq:cross_entropy_cost}
\end{equation}

\noindent Given a cost function, the goal of the optimization/training procedure is then to minimize the cost function w.r.t. the set of the weight parameters. The change in the weight parameters, $\Delta w_{ij}$ is calculated in an iterative gradient descent procedure:

\begin{equation}
\Delta w_{i j}=-\eta_{L} \frac{\partial E}{\partial w_{i j}}=-\eta_{L} y_{i} \delta_{j},
\label{eq:gradient_descent}
\end{equation}

\noindent where $\eta_{L}$ is the learning rate parameter, $y_{i}$ is the output of the layer $i$ and $\delta_{j}$ is the gradient at the layer $j$. Namely, the aim of the procedure is then to calculate the partial derivative of the cost function term w.r.t. the weight parameters, which is done by backpropagation, i.e. evaluating the gradient terms for each layer starting with the final layer. Assuming the cost and the activation functions are well-behaved and their derivatives can be calculated, each gradient can be evaluated and the weight parameters can be updated in a way that reduces the value of the cost function. If the training procedure converges, gradient descent finds the minimum (this is usually one of the local minima). After the training procedure, assuming there is enough data and the neural network architecture is well-chosen, the network is capable of making accurate predictions in classification, regression and other tasks. Modern software packages, such as \textit{TensorFlow} allow performing gradient descent quickly and efficiently for a pre-defined architecture \citep{Tensorflow2015}.

\section{Convolutional Neural Networks}

Another type of artificial neural networks that are important to discuss are CNNs. CNNs share a lot of the features with the previously discussed multilayer perceptrons with one major difference being that they extract useful features from the data using convolutions. To put it simply, the convolution procedure refers to convolving a filter (kernel) with different parts of an image, which allows extracting visual features from that image. In this respect, CNNs draw inspiration from the human visual cortex. The extracted visual features (edges, corners, main shapes etc.) are then processed and combined in order to make a prediction in a classification or another kind of machine learning task. 

\begin{figure}[ht!]
  \centering
    \includegraphics[width=1.0\columnwidth]{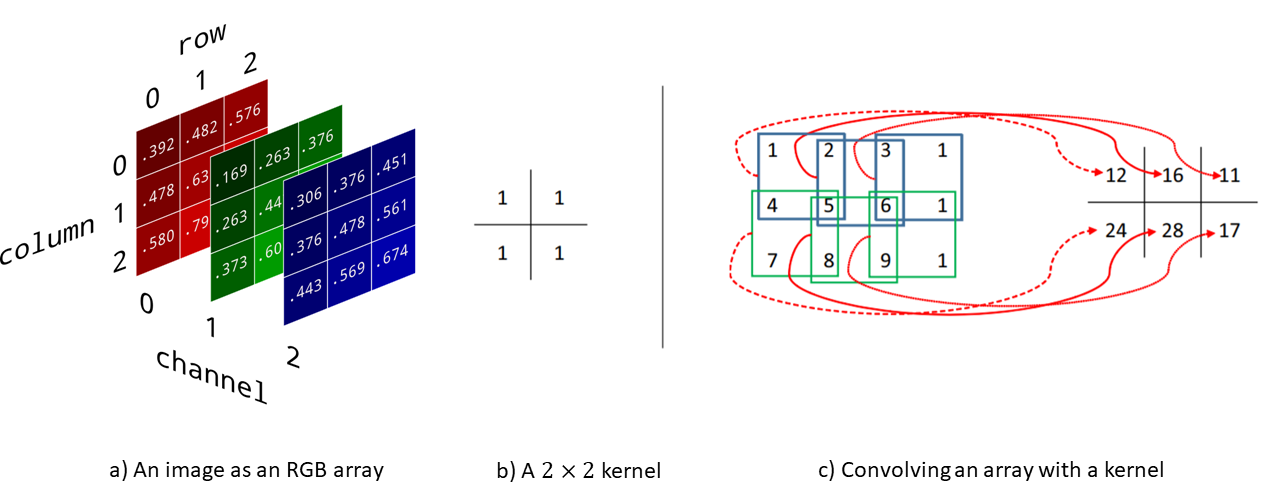}
    \caption[RGB images and the procedure of convolution]{\textbf{a:} representing a coloured image as an RGB array; \textbf{b:} a $2 \times 2$ kernel; \textbf{c:} the procedure of convolving a kernel with a single plane of an image. Image adapted from \citet{Wu2017, machine_learning2020}. }
\label{figure:convolving_images}
\end{figure}

\begin{figure}[ht!]
  \centering
    \includegraphics[width=0.9\columnwidth]{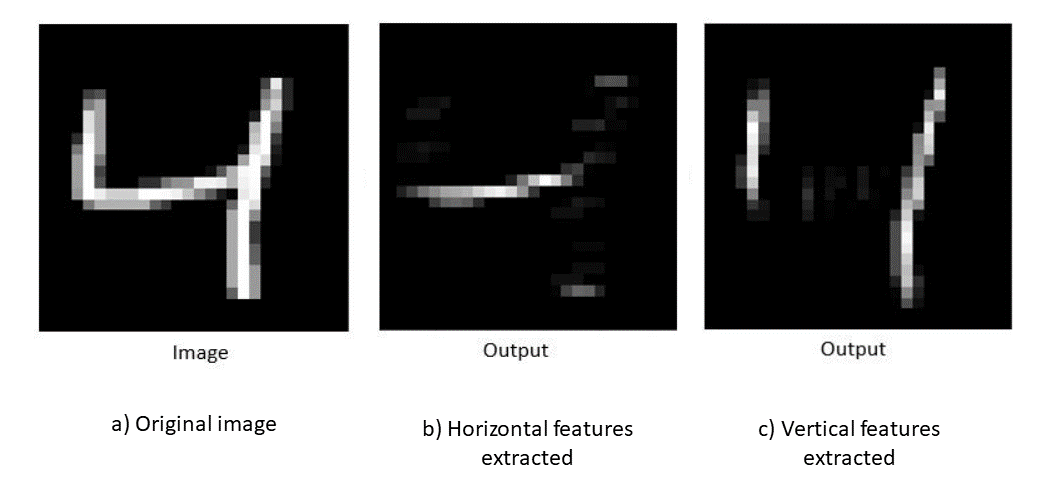}
    \caption[Edge extraction from an image]{\textbf{a:} a greyscale image representing a hand-written digit; \textbf{b:} horizontal features extracted by convolving the image with a horizontal edge kernel and passing it through a rectifier (ReLU) layer (i.e. $f(x) = \mathrm{max}(0,x) $); \textbf{c:} vertical features extracted by convolving the image with a vertical edge kernel and passing it through a rectifier layer. Image adpated from: \citet{Karkare2019}.
    }
\label{figure:edge_extraction}
\end{figure}

Mathematically, the procedure of convolving a kernel with an image can be described as: 

\begin{equation}
I^{'}(x, y)=k(x,y) * I(x, y)=\sum_{d x=-a}^{a} \sum_{d y=-b}^{b} k(d x, d y) I(x+d x, y+d y),
\label{eq:convolution}
\end{equation}

\noindent where $k(x,y)$ refers to the kernel (filter) matrix, $I(x,y)$ is the original image and $I^{'}(x,y)$ is the convolved image with the matrix values spanning $-a \leq d x \leq a$ and $-b \leq d y \leq b$. Figure \ref{eq:convolution} illustrates the convolution procedure for a simple $2 \times 2$ kernel with trivial values, which simply sums all the values in the corresponding section of an image. Choosing different values for the kernel matrix allows extracting different visual features from a given image. Figure \ref{figure:edge_extraction} illustrates the results of this procedure for a simple greyscale image. 

CNNs work most naturally with full-colour RGB images (figure \ref{figure:convolving_images}) which can be represented as 3-dimensional arrays. More generally, any dataset can be represented as a $N$-dimensional array (tensor\footnote{Note regarding the terminology: \textit{tensors} in machine learning and computer science literature often simply refer to $N$-dimensional arrays, rather than algebraic objects with specific transformation properties. }). In particular, we can denote the input data to the $l$-th CNN layer as: $\boldsymbol{x}^{l} \in \mathbb{R}^{H^{l} \times W^{l} \times D^{l}}$, where $H$, $W$, $D$ refer to the height, width and the number of channels in the input array (see figure \ref{figure:convolving_images}). During the training procedure CNNs are often trained on batches of input images, which can be represented as 4-dimensional arrays: $\mathbb{R}^{H^{l} \times W^{l} \times D^{l} \times N}$, where $N$ is the number of the images in a batch. Hence a CNN receives an input $\boldsymbol{x}^{l}$, transforms it (by convolving it or applying some other operation) all the way till the final layer $\boldsymbol{x}^{L} \in \mathbb{R}^{C}$, which corresponds to an array of values representing probabilities of the input image belonging to some class (classification task). Alternatively, the output of CNN could be another image  $\boldsymbol{x}^{L} \in \mathbb{R}^{H \times W}$.

Convolution kernels can be easily described as 2-dimensional or more generally as multiple $N$-dimensional arrays. In addition, in real CNNs kernels do not necessarily have to convolve images by covering every pixel and instead can skip every $n$-th pixel. This behaviour is summarized by the \textit{stride} parameter. Putting everything together, the output of the $l$-th layer can be denoted as: 
  
\begin{equation}
y_{i^{l+1}, j^{l+1}, d}=\sum_{i=0}^{H} \sum_{j=0}^{W} \sum_{d^{l}=0}^{D^{l}} k_{i, j, d^{l}, d} \times x_{i^{l+1}+i, j^{l+1}+j, d^{l}}^{l},
\label{eq:convolution2}
\end{equation}  

\noindent where $k$ as before denotes the kernel matrix. Equation \ref{eq:convolution2} looks complex, but it simply generalizes equation \ref{eq:convolution} for any input image shape and multiple kernels of custom size. 

As mentioned the training procedure of CNNs is in principle the same as for the multilayer perceptron networks. In particular, the arrays in the network can be vectorized and each value weighted, making the goal of the training procedure, as before, to find the optimal values for those weight parameters (see equation \ref{eq:gradient_descent}). Using the chain rule, for the $(i + 1)$-th layer, the main part of the optimization procedure is calculating the dependence of the loss/error function on the input and the weight parameters \citep{Wu2017}: 

\begin{equation}
\frac{\partial E}{\partial\left(\operatorname{vec}\left(\boldsymbol{w}^{i}\right)^{T}\right)}=\frac{\partial E}{\partial\left(\operatorname{vec}\left(\boldsymbol{x}^{i+1}\right)^{T}\right)} \frac{\partial \operatorname{vec}\left(\boldsymbol{x}^{i+1}\right)}{\partial\left(\operatorname{vec}\left(\boldsymbol{w}^{i}\right)^{T}\right)},
\label{eq:gradient_descent2}
\end{equation}

\begin{equation}
\frac{\partial E}{\partial\left(\operatorname{vec}\left(\boldsymbol{x}^{i}\right)^{T}\right)}=\frac{\partial E}{\partial\left(\operatorname{vec}\left(\boldsymbol{x}^{i+1}\right)^{T}\right)} \frac{\partial \operatorname{vec}\left(\boldsymbol{x}^{i+1}\right)}{\partial\left(\operatorname{vec}\left(\boldsymbol{x}^{i}\right)^{T}\right)},
\label{eq:gradient_descent3}
\end{equation}

\noindent where vec represents the vectorization operation. Calculating these terms is more challenging than in the case of ordinary multilayer networks, but can be done quite efficiently with modern software and graphical processing unit (GPU) support. 

Finally, an important topic to discuss in the context of the architecture of CNNs is the variety of the types of layers that can be used. Here some of the most important types of layers are listed and discussed:

\begin{itemize}
    \item \textbf{Batch normalisation layers}. These layers basically normalize the input data to have a zero mean and unit variance (or other specified values). This has been demonstrated to increase the efficiency of the training procedure, while also adding stability.
    \item \textbf{Pooling layers}. These layers reduce the dimensionality of the input data by using averaging, summing or maximization operations. For instance, this can simply refer to extracting the maximum values in each section of an input array by a simple kernel. 
    \item \textbf{Softmax layer}. A layer of this type is used as a final layer to provide the output values corresponding to probabilities of an object belonging to one of the $K$ classes in a classification problem.
    \item \textbf{Flatten layers}. These layers vectorize the $N$-dimensional input arrays to 1-dimensional vectors. 
\end{itemize}

\section{Generative Adversarial Networks}
\label{section:introduction_to_GANs}

As mentioned, generative models form an important class of machine learning algorithms that have been applied to solve a wide variety of problems in science. With the discovery of GANs in \citet{Goodfellow2014}, an entirely new way of using artificial neural networks has been discovered. The GAN algorithm refers to a system of two neural networks (these can be convolutional, but it is not necessary for the algorithm to work), a \textit{generator} and a \textit{discriminator} that are trained adversarially to produce novel statistically realistic data. In particular, the two neural networks compete in an adversarial fashion during the training process -- the generator is optimized to produce realistic datasets statistically identical to the training data and hence to fool the discriminator. Mathematically, such an optimization corresponds to minimizing the cost function $E$: 

\begin{equation}
 \min_{G_{\theta}} \max_{D_{\phi}}E(D_{\theta},G_{\phi}) = - \mathbb{E}_{X \sim p_{data}}\log(D_{\theta}(X)) - \mathbb{E}_{Z \sim p_{g}}\log(1 - D_{\theta}(G_{\phi}(Z))),  
\end{equation}

\noindent where $\mathbb{E}$ refers to the expectation function, $D_{\theta}$ to the discriminator with weights $\theta$, $G_{\phi}$ to the generator with weights $\phi$, $p_{r}$ to the distribution of the data we are aiming for, $p_{g}$ to the generated distribution, $X$ to the data (real or generated) analyzed by the discriminator and $Z$ to the random noise vector input to the generator.

Such an optimization procedure is a nice example of game theory where the two agents (the generator and the discriminator) compete in a two player zero sum game and adjust their \textit{strategies} (neural network weights) based on the common cost function. In case of perfect convergence, the GAN would reach Nash equilibrium, i.e. the generator and the discriminator would reach optimal configurations (optimal sets of weights). In practice, however, reaching convergence is difficult and the training procedure is often unstable and prone to mode collapse\footnote{This refers to the generator overpowering the discriminator, which results in the generator getting stuck in producing a small subset of identical or nearly identical realistic outputs.} \citep{farnia2020}. 

The two neural networks, the discriminator and the generator, have two different training procedures. In particular, the discriminator classifies the datasets into real (coming from the training dataset) or fake (produced by the generator) and is penalized for misclassification via the discriminator loss term. The discriminator weights are updated through backpropagation as usual. The generator, on the other hand, samples random noise, produces an image, gets the classification of that image from the discriminator and updates its weights accordingly via backpopagation using the generator loss function term. The full training procedure is done by alternating between the discriminator and the generator training cycles.

The described training procedure can be summarized more formally in pseudo-code (this is the original algorithm for the simplest version of a GAN described in great detail in \citet{Goodfellow2014}):

\begin{enumerate}
    \item For training iterations $n = 1$ to $N$ \textbf{do}:
    
    \begin{enumerate}[label=\roman*)]
        \item for $k$ steps \textbf{do}:
        
            \begin{itemize}
                \item Sample a minibatch of $m$ random noise samples $\left\{Z^{(1)}, \ldots, Z^{(m)}\right\}$ from the noise prior distribution $p_{g}(Z)$.
                \item Sample a minibatch of $m$ samples $\left\{X^{(1)}, \ldots, X^{(m)}\right\}$ from the training data distribution $p_{data}(X)$.
                \item Update the discriminator neural network by evaluating its stochastic gradient:
                
                \begin{equation}
                    \nabla_{\theta} \frac{1}{m} \sum_{i=1}^{m}\left[\log D_{\theta}\left(X^{(i)}\right)+\log \left(1-D_{\theta}\left(G_{\phi}\left(Z^{(i)}\right)\right)\right)\right].
                \label{eq:discriminator_gradient}
                \end{equation}
            \end{itemize}
            
        \textbf{end for}
        
        \item Sample a minibatch of $m$ random noise samples $\left\{Z^{(1)}, \ldots, Z^{(m)}\right\}$ from the noise prior distribution $p_{g}(Z)$.
        \item Update the generator neural network by evaluating its stochastic gradient: 
        
        \begin{equation}\nabla_{\phi} \frac{1}{m} \sum_{i=1}^{m} \log \left(1-D_{\theta}\left(G_{\phi}\left(Z^{(i)}\right)\right)\right).
        \label{eq:generator_gradient}
        \end{equation}
        
        \textbf{end for}
        
    \end{enumerate}
    
\item Output the updated discriminator and the generator neural networks: $D_{\theta}$, $G_{\phi}$.
    
\end{enumerate}

\begin{figure}[ht!]
  \centering
  \makebox[\textwidth][c]{\includegraphics[width=1.2\textwidth]{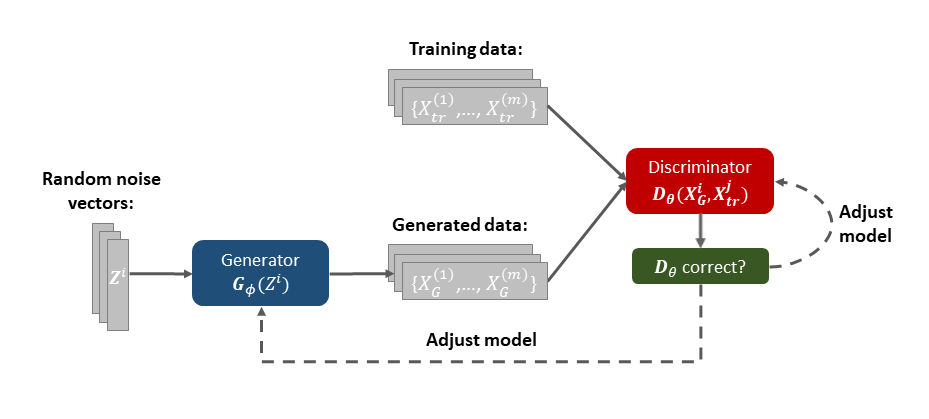}}
    \caption[GAN training procedure]{The training procedure for the simplest GAN algorithm as described in \citet{Goodfellow2014}. The figure is author's own.
    }
\label{figure:GAN_training}
\end{figure}

The training procedure described in the algorithm above is illustrated visually in figure \ref{figure:GAN_training}. Assuming the adversarial training is successful, the generator $G_{\phi}(Z)$ can then be used separately to produce realistic synthetic data from a randomized input vector $Z$. 

\begin{figure}[ht!]
  \centering
  \makebox[\textwidth][c]{\includegraphics[width=0.9\textwidth]{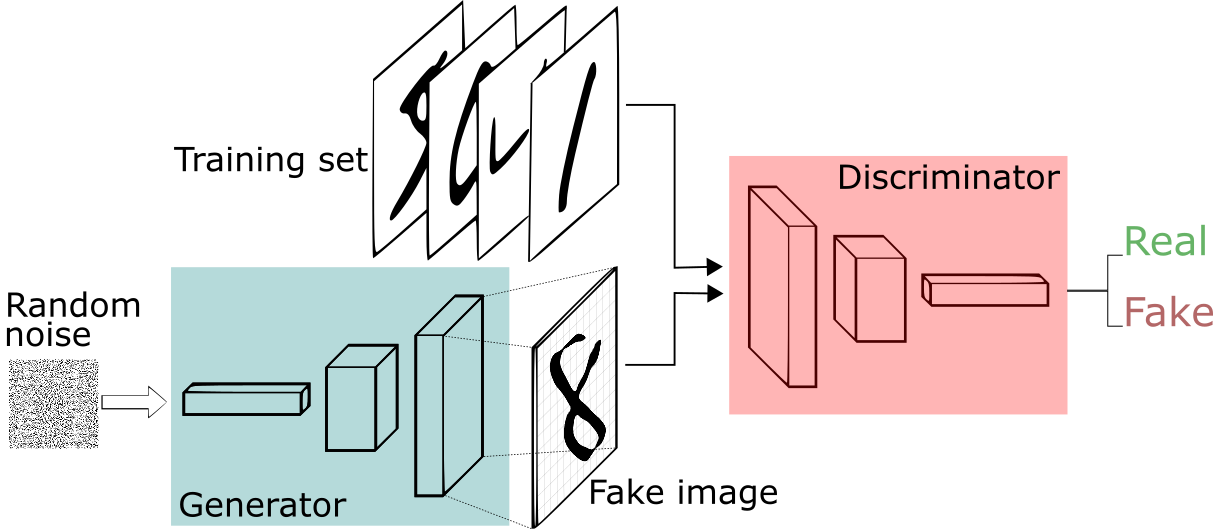}}
    \caption[GAN training procedure for MNIST images]{The training procedure for the simplest GAN algorithm as described in \citet{Goodfellow2014} for the MNIST dataset. Image from \cite{Gan_mnist}.
    }
\label{figure:GAN_mnist}
\end{figure}

A useful pedagogical example to examine is that of using a GAN to produce realistic hand-written digits. In this case the training dataset consists of 60,000  hand-written digits represented as $28 \times 28$ px images (the MNIST dataset described in \cite{MNIST-2010}). The training procedure then consists of the generator producing a batch of hand-written images from a random noise vector. The produced and the training image batches are used by the discriminator for updating the discriminator cost function, which in turn is passed to the generator for updating the corresponding error function. During the initial stages of training the generated images are not-realistic, however, after a few epochs, the generator weights are updated sufficiently in order to produce high quality images. Once the training is finished (provided that the common problem of mode collapse is avoided), the generator neural network can be used to produce high quality realistic hand-written images from a batch of noise vectors. The training procedure for the MNIST dataset is illustrated pictorially in figure \ref{figure:GAN_mnist}. 

\chapter{Using GANs for Emulating Cosmological Simulation Data}
\label{ch:GAN_emulators}

This chapter contains a novel GAN algorithm that is used for emulating cosmological simulation data. More specifically, a brief summary of the motivations and approaches to cosmological simulation emulators is given. In addition, various properties of GAN algorithms are discussed. And finally, an approach of emulating simulations of different cosmological parameters, redshifts and modified gravity parameters is introduced. The algorithm is a modified version of the \textit{cosmoGAN} code described in great detail in \citet{Mustafa2019}. Most figures are from \citet{Tamosiunas2020} unless otherwise specified. The calculations, coding and the result analysis was done by the author with consultation and the supervision by the supervisors and the co-authors. The L-PICOLA and MG-PICOLA simulation data was produced primarily by Hans Winther. The other datasets are given in appendix \ref{data_code_availability}.  

\section{The Need for Cosmological Emulators}

In the era of precision cosmology an important tool for studying the evolution of large scale structure is \textit{N}-body simulations. Such simulations evolve a large number of particles under the influence of gravity (and possibly other forces) throughout cosmic time and allow detailed studies of the non-linear structure formation. Modern cosmological simulations are highly realistic and extremely complex and may include galaxy evolution, feedback processes, massive neutrinos, weak lensing and many other effects. Such complexity however comes at a price in terms of computational resources and large simulations may take several days or even weeks to run. In addition, to fully account for galaxy formation and other effects various simplification schemes and semi-analytical models are required. To address these issues a variety of emulation techniques have been discussed in the literature \citep{ Kwan2015, Winther2019, Knabenhans2019}. In light of upcoming surveys like Euclid, such emulators will be an invaluable tool for producing mock data quickly and efficiently.

Lately, machine learning techniques have also been applied as an alternative to the traditional emulation methods. For instance, deep learning has been used to accurately predict non-linear structure formation \citep{He2019}. Similarly GANs and variational autoencoders have been used to produce novel realistic cosmic web 2-D projections, weak lensing maps and to perform dark energy model selection \citep{Rodriguez2018, Mustafa2019, Li2019C}. In addition the GAN approach has also been used to produce realistic cosmic microwave background temperature anisotropy 2-D patches as well as deep field astronomical images \citep{Mishra2019, Smith2019}. Finally, generating full 3-D cosmic web data has been discussed in \citet{Perraudin2019, Ramanah2020}. The cited works show that GANs are capable of reproducing a variety of cosmological simulation outputs efficiently and with high accuracy. 

However, certain challenges remain: the training process of the GAN algorithm is complicated and prone to failure and producing full scale 3-D results is computationally expensive. A common problem when training GANs is \textit{mode collapse}, when the generator neural network overpowers the discriminator and gets stuck in producing a small sample of identical outputs. Mode collapse can be addressed in multiple ways -- modern GAN architectures introduce label flipping or use different loss functions, such as the Wasserstein distance, which has been shown to reduce the probability of mode collapse \citep{Arjovsky2017}. 
In this chapter I address some of these issues and present the results on extending some of the currently existing GAN algorithms. In particular, as mentioned, a modified version of the \textit{cosmoGAN} algorithm (introduced in \cite{Mustafa2019}) is used to produce weak lensing convergence maps and 2-D cosmic web projections of different redshifts and multiple cosmologies, including dark matter, gas and internal energy data. Furthermore, other techniques from contemporary research in the field of deep learning are explored, such as latent space interpolation, which offers a way to control the outputs of the algorithm. This, to my best knowledge, is a novel approach that in the context of cosmology has not been explored in the literature so far. Finally, a discussion of GANs in the framework of Riemannian geometry is given in order to put the problem on a more theoretical footing and to explore the feature space learnt by the algorithm. Ultimately, the goal of the research described in this chapter is to adapt the existing algorithms towards becoming fully-controllable, universal emulators capable of producing both novel large scale structure data as well as other datasets, such as weak lensing convergence maps.

\section{DCGAN Architecture for Emulating Cosmological Simulation Data}

As outlined in section \ref{section:introduction_to_GANs}, the GAN algorithm can be used to generate novel, statistically realistic data based on some training dataset. In our case three types of training datasets are used: 2-D cosmic web slices, 2-D weak lensing convergence maps and stacks of cosmic-web slices for dark matter and baryonic simulation data along with internal energy data. Here it is important to clarify what exactly is meant by cosmic web slice and weak lensing convergence data. Cosmic web slices, in particular, refer to the 2-D discrete dark matter overdensity field $1 + \delta(x)$ data, where the value at each position $x$ refers to the density relative to the average density (i.e. $\delta(x) = (\rho(x) - \bar{\rho} )/\bar{\rho}$). Similarly, weak lensing maps refer to the discrete 2-D corvengence fields, where the numerical value at each position $x$ simply refers to the value of convergence $\kappa_{c}(x)$. Both types of datasets can be represented as 2-D arrays, where each entry of the array corresponds to the value of the overdensity or convergence. Analogously, these arrays can be represented visually as images, with each pixel value corresponding to the mentioned quantities (note that in some cases the pixel values are rescaled for visualization purposes).

Since we are dealing with data that can be naturally represented in a visual form, a straightforward choice is to pick a GAN architecture that works well with image data. In this regard, deep convolutional generative adversarial networks (DCGANs) have shown good results in producing statistically realistic novel visual data. Thus, in order to generate cosmic web and weak lensing convergence data, we chose to use the DCGAN architecture, as described in \citep{Mustafa2019}. More specifically, as a starting point the DCGAN implementation publicly available in \citet{cosmoGAN} was used.

\begin{table}[ht!]
\centering
\begin{tabular}{lccc}
\hline & \textbf{Activ.} & \textbf{Output shape} & \textbf{Params.} \\
\hline Latent & $-$ & 64 & $-$ \\
\hline Dense & $-$ & $512 \times 16 \times 16$ & $8.5 \mathrm{M}$ \\
BatchNorm & $\mathrm{ReLU}$ & $512 \times 16 \times 16$ & 1024 \\
\hline TConv $5 \times 5$ & $-$ & $256 \times 32 \times 32$ & $3.3 \mathrm{M}$ \\
BatchNorm & $\mathrm{ReLU}$ & $256 \times 32 \times 32$ & 512 \\
\hline TConv $5 \times 5$ & $-$ & $128 \times 64 \times 64$ & $819 \mathrm{K}$ \\
BatchNorm & $\mathrm{ReLU}$ & $128 \times 64 \times 64$ & 256 \\
\hline TConv $5 \times 5$ & $-$ & $64 \times 128 \times 128$ & $205 \mathrm{K}$ \\
BatchNorm & $\mathrm{ReLU}$ & $64 \times 128 \times 128$ & 128 \\
\hline TConv $5 \times 5$ & Tanh & $1 \times 256 \times 256$ & 1601 \\
\hline \multicolumn{2}{l} { Total trainable parameters } &   &$\mathbf{1 2 . 3 M}$ \\
\hline
\end{tabular}
\caption[Architecture of the generator neural network]{The architecture of the generator neural network as described in \citet{Mustafa2019}. \textit{TConv} corresponds to the transposed convolutional layer with $\mathrm{stride}=2$ (and the kernel size given by the shown numerical values). \textit{ReLU} corresponds to the rectified linear unit activation function.   }
\label{table:generator_network}
\end{table}

\noindent Tables \ref{table:generator_network} and \ref{table:discriminator_network} describe the key features of the architecture. Both the discriminator and the generator are standard convolutional neural networks using primarily ReLU and leaky ReLU activation functions along with transposed convolutional and standard convolutional layers.

\begin{table}[ht!]
\centering
\begin{tabular}{lccc}
\hline & \textbf{Activ.} & \textbf{Output shape} & \textbf{Params.} \\
\hline Input map & $-$ & $1 \times 256 \times 256$ & $-$ \\
\hline Conv $5 \times 5$ & LReLU & $64 \times 128 \times 128$ & $1664$ \\
\hline Conv $5 \times 5$ & $-$ & $128 \times 64 \times 64$ & $205 \mathrm{K}$ \\
BatchNorm & LReLU & $128 \times 64 \times 64$ & 256 \\
\hline Conv $5 \times 5$ & $-$ & $256 \times 32 \times 32$ & $819 \mathrm{K}$ \\
BatchNorm & LReLU & $256 \times 32 \times 32$ & 512 \\
\hline Conv $5 \times 5$ & $-$ & $512 \times 16 \times 16$ & $3.3 \mathrm{M}$ \\
BatchNorm & LReLU & $512 \times 16 \times 16$ & 1024 \\
\hline Linear& Sigmoid & $1$ & $131 \mathrm{K}$ \\
\hline \multicolumn{2}{l} { Total trainable parameters } &   &$\mathbf{4.4 M}$ \\
\hline
\end{tabular}
\caption[Architecture of the discriminator neural network]{The architecture of the discriminator neural network as described in \citet{Mustafa2019}. \textit{Conv} stands for convolutional layers with $\mathrm{stride}=2$ the kernel size given by the numerical value. \textit{LReLU} stands for the leaky rectified linear unit activation function with the leakiness parameter $=0.2$.  }
\label{table:discriminator_network}
\end{table}

To adapt the outlined architecture to the problem at hand, I experimented with different activation functions, different strides and different sizes of the convolutional layers. The results indicated that the architecture used in \citet{Mustafa2019} with minor variations generally worked well for producing realistic cosmic web and weak lensing data as well as the combined dark matter and baryonic data samples. More specifically, for  the cosmic web data, the input shape (i.e. the size of the random noise vector) was changed from $64$ to $256$ to account for the higher complexity of the cosmic web images when compared to the weak lensing maps. In the case of emulating dark matter, gas and internal energy slices, the public code was adapted to work with multi-channel data (i.e. RGB input arrays). In addition, the default batch size (number of input arrays that the algorithm uses during an iteration of training) was changed from $32$ to $64$. Extra functions were also added to the code to allow performing the latent space interpolation easier (see section \ref{latent_space_interpolation} and appendix \ref{data_code_availability}). Finally, when training on weak lensing convergence maps, the architecture was left unchanged as one of the key goals was to reproduce the results described in \citet{Mustafa2019}.

\begin{figure}[ht!]
  \centering
    \includegraphics[width=1.05\columnwidth]{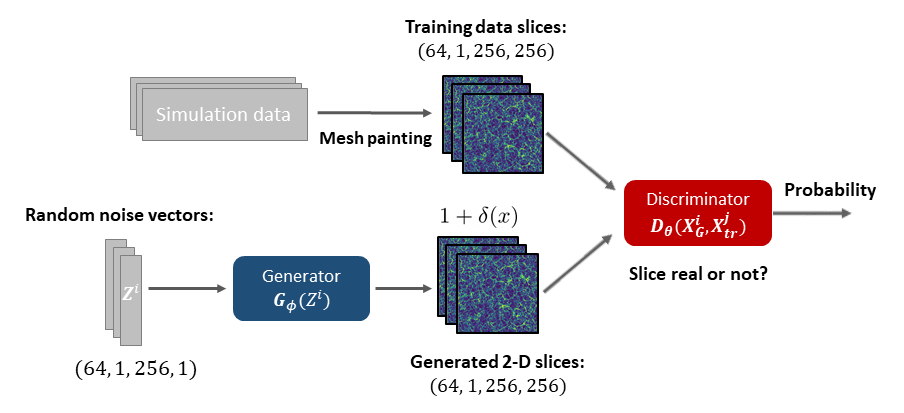}
    \caption[The pipeline of training a GAN on cosmic web slice data]{The pipeline of training a GAN on cosmic web slice data. The numbers in the brackets are the sizes of the input/output arrays and correspond to the number of images in the batch along with the channel number, height and the width of the images.  
    }
\label{figure:GAN_pipeline}
\end{figure}

In summary, if we choose convolutional neural networks (rather than simple multi-layer perceptrons) for the generator and the discriminator, the training procedure described in section \ref{section:introduction_to_GANs} essentially remains the same. In fact, any algorithm that is capable of producing and classifying data could be used as the generator and the discriminator. And then, if we represent the cosmic web and the weak lensing data as 2-D arrays, the pipeline of training a GAN on our data is summarized in figure \ref{figure:GAN_pipeline}. In particular, the simulation data is used to produce the 2-D cosmic web slices, which are then used as a training dataset. More specifically, during the training procedure the generator network produces a batch of $64$ images from a batch of $64$ random noise vectors (drawn from a Gaussian distribution centered around $0$), which is then sent to the discriminator network, where it is combined with a batch of $64$ training images for classification. In the case of a training dataset with multi-channel images (RGB images), the generator and the discriminator process batches of images of the following shape: $(64,3,256,256)$, where the second value corresponds to the number of channels.

To summarize, the problem at hand is to emulate cosmological simulation data using GANs. More specifically, the aim is to emulate novel dark matter overdensity fields, represented by 2-D arrays produced via mesh painting from the raw simulation output data. In addition, we also aim to produce realistic convergence field and gas density data from hydrodynamic simulations. In all cases the training data consists of batches of 2-D arrays, which, when plotted visually, represent the mentioned fields (projected to 2-D). A more technical discussion of the datasets is given in section \ref{section:datasets}. Once the training is completed, the generator neural network can be used to produce realistic novel 2-D arrays in the same format as the training dataset. Note that the problem at hand is fundamentally analogous to that of generating hand-written images (i.e. figure \ref{figure:GAN_mnist}) the only difference being the size of the dataset arrays and the architecture of the neural networks used.  

\section{Latent Space Interpolation}
\label{latent_space_interpolation}

Before discussing the different datasets and the training procedure, it is important to review another important feature of the GAN algorithms. Latent space interpolation refers to the procedure of interpolating between a pair of outputs produced by a GAN. This procedure not only allows us to study the feature space learnt by the algorithm, but also allows us to control which outputs the algorithm produces. Here a review of the procedure and its various uses is given.  

If the training procedure is successful, the generator $G_{\phi}(Z^{i})$ learns to map the values of a random vector $Z^{i}$ to the values of a statistically realistic output vector $X^{i}_{G}$, which can be reshaped to the original 2-D array shape representing an image $X^{jk}_G$ (a cosmic web slice or a convergence map in our case). This can be viewed as mapping from a low-dimensional \textit{latent} space $Z \subseteq \R^{d}$ to a higher-dimensional data (pixel) space $X \subseteq \R^{D}$ (where $d$ is the size of the noise input vector and $D$ is the total number of the of the output image pixels; for more details see \cite{shao2017}). For a generator neural network $d \ll D$ (in our case $d = 256$ or $64$, while $D = 256^{2}$).

The training procedure can be viewed as the generator learning to map clusters in the $Z$ space to the clusters in the $X$ space. Hence, if we treat the random input vectors\footnote{Note regarding notation: here a superscript refers to input/output vectors, while a subscript refers to the corresponding point in the latent/output data space.} $Z^{i}$ as points in a $d$-dimensional space, we can interpolate between multiple input vectors and produce a transition between the corresponding outputs. In particular, if we choose two input vectors that correspond to points $Z_{1}$ and $Z_{2}$ and find a line connecting them, sampling intermediate input points along that line leads to a set of outputs that correspond to an almost smooth transition between the output points $X_{1}$ and $X_{2}$. 

\begin{figure}
  \centering
    \includegraphics[width=0.98\textwidth]{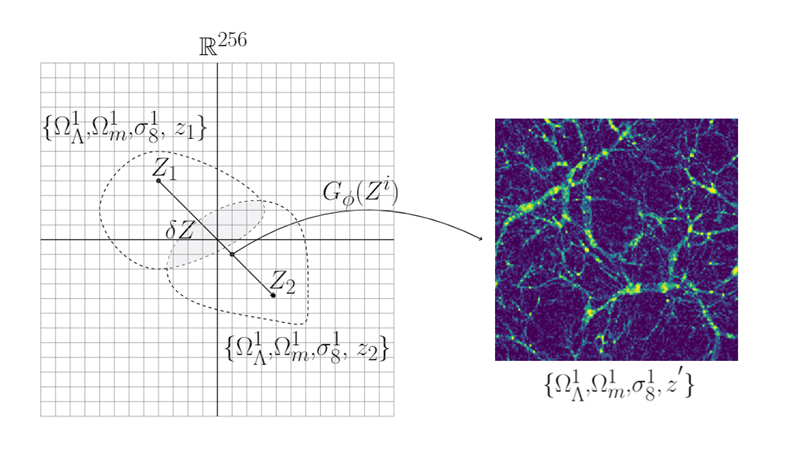}
    \caption[Illustration of the linear latent space interpolation procedure]{Illustration of the latent space interpolation procedure. Training the GAN algorithm on the cosmic web slices of two different redshifts encodes two different clusters in the latent space (which is a subset of a 256-dimensional space, i.e. the size of the random noise input vector). Sampling a point from the line connecting two input points $Z_{1}$ and $Z_{2}$ in this space produces an output with redshift  $z^{'}$. In the case of our dataset with $z_{1} = 1.0$ and $z_{2} = 0.0$, several points near the centre of this line correspond to outputs approximately emulating $z^{'} \approx 0.5$. }
    \label{figure 2}
\end{figure}

As an example, if we train the generator to produce cosmic web slices of two different redshifts, we can produce a set of outputs corresponding to a transition between those two redshifts by linearly interpolating between the input points $Z_{1}$ and $Z_{2}$ (see figure \ref{figure 2}). More concretely, if we train the algorithm on cosmic web slices of redshifts $\{0.0,1.0\}$, somewhere between the two input points, one can find a point $Z^{'}$, which produces an output that has a matter power spectrum approximately corresponding to a redshift  $z^{'}\approx 0.5$. This is fascinating given that the training dataset did not include intermediate redshift data. Here it is important to note that such an interpolation procedure does not necessarily produce a perfectly smooth transition in the data space, i.e. the produced outputs corresponding to the points $\{X_{i}\}$ between $X_{1}$ and $X_{2}$ are not always realistic (in terms of the matter power spectrum and other statistics; see figure \ref{figure11} and section \ref{section:latent_results} for further details). Also, one might naively think that the point $Z^{'}$ lies in the middle of the line connecting $Z_{1}$ and $Z_{2}$, but in general we found it not to be the case (as the middle of the mentioned line does not necessary correspond to the middle between $X_{1}$ and $X_{2}$ in the data space, which is known to be non-Euclidean (see section \ref{riemannian_geometry})). In the upcoming chapters I investigate whether the latent space interpolation procedure can be used to map between outputs of different redshifts and cosmologies and whether the produced datasets are physically realistic. 

The latent space interpolation technique was performed by randomly choosing two input points $Z_{1}$ and $Z_{2}$, finding the line connecting the two points in the 256 (64)-dimensional space (256 (64) is the size of the corresponding input vectors) and then sampling 64 equally spaced points along that line. The outputs of the generator neural network of those intermediate input points $G_{\phi}(\{Z_{i}^{int}\})$ then correspond to cosmic web slices and weak lensing maps that represent a transition between the two outputs $G_{\phi}(Z_{1})$ and $G_{\phi}(Z_{2})$.

In order to perform linear latent space interpolation it is crucial to have the ability to distinguish between different data classes produced by the GAN (e.g. cosmic web slices of different redshifts). This was resolved by employing a combination of the usual summary statistics like the power spectrum and the Minkowski functionals along with two different machine learning algorithms. In particular, deep convolutional neural network and gradient boosted decision trees were used for distinguishing the different classes of datasets produced by the GAN \citep{chen2016_xgboost}.  

\section{Riemannian Geometry of GANs}
\label{riemannian_geometry}

The latent space interpolation procedure described in the previous section is a good example of how Riemannian geometry can be employed to describe certain features of the GAN algorithm. Recently various connections between GANs and Riemannian geometry have been explored in the machine learning literature in a more general context. Such connections are important to explore not only for the sake of curiosity, but also because they allow us to describe GANs and their optimization procedure in a language more familiar to physicists. A Riemannian geometry description of GANs is also powerful when exploring the latent space of a trained generator neural network and the outputs that it produces. Finally, a differential geometry description could shine some light on the connections between generative models and information geometry, which is a well-established field and could offer some new insights into training and analyzing the outputs of such models.

\begin{figure}[!ht]
  \centering
    \includegraphics[width=0.85\textwidth]{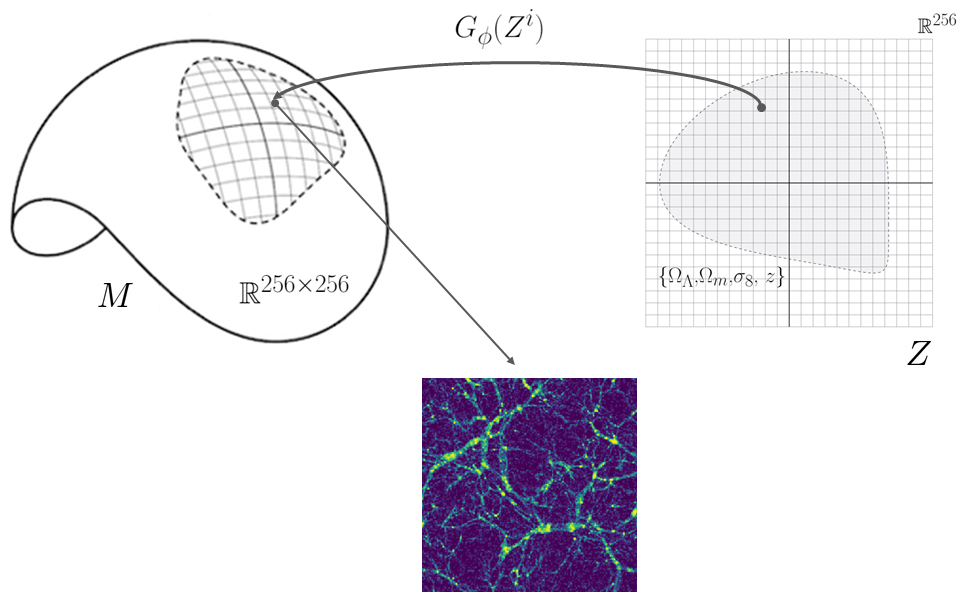}
    \caption[Riemannian geometry of generative adversarial networks]{Riemannian geometry of generative adversarial networks. The generator $G_{\phi}(Z^{i})$ can be treated as a mapping from the lower dimensional Euclidean latent space $Z$ (corresponding to the random noise input) to a high dimensional data (pixel) space $M$ (in general non-Euclidean). Each point on $M$ corresponds to a cosmic web slice (or a weak lensing map). }
    \label{riemanian_geometry}
\end{figure}

Recent work in \cite{shao2017} proposes treating the trained generator neural network as a mapping from a lower dimensional latent space $Z$ to the higher dimensional data space $X$: $G_{\phi}: Z \rightarrow X$ (see fig. \ref{riemanian_geometry}). More specifically, the generator $G_{\phi}(Z^{i})$ maps the latent space vectors of size $n$ (in our case $n = 256$ or $64$) to a manifold $M$ of dimensionality $m$ ($256 \times 256$, i.e. the number of pixels in the output images). Manifold $M$ here simply refers to a subset of the data space (all possible combinations of pixel values), which correspond to realistic images of weak lensing/cosmic web slices. The existence of such a manifold is postulated by the \textit{manifold hypothesis} in deep learning, which states that high-dimensional data can be encoded on a manifold of a much lower dimension \citep{Fefferman2013}. 

Hence if we treat the generator neural network $G_{\phi}$ as a mapping for the latent space to the data space manifold, one can naturally define an induced metric $g$, which then allows to quantify the distance between the points on the manifold and the length of curves. For a mapping described by the generator neural network, the metric is simply equal to a product of the Jacobian and the transposed Jacobian \citep{shao2017}: 

\begin{equation}
    g = J(Z)^{T}J(Z). 
\end{equation}

\noindent The Jacobian in our case refers to the partial derivative of each output value w.r.t. to each input value, i.e.: 

\[
J = \begin{bmatrix} 
    \frac{\partial X^{1}}{\partial Z^{1}} & \frac{\partial X^{1}}{\partial Z^{2}} & \dots  &\frac{\partial X^{1}}{\partial Z^{n}} \\
    \vdots & \vdots &\ddots & \\
    \frac{\partial X^{m}}{\partial Z^{1}}& \frac{\partial X^{m}}{\partial Z^{2}} & \dots        & \frac{\partial X^{m}}{\partial Z^{n}} 
    \end{bmatrix}.
\]

\noindent Once a metric is defined, one can use the usual tools to describe geodesics on the manifold $M$. For instance, one can define a curve $\kappa_{Z}$ between two points $a$ and $b$ in the latent space $Z$ parametrized by some parameter $t$. Using the mapping $G_{\phi}$, the corresponding curve on the manifold $M$ is then: $G_{\phi}(\kappa_{Z}(t)) \in M$. To find a curve that corresponds to a geodesic on the manifold one has to solve the Euler-Lagrange equation, which gives: 

\begin{equation}
    \frac{d^{2}\kappa_{Z}^{\alpha}}{dt^{2}} = - \Gamma^{\alpha}_{\beta \gamma} \frac{d\kappa_{Z}^{\beta}}{dt} \frac{d\kappa_{Z}^{\gamma}}{dt},
\end{equation}

\noindent where $\Gamma$ is the usual Christoffel symbol, given by:

\begin{equation}
    \Gamma^{\alpha}_{\beta \gamma} = \frac{1}{2} g^{\alpha \delta} \bigg(\frac{\partial g_{\delta \beta}}{\partial X^{\gamma}} + \frac{\partial g_{\delta \gamma}}{\partial X^{\alpha}} - \frac{\partial g_{\alpha \beta}}{\partial X^{\delta}} \bigg).
\end{equation}

As discussed in \cite{shao2017} geodesics between points on the manifold are of special importance, as they give the smoothest possible transition between multiple outputs. One of the main findings in \cite{shao2017} was that the Riemannian curvature of the manifold corresponding to the their data was surprisingly small and, hence, linear interpolation produced realistic results comparable to the results produced by calculating a geodesic curve between the outputs. In our work we also found that linear interpolation generally produced realistic results. However, to ensure that the outputs produced via the latent space interpolation are indeed realistic, one would have to interpolate on a curve in the latent space (corresponding to the geodesic connecting the needed outputs on the data manifold $M$) rather than a line.

Another important connection to Riemannian geometry comes in the context of the discriminator neural network. The discriminator can be viewed as a mapping from the data manifold to a probability manifold $P$, where each point on the manifold corresponds to the probability of a given data sample being real (i.e. belonging to the training dataset). Such a manifold looks remarkably similar to the statistical manifolds studied in the field of information geometry. Insights from information geometry have a long tradition of being used in neural network optimization (e.g. \cite{Hauser2017}). Exploring such connections could lead to deeper insights into the GAN training process, which is an interesting direction for future work.

\section{Datasets and the Training Procedure}
\label{section:datasets}

This section contains a detailed introduction to the datasets that were used to train the GAN algorithm described in the previous sections. In each case the simulations that were used to produce the dataset are described as well. Finally, the cosmological parameters and the used smoothing techniques are described as well.   

\subsection{Weak Lensing Convergence Map Data}

Gravitational potentials influence the path of photons in such a way that they introduce coherent distortions in the apparent shape (shear) and position of light sources. Weak gravitional lensing introduces ellipticity changes in objects of the order of $\approx$ 1\% and can be measured across the sky, meaning that maps of the lensing distortion of objects can be made and related to maps of the mass distribution in the Universe. The magnitude of the shear depends upon the combined effect of the gravitational potentials between the source and the observer. An observer will detect this integrated effect and maps of the integrated mass, or convergence, can be made. Gravitational lensing has the significant advantage that it is sensitive to both luminous and dark matter, and can therefore directly detect the combined matter distribution. In addition, weak lensing convergence maps allow for detecting the growth of structure in the Universe and hence they can also be used for probing statistics beyond two point correlation functions, such as in the higher moments of the convergence field or by observing the topology of the field with Minkowski functionals and peak statistics \citep{Dietrich2010, Mawdsley2020}. As future surveys attempt to further probe the non-linear regime of structure growth, the information held in these higher order statistics will become increasingly important, and will also require accurate simulations in order to provide cosmological constraints. This requirement for large numbers of simulations that also model complex physical phenomena means that more computationally efficient alternatives to \textit{N}-body simulations, such as the GAN approach proposed in this work, are required.

In order to train the GAN algorithm to produce realistic convergence maps, publicly available datasets were used. In particular, to test whether we could reproduce the original results from \citet{Mustafa2019} the publicly available data from \citet{cosmoGAN} was used. The dataset consists of 8000 weak lensing maps that were originally produced by running a Gadget2 \citep{Volker2005} simulation with $512^{3}$ particles in a $240$ $\mathrm{Mpc}/h$ box. To perform ray tracing the Gadget weak lensing simulation pipeline was used. The simulation box was rotated multiple times for each ray tracing procedure, resulting in 1000 12 sq. degree maps per simulation box. 

In order to train the GAN algorithm on convergence maps of different cosmologies and redshifts, the dataset publicly available at \citep{Matilla2016, Gupta2018, Columbia} was used. The available dataset contains weak lensing convergence maps covering a field of view of 3.5 deg $\times$ 3.5 deg, with resolution of 1024 $\times$ 1024 pixels. The maps were originally produced using Gadget2 DM-only simulation data with 240 Mpc$/h$ side box and $512^{3}$ particles. The dataset includes 96 different cosmologies (with varying $\Omega_{m}$ and $\sigma_{8}$ parameters). The values of $\Omega_{m} = 0.260$ and $\sigma_{8} = 0.8$ were used as the fiducial cosmology. In this work only a small subset of this dataset was used, namely, the maps where only one of the two cosmological parameter varies. In particular, the dataset consisting of the maps with $\sigma_{8} = \{0.436, 0.814 \}$ with a common value of $\Omega_{m} = 0.233$ was used. This was done in order to simplify the latent space analysis.

For the weak lensing map data the same architecture as described in tables \ref{table:generator_network} and \ref{table:discriminator_network} was used. In fact the same basic architecture with minor variations was used for training all the datasets described later on. The key parameter in terms of the training procedure is the learning rate. For all the cosmic web slice datasets, I found the learning rate value of $R_{L} = 3 \times 10^{-5}$ to work well. In the case of all the considered weak lensing datasets $R_{L} = 9 \times 10^{-6}$ was used. The training procedure and all the key parameters are described in great detail in the publicly available code (see appendix \ref{data_code_availability} for more information). 

\subsection{Cosmic Web Slice Data}
\label{cosmic-web data}

The cosmic web or the dark matter overdensity field refers to the intricate network of filaments and voids as seen in the output data of \textit{N}-body simulations. The statistical features of the cosmic web contain important information about the underlying cosmology and could hide imprints of modifications to the standard laws of gravity. In addition, emulating a large number of overdensity fields is important for reliable estimation of the errors of cosmological parameters. Hence, emulators, such as the one proposed in this work, are of special importance for the statistical analysis in the context of the upcoming observational surveys.      

The cosmic web training dataset was produced by employing a similar procedure to the one outlined in \citet{Rodriguez2018}. In particular, we ran L-PICOLA \citep{howlett2015} to produce a total of 15 independent simulation boxes with different cosmologies. Initially, the same cosmology as described in \citet{Rodriguez2018} was used with $h = 0.7$, $\Omega_{\Lambda} = 0.72$ and $\Omega_{m} = 0.28$. Subsequently, the effects of varying one of the cosmological parameters, namely the $\sigma_{8}$ parameter, was studied. The values of $\sigma_{8} = \{0.7,0.8,0.9\}$ along with $\Omega_{\Lambda} = 0.7$, $\Omega_{m} = 0.3$ and $h = 0.67$ were explored. For each different set of simulations, snapshots at 3 different redshifts: $z = \{0.0, 0.5,1.0\}$ were saved. For each simulation, a box size of 512 Mpc/$h$ was used with $512^{3}$ particles. For the latent space interpolation procedure, the GAN was trained on slices with redshifts $\{0.0,1.0\}$, with a common value of $\sigma_{8} = 0.8$.

To produce the slices for training the GAN, I used \textit{nbodykit} \citep{nbodykit}, which allows painting an overdensity field from a catalogue of simulated particles. To obtain the needed slices, the simulation box was cut into sections of 2 Mpc width in $x,y,z$ directions and for each section a mesh painting procedure was done. This refers to splitting the section into cells, where the numerical value of each cell corresponds to the dark matter overdensity $1 + \delta(x)$. Finally, after a 2-D projection of each slice, a $256^{2}$ px image was obtained, with each pixel value corresponding to the overdensity field. To emphasize the features of the large scale structure, I applied the same non-linear transformation as described in \citet{Rodriguez2018}: $s(x) = 2x/(x+a) - 1$, with $a = 250$, which rescales the overdensity values to $[-1,1]$ and increases the contrast of the images.

In order to emulate modified gravity effects the MG-PICOLA code was used. MG-PICOLA extends the original L-PICOLA code in order to allow simulating theories that exhibit scale-dependent growth \citep{Scoccimarro2012, Tassev2013, Winther2017, mg-picola}. This includes models such as $f(R)$ theories, which replace the Ricci scalar with a more general function in the Einstein-Hilbert action (see \citet{Li2019B} and chapter \ref{ch:modified_gravity} for an overview of the phenomenology of such models). In particular, multiple runs of MG-PICOLA were run with the following range of the $f_{R0}$ parameter: $[10^{-7}, 10^{-1}]$. Such a wide range was chosen to make the latent space interpolation procedure easier. The $f(R)$ simulations were also run with the same seed as the corresponding $\Lambda$CDM simulations, making the two datasets described above directly comparable.   

\subsection{Dark Matter, Gas and Internal Energy Data}

Simultaneously generating dark matter and the corresponding baryonic overdensity field data is a great challenge from both the theoretical and the computational perspectives. Namely, generating the baryonic distribution requires detailed hydrodynamical simulations that account for the intricacies of galaxy formation and feedback processes, which leads to a major increase in the required computational resources. For this reason, emulating large amounts of hydrodynamical simulation data is of special importance.   

To produce the dark matter, baryonic matter and the internal energy distribution slices I used the publicly available Illustris-3 simulation data \citep{vogelsberger2014,nelson2015}. Illustris-3 refers to the low resolution Illustris run including the full physics model with a box size of 75000 $\textrm{kpc}/h$ and over $9 \times 10^{7}$ dark matter and gas tracer particles. The cosmology of the simulation can be summarized by the following parameters: $\Omega_{m} = 0.2726$, $\Omega_{\Lambda} = 0.7274$, $h =0.704$. The simulation included the following physical effects: radiative gas cooling, star formation, galactic-scale winds from star formation feedback, supermassive black hole formation, accretion, and feedback. 

\begin{figure}[!ht]
\centering
\captionsetup[subfigure]{justification=centering}
  \begin{subfigure}[b]{0.37\textwidth}
    \includegraphics[width=\textwidth]{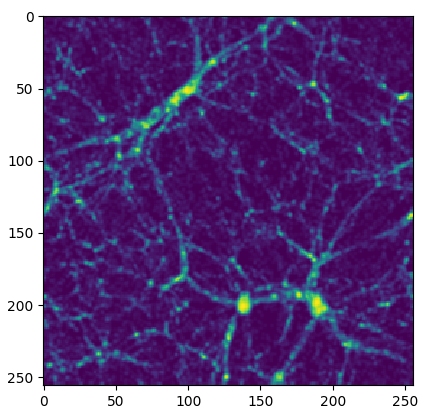}
    \caption{DM overdensity field}
    \label{fig3:a1}
    
  \end{subfigure}
  \begin{subfigure}[b]{0.364\textwidth}
    \includegraphics[width=\textwidth]{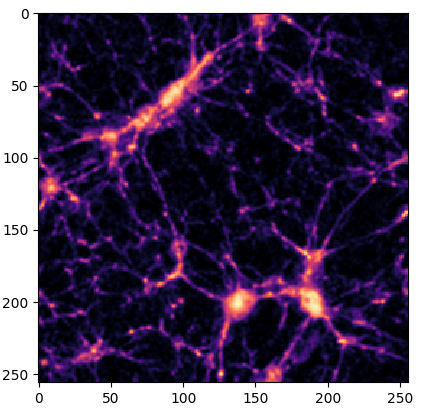}
    \caption{Gas overdensity field}
    \label{fig3:a2}
  \end{subfigure}
    \begin{subfigure}[b]{0.37\textwidth}
    \includegraphics[width=\textwidth]{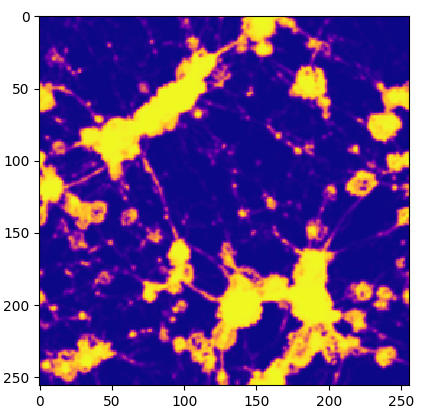}
    \caption{Internal energy field}
    \label{fi3:a3}
  \end{subfigure}
    \begin{subfigure}[b]{0.37\textwidth}
    \includegraphics[width=\textwidth]{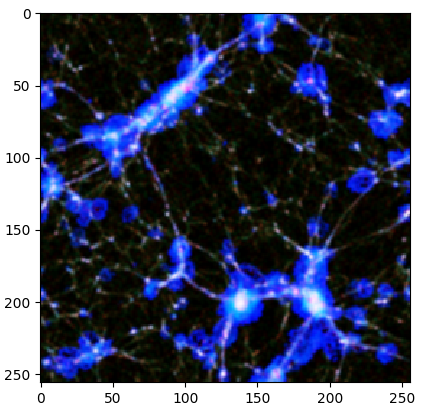}
    \caption{All components combined}
    \label{fig3:a4}
  \end{subfigure}
  \caption[Samples of the Illustris dataset]{Samples from the Illustris simulation dataset used to train the GAN algorithm: 2-D slices of the different simulation components (box size: 75000 $\textrm{kpc}/h$).}
  \label{figure3}
\end{figure}

To form the training dataset I used an analogous procedure to the one used for the cosmic web slices in section \ref{cosmic-web data}. In particular, the full simulation box was cut into slices of 100 $\textrm{kpc}/h$ and for each slice mesh painting was done to obtain the overdensity field. This was done for the dark matter and gas data. In addition, the available internal energy (thermal energy in the units of $(\rm km/s)^{2}$) distribution data was used as well. Figure \ref{figure3} shows a few samples from the dataset.

To investigate whether the GAN algorithm could be trained on multidimensional array data, the DM, gas and energy distribution 2-D slices were treated as RGB planes in a single image. In particular, a common way of representing colors in an image is forming a full color image out of three planes, each corresponding to the pixel values for red, green and blue colours (see figure \ref{figure:convolving_images}). In this framework, a full-color image corresponds to a 3-D array. Convolutional neural networks, including the one that the \textit{cosmoGAN} algorithm is based on are originally designed to be trained on such RGB images. Hence we combined the mentioned DM, gas and internal energy slices into a set of RGB arrays that were used as a training set.  

\subsection{The Training Procedure}

The initial stages of training (i.e. reproducing the results in \citep{Rodriguez2018, Mustafa2019} were done using the Google Cloud computing platform. The following setup was used: 4 standard vCPUs with 15 GB memory, 1 NVIDIA Tesla K80 GPU and 2TB of SSD hard drive space. 

Later stages of training (i.e. training the GAN on different cosmology, modified gravity and redshift data) were done using the local Sciama HPC cluster, which has 3702 cores of 2.66 GHz Intel Xeon processors with 2 GB of memory per core.

Given how unstable the GAN training procedure is, a simple way of evaluating the best checkpoint was used: I calculated the mean square difference between the mean values of the GAN-produced and the training dataset power spectra, pixel histograms and the Minkowski functionals. The set of GAN weights that minimizes this value was used for the plots displayed in the result section.  

\section{Diagnostics}

A key aspect of the analysis of the produced samples is being able to quantify how realistic the GAN-generated data is. This was done at an ensemble level -- i.e. we generated multiple batches of data (see figures in section \ref{section:results}) and calculated the average summary statistics, which were then compared against analogous results produced using the training dataset. 

The results produced by the algorithm were investigated using the following diagnostics: the 2-D matter power spectrum, overdensity (pixel) value histogram and the three Minkowski functionals. In addition, the cross and the auto power spectrum were computed in order to investigate the correlations between the datasets on different scales. The cross-power spectrum was calculated using: 

\begin{equation}
    \langle \tilde{\delta_{1}}(l)\tilde{\delta_{2}^{*}}(l^{'}) \rangle = (2 \pi)^{2}\delta_{D}(l-l^{'})P_{\times}(l),
    \label{eq:power_spectra}
\end{equation}

\noindent where $\tilde{\delta_{1}}$ and  $\tilde{\delta_{2}^{*}}$ are the Fourier transforms of the two overdensity fields at some Fourier bin $l$ and $\delta_{D}$ is the Dirac delta function.

The Minkowski functionals are a useful tool in studying the morphological features of fields that provide not only the information of spatial correlations but also the information on object shapes and topology. For some field $f(x)$ in 2-D we can define the three Minkowski functionals as follows: 

\begin{equation}
    V_{0}(\nu) = \int^{}_{Q_{\nu}} d \Omega,
    \quad\text{}\quad
    V_{1}(\nu) = \int^{}_{\partial Q_{\nu}} \frac{1}{4}dl,
    \quad\text{}\quad
    V_{2}(\nu) = \int^{}_{\partial Q_{\nu}} \frac{1}{2\pi} \kappa_{b}dl.
    \label{minkowski}
\end{equation}

Where $Q_{\nu} \equiv \{x \in \R^{2} | f(x) > \nu \} $ is the area and $\partial Q_{\nu} \equiv \{x \in \R^{2} | f(x) = \nu \} $ is the boundary of the field at the threshold value $\nu$. The integrals $V_{0}$, $V_{1}$, $V_{2}$ correspond to the area, boundary length and the integrated geodesic curvature $\kappa_{b}$ along the boundary. To put it simply, the procedure of measuring the Minkowski functionals refers to taking the values of the field at and above a given threshold $\nu$, evaluating the integrals in eq. \ref{minkowski} and then changing the threshold for a range of values. In the case of the 2-D fields one can imagine the field values at different positions as \textit{height} in the third dimension. Then a 2-D convergence map or an overdensity field can be visualised as a 3-D surface. And the Minkowski functionals then correspond to taking slices of the 3-D surface at and above the different \textit{heights} and measuring the area, curve length and the geodesic curvature as described in equation \ref{minkowski}. In this way Minkowski functionals allow to capture detailed morphological features of the generated field data which can then be directly compared against the training dataset.  

Minkowski functionals are also a useful tool in weak lensing convergence map studies as they allow us to capture non-Gaussian information on the small scales, which is not fully accessed by the power spectrum alone. In addition, Minkowski functionals have been used to detect different cosmologies, modified gravity models and the effects of massive neutrinos in weak lensing convergence maps \citep{petri2013, Ling2015L, marques2019}. Given the usefulness of Minkowski functionals in accessing the non-Gaussian information on the small scales, the functionals were chosen for studying the produced cosmic web data as well. To calculate the Minkowski functionals properly on a 2-D grid I used the \textit{minkfncts2d} algorithm, which utilizes a marching square algorithm as well as pixel weighting to capture the boundary lengths correctly \citep{mantz2008,minkfncts2d}. 

Minkowski functionals are sensitive to the Gaussian smoothing applied to the GAN-produced images and the training data. Hence, it is important to study the effects of Gaussian smoothing as it might give a deeper insight into the detected differences between the datasets. The procedure of smoothing refers to a convolution between a chosen kernel and the pixels of an image. In more detail, a chosen kernel matrix is centered on each pixel of an image and each surounding pixel is multiplied by the values of the kernel and subsequently summed. In the simplest case, such a procedure corresponds to averaging a chosen number of  pixels in a given image. In the case of Gaussian filtering, a Gaussian kernel is used instead.   

\begin{figure}[!ht]
\centering
\captionsetup[subfigure]{justification=centering}
  \begin{subfigure}[b]{0.327\textwidth}
    \includegraphics[width=\textwidth]{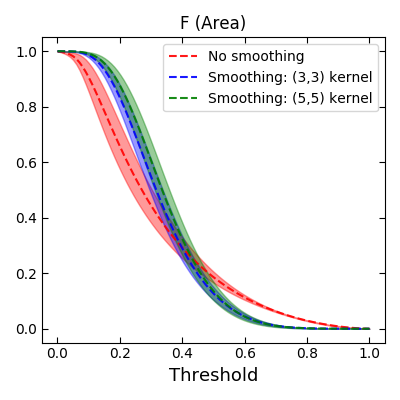}
    \label{MF_smoothing:1}
    
  \end{subfigure}
  \begin{subfigure}[b]{0.327\textwidth}
    \includegraphics[width=\textwidth]{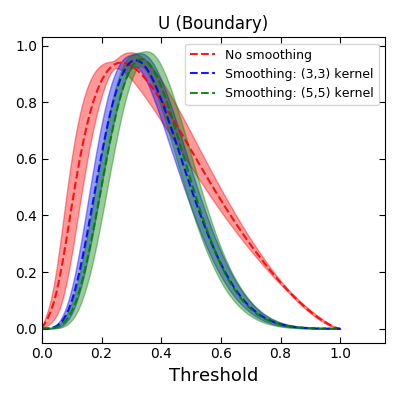}
    \label{MF_smoothing:2}
  \end{subfigure}
    \begin{subfigure}[b]{0.327\textwidth}
    \includegraphics[width=\textwidth]{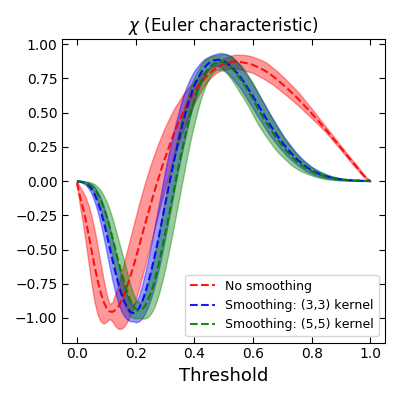}
    \label{MF_smoothing:3}
  \end{subfigure}
  \caption[The effects of Gaussian smoothing on Minkowski functionals]{An illustration of the effects of Gaussian smoothing on the Minkowski functionals calculated using cosmic web slices from the training data with redshift $z = 0.0$. The colored bands correspond to the mean and the standard deviation of the functionals calculated using different sizes of Gaussian smoothing kernels on a batch of 64 images.}
  \label{MF_smoothing_effects}
\end{figure}

To filter the noise we used Gaussian smoothing with a $3 \times 3$ kernel window and a standard deviation of 1 px. The Minkowski functionals were found to be especially sensitive to any kind of smoothing. For instance, the position and the shape of the trough of the third Minkowski functional is highly sensitive to the existence of any small-scale noise. Figure \ref{MF_smoothing_effects} illustrates the effects of Gaussian smoothing with different kernel sizes on the three Minkowski functionals.

\section{Results}
\label{section:results}
\subsection{Weak Lensing Map Results}
\label{cosmoGAN_original_results}

After around 150 epochs (corresponding to around 96 hours on a local HPC) the GAN started producing statistically realistic convergence maps as measured by the power spectrum and the Minkowski functionals. The diagnostics were computed at an ensemble level -- 100 batches of 64 convergence maps were produced by the GAN and the mean values along with the standard deviation were computed and compared with the training data. An analogous procedure was done when calculating the pixel intensity distribution histograms.

\begin{figure}[!ht]
\centering
\captionsetup[subfigure]{justification=centering}
  \begin{subfigure}[b]{0.49\textwidth}
    \includegraphics[width=\textwidth]{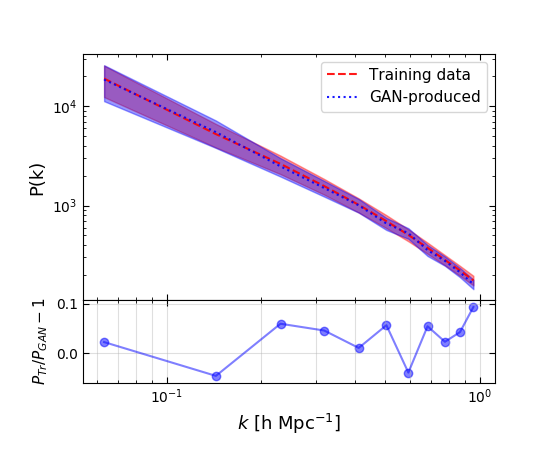}
    \caption{Power spectrum}
    \label{fig4:a1}
    
  \end{subfigure}
  \begin{subfigure}[b]{0.49\textwidth}
    \includegraphics[width=\textwidth]{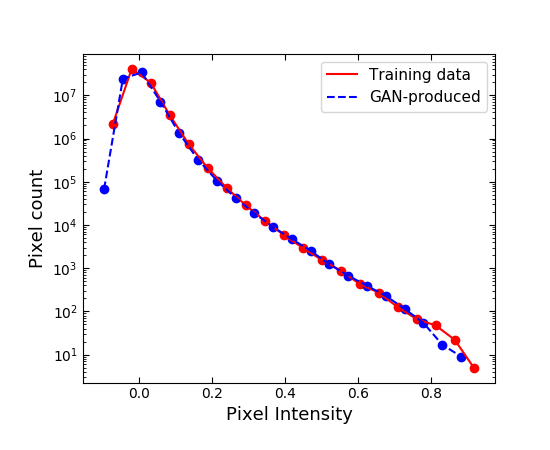}
    \caption{Pixel intensity histogram}
    \label{fig4:a2}
  \end{subfigure}
  \caption[The power spectrum and the pixel intensity histogram for the weak lensing convergence maps ($\Lambda$CDM)]{The matter power spectrum (with the relative difference) and the pixel intensity histogram for an ensemble of 6400 weak lensing convergence maps. The dashed lines correspond to the mean values, while the contours correspond to the standard deviation. Note that the pixel intensity values were normalized to the range of $[-1,1]$. }
  \label{figure4}
\end{figure}

The power spectra agree well between the GAN-produced and the training data, with minor differences on the small scales (see figure \ref{figure4}). In particular, the difference between the training and the GAN-produced dataset power spectra is around 5\% or lower for most values of $k$. Only at the smallest scales a significant difference of 10\% is reached. Similarly, the pixel intensity histogram in general shows a good agreement with significant differences appearing only for the highest and the lowest pixel intensity values (which is also detected in the original work in \cite{Mustafa2019}). A selection of GAN-produced maps are presented for visual inspection in figure \ref{cosmoGAN_samples}.

\begin{figure}[!ht]
\centering
\captionsetup[subfigure]{justification=centering}
  \begin{subfigure}[b]{0.327\textwidth}
    \includegraphics[width=\textwidth]{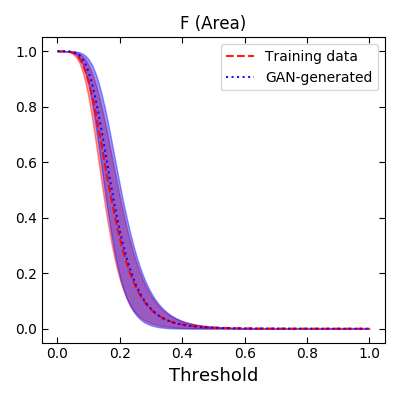}
    \label{fig5:a1}
    
  \end{subfigure}
  \begin{subfigure}[b]{0.327\textwidth}
    \includegraphics[width=\textwidth]{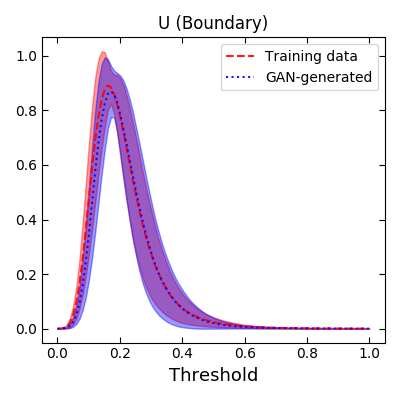}
    \label{fig5:a2}
  \end{subfigure}
    \begin{subfigure}[b]{0.327\textwidth}
    \includegraphics[width=\textwidth]{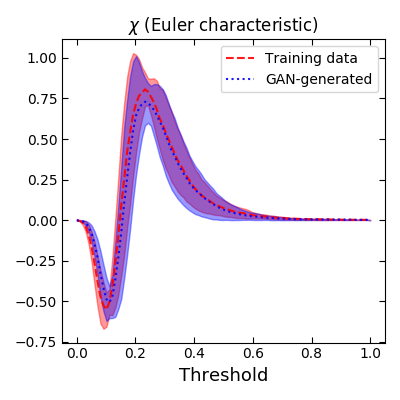}
    \label{fig5:a3}
  \end{subfigure}
  \caption[Minkowski functional analysis for the weak lensing convergence maps ($\Lambda$CDM)]{A comparison of the Minkowski functionals evaluated using 100 batches of 64 ramdomly selected maps for both datasets.}
  \label{figure5}
\end{figure}

Minkowski functionals were also calculated for the GAN-produced and the training datasets. The results are shown in figure \ref{figure5}. In general there is a good agreement between the training data and the GAN-produced maps, given the standard deviation, however, some minor differences can be detected in the Euler characteristic and the boundary functional, likely resulting from noise. 

\subsection{Weak Lensing Maps of Multiple Cosmologies}
\label{section_weak_lensing_multiple_cosmologies}

The results also indicate that the GAN is capable of producing realistic weak lensing maps for multiple cosmologies. This is an important result as it shows that the algorithm is able to pick up on the various subtle statistical differences between different cosmologies that usually requires a detailed study of the power spectrum, Minkowski functionals and other statistics.
 
However, the training procedure was found to be highly prone to mode collapse. A wide hyperparameter search had to be performed to find an optimal set of parameters that did not lead to full or partial mode collapse. The most important parameter in this context was found to be the learning rate. As a rule of thumb, decreasing the learning rate led to mode collapse happening later in the training procedure. When the learning rate was reduced below a certain value (discussed further in the analysis section), mode collapse was avoided altogether. As in the case with the cosmic web slice data, applying a transformation to each pixel of the image in order to increase the contrast had a positive effect in reducing the probability of mode collapse as well.  

\begin{figure}[!ht]
\centering
\captionsetup[subfigure]{justification=centering}
  \begin{subfigure}[b]{0.495\textwidth}
    \includegraphics[width=\textwidth]{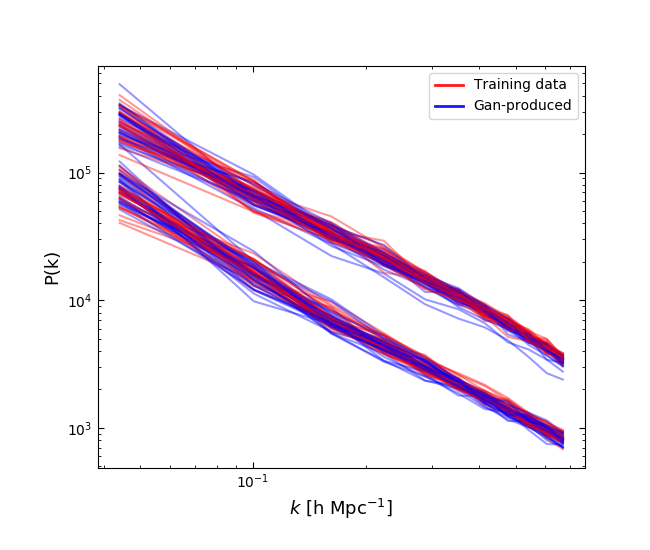}
    \label{shear_maps_sigma:1}
    
  \end{subfigure}
  \begin{subfigure}[b]{0.495\textwidth}
    \includegraphics[width=\textwidth]{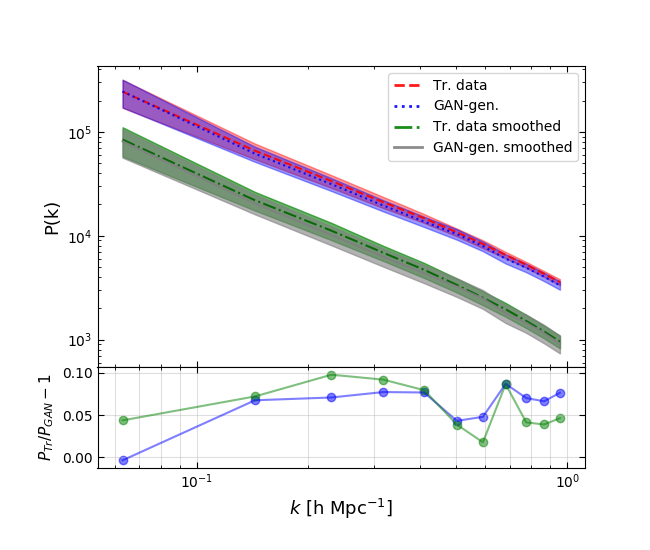}
    \label{shear_maps_sigma:2}
  \end{subfigure}
    \begin{subfigure}[b]{0.495\textwidth}
    \includegraphics[width=\textwidth]{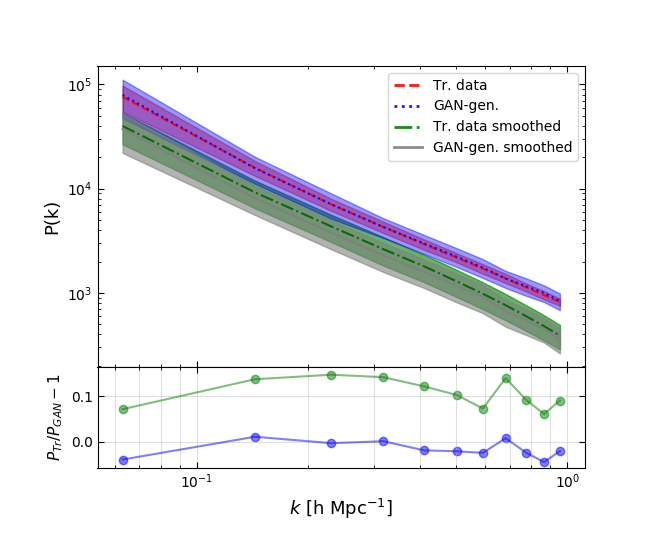}
    \label{shear_maps_sigma:3}
  \end{subfigure}
    \begin{subfigure}[b]{0.495\textwidth}
    \includegraphics[width=\textwidth]{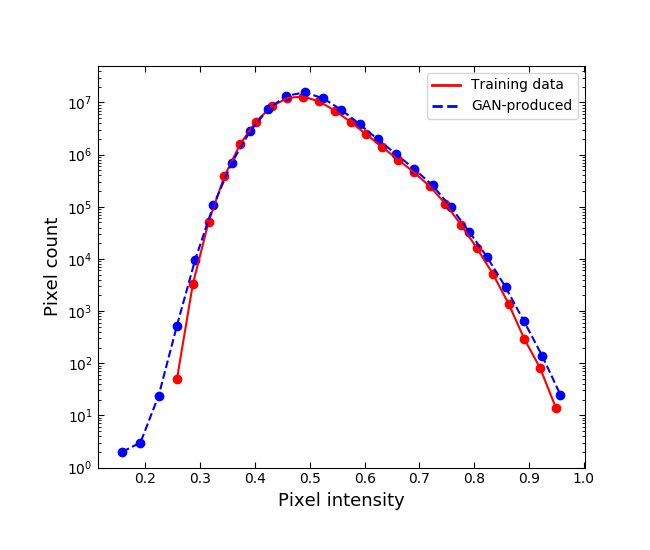}
    \label{shear_maps_sigma:4}
  \end{subfigure}
  \caption[The power spectrum and the pixel intensity histogram for the weak lensing convergence maps ($\sigma_{8} = \{0.436,0.814\}$)]{A selection of diagnostics to compare the training and the GAN-produced weak lensing convergence maps for $\sigma_{8}$ = $\{0.436,0.814\}$ with $\Omega_{m} = 0.233$. \textbf{Top left}: power spectra for an ensemble of 64 randomly chosen shear maps; \textbf{top right}: power spectra (mean and standard deviation) with and without Gaussian smoothing produced using 1000 randomly chosen shear maps with $\sigma_{8} = 0.814$; \textbf{bottom left}: same as top right, but for $\sigma_{8} = 0.436$; \textbf{bottom right}: the pixel intensity distribution (for both datasets combined). The blue and the green dots give $P_{tr}/P_{GAN} - 1$ with and without Gaussian smoothing applied correspondingly.   }
  \label{shear_maps_sigma}
\end{figure}

Figure \ref{shear_maps_sigma} summarizes the results of training the GAN on shear maps with different $\sigma_{8}$ values. The results indicate an agreement of the power spectra in the range of 5-10\% for $k > 10^{-1}$ $\mathrm{h}$ $\mathrm{Mpc^{-1}}$ for $\sigma_{8} = 0.814$. In the case of $\sigma_{8} = 0.436$ the agreement is significantly better, ranging between 1-3\% on most scales. Interestingly, Gaussian smoothing  increases the difference to around 5-15\% in this particular case. This shows that for this dataset Gaussian noise is not the major source of the statistical differences between the training and the GAN-generated datasets.   

\begin{figure}[!ht]
\centering
\captionsetup[subfigure]{justification=centering}
  \begin{subfigure}[b]{0.327\textwidth}
    \includegraphics[width=\textwidth]{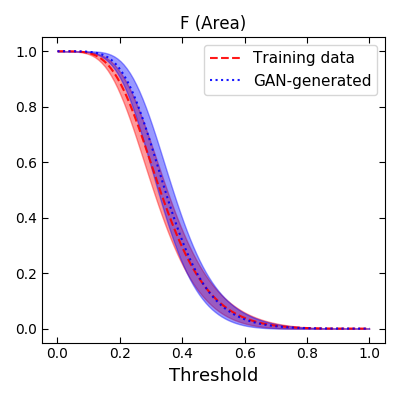}
    \label{shear_map_sigma_MFs:1}
    
  \end{subfigure}
  \begin{subfigure}[b]{0.327\textwidth}
    \includegraphics[width=\textwidth]{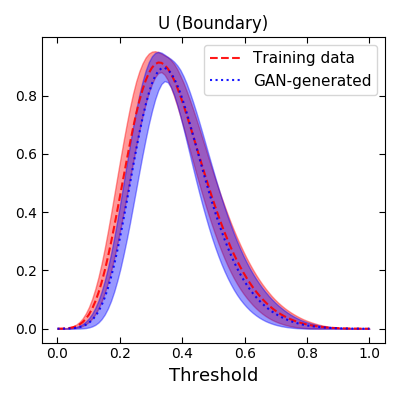}
    \label{shear_map_sigma_MFs:2}
  \end{subfigure}
    \begin{subfigure}[b]{0.327\textwidth}
    \includegraphics[width=\textwidth]{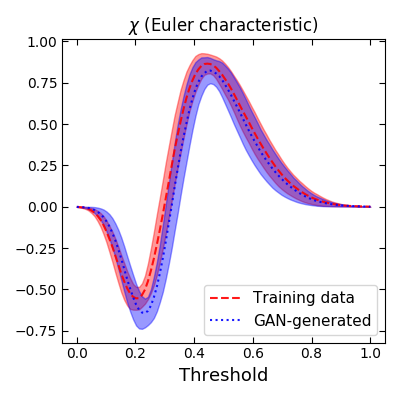}
    \label{shear_map_sigma_MFs:3}
  \end{subfigure}
  \caption[Minkowski functional analysis for the weak lensing convergence maps ($\sigma_{8} = \{0.436,0.814\}$)]{ A comparison of the Minkowski functionals evaluated using 1000 randomly selected weak lensing convergence maps with $\sigma_{8} = \{0.436,0.814\}$. Gaussian smoothing is applied for all datasets. }
  \label{shear_map_sigma_MFs}
\end{figure}

Figure \ref{shear_map_sigma_MFs} compares the Minkowski functionals calculated using the training and the GAN-produced datasets. Given the standard deviation in both datasets, the results overlap for all threshold values. However, for thresholds in the range of $[0.0,0.4]$ there is a significant difference between the training and the GAN-generated datasets. We found that this is partially due to small-scale noise in the GAN-produced data (see figure \ref{MF_smoothing_effects}). However, after experimenting with adding artificial noise to the training dataset images, it is clear that the noise alone cannot fully account for the observed differences in the Minkowski functionals. Another reason for the observed differences could be a relatively small size of the used dataset consisting of a few thousand weak lensing maps. It is likely that having more training data samples could significantly improve the results.    

\subsection{Cosmic Web for Multiple Redshifts}
\label{section_cosmic_web_redshifts}

\begin{figure}[!ht]
\centering
\captionsetup[subfigure]{justification=centering}
  \begin{subfigure}[b]{0.49\textwidth}
    \includegraphics[width=\textwidth]{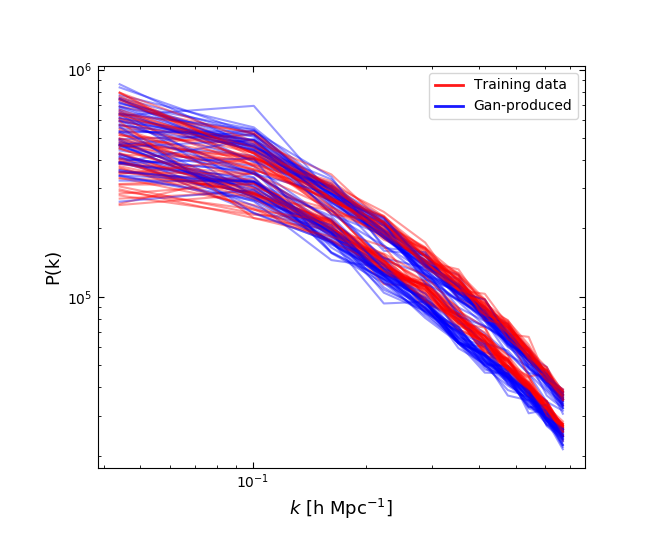}
    \label{fig6:a1}
  \end{subfigure}
  \begin{subfigure}[b]{0.49\textwidth}
    \includegraphics[width=\textwidth]{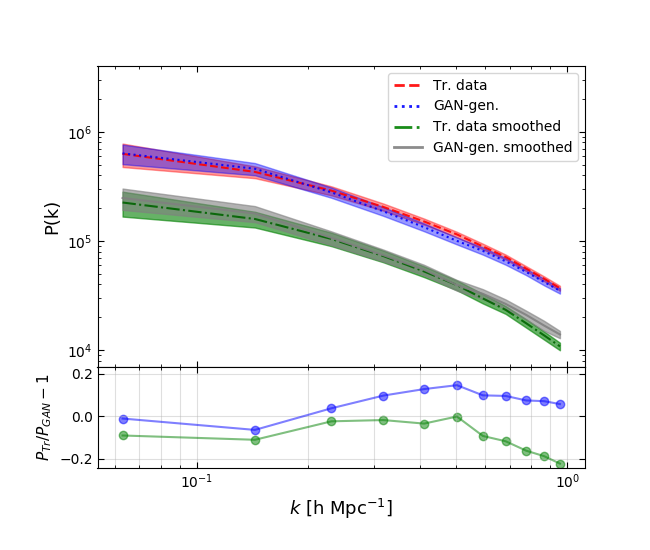}
    \label{fig6:a2}
  \end{subfigure}
    \begin{subfigure}[b]{0.49\textwidth}
    \includegraphics[width=\textwidth]{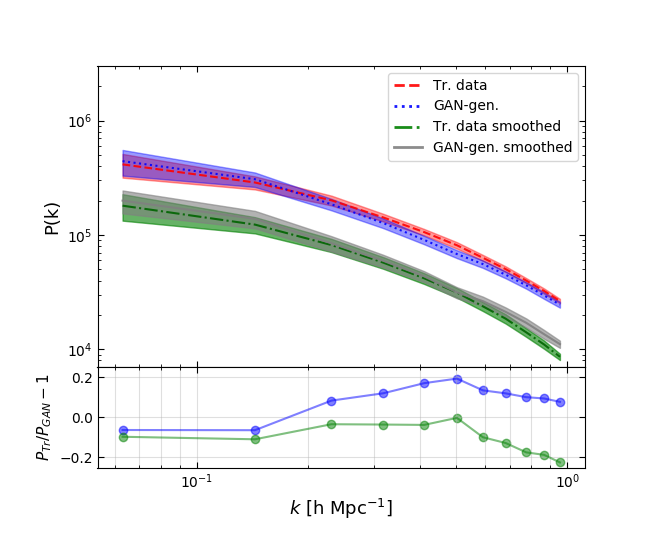}
    \label{fig6:a3}
  \end{subfigure}
    \begin{subfigure}[b]{0.49\textwidth}
    \includegraphics[width=\textwidth]{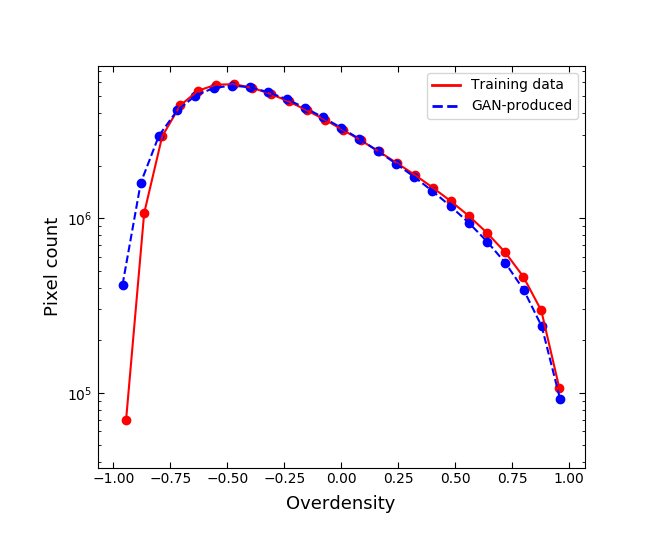}
    \label{fig6:a4}
  \end{subfigure}
  \caption[The power spectra and the overdensity histogram for the cosmic web slices with different redshifts: $z = \{0,1\}$]{A selection of diagnostics to compare the training and the GAN-produced cosmic web slices for redshifts $z=0.0$ and $z=1.0$ with $\sigma_{8}= 0.8$. \textbf{Top left}: power spectra for an ensemble of 64 randomly chosen slices for two different redshifts; \textbf{top right}: mean and standard deviation of the power spectra produced using 1000 randomly chosen slices with $z=0.0$; \textbf{bottom left}: same as top right, but for $z=1.0$; \textbf{bottom right}: the overdensity histogram (no smoothing). The blue and the green dots give $P_{tr}/P_{GAN} - 1$ with and without Gaussian smoothing applied correspondingly. }
  \label{figure6}
\end{figure}

\begin{figure}[!ht]
\centering
\captionsetup[subfigure]{justification=centering}
  \begin{subfigure}[b]{0.327\textwidth}
    \includegraphics[width=\textwidth]{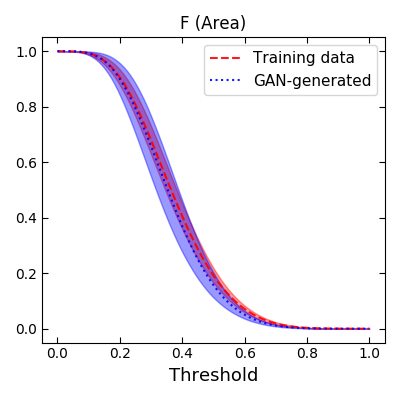}
    \label{fig7:a1}
    
  \end{subfigure}
  \begin{subfigure}[b]{0.327\textwidth}
    \includegraphics[width=\textwidth]{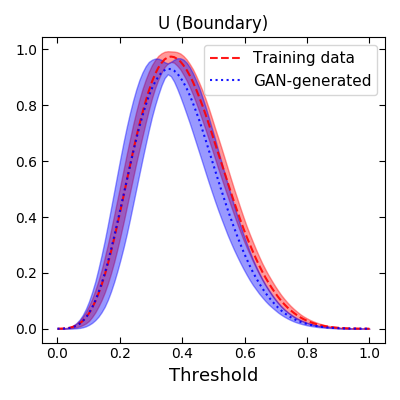}
    \label{fig7:a2}
  \end{subfigure}
    \begin{subfigure}[b]{0.327\textwidth}
    \includegraphics[width=\textwidth]{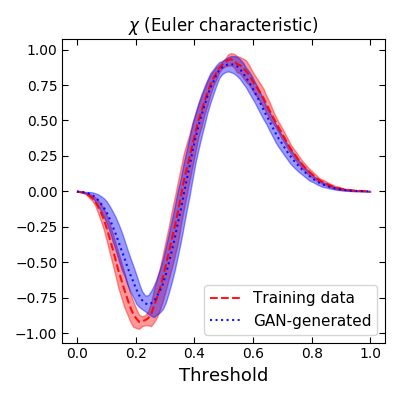}
    \label{fig7:a3}
  \end{subfigure}
  \caption[Minkowski functional analysis for the cosmic web slices with different redshifts: $z = \{0,1\}$]{A comparison of the Minkowski functionals evaluated using 1000 randomly selected cosmic web slices of redshifts $z = \{0.0,1.0\}$ for both datasets. Gaussian smoothing is applied for all datasets. }
  \label{figure7}
\end{figure}

The results also indicate that the GAN approach is capable of producing realistic cosmic web slices for different redshifts. As before with the weak lensing maps of different cosmologies, this illustrates that the algorithm in general does not get \textit{confused} between the two different redshifts and is capable of detecting subtle statistical differences between the different datasets (figure \ref{figure6}). In addition, I found that using Gaussian smoothing, as before, led to a better agreement between the training and the GAN-produced datasets. The effect is especially noticeable in the Minkowski functional analysis (figure \ref{figure7}). Visual samples of the produced cosmic web slices are shown in figure \ref{cosmoGAN_cw_samples}. 

The power spectra results for both redshift values were found to be very similar. Namely, for the non-smoothed case the difference between the training and the GAN-produced power specta ranges between 5-10\%. The results are similar for the smoothed case, with exception of $k$ values around 1 $\mathrm{h}$ $\mathrm{Mpc^{-1}}$ where the difference reaches 20\%. 

The effects of the Gaussian smoothing on both the power spectra and the Minkowski functionals illustrate that one of the reasons for the differences between the GAN-generated and the training datasets is noise appearing on different scales in the GAN-produced images. Applying Gaussian smoothing, in general, filters the majority of such noise, however, it cannot fully account for all the differences appearing in the different statistical diagnostics. In addition, smoothing can improve the results on some scales, while worsening them on others. As an example, in figure \ref{figure6}, Gaussian smoothing increases the difference between the GAN-produced and the training dataset power spectra on the smallest scales.

\subsection{Cosmic Web for Multiple Cosmologies and Modified Gravity Models}

Training the GAN on the cosmic web slices of different cosmologies and modified gravity models offered another way of testing whether the algorithm would pick up on the subtle statistical differences between the different datasets. In addition, the classification task for the discriminator neural network is more difficult when training on datasets with multiple cosmologies leading to longer training times.  

\begin{figure}[!ht]
\centering
\captionsetup[subfigure]{justification=centering}
  \begin{subfigure}[b]{0.49\textwidth}
    \includegraphics[width=\textwidth]{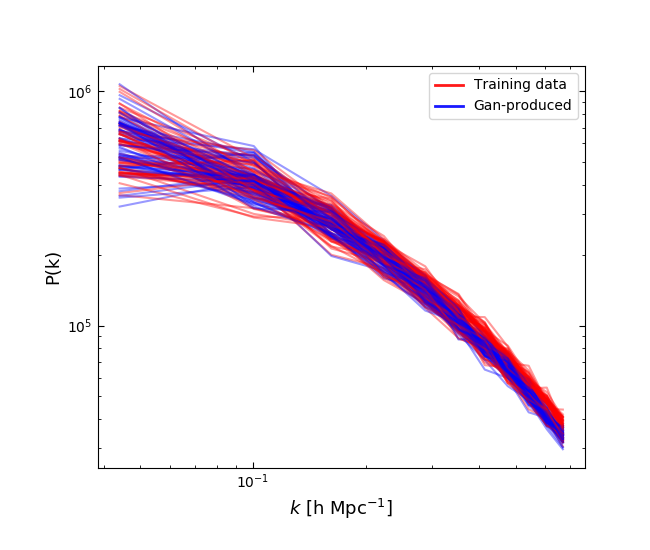}
    \label{fig8:a1}
    
  \end{subfigure}
  \begin{subfigure}[b]{0.49\textwidth}
    \includegraphics[width=\textwidth]{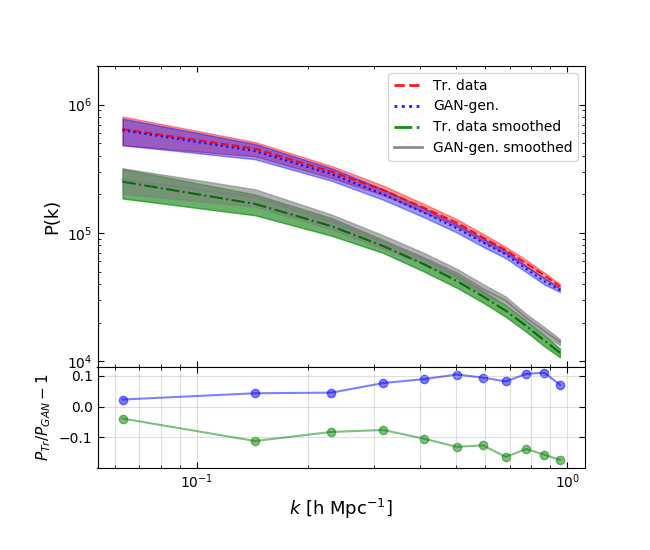}
    \label{fig8:a2}
  \end{subfigure}
    \begin{subfigure}[b]{0.48\textwidth}
    \includegraphics[width=\textwidth]{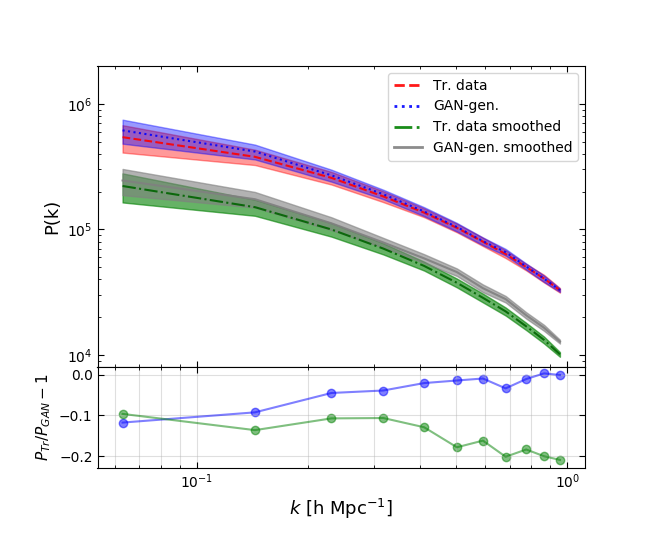}
    \label{fig8:a3}
  \end{subfigure}
    \begin{subfigure}[b]{0.48\textwidth}
    \includegraphics[width=\textwidth]{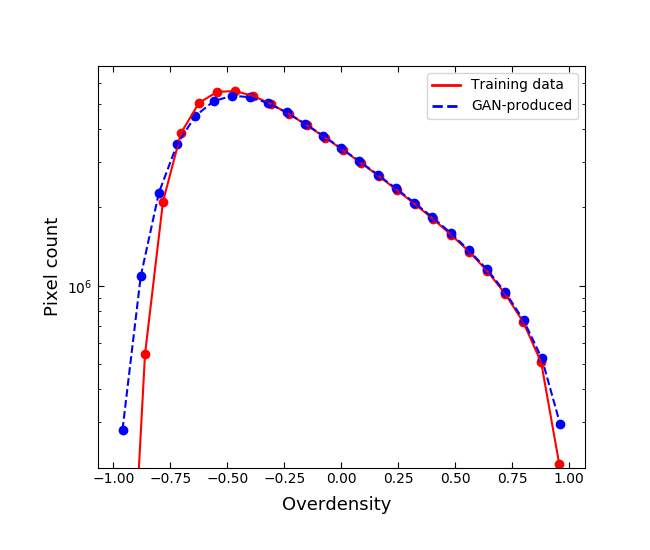}
    \label{fig8:a4}
  \end{subfigure}
  \caption[The power spectra and the overdensity histogram for the cosmic web slices with $\sigma_{8} = \{0.7,0.9\}$]{A selection of diagnostics to compare the training and the GAN-produced cosmic web slices for $\sigma_{8} = 0.7$ and $\sigma_{8} = 0.9$ at $z = 0.0$. \textbf{Top left}: power spectra for an ensemble of 64 randomly chosen slices for both datasets; \textbf{top right}: mean and standard deviation of the power spectra computed using 1000 randomly chosen slices of $\sigma_{8} = 0.9$; \textbf{bottom left}: same as top right, but for $\sigma_{8} = 0.7$; \textbf{bottom right}: the overdensity histogram (no smoothing). The blue and the green dots give $P_{tr}/P_{GAN} - 1$ with and without Gaussian smoothing applied correspondingly. }
  \label{figure8}
\end{figure}

The results indicate that the GAN is indeed capable of producing statistically realistic cosmic web data of different cosmologies and modified gravity models. With no Gaussian smoothing applied, the relative agreement between the power spectra is 1-10\% (see figure \ref{figure8}). Applying smoothing in this case resulted in increasing the relative power spectrum difference to over 10\% on average. In the case of cosmic web slices for different $f_{R0}$ values, the agreement between the two datasets was good, ranging between 1-10\% on all scales. Smoothing improved the situation only in the mid-range of the covered $k$ values, reducing the agreement on the smallest scales (see figure \ref{cw_fR_plots}). 

\begin{figure}[!ht]
\centering
\captionsetup[subfigure]{justification=centering}
  \begin{subfigure}[b]{0.327\textwidth}
    \includegraphics[width=\textwidth]{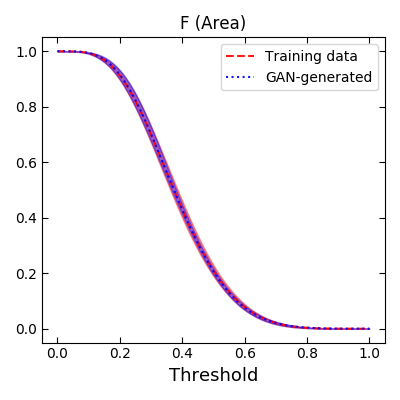}
    \label{fig9:a1}
    
  \end{subfigure}
  \begin{subfigure}[b]{0.327\textwidth}
    \includegraphics[width=\textwidth]{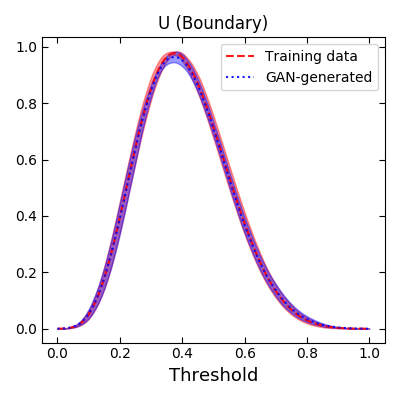}
    \label{fig9:a2}
  \end{subfigure}
    \begin{subfigure}[b]{0.327\textwidth}
    \includegraphics[width=\textwidth]{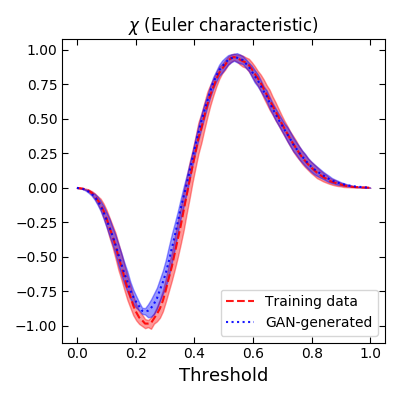}
    \label{fig9:a3}
  \end{subfigure}
  \caption[Minkowski functional analysis for the cosmic web slices with $\sigma_{8} = \{0.7,0.9\}$]{A comparison of the Minkowski functionals evaluated using 1000 randomly selected cosmic web slices from the dataset with two different values of $\sigma_{8} = \{0.7,0.9\}$. Gaussian smoothing is applied for both datasets. }
  \label{figure9}
\end{figure}

\begin{figure}[!ht]
\centering
\captionsetup[subfigure]{justification=centering}
  \begin{subfigure}[b]{0.49\textwidth}
    \includegraphics[width=\textwidth]{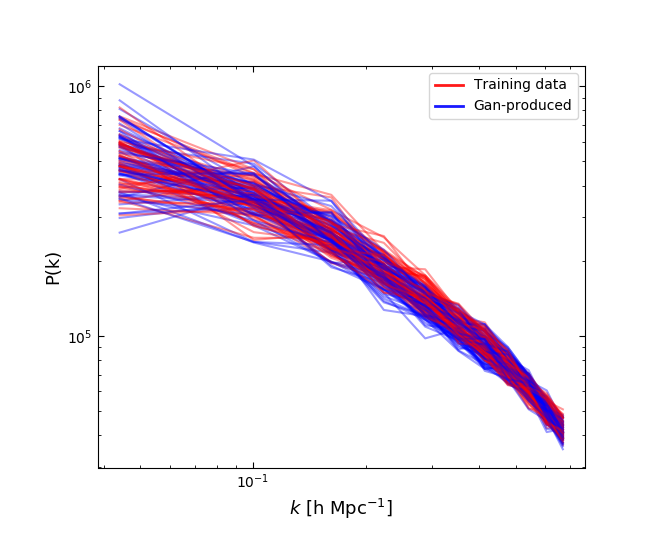}
    \label{cw_fR_plots:1}
  \end{subfigure}
  \begin{subfigure}[b]{0.49\textwidth}
    \includegraphics[width=\textwidth]{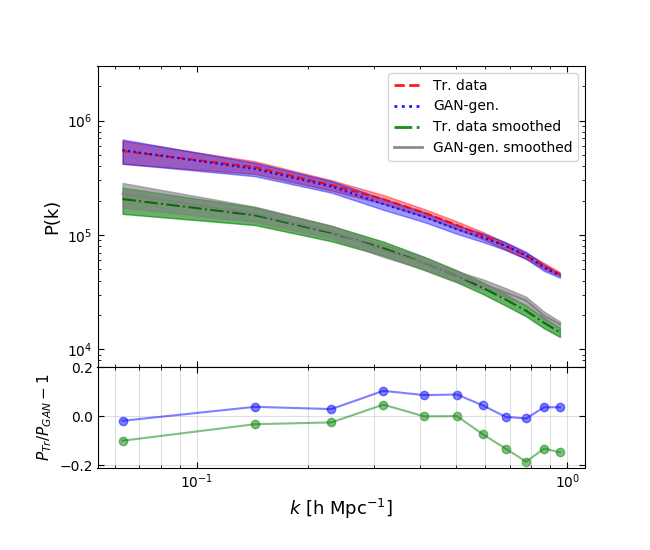}
    \label{cw_fR_plots:2}
  \end{subfigure}
    \begin{subfigure}[b]{0.49\textwidth}
    \includegraphics[width=\textwidth]{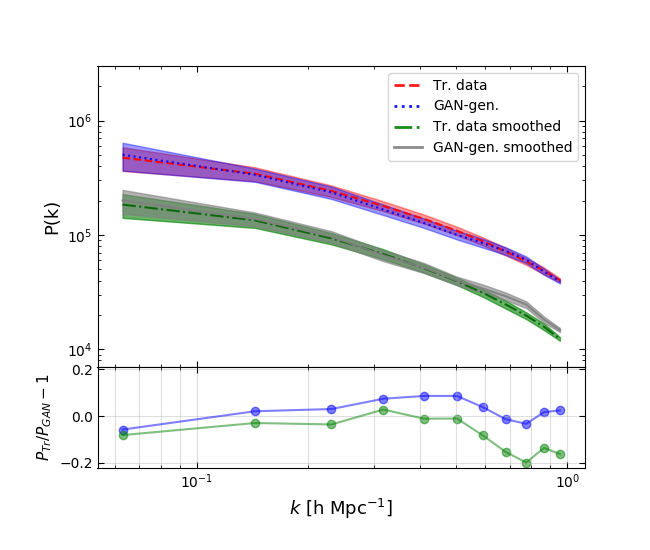}
    \label{cw_fR_plots:3}
  \end{subfigure}
    \begin{subfigure}[b]{0.49\textwidth}
    \includegraphics[width=\textwidth]{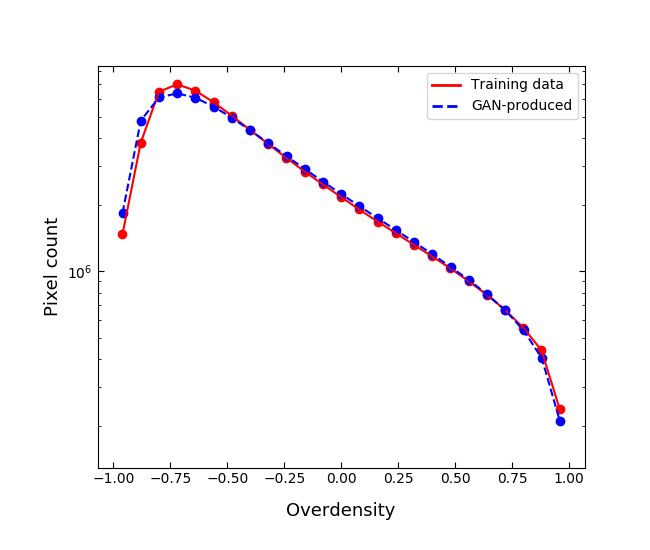}
    \label{cw_fR_plots:4}
  \end{subfigure}
  \caption[The power spectra and the overdensity histogram for the cosmic web slices with $f_{R0} = \{10^{-7}, 10^{-1} \}$]{A selection of diagnostics to compare the training and the GAN-produced cosmic web slices for $f_{R0} = \{10^{-7}, 10^{-1} \}$ (with $\sigma_{8} = 0.8$ and $z = 0.0$). \textbf{Top left}: power spectra for an ensemble of 64 randomly chosen slices for both datasets; \textbf{top right}: mean and standard deviation of the power spectra produced using 1000 randomly chosen slices with $f_{R0} = 10^{-1}$; \textbf{bottom left}: same as top right, but for $f_{R0} = 10^{-7}$; \textbf{bottom right}: the overdensity histogram (no smoothing). The blue and the green dots give $P_{tr}/P_{GAN} - 1$ with and without Gaussian smoothing applied correspondingly.  }
  \label{cw_fR_plots}
\end{figure}

Figure \ref{figure9} shows the Minkowski functional analysis. In this case, very little deviation is observed. In general, there is a good agreement between the GAN-produced and the training datasets, especially for the first and the second Minkowski functionals. For the third Minkowski functional, the results diverge around the lower trough area, which is also observed for other datasets. This is at least in part related to small-scale noise as indicated by the previous analysis. 

\begin{figure}[!ht]
\centering
\captionsetup[subfigure]{justification=centering}
  \begin{subfigure}[b]{0.327\textwidth}
    \includegraphics[width=\textwidth]{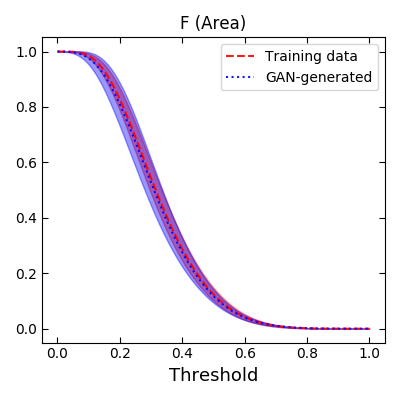}
    \label{cw_fR_MF:1}
  \end{subfigure}
  \begin{subfigure}[b]{0.327\textwidth}
    \includegraphics[width=\textwidth]{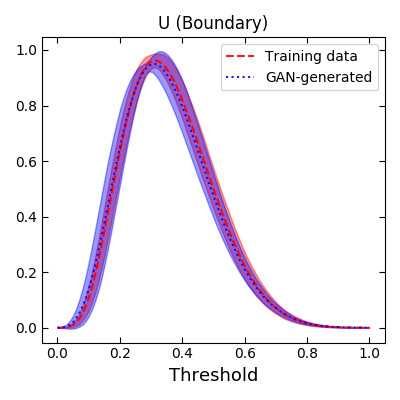}
    \label{cw_fR_MF:2}
  \end{subfigure}
    \begin{subfigure}[b]{0.327\textwidth}
    \includegraphics[width=\textwidth]{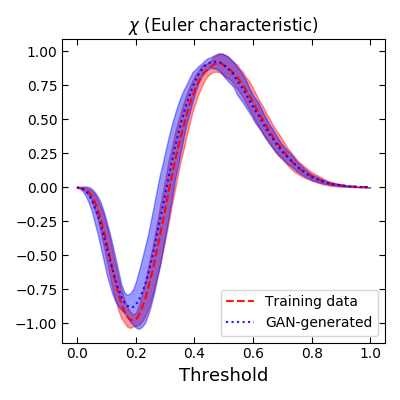}
    \label{cw_fR_MF:3}
  \end{subfigure}
  \caption[Minkowski functional analysis for the cosmic web slices with $f_{R0} = \{10^{-7},10^{-1}\}$]{A comparison of the Minkowski functionals evaluated using 1000 randomly selected cosmic web slices from the dataset with two different values of $f_{R0} = \{10^{-7},10^{-1}\}$. Gaussian smoothing is applied for both datasets. }
  \label{cw_fR_MF}
\end{figure}

The results are similar for the GAN trained on cosmic web slices corresponding to different $f(R)$ models (figure \ref{cw_fR_MF}). In general, a good agreement between the datasets was found (given the standard deviation of the data and the GAN-produced results). Gaussian smoothing, in this case, was more effective in reducing some of the offset observed in the power spectrum analysis. However, it increased the offset on the smallest scales.

\subsection{Dark Matter, Gas and Internal Energy Results}

In the case of training the GAN algorithm on multiple components at the same time, the training procedure was relatively quick and efficient (around 1.3 times quicker compared to the datasets discussed previously) despite the training dataset being 3 times bigger. This is most likely due to the fact that the cosmic web slices in this particular dataset corresponded to a much larger simulation box and hence were not as detailed on the smallest scales. 

\begin{figure}[!ht]
\centering
\captionsetup[subfigure]{justification=centering}
  \begin{subfigure}[b]{0.49\textwidth}
    \includegraphics[width=\textwidth]{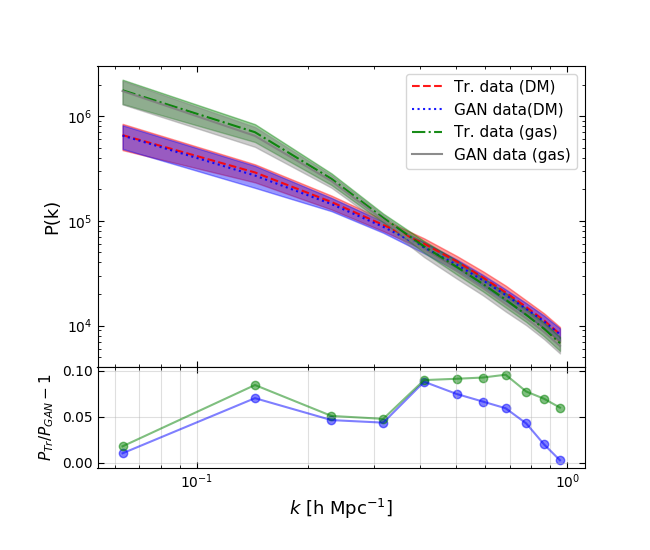}
    \label{cw_illustris:1}
  \end{subfigure}
  \begin{subfigure}[b]{0.49\textwidth}
    \includegraphics[width=\textwidth]{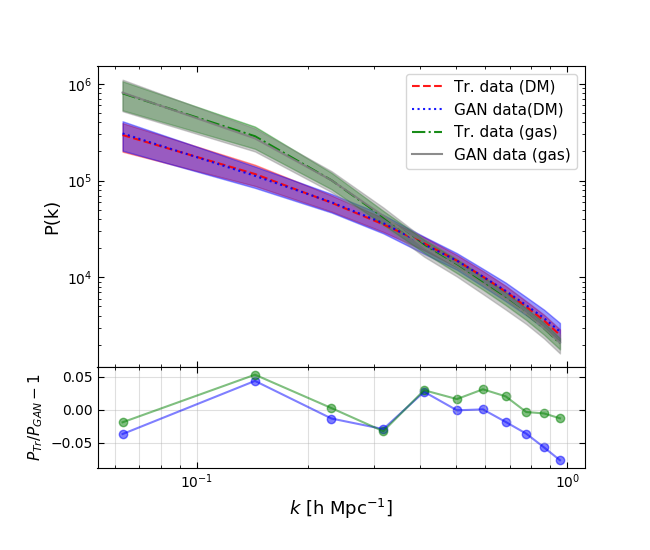}
    \label{cw_illustris:2}
  \end{subfigure}
    \begin{subfigure}[b]{0.49\textwidth}
    \includegraphics[width=\textwidth]{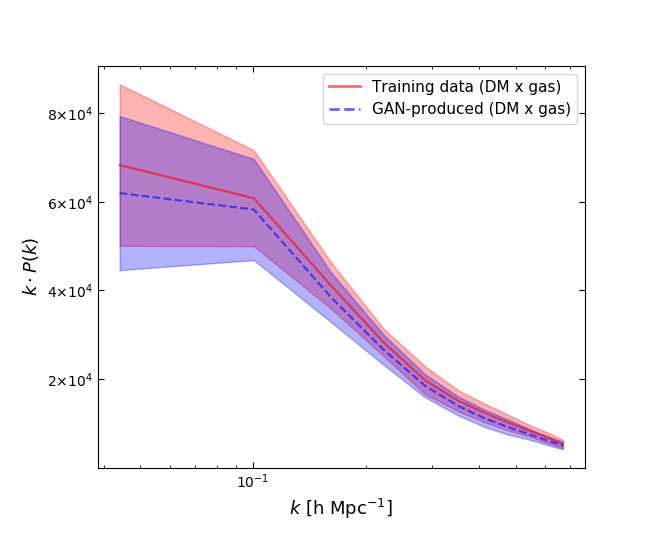}
    \label{cw_illustris:3}
  \end{subfigure}
    \begin{subfigure}[b]{0.49\textwidth}
    \includegraphics[width=\textwidth]{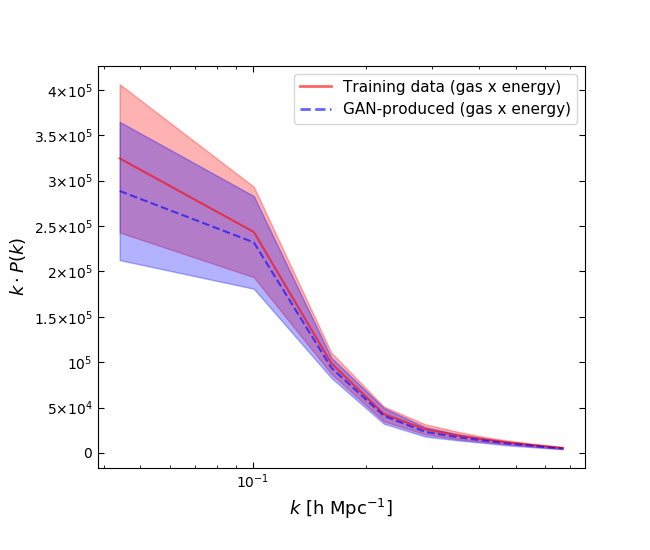}
    \label{cw_illustris:4}
  \end{subfigure}
  \caption[Diagnostics for the dark matter, gas and internal energy results]{A selection of diagnostics to compare the training and the GAN-produced multi-component cosmic web slices. \textbf{Top left}: the mean and the standard deviation of the power spectrum for 1000 randomly chosen slices for both datasets along with the corresponding relative difference between the datasets (green for $P_{Tr}^{gas} / P_{GAN}^{gas} - 1$ and blue for $P_{Tr}^{DM} / P_{GAN}^{DM} - 1$); \textbf{top right}: same as top left, but with Gaussian smoothing applied; \textbf{bottom left}: the cross-power spectrum calculated between 1000 randomly chosen dark matter and the corresponding gas cosmic web pairs for both the training and the GAN-produced datasets; \textbf{bottom right}: same as bottom left, but for the gas-energy cross-power. }
  \label{cw_illustris}
\end{figure}

As before, the relative difference between the GAN-produced and the training datasets was calculated. The internal energy slices were analysed using Minkowski functionals as well as the cross-power spectrum (figure \ref{cw_illustris}). The analysis was done for both dark matter and the gas components. The relative difference between the power spectra for both DM and gas cosmic web slices was found to be at around 5\% level for all the covered range. Gaussian smoothing reduced this value to 1-5\%. In addition, the cross-power spectrum was calculated for all the components. For both the dark matter-gas and the gas-energy pairs there is a good agreement between the training and the GAN-produced datasets given the large standard deviation. Both plots show values well above zero for most $k$ values, indicating a significant correlation between the dark matter and the corresponding gas as well as the internal energy distributions on all scales as expected. 

\begin{figure}[!ht]
\centering
\captionsetup[subfigure]{justification=centering}
  \begin{subfigure}[b]{0.327\textwidth}
    \includegraphics[width=\textwidth]{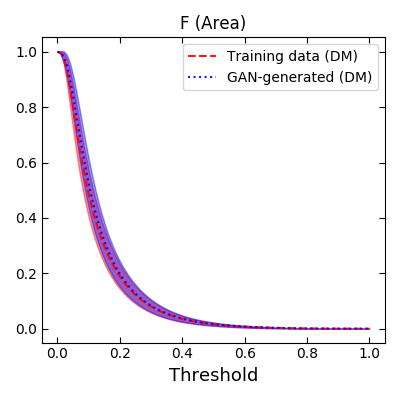}
    \label{cw_dm_gas_MFs:1}
  \end{subfigure}
  \begin{subfigure}[b]{0.327\textwidth}
    \includegraphics[width=\textwidth]{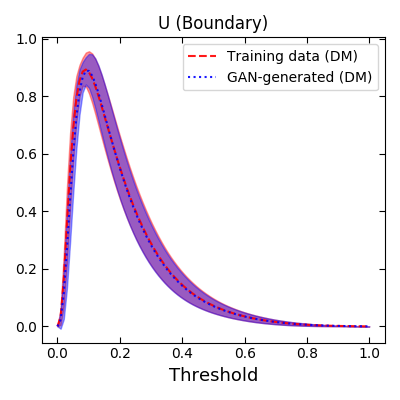}
    \label{cw_dm_gas_MFs:2}
  \end{subfigure}
    \begin{subfigure}[b]{0.327\textwidth}
    \includegraphics[width=\textwidth]{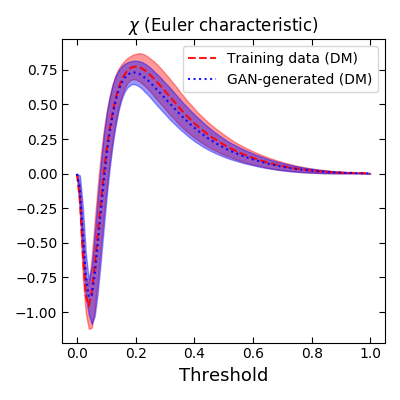}
    \label{cw_dm_gas_MFs:3}
  \end{subfigure}
    \begin{subfigure}[b]{0.327\textwidth}
    \includegraphics[width=\textwidth]{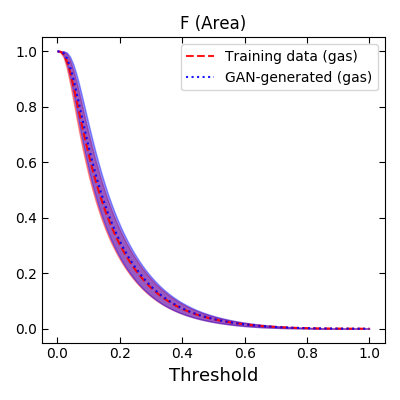}
    \label{cw_dm_gas_MFs:4}
  \end{subfigure}
    \begin{subfigure}[b]{0.327\textwidth}
    \includegraphics[width=\textwidth]{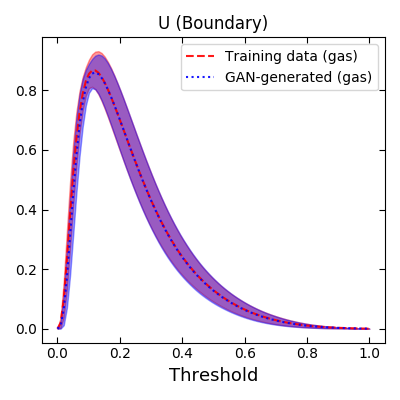}
    \label{cw_dm_gas_MFs:5}
  \end{subfigure}
    \begin{subfigure}[b]{0.327\textwidth}
    \includegraphics[width=\textwidth]{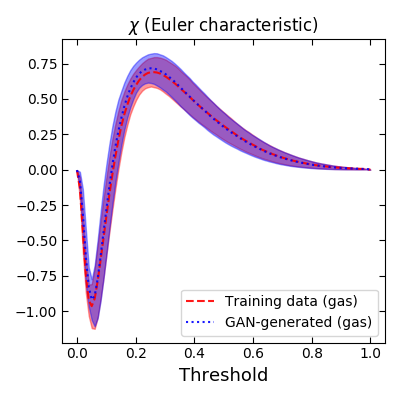}
    \label{cw_dm_gas_MFs:6}
  \end{subfigure}  
    \begin{subfigure}[b]{0.327\textwidth}
    \includegraphics[width=\textwidth]{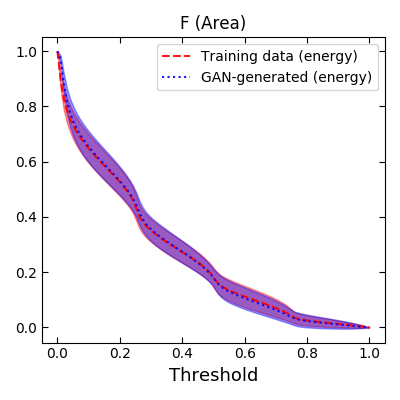}
    \label{cw_dm_gas_MFs:7}
  \end{subfigure}  
    \begin{subfigure}[b]{0.327\textwidth}
    \includegraphics[width=\textwidth]{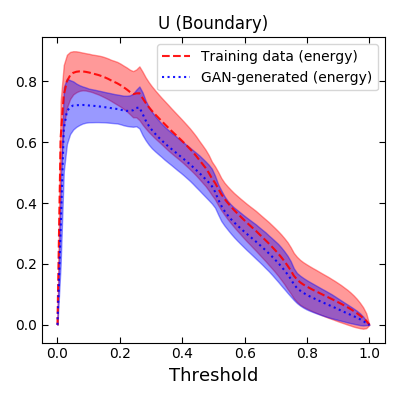}
    \label{cw_dm_gas_MFs:8}
  \end{subfigure}  
    \begin{subfigure}[b]{0.327\textwidth}
    \includegraphics[width=\textwidth]{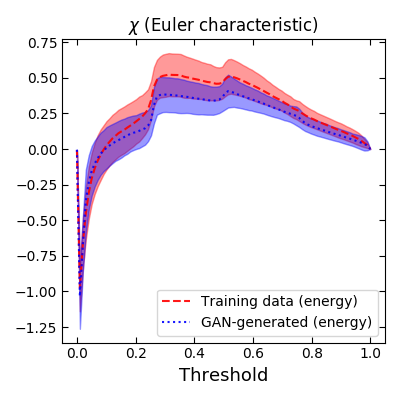}
    \label{cw_dm_gas_MFs:9}
  \end{subfigure} 
  \caption[Minkowski functional analysis for the dark matter, gas and internal energy results]{Results of the Minkowski functional analysis for the GAN trained on the DM, gas and the internal energy data. \textbf{Top row:} Minkowski functionals for the DM cosmic web slices; \textbf{middle row:} Minkowski functionals for the gas overdensity slice data; \textbf{bottom row:} the corresponding Minkowski functionals for the internal energy data. In all cases Gaussian smoothing is applied. }
  \label{cw_dm_gas_MFs}
\end{figure}

The Minkowski functional analysis (figure \ref{cw_dm_gas_MFs}) revealed a generally good agreement between the two datasets, with significant differences appearing only in the boundary and the Euler characteristic Minkowski functionals for the energy cosmic web slices. This is somewhat surprising as the internal energy slices, in general, are significantly less complex on the smallest of scales when compared to the corresponding dark matter and gas data (see figure \ref{figure3}), hence we expected the GAN to easily learn to reproduce the named dataset. However, we also found that the internal energy data and the corresponding Minkowski functionals are especially sensitive to adding any small scale artificial noise. A more detailed Minkowski functional analysis is required to determine the reason for this divergence.    

\subsection{Latent Space Interpolation Results}
\label{section:latent_results}

To perform the latent space interpolation procedure I trained the GAN to produce cosmic web slices of two different redshifts along with weak lensing maps of different $\sigma_{8}$ values. Once trained, a batch of outputs was produced and in each case a pair of slices/maps corresponding to different redshifts or $\sigma_{8}$ values was chosen. Subsequently, I interpolated between the input points $Z_{1}$ and $Z_{2}$ corresponding to the outputs with different redshifts and $\sigma_{8}$ values (see figure \ref{figure 2}).   

\begin{figure}[!ht]
\centering
\captionsetup[subfigure]{justification=centering}
  \begin{subfigure}[b]{0.49\textwidth}
    \includegraphics[width=\textwidth]{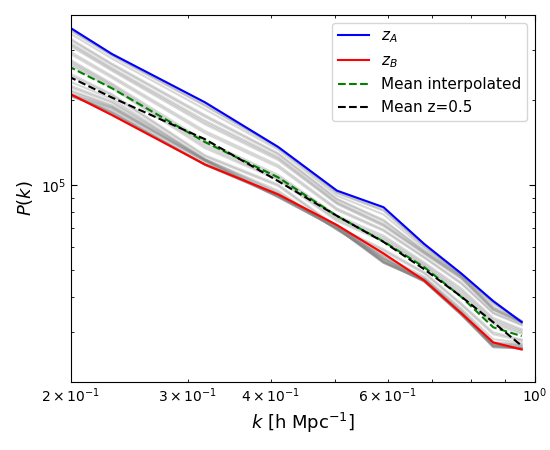}
    \caption{CW slice redshift interpolation}
    \label{fig10:a1}
  \end{subfigure}
  \begin{subfigure}[b]{0.49\textwidth}
    \includegraphics[width=\textwidth]{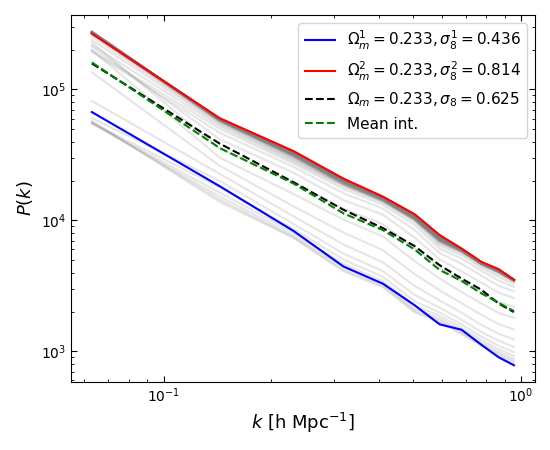}
    \caption{WL $\sigma_{8}$ interpolation}
    \label{fig10:a2}
  \end{subfigure}
  \caption[Latent space interpolation results]{The results of the linear latent space interpolation technique. \textbf{Left:} the matter power spectrum corresponding to a linear interpolation between two cosmic web slices of redshifts $z=0.0$ and $z=1.0$. The lines in grey are the intermediate output slices generated by the procedure, while the black line corresponds to the mean value of the power spectrum calculated by choosing 100 random (training data) slices of redshift $z=0.5$. The green dashed line corresponds to the mean of 100 outputs produced using latent space points lying close to the centre of the line connecting the two clusters of redshifts $z=0.0$ and $z=1.0$. More specifically, 100 points from a region equal to 1/4 of the total length of the line centered at the middle point was sampled. \textbf{Right:} interpolating between two randomly chosen weak lensing maps with different values of $\sigma_{8}$. As before, the black line corresponds to the mean power spectrum produced from 100 random maps with $\sigma_{8} = 0.625$. The green line is the mean power spectrum of 100 outputs generated using latent space points lying close to the centre of the line connecting the two clusters corresponding to $\sigma_{8}=0.436$ and $\sigma_{8}=0.814$
  }
  \label{figure10}
\end{figure}

Figure \ref{figure10} illustrates the results of the latent space interpolation procedure. In particular, it shows that the technique does indeed produce intermediate power spectra. However, the transition is not linear -- the power spectra lines corresponding to equally spaced inputs (in the latent space) are not equally spaced in the power spectrum space. This is the case as the produced data samples can be described as points on a Riemannian manifold, which in general has curvature (see appendix \ref{riemannian_geometry} for more details).

\begin{figure}[!ht]
  \centering
    \includegraphics[width=1.0\textwidth]{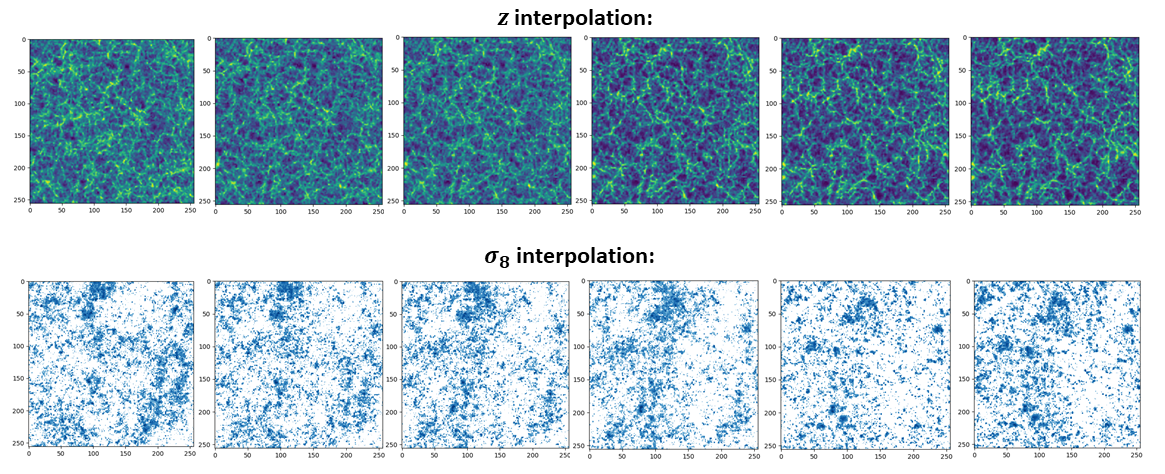}
    \caption[Samples of data produced by the latent space interpolation procedure]{The results of the latent space interpolation procedure for cosmic web slices of redshifts $z=0.0$ (far right) and $z=1.0$ (far left) and weak lensing convergence maps of $\sigma_{8}^{1} = 0.436$ (far left) and $\sigma_{8}^{1} = 0.814$ (far right).  }
    \label{figure11}
\end{figure}

Figure \ref{figure10} and \ref{figure11} show the results of interpolating between cosmic web slices with redshifts $z=0.0$ and $z=1.0$ and weak lensing maps with $\sigma_{8} =0.436$ and $\sigma_{8} = 0.814$. The interpolated samples are statistically realistic and the transition is nearly smooth. The power spectrum analysis was done by comparing 100 latent space points drawn from the central region (equal in length to 1/4 of the total length of the line) of the line connecting the two latent space clusters corresponding to the different redshifts and $\sigma_{8}$ values against 100 training data samples (see figure \ref{figure 2} and \ref{figure10} for more information). The intermediate power spectra was found to be in good agreement. 

An important part of the latent space interpolation procedure is being able to distinguish between the GAN-generated cosmic web slices and weak lensing maps of different redshifts, cosmologies and modified gravity parameters. In this regard, I have tested two machine learning algorithms: a convolutional neural network and gradient boosted decision trees. Initially a convoluational neural network architecture described in table \ref{table:discriminator_network} was used, as we already knew that such neural networks are effective in classifying cosmic web slices and convergence maps. This resulted in accuracy of around $90$\% when classifying unseen data samples. However, the training procedure was prone to overfitting, requiring a thorough hyperparameter optimization. In addition, I found that the small scale noise appearing would highly reduce the prediction accuracy of the CNN. This is a known problem in the deep learning literature and can be mitigated to a certain degree by adding artificial noise to the training dataset \citep{Liu2017}. However, finding the right amount of noise needed to mimic the noise appearing in the GAN-generated outputs is difficult.  

The gradient boosted decision tree algorithm (\textit{XGBoost} \citep{chen2016_xgboost}) was found to be faster and more accurate in predicting the dataset class. In particular, 95-98\% accuracy was reached (depending on the dataset and hyperparameters used), when predicting the dataset class of unseen test samples. Table \ref{table:XGBoost_parameters} summarizes the parameters used when training the \textit{XGBoost} algorithm. 

\begin{table}[ht!]
\noindent\makebox[\textwidth]{%
\begin{tabular}{ |c||c|c|c|c| } 
 \hline
 \textbf{Parameter:} & Learning rate & Max. tree depth & Training step & Objective\\ 
 \hline 
 \textbf{Value:} & 0.08 & 2  & 0.3  & \textit{multi:softprob}\\ 
 \hline
\end{tabular}}
\caption[The used \textit{XGBoost} settings]{The \textit{XGBoost} parameters used for classifying the cosmic web slices with redshifts $z=\{0.0,1.0\}$ and the weak lensing maps with $\sigma_{8} = \{ 0.436,0.814 \}$. }
\label{table:XGBoost_parameters}
\end{table}

Combining such a machine learning approach with a power spectrum analysis allowed us to distinguish between the different classes of the GAN-produced outputs reliably. 

The latent space interpolation results illustrate a number of interesting features of GANs. Firstly, the results illustrate that the GAN training procedure tightly encodes the various features discovered in our training dataset in the high-dimensional latent space. By finding clusters in this latent space, corresponding to outputs of different redshifts or cosmology parameters, and linearly interpolating between them, we can produce outputs with intermediate values of the mentioned parameters. This allows us to control the outputs produced by the generator.    

\section{Analysis and Conclusions}

The main goal of this work was to investigate whether GANs can be used as a universal, fast and efficient emulator capable of producing realistic and novel mock data. The results of this work are encouraging, illustrating that GANs are indeed capable of producing realistic mock datasets. In addition, I have shown that GANs can be used to emulate dark matter, gas and internal energy distribution data simultaneously. This is a key result, as generating realistic gas distributions requires complex and computationally expensive hydrodynamical simulations. Hence, producing vast amounts of realistic multi-component mock data quickly and efficiently will be of special importance in the context of upcoming observational surveys.    

The GAN-produced data in general cannot be distinguished from the training dataset visually. In terms of the power spectrum analysis, the relative difference between the GAN-produced and the training data ranges between 1-20\%  depending on the dataset and whether Gaussian smoothing was applied. The Minkowski functional analysis revealed a generally good agreement between the two datasets with an exception of the third Minkowski functional corresponding to curvature, which showed subtle differences for all studied datasets. In addition, greater differences were observed when training the GAN on datasets with multiple data classes. This is somewhat expected, as the training task becomes more difficult. In general, these differences can be partially accounted for as a result of small-scale noise in the GAN-generated images. Gaussian smoothing with a $3 \times 3$ pixel kernel size was found to be effective in filtering away most of such noise. In addition, the training datasets used in this work are smaller than those used in \citep{Rodriguez2018,Mustafa2019}, which, at least partially, accounts for the differences between our and their corresponding results.

A commonly used technique of latent space interpolation was also investigated as a tool for controlling the outputs of the generator neural network. Interestingly, the results indicated that such a procedure allows us to generate samples with intermediate redshift/cosmology/$f_{R0}$ parameter values, even if our model had not been explicitly trained on those particular values. In general, the latent space interpolation procedure offers a powerful way of controlling the outputs of the GAN as well as a tool for investigating the feature space of the generator neural network. However, it is important to point out some of the drawbacks of this procedure. Namely, as pointed out in machine learning literature, the latent space of a convolutional GAN is known to be \textit{entangled}. In other words, moving in a different direction in the latent space necessarily causes multiple changes to the outputs of the GAN. As a concrete example, finding a latent space line that induces a change in redshift of a given output necessarily also introduces other subtle changes to the output (e.g. the depth of the voids or the distribution of the filaments). So if we take a random output of redshift $z = 1.0$ and perform the linear interpolation procedure to obtain a cosmic web slice of $z = 0.0$, the obtained slice will correspond to a realistic but \textit{different} distribution of the required redshift. This is a drawback as in an ideal case we would love to have full control of individual parameters, while not affecting other independent features of a dataset. There are however other generative models discussed in the literature that allow such manipulation of the latent space. Namely, the $\beta$-VAE variational autoencoder and the InfoGAN algorithms, allow encoding features into the latent space in a special way that allows full control of individual key parameters without affecting the other features of the dataset (latent space disentanglement) \citep{Chen2016, Higgins2017, Burgess2018}. 

Another important pitfall to discuss is the problem of mode collapse. As is widely discussed in the literature, the generator neural network is prone to getting stuck in producing a very small subsample of realistic mock datapoints that fool the discriminator neural network. Resolving mode collapse is an important open problem in the field of deep learning, with a variety of known strategies ranging from choosing a particular GAN architecture, to altering the training procedure or the cost function \citep{Srivastava2017, Hong2019}. Mode collapse was encountered multiple times in our training procedure as well. As a rule of thumb, I found that reducing the learning rate parameter had the biggest effect towards resolving mode collapse for all studied datasets. Learning rates around the values of $R_{L} = 3 \times 10^ {-5}$ for the cosmic web data and $R_{L} = 9 \times 10^{-6}$ for the weak lensing maps were found to be the most effective in avoiding any mode collapse. 

As indicated by the results, GANs can be used to generate novel 2-D data efficiently. A natural question to ask is whether this also applies to 3-D data. As an example, an analogous emulator capable of generating 3-D cosmic web data, such as that produced by state of the art hydrodynamic and DM-only simulations would be very useful. In principle there is no limit on the dimensionality of the data used for training a GAN, however, in practice, going from 2-D to 3-D data leads to a significant increase of the generator and the discriminator networks. In addition, in the case of 3-D cosmic web data, forming a big enough training dataset would become an issue, as running thousands of simulations would be required. However, as previously mentioned, there are sophisticated ways of emulating 3-D cosmic web data as shown in \citep{Perraudin2019}, where a system of GANs is used to upscale small resolution comic web cubes to full size simulation boxes. Note that the techniques introduced in this work (e.g. latent space interpolation) can be readily combined with the mentioned 3-D approach. 

A number of interesting directions can be explored in future work. Namely, it would be interesting to further investigate the latent space interpolation techniques in the context of more advanced generative models, such as the InfoGAN algorithm. In addition, a more detailed investigation into the Riemannian geometry of GANs could lead to a better understanding of the feature space of the algorithm. Finally, many other datasets could be explored. With upcoming surveys such as Euclid generating mock galaxy and galaxy cluster data quickly and efficiently is of special interest. A GAN could be used to generate galaxies with realistic intrinsic alignments, density distributions and other properties. Similarly, GANs could be used to quickly emulate realistic galaxy cluster density distributions at a fraction of the computational cost required to run full hydrodynamic simulations. 

To conclude, GANs offer an entirely new approach for cosmological data emulation. Such a game theory based approach has been demonstrated to offer a quick and efficient way of producing novel data for a low computational cost. As we have shown in this work, the trade-off for this is a 1-20\% difference in the power spectrum, which can be satisfactory or not depending on what application such an emulator is used for. Even though a number of questions remain to be answered regarding the stability of the training procedure and training on higher dimensional data, GANs will undoubtedly be a useful tool for emulating cosmological data in the era of modern \textit{N}-body simulations and precision cosmology.

\chapter{Conclusions and Future Work}
\label{ch:conclusions}

The main goal of this thesis was to introduce tools and techniques for studying modified gravity. In summary, a method of testing modified gravity, first introduced in \citet{Terukina2013} and \citet{Wilcox2015}, was extended by generating a new more accurate dataset and by testing a new theory. In addition, machine learning techniques were explored in the context of emulating $\Lambda$CDM and modified gravity $N$-body simulations. 

In terms of all the mentioned projects, a lot of work remains to be done. In particular, the outlined technique of testing modified gravity relies on stacking multiple galaxy clusters. Stacking clusters, in some sense, produces an idealized galaxy cluster by averaging out the various irregularities that individual clusters possess. This is a powerful technique capable of producing competitive modified gravity constraints, however, the produced stack dataset is only an approximation of real galaxy clusters. In nature, no cluster is exactly spherical, hence it is important to understand what effects deviations from spherical symmetry would have in the context of chameleon gravity. Deviations from spherical symmetry must be better understood on two fronts. Firstly, the hydrostatic equilibrium equations used in our work must be generalized for arbitrary 3-D mass, pressure/temperature and surface brightness distributions. Alternatively, the bias due to non-spherical mass distributions can be quantified and accounted for in the mass calculations. Such bias has been studied both observationally and in the context of hydrodynamic simulations, e.g. see \citet{Morandi2010,Martizzi2016}. Secondly, it is important to understand how the theoretical predictions are affected by breaking the assumption of spherical symmetry. This is of special importance to the model of EG, the main predictions of which were derived under the key assumption of spherical symmetry. Hence, the predictions of the model for the relationship between the baryonic and the apparent dark matter distributions must be generalized for arbitrary mass distributions. This might be easier to accomplish in different models of EG, such as Hossenfelder's covariant EG.

The original motivation for stacking galaxy clusters is mainly due to the dominant weak lensing profile errors. This particular issue will be possible to address when the newest data from DES and future surveys such as Euclid becomes available. Another approach is to employ different data. As shown in \citet{Terukina2013}, stringent constraints can be calculated using the data from a single cluster. However, the main difference in that work when compared to our approach is that the SZ effect and temperature profile data is used in addition to the surface brightness and weak lensing profiles. As the mentioned results show, including these extra datasets leads to constraints comparable to the results presented in this work, even if a single cluster with high quality is used rather than a stack of clusters. Hence, a straightforward extension of our work is to introduce extra datasets, which could significantly improve the constraints. 

In addition to the previously mentioned shortcomings of EG, it is important to mention the lack of a rigid description of cosmology and weak lensing. More concretely, the original formulation of EG as described in \citet{Verlinde2017} is only valid for redshifts $z \sim 0$ and $H(z) \sim H_{0}$. In addition, a thorough description of lensing would require deriving a geodesic equation, which, in turn, requires a full covariant description of the theory. A covariant description of the theory does exist, as described in \citet{Hossenfelder2017}, however, the lensing equations have not been derived yet. Hence, on the theoretical front, the key issues to tackle are related to extending the original Verlinde's EG framework to account for cosmological effects and weak lensing. Alternatively, Hossenfelder's framework can be extended by deriving the geodesic equation. Here, however, it is important to point out that, strictly speaking, these two theories are not identical and agree only in the non-relativistic, stationary spherical mass distribution limit. The exact relationship between the two approaches deserves a more detailed investigation as well. 

In summary, EG undoubtedly suffers from certain theoretical and observational shortcomings. However, the theory has been successful in encouraging further studies of the various connections between thermodynamics and gravity. Multiple recent studies have extended Verlinde's ideas in different contexts, e.g. \citet{Vacaru2019, Peach2019}. Finally, Verlinde's work has also contributed to the current resurgence of the rather unique approach of treating gravity as spacetime elasticity previously studied in \citet{Visser2002,Padmanabhan2004}.    

In terms of the machine learning algorithms discussed in chapters \ref{ch:machine_learning} and \ref{ch:GAN_emulators} a lot of work remains to be done as well. As mentioned, the results in \citet{Tamosiunas2020} are encouraging, however, multiple issues remain to be addressed. Firstly, using GANs for emulating 3-D simulation data remains an issue in terms of the memory issues and the availability of large 3-D training datasets. These and other issues have already been partially addressed in the literature. As an example, the training procedure of the GAN can be modified such that small 3-D overdensity cubes are patched together to form a full-size 3-D overdensity field as shown in \citet{Perraudin2019}. An alternative approach is to start with a low resolution mock dataset, which is gradually upscaled during a multi-stage training procedure (e.g. see \citet{Ledig2016}). These sort of approaches combined with modern GPU training will likely make emulating 3-D datasets easy and efficient in the near future. 

Another key issue encountered in our work was controlling the outputs. In particular, when training the algorithm on a dataset consisting of different data classes, the simple DCGAN architecture does not allow to control which outputs will be produced. In other words, one cannot choose the output class due to the inherent randomness of the generation procedure. The proposed workaround discussed in this work was to use the latent space interpolation procedure, which resolves the problem partially. However, the key issue with the procedure is that it is not efficient as it requires one to manually detect interesting regions of the latent space that can be used in the interpolation. A much more elegant approach is to use a different GAN architecture that is specifically designed for producing multi-class data. In particular, the conditional GAN architecture (CGAN) allows to take full control of the generation procedure. In this architecture the neural networks are trained using labeled data and hence the outputs of the generator can be controlled by directly specifying the label of the class to be generated. As a concrete example, this type of architecture has been recently used to generate convergence maps with different cosmological parameters as described in \citet{Perraudin2020}. Hence a natural extension of our work would be to apply the techniques described in this thesis on different GAN architectures. This would lead not only to more control of the data generation, but also to a better understanding of the latent space. Finally, alternative architectures, such as the mentioned CGAN algorithm, would make it easier to explore the Riemannian geometry of the latent space produced during the training procedure.

As shown in this work, generative models offer a completely new approach of generating mock data. Like any algorithm such models come with a set of shortcomings. However, the ability to generate large multi-class mock datasets quickly and efficiently undoubtedly makes such algorithms useful. This is especially true in the context of the upcoming large scale surveys such as Euclid and SKA, which will require accurate and fast emulators. 

In conclusion, observational tests of gravity have traditionally played an important role in the theoretical development of theories of modified gravity and dark energy. Starting with the initial tests of GR and ending with cutting edge gravitational wave tests, viable theories of gravity have always been firmly constrained by the most recent observational data. The set of techniques described in this work draws an optimistic picture for the near future of cosmological tests of modified gravity. With the next generation of observational surveys and new high quality data becoming available, constraints of unprecedented accuracy will become possible. Similarly, with the new machine learning techniques becoming available, new ways of emulating modified gravity will become possible as well. And hence the techniques described in this work will hopefully play an important role in forming a better understanding of gravity.

\begin{appendices}
\chapter{}
\section{Availability of Data and Codes}
\label{data_code_availability}

The key datasets generated and analysed in this work are available at the following GitHub repository: \href{https://github.com/AndriusT/cw_wl_GAN}{https://github.com/AndriusT/cw\_wl\_GAN}. The link also contains detailed instructions on how to produce the data samples from the publicly available Illustris data. The full Illustris datasets can be found at: \href{https://www.illustris-project.org/data/}{https://www.illustris-project.org/data/}.

The used weak lensing data can be accessed at: \href{http://columbialensing.org/}{http://columbialensing.org/}.

\section{Samples of the GAN-produced Data}
\label{samples}

\begin{figure}[!ht]
  \centering
    \includegraphics[width=0.99\textwidth]{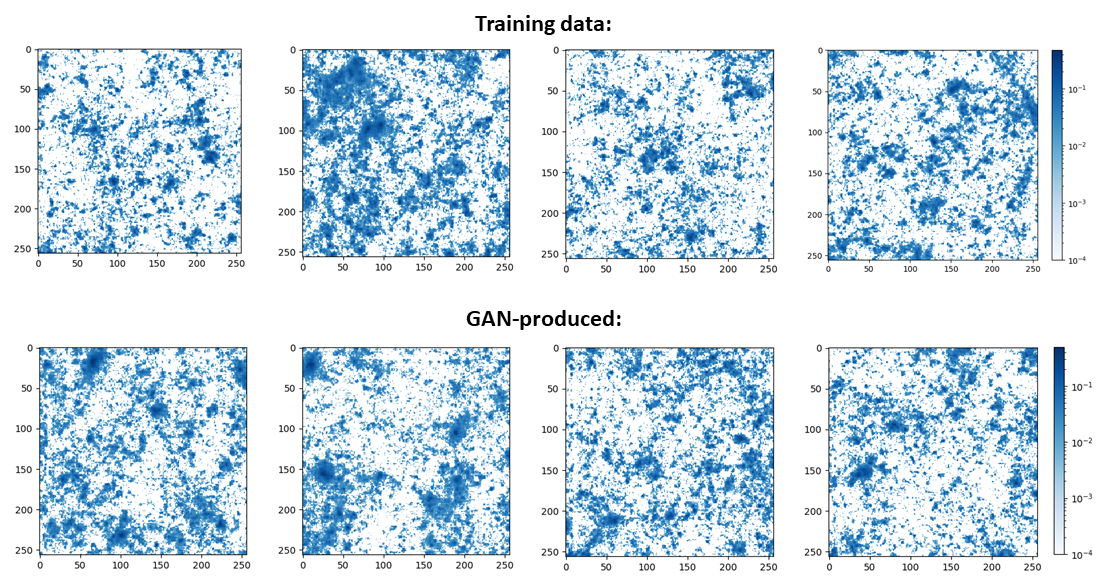}
    \caption[A comparison of 4 randomly selected weak lensing convergence maps.]{ A comparison of 4 randomly selected weak lensing convergence maps. The colors are log-normalized to emphasize the main features and to allow a direct comparison with the results in \cite{Mustafa2019}.}
    \label{cosmoGAN_samples}
\end{figure}

This section contains a selection of GAN-produced samples for visual inspection. Fig. \ref{cosmoGAN_samples} contains randomly selected weak lensing convergence maps produced by the GAN algorithm (these are the samples described in sections \ref{section_weak_lensing_multiple_cosmologies} and \ref{section_cosmic_web_redshifts}). 

\begin{figure}[!ht]
  \centering
    \includegraphics[width=0.99\textwidth]{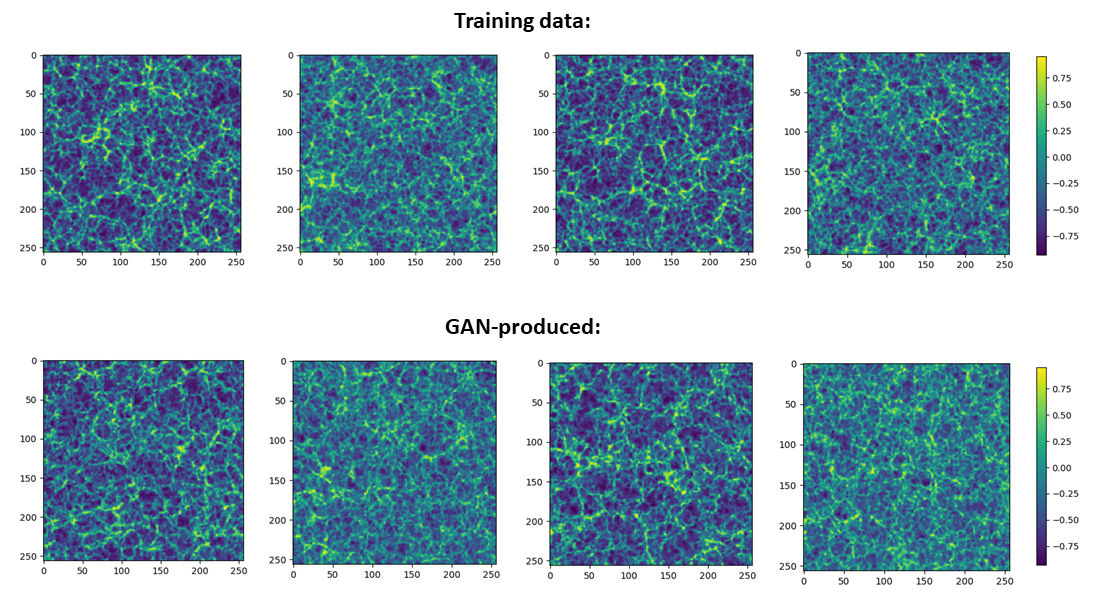}
    \caption[A comparison of 4 randomly selected cosmic web slices.]{ A comparison of 4 randomly selected cosmic web slices. Columns 1 and 3 correspond to redshift 0.0 while columns 2 and 4 are redshift 1.0.}
    \label{cosmoGAN_cw_samples}
\end{figure}

Fig. \ref{cosmoGAN_cw_samples} shows a selection of randomly selected cosmic web 2-D slices for two different redshifts. Both the training data and the produced slices have been Gaussian-smoothed. 

\section{MCMC Contours}

This section contains the MCMC contours along with the corresponding likelihood distributions for the key datasets used in chapter \ref{ch:modified_gravity}. In particular, the MCMC results from \cite{Terukina2013}, \cite{Wilcox2015} and our results produced using the newest dataset consisting of 77 galaxy clusters. In all cases, the light gray contours correspond to the 99\% CL, while the dark gray contours are the 95\% CL for each best-fit parameter. Notice also that the colors in the individual modified gravity parameter plots appearing in chapter \ref{ch:modified_gravity} are inverted for the sake of clarity when comparing the results from different papers. 

\begin{figure}[!ht]
  \centering
    \includegraphics[width=0.73\textwidth]{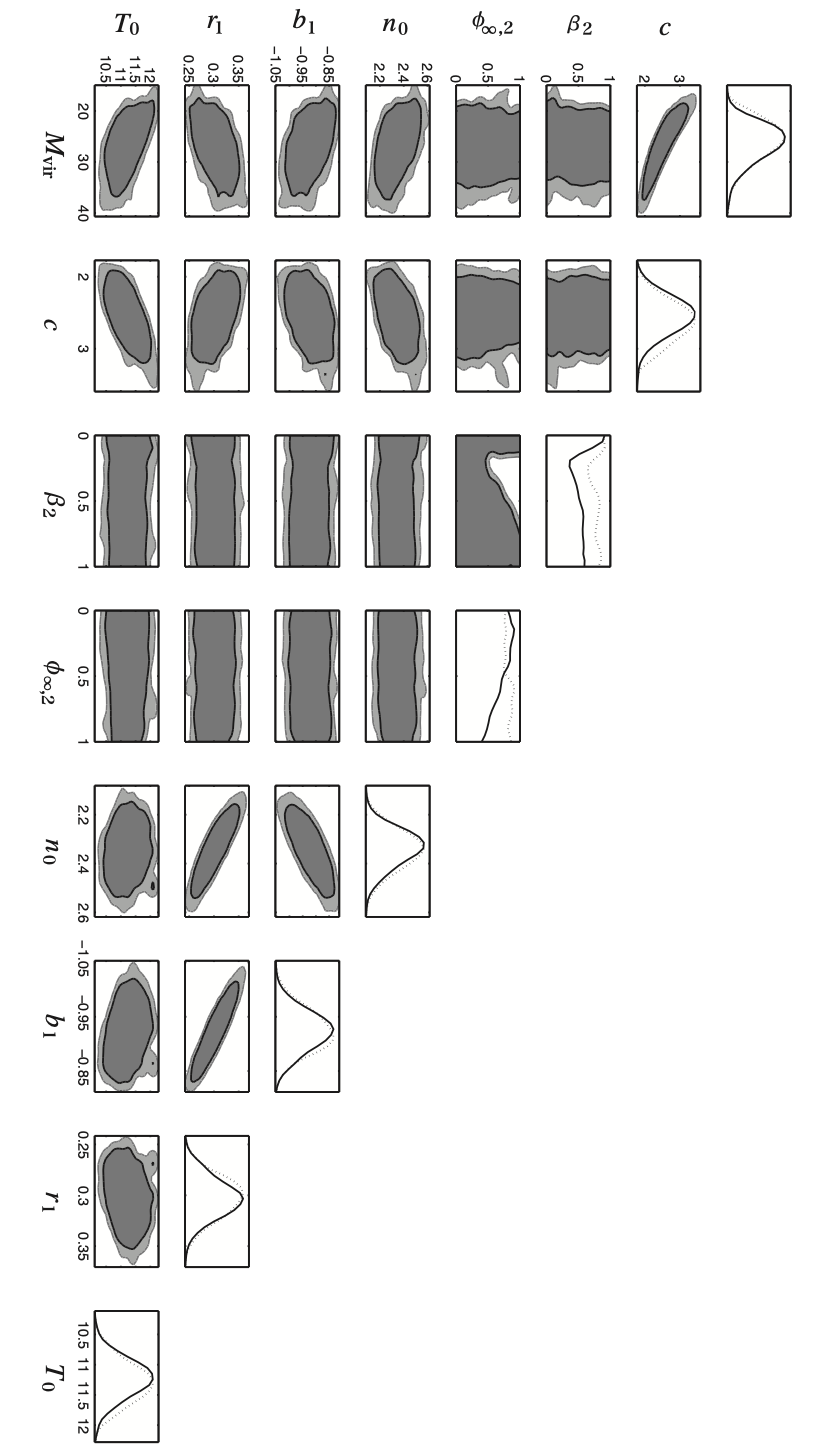}
    \caption[The MCMC contours from \cite{Terukina2013}]{The MCMC marginalised contours of the 6 model parameters used in \cite{Terukina2013} to fit the data. The light gray and the dark gray contours show the 95\% and the 99\% CL correspondingly. The rightmost panels show the marginalised 1-dimensional constraints (solid) and the likelihood distribution (dotted).}
    \label{MCMC_contours_terukina}
\end{figure}

\begin{figure}[!ht]
  \centering
    \includegraphics[width=0.99\textwidth]{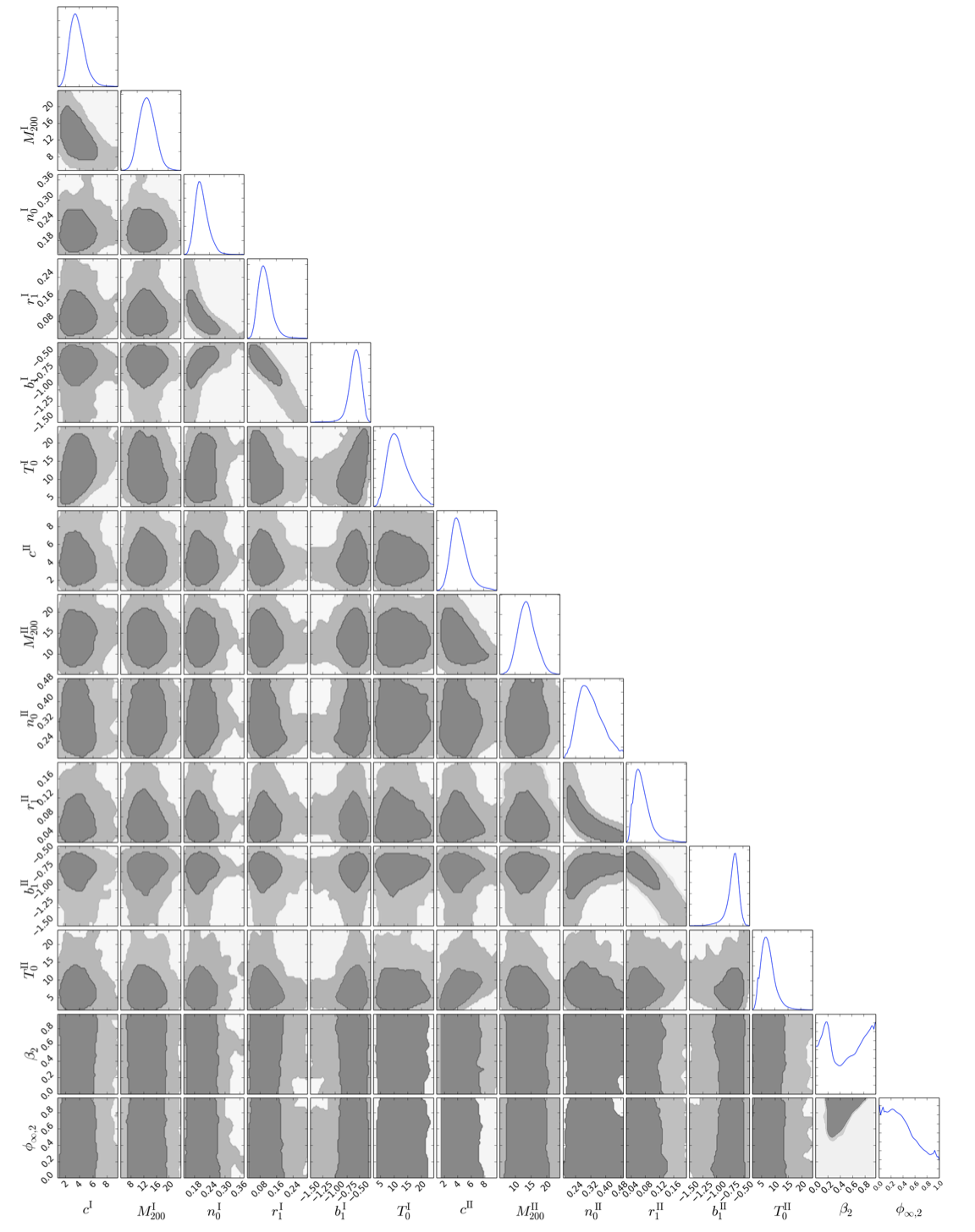}
    \caption[The MCMC contours from \cite{Wilcox2015}]{The MCMC marginalised contours of the 14 model parameters used in \cite{Wilcox2015} to fit the data. The light gray and the dark gray contours show the 95\% and the 99\% CL correspondingly. The rightmost panels show the likelihood distribution.}
    \label{MCMC_contours_wilcox}
\end{figure}

\begin{figure}[!ht]
  \centering
    \includegraphics[width=0.99\textwidth]{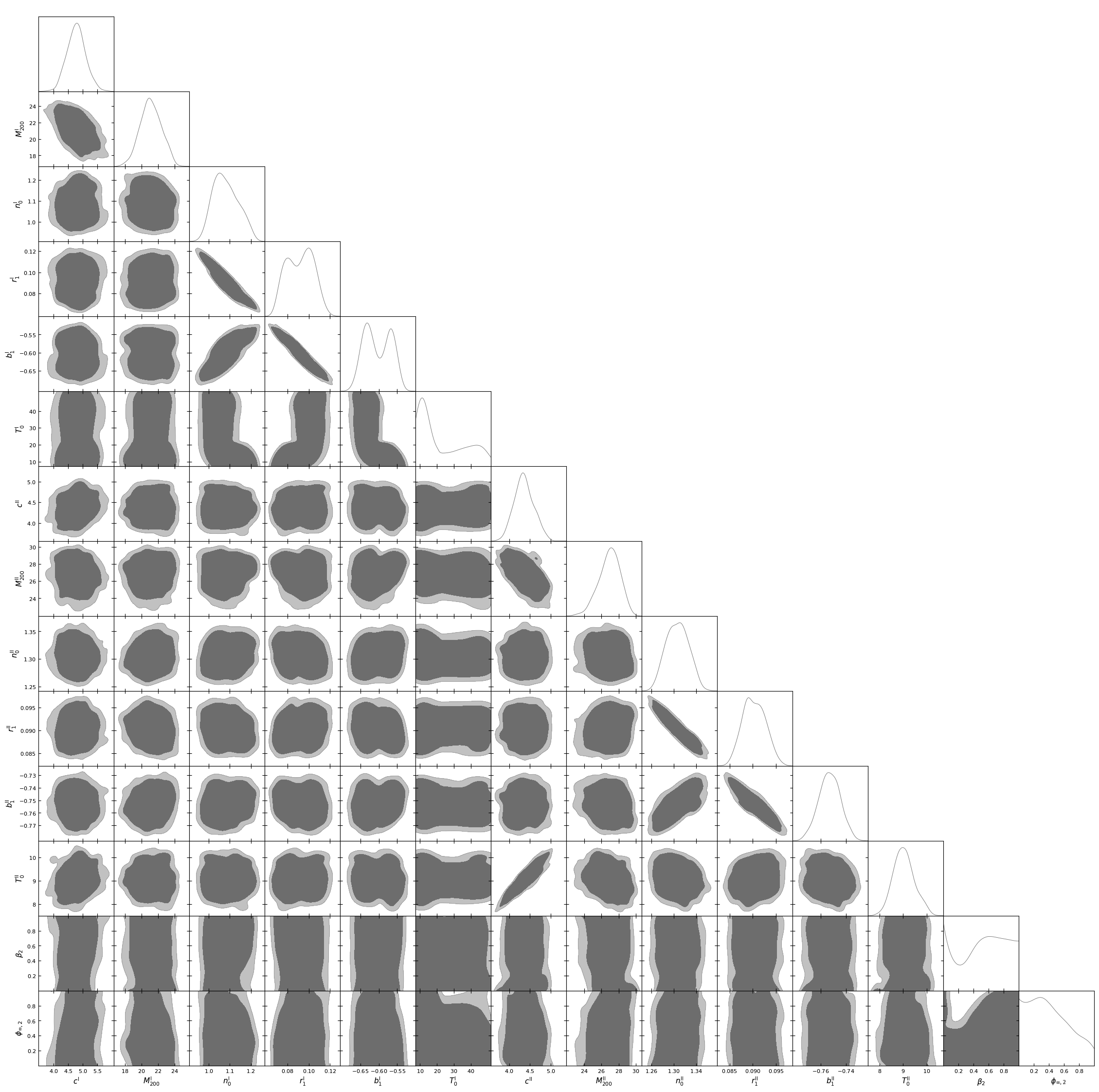}
    \caption[The MCMC contours of the best-fit parameters determined in this work]{The MCMC marginalised contours of the 14 model parameters determined using the newest dataset of 77 galaxy clusters. The light gray and the dark gray contours show the 95\% and the 99\% CL correspondingly. The rightmost panels show the likelihood distribution.}
    \label{MCMC_contours_77_clusters}
\end{figure}

\clearpage



\end{appendices}

\newpage
\phantomsection
\addcontentsline{toc}{chapter}{Bibliography}
\bibliography{ThesisRefs}

\end{document}